\documentclass[prb,reprint,a4paper,citeautoscript]{revtex4-2}

\pdfoutput=1						
\usepackage[charter]{mathdesign}	
\usepackage{amsmath}				

\usepackage{mathrsfs}					

\usepackage[pdftex]{graphicx}
\usepackage{microtype}
\usepackage{bm}\let\vec\bm
\usepackage{mathtools}
\usepackage[]{stackengine}
\DeclareMathOperator{\Pint}{\ensurestackMath{\stackinset{c}{0pt}{c}{0pt}{\mathscr P}{\displaystyle\int}}}
\newlength\correct
\usepackage[svgnames]{xcolor}
\usepackage[colorlinks=True,linkcolor=DarkRed,citecolor=ForestGreen,urlcolor=MediumBlue,pdfstartview=FitH,bookmarks=False,pdfpagemode=UseNone]{hyperref}

\newif\ifshowcomments\showcommentstrue

\begin{document}

\title{Optical signatures of shear collective modes in strongly interacting Fermi-liquids}

\author{D.~Valentinis$^{1,2}$}
\email{davide.valentinis@kit.edu}
\affiliation{$^1$ Department of Quantum Matter Physics, University of Geneva, 24 quai Ernest-Ansermet, 1211 Geneva 4, Switzerland}
\affiliation{$^2$ Institute for Theory of Condensed Matter, Karlsruhe Institute of Technology, Wolfgang-Gaede Stra{\ss}e 1, 76131 Karlsruhe, Germany}

\date{April 27, 2021}

\begin{abstract}

The concept of Fermi liquid lays a solid cornerstone to the understanding of electronic correlations in quantum matter. This ordered many-body state rigorously organizes electrons at zero temperature in progressively higher momentum states, up to the Fermi surface. As such, it displays rigidity against perturbations. Such rigidity generates Fermi-surface resonances which manifest as longitudinal and transverse collective modes. Although these Fermi-liquid collective modes have been analyzed and observed in electrically neutral liquid helium, they remain unexplored in charged solid-state systems up to date. In this paper I analyze the transverse shear response of charged three-dimensional Fermi liquids as a function of temperature, excitation frequency and momentum, for interactions expressed in terms of the first symmetric Landau parameter. I consider the effect of momentum-conserving quasiparticle collisions and momentum-relaxing scattering in relaxation-time approximation on the coupling between photons and Fermi-surface collective modes, thus deriving the Fermi-liquid optical conductivity and dielectric function. In the high-frequency, long-wavelength excitation regime the electrodynamic response entails two coherent and frequency-degenerate polaritons, and its spatial nonlocality is encoded by a frequency- and interaction-dependent generalized shear modulus; in the opposite high-momentum low-frequency regime anomalous skin effect takes place. I identify observable signatures of propagating shear collective modes in optical spectroscopy experiments, with applications to the surface impedance and the optical transmission of thin films. 
\end{abstract}

\maketitle

\section{Introduction}\label{Intro}

The Fermi liquid represents a ``Rosetta stone'' for electronic correlations in weakly interacting electron systems. It translates (maps) the complexity of many-body interactions into a simpler description built in terms of nearly independent constituents, the electron-hole quasiparticles \cite{Nozieres-1999ql,Abrikosov-1959}. Such conception, introduced by Landau in the 1960s, fostered profound insight into the phenomenology of electrical and thermal conduction in metals throughout the twentieth century, and it still reserves unexpected surprises in the application to modern-day materials. 

The phenomenology of nearly free quasiparticles in a Fermi liquid actually stems from a rigorously ordered microscopic state: at temperature $T=0$, long-ranged entanglement associated with the Pauli exclusion principle arranges electrons in a hierarchy of progressively higher momentum states, with the uppermost states composing the Fermi surface. Electron-hole quasiparticles are created by promoting an occupied state from below the Fermi surface to unoccupied levels above: hence, they are the single-particle elementary excitations of the system, in direct correspondence with the original interacting electrons. The energy $\hbar \omega$ required to generate such excitations increases by going deeper below the Fermi level $E_F$, so that states at $E_F$ essentially dictate all thermal and electrical properties, while lower-energy states remain largely unperturbed. The Fermi liquid thus forms a stable, cohesive state of zero-temperature matter. 

The definition of Fermi-liquid order does not straightforwardly extend to finite temperatures: it is not captured by the conventional language of phase transitions, whereby a generalized rigidity emerges from spontaneous symmetry breaking of a high-energy disordered phase. 
In fact, for energies $\hbar \omega <k_B T$ the Fermi liquid is adiabatically connected to a classical fluid, in which thermal fluctuations ``disorder'' the system through quasiparticle excitations across the Fermi surface but without any thermodynamic singularity. 
However, at higher energies $\hbar \omega > k_B T$ Fermi-surface rigidity resists thermal fluctuations, so that one recovers substantial remnants of the $T=0$ physics. All this wisdom is conventionally parametrized in terms of a quasiparticle collision time $\tau_c \propto (\hbar E_F)/\left[(\hbar \omega)^2+ (k_B T)^2\right]$, stemming from the phase-space restriction for collision processes entailed by the Pauli principle \cite{Tokatly-2000, Landau-2013stat}. 
Multiple experiments confirm that the low-temperature ground state of many metals indeed complies with the Fermi-liquid picture: notable examples include Sr$_2$RuO$_4$ \cite{Maeno-1997,Bergemann-2003,Stricker-2014} and electron-doped BaFe$_{2-x}$Co$_x$As$_2$ \cite{Tytarenko-2015}. 

Much of the contemporary literature on Fermi-liquid transport focuses on the classical-fluid regime $\hbar \omega <k_B T$. There, if the collision time $\tau_c$ provides the smallest timescale in the system, local equilibrium is established among quasiparticles. This condition allows for an effective hydrodynamic description of quasiparticle flow \cite{Conti-1999,Principi-2016}, whereby momentum and energy dissipation are encoded in the viscosity tensor \cite{Bradlyn-2012}, while dissipationless deformations of the fluid can be formulated in terms of ``generalized elasticity'' \cite{Conti-1999,Landau-1957,Landau-1986el, Link-2018} and quantified by elastic moduli \cite{Conti-1999,Vignale-2005}. 

However, electron viscosity has eluded experimental observation in standard three-dimensional (3D) metals so far. 
One main reason is the presence of multiple scattering channels, which inevitably relax momentum and hinder the establishment of local equilibrium \cite{Lucas-2018b}. In fact, even in ideal crystalline lattices, momentum conservation is already destroyed on the scale of the lattice periodicity, due to momentum-relaxing scattering of quasiparticles on lattice ions. On top of this, impurities and defects provide additional relaxation channels in real crystals.

Besides, Fermi liquids are intrinsically less viscous than strongly interacting electron systems, which makes viscosity effects more difficult to observe in practice. On the contrary, strongly interacting systems with reduced dimensionality provide fruitful platforms for electron hydrodynamics, as testified by applications to graphene \cite{Andreev-2011, Briskot-2015, Levitov-2016, Narozhny-2017,Narozhny-2019}, delafossites \cite{Moll-2016,Cook-2019}, Weil semimetals \cite{Gorbar-2018, Gorbar-2019}, the two-dimensional electron gas in (Al,Ga)As heterostructures \cite{Molenkamp-1994,deJong-1995}, and bad metals \cite{Lucas-2017b}. Quantum critical (QC) electrons are especially suitable for hydrodynamic descriptions: their strong interactions imply an extremely low ``Planckian'' collision time $\tau_c \propto \hbar/(k_B T)$ --- with $\hbar=1.05 \cdot 10^{-34}$ J s reduced Planck's constant --- in an extended range of temperatures $T$ above the $T=0$ QC point. Such a high collision rate invalidates the notion of quasiparticles \cite{Schofield-1999, Zaanen-2015holo}, but it also favors local equilibrium. Indeed, experimental signatures of hydrodynamic flow of electrons have been retrieved in QC systems like graphene \cite{Crossno-2016,Krishna-Kumar-2017,Ella-2019}, PdCoO$_2$ \cite{Moll-2016,Mackenzie-2017,Nandi-2018}, PtCoO$_2$ \cite{Nandi-2018}, WP$_2$ \cite{Gooth-2018,Jaoui-2018}, and PtSn$_4$ \cite{Fu-2018}. 

On the other hand, the dynamical regime $\hbar \omega >k_B T$, where Fermi-liquid order survives thermal effects, entails phenomena linked to Fermi-surface rigidity against perturbations \cite{Tokatly-2000,Link-2018}. A striking example of such phenomena is the propagation of collective modes of electron density and current. These modes are coherent vibrations of the Fermi surface in space and time, which are analogous quantum counterparts of classical waves on the surface of an elastic carpet. Nevertheless, it would be inaccurate to label Fermi-liquid collective modes as elastic phenomena, since in the strict sense elasticity refers to \emph{static} reactive properties of solids, which are absent in the Fermi liquid. In fact, from a collective excitation viewpoint, the \emph{dynamical} shear deformation of the Fermi surface behaves like a spin-1 object, akin to a ``transverse phonon'', while shear rigidity resides in the static spin-2 channel \cite{Zaanen-2015holo}: in other words, reactive shear in the Fermi liquid is disconnected from elasticity, as in the static limit the system rather behaves as a viscous liquid. Perhaps water offers a poignant analogy: plunging slowly into a pond produces a viscous, dissipative response typical of a fluid, while diving at high speed from above generates much more resistance at impact, whereby the liquid reacts almost like a solid. 

Among such coherent resonances is the so-called ``zero sound'', occurring in both the longitudinal and transverse (shear) channels. The observation of zero sound in liquid $^3$He, the archetype of electrically neutral Fermi liquid, stroke a landmark of twentieth-century low-temperature physics \cite{Lea-1973,Roach-1976}. Nevertheless, the physics of Fermi-liquid collective modes remains hitherto unexplored in solid-state settings \cite{Tokatly-2000,Link-2018}.

At first sight, the analysis of charge collective modes in Fermi liquids formed in crystalline solids bears additional challenges with respect to electrically neutral systems, due to the presence of the long-ranged Coulomb interaction \cite{Silin-1958a,Silin-1958b} and of momentum relaxation imposed by the breaking of translation invariance. The first conundrum is solved by the Silin theory \cite{Silin-1958a,Silin-1958b}, which prescribes a proper redefinition of the Landau quasiparticle interaction parameters, to take into account Coulomb repulsion. Still, as we will describe later, electric charge qualitatively modifies the dispersion of transverse shear collective modes \cite{Nozieres-1999ql,Beekman-2017}. The second issue of momentum relaxation is more serious, as it prevents hydrodynamics to occur in standard 3D metals as previously mentioned. However, the reactive shear response is more robust than hydrodynamics with respect to relaxation, because the former depends mainly on quasiparticle interactions and does not rely on $\tau_c$ being the smallest timescale in the system. Hence, propagating Fermi-liquid modes might be detectable even in the presence of momentum-relaxing processes.

How is it possible to observe shear rigidity in solid-state Fermi liquids? The photon is an ideal candidate probe, as it exerts transverse perturbations on electrons and it couples to electric charge. However, due to the interactions among quasiparticles in the Fermi liquid, the electromagnetic response is \emph{spatially nonlocal}: applying a local electromagnetic field results in a perturbation spread out in an extended region of space surrounding the probe, by virtue of the nonlocal character of the optical response. Such nonlocality entails a precise relation between the transverse electric field $\vec{E}(\vec{r},t)$ of an incident electromagnetic wave, depending on space coordinate $\vec{r}$ and time $t$, and the induced dielectric displacement field $\vec{D}_T(\vec{r},t)$ in the material. In linear and isotropic media (the focus of this paper) and in the presence of translation invariance we have 
\begin{equation}\label{eq:epsilon_gen}
\vec{D}_T(\vec{r},t)=\int d\vec{r}' \int dt' \epsilon_T(\vec{r}-\vec{r}',t-t') \vec{E}(\vec{r}',t'). 
\end{equation}
The relation (\ref{eq:epsilon_gen}) defines the dielectric function $\epsilon_T(\vec{r}-\vec{r}',t-t')$, which is manifestly nonlocal in space: applying the electric field at the coordinate $\vec{r}'$ produces a dielectric response also at other coordinates $\vec{r} \neq \vec{r}'$. A relation analogous to Eq.\@ (\ref{eq:epsilon_gen}) between the transverse current $\vec{J}_T(\vec{r},t)$ and $\vec{E}(\vec{r},t)$ defines the transverse optical conductivity $\sigma_T(\vec{r}-\vec{r}',t-t')$. Fourier-transforming Eq.\@ (\ref{eq:epsilon_gen}) to reciprocal space of wave vectors $\vec{q}$ and excitation frequency $\omega$ results in
\begin{equation}\label{eq:epsilon_gen_q_om}
\vec{D}_T(\vec{q},\omega)=\epsilon_T(\vec{q},\omega) \vec{E}(\vec{q},\omega),
\end{equation} 
i.e.\@, the dielectric function $\epsilon_T(\vec{q},\omega)$ depends on wave vector $\vec{q}$ due to spatial nonlocality. Physically, Eq.\@ (\ref{eq:epsilon_gen_q_om}) represents the transverse dielectric response of the electron ensemble to perturbations with momentum $\hbar \vec{q}$ and frequency $\omega$. 
Knowing the exact functional dependence of $\epsilon_T(\vec{q},\omega)$ on $\vec{q}$ requires microscopic models of the electronic system, such as the Fermi liquid. 

Textbook treatments of optical spectroscopy often neglect the $\vec{q}$ dependence of Eq.\@ (\ref{eq:epsilon_gen}) by calculating the optical properties at $q=0$, e.g.\@, $\epsilon_T(0,\omega)$. The $q=0$ response would be exact if the dielectric response was entirely local in space, i.e.\@, if it did not depend on $\vec{r}$. In the presence of spatial nonlocality, the limit $q \rightarrow 0$ is equivalent to the spatial average of the response function $\epsilon_T(\vec{r}-\vec{r}',t-t')$. The common justification for considering the $q=0$ limit resides in the smallness of the momentum transferred by radiation to the solid, due to the high velocity of light with respect to the Fermi velocity. This generally suffices to describe optical phenomena in standard metals at room temperature \cite{Jackson-1962,Dressel-2001,Nozieres-1999ql}. 

However, notable known exceptions like anomalous skin effect show that, even with a perfectly local probe, the electromagnetic response becomes spatially nonlocal at low temperatures due to the increase of the electronic mean-free path \cite{Dressel-2001,Sondheimer-2001}. The influence of Fermi-liquid collective modes on the nonlocal dielectric response, and distinctive optical signatures of such influence, are not comprehensively understood to the best of my knowledge. Hence, this paper focuses on how to observe reactive shear collective modes in solid-state bulk Fermi liquids using low-temperature optical spectroscopy.  

I use the Landau kinetic equation approach \cite{Abrikosov-1959} to calculate the Fermi-liquid optical properties, in particular the dielectric function $\epsilon_T(\vec{q},\omega)$, and to analyze the coupling of Fermi-surface oscillations to electromagnetic radiation. The results of this analysis can be compared with similar kinetic equation approaches \cite{Tokatly-2000,Levitov-2013, Sun-2018,Levchenko-2020,Khoo-2021_preprint}, with the general formulation in terms of the stress and viscosity tensors \cite{Bradlyn-2012,Link-2015}, and with optical conductivities derived from the AdS/CFT correspondence \cite{Zaanen-2015holo,Delacretaz-2017,Inkof-2019}. Moreover, the outcomes of this paper may be linked with studies of negative refraction \cite{FZVM-2014}, which occurs in the presence of spatial nonlocality and dissipation \cite{Veselago-1968,FZVM-2014,Forcella-2016} and is usually realized in artificial metamaterials \cite{Veselago-2006}. Furthermore, recent studies analyzed the propagation of Fermi-liquid shear sound in two-dimensional (2D) systems \cite{Khoo-2019} and identified their signature in AC conductivity dips in narrow strips \cite{Khoo-2020}. Interestingly, the general form of the static transverse Fermi-liquid conductivity in two dimensions has been recently shown to equally apply to spinon Fermi-surface states \cite{Khoo-2021_preprint}.

The paper is organized as follows. Section \ref{Summary} provides a concise summary of the methods and the new results presented in this paper, and of how such results fit into the existing literature. This summary helps the reader navigating through all subsequent sections. In particular, it contains the definition of the generalized shear modulus $\nu(\omega)$, which will quantify Fermi-surface rigidity effects --- and the ensuing spatially nonlocal dielectric response ---  throughout this work. Section \ref{Kin_FL_neutral} hosts a recollection of some previous fundamental milestones in the analysis of the transverse shear response in electrically neutral Fermi liquids, with application to transverse sound in liquid $^3$He. Section \ref{Charged_FL_XT} extends aforementioned analysis to charged Fermi liquids, and it presents the calculation of the transverse susceptibility in the kinetic equation approach. The transverse susceptibility is then used in Sec.\@ \ref{FL_epsT_coll} to derive the transverse dielectric function and the optical conductivity. Section \ref{FL_DC} deals with the zero-frequency (DC) limit of the optical conductivity and its relation to hydrodynamic effects in transport experiments. Section \ref{Scatter} describes the microscopic models for temperature- and frequency-dependent scattering rates, due to electron collisions and momentum relaxation, assumed in later calculations. In particular, I use the phenomenological Matthiessen's rule to sum independent contributions from acoustic phonons, impurities and Umklapp processes to the momentum-relaxation rate. In Sec.\@ \ref{Opt} I propose specific experimental setups to detect Fermi-surface rigidity in optical spectroscopy, namely the surface impedance and the thin-film transmission, and I compare numerical and analytical results for such experimental configurations with the standard predictions in the absence of spatial nonlocality, i.e.\@, for an Ohmic (Drude) conductor. The main messages of this paper are summarized and discussed in Sec.\@ \ref{Concl}.

\subsection{Summary of main results}\label{Summary}

The dielectric function $\epsilon_T(\vec{q},\omega)$ of a charged Fermi liquid in the presence of both short-ranged interactions and quasiparticle collisions is Eq.\@ (\ref{eq:epsilon_T_FL}), which is the first main result of this paper. It descends from the Kubo formalism, which relates the optical conductivity with the Fermi-liquid transverse susceptibility calculated from the Landau kinetic equation.

Specifically, for an interaction comprising the symmetric Landau parameters $F_0^S$ and $F_1^S$, $\epsilon_T(\vec{q},\omega)$ depends on the combination of variables $s=\omega/(v_F^{*} q)$ and $\omega \tau_c$, where $v_F^{*}=v_F/(1+F_1^S/3)$ is the Fermi velocity $v_F$ renormalized by the Landau interaction parameter $F_1^S$. The collision time $\tau_c$ stems from a single-time approximation for the quasiparticle collision integral in the kinetic equation, taking into account the conservation of particle number, momentum and energy in collision processes \cite{Abrikosov-1959,Lea-1973,Roach-1976}.
The pole in $\epsilon_T(\vec{q},\omega)$ signals the existence of a transverse collective mode in the Fermi liquid: physically, this means that applying a perturbation with specific wave vector $\vec{q}$ and frequency $\omega$ excites a Fermi-surface shear resonance with dispersion relation $\vec{q}(\omega)$, for which $\epsilon_T(\vec{q},\omega)$ diverges \footnote{The dispersion relation can equivalently be written in the form $\omega(\vec{q})$, as commonly done in solid-state textbooks. In this paper, we adopt the viewpoint of $\vec{q}(\omega)$, consistently with the picture of an electromagnetic wave of real-valued frequency $\omega \in \mathbb{R}$ which excites a Fermi-liquid shear mode with complex-valued momentum $\vec{q} \in \mathbb{C}$.}. The real part $\mathrm{Re}\left\{\vec{q}\right\}$ describes the spatial dispersion of the excited transverse mode, while the imaginary part $\mathrm{Im}\left\{\vec{q}\right\}$ is connected to mode damping. Assuming that $\omega$ is the frequency of an incident electromagnetic field, the pole in $\epsilon_T(\vec{q},\omega)$ shows that radiation can couple to shear resonances of the Fermi surface.

The dispersion $\vec{q}(\omega)$ of the pole in $\epsilon_T(\vec{q},\omega)$ for the charged Fermi liquid is formally equivalent to the dispersion relation of transverse sound in electrically neutral systems, previously analyzed and experimentally detected in $^3$He \cite{Lea-1973,Roach-1976}, and recalled in Sec.\@ \ref{FL_Tmode_coll}. Hence, Sec.\@ \ref{FL_epsT_coll} connects the electrodynamics of bulk charged Fermi liquids with the theory of transverse sound in neutral systems like liquid helium.

Conversely, the dielectric response of the system is qualitatively modified by the coupling between transverse collective modes and radiation. In particular, the propagation of photons into the Fermi liquid is affected by the presence of electronic shear modes like transverse sound. Physically, as radiation couples to the charged Fermi liquid, the resulting optical modes are mixtures of the incident electromagnetic wave with the excited electronic Fermi-surface resonance, i.e.\@, polaritons. Technically, the polaritons are self-consistent solutions of Maxwell's equations in the dielectric medium with dielectric function $\epsilon_T(\vec{q},\omega)$, satisfying $\epsilon_T(\vec{q},\omega)=\left(q c/\omega\right)^2$. The sensitivity of photons to electronic shear modes suggests a way to probe transverse sound in charged Fermi liquids using electromagnetic waves, as the optical properties of the system bear potentially observable traces of underlying Fermi-surface shear resonances. This strategy will be pursued in later sections. 

In the high-frequency, low momentum regime, the dielectric function behaves as
\begin{multline}\label{eq:eps_nu_qualit}
\epsilon_T(\vec{q},\omega)=\epsilon_T\left[i \omega \nu(\omega) q^2,\omega\right]+o(q^4) \\
=1-\frac{(\omega_p)^2}{\omega^2+ i \omega \nu(\omega) q^2}
\end{multline}
where $\nu(\omega)$ is the \emph{generalized shear modulus} of the isotropic Fermi liquid \cite{Conti-1999,Vignale-2005,Bedell-1982}
\begin{equation}\label{eq:nu_q0_vel}
\nu(\omega)=\left(1+\frac{F_1^S}{3}\right) \frac{ (v_F^*)^2 \tau_c}{5 \left(1-i \omega \tau_c\right)}=\frac{\nu(0)}{1-i \omega \tau_c},
\end{equation}
and
\begin{equation}\label{eq:nu_0}
\nu(0)= \frac{1}{5} (v_F)^2 \tau_c \frac{1}{1+\frac{F_1^S}{3}}
\end{equation}
is the static viscosity coefficient \cite{Conti-1999, Nozieres-1999ql, Abrikosov-1959}. The term $i\omega\nu(\omega)$ is imaginary for $\omega\tau_c\ll1$ and it becomes real for $\omega\tau_c\gg1$, reflecting the transition from a dissipative to a reactive Fermi-surface shear response. Hence, the result (\ref{eq:eps_nu_qualit}) directly links microscopic Fermi-liquid theory with the phenomenologies of viscous charged fluids \cite{FZVM-2014} and of elasticity in electron liquids \cite{Conti-1999, Tokatly-2000, Vignale-2005}. 

Coupling the Fermi liquid to photons in this regime produces \emph{two} degenerate polariton modes for each frequency $\omega$. The ``plasmon-polariton'' is weakly affected by $\nu(\omega)$ and resembles the usual optical plasmon, while the ``shear-polariton'' is much more sensitive to $\nu(\omega)$. The shear-polariton has quartic dispersion $q \propto \omega^{1/4}$ for $\omega \tau_c \rightarrow 0$, and it emerges from the electron-hole continuum above a finite frequency $\omega \sim \omega_p/c$ \cite{Nozieres-1999ql}. 

In the low-frequency, high-momentum regime, where Landau damping prevents the formation of polariton modes, a real-valued optical conductivity $\sigma_T(\vec{q},\omega)\propto(v_F q)^{-1}$ is found, from which the phenomenology of the anomalous skin effect is recovered \cite{Nozieres-1999ql}. 

Finally, at leading-order in $\omega \tau_c \ll 1$, the dielectric function approaches an expression given by Eq.\@ (\ref{eq:eps_nu_qualit}) with $\nu(\omega)$ replaced by $\nu(0)$. This is the hydrodynamic regime, where local equilibrium among quasiparticles is ensured by the collision time $\tau_c$ being the smallest timescale in the system. Here, the nonlocal part of the Fermi-liquid response is dissipative, in analogy with a viscous charged fluid \cite{FZVM-2014}, and the shear-polariton is critically damped, with equal real and imaginary parts of $\vec{q}$. 
The different regimes analyzed above compose a unified picture of the Fermi-liquid shear response, summarized in Fig.\@ \ref{fig:regimes} below.

The effect of momentum relaxation on the Fermi-liquid dielectric function is analyzed in Sec.\@ \ref{VE_tauk}, by introducing the associated time $\tau_K$ in the Landau kinetic equation in single-time approximation. The Kubo formalism then leads to $\epsilon_T(\vec{q},\omega)$ with momentum relaxation, momentum-conserving collisions and interactions. Due to momentum relaxation, $\epsilon_T(\vec{q},\omega)$ depends on $i \omega/\tau_K$ explicitly, as well as on the variable $\zeta=s\left[1+i/\left(\omega \tau_m\right)\right]$ where $(\tau_m)^{-1}=\left(\tau_c\right)^{-1}+\left(\tau_K\right)^{-1}$. This result serves as the starting point for all applications in subsequent sections. 

The high-frequency, low-momentum dielectric function (\ref{eq:eps_nu_qualit}) is affected by momentum-relaxing processes according to
\begin{multline}\label{eq:eps_nu_qualit_tauK}
\epsilon_T(\vec{q},\omega)=\epsilon_T\left[i \omega \tilde{\nu}(\omega) q^2,\omega\right]+o(q^4) \\
=1-\frac{(\omega_p)^2}{\omega^2+i \frac{\omega}{\tau_K} +i \omega \tilde{\nu}(\omega) q^2},
\end{multline}
where
\begin{equation}\label{eq:nu_tilde}
\tilde{\nu}(\omega)=\left(1+\frac{F_1^S}{3}\right)  \frac{(v_F^{*})^2 \tau_c}{5\left(1+\tau_c/\tau_K-i \omega \tau_c\right)}
\end{equation}
analytically shows the effect of momentum relaxation on the generalized shear modulus.
In the regime $\tau_c \ll \tau_K$ (weak momentum relaxation), the dielectric function (\ref{eq:eps_nu_qualit_tauK}) depends on the relaxationless shear modulus (\ref{eq:nu_q0_vel}), and it is obtained from Eq.\@ (\ref{eq:eps_nu_qualit}) upon substituting $\omega^2 \mapsto \omega^2+i \omega/\tau_K$ at the denominator. This expression is identical to the phenomenological dielectric function of viscous charged fluids with momentum damping \cite{FZVM-2014} from microscopic Fermi-liquid theory: physically, the Fermi liquid macroscopically responds to shear stresses like a visco-elastic substance in the low-momentum, high-frequency regime. The dispersion of the shear-polariton is robust against momentum relaxation, as it is negligibly affected by $\tau_K$ unless $\tau_K < \tau_c$. This result corroborates the idea that shear-propagation effects might be visible in solid-state Fermi liquids despite relaxation imposed by the breaking of translation invariance.

In the DC limit $\omega \tau_c \rightarrow 0$ with $\nu(0) q^2 \tau_K \gg 1$ (strong spatial nonlocality with respect to relaxation) the nonlocal optical conductivity $\sigma_{DC}(\vec{q})= 1/\left[(\tau_K)^{-1}+\nu(0)q^2\right]$ is limited by the Fermi-liquid viscosity (\ref{eq:nu_0}), which governs diffusive electron flow through narrow channels \cite{Gurzhi-1963,Jaggi-1990,Alekseev-2016, Scaffidi-2017, Levchenko-2020}. Thus, Sec.\@ \ref{FL_DC} links the Fermi-liquid results to the phenomenology of hydrodynamic electron transport. 

The propagation of the shear-polariton qualitatively affects both the surface impedance and the thin-film transmission of solid-state Fermi liquids, while the boundary conditions at vacuum-sample interfaces have relatively minor influence on the results even at quantitative level. 
In the low-frequency, high-momentum regime we retrieve the relaxationless limit of the surface resistance (real part of the surface impedance) of anomalous skin effect \cite{Reuter-1948,Sondheimer-2001,Dressel-2001}. Accordingly, the characteristic penetration depth of electromagnetic fields is the anomalous skin depth $\delta_s^{\rm an} \propto \omega^{-1/3}$. 
In the high-frequency, low-momentum regime the surface resistance saturates to a $\nu(\omega)$-dependent analytical limit for $\tau_K \rightarrow +\infty$ with $\tau_c \ll \tau_K$. The skin depth is determined by $\lambda_L \delta_\nu$, where $\lambda_L=c/\omega_p$ is the London penetration depth and the complex-valued length scale is $\delta_\nu=\sqrt{\nu(\omega)/\omega}$: this scale characterizes the crossover between the hydrodynamic regime $\omega \tau_c \rightarrow 0$, where the skin depth is $\delta_s^{\rm hd}=\sqrt{\lambda_L \delta_\nu^0} \propto \omega^{-1/4}$ with $\delta_\nu^0=\sqrt{\nu(0)/\omega}$ \cite{Levchenko-2020}, and the collisionless regime $\omega \tau_c \rightarrow +\infty$, where the skin depth is $\delta_s^{\rm el}=\sqrt{\lambda_L \delta_s^{\rm sh}} \propto \omega^{-1/2}$ with $\delta_s^{\rm sh}=v_F^{*} \sqrt{(1+F_1^S/3)/5}/\omega$. 
The $\nu(\omega)$-dependent surface impedance is also analytically expressed in terms of the refractive indexes for the plasmon-polariton and the shear-polariton. These results for the surface impedance and the skin depth allow one to use anomalous skin effect in strongly interacting Fermi liquids as a way to measure $\nu(\omega)$. 

In the low-momentum regime where the shear-polariton propagates, the absolute value of the thin-film transmission coefficient is amplified by spatial nonlocality through $\nu(\omega)$, and it is accompanied by oscillations as a function of $\omega$ due to the interference of the two mutually coherent polariton modes \cite{FZVM-2014}. The characteristic wavelength of these oscillations depends on $\nu(\omega)$ through the difference between the real parts of the polariton refractive indexes. Such dependence offers another way of extracting the generalized shear modulus from experiments, by fitting the theory to optical data on Fermi-liquid slabs.
Microscopic models for the evolution of $\tau_c$ and $\tau_K$ with temperature and frequency are considered, to qualitatively simulate temperature-dependent optical experiments. The collision rate is $1/\tau_c \propto (\hbar \omega)^2 +(\pi k_B T)^2$ from phase-space restriction of Fermi-liquid quasiparticles, and the momentum-relaxation rate $1/\tau_K$ includes independent contributions from Umklapp scattering \cite{FZVM-2014}, impurities, and acoustic phonons due to electron-phonon coupling. The oscillations in the thin-film transmission are shown to resist the detrimental effect of momentum relaxation at cryogenic temperatures $T \sim 1$ K in pure samples with sufficiently long impurity scattering times $\tau_{i} \omega_p \gtrapprox 1000$.

\section{Transverse response of neutral Fermi liquids}\label{Kin_FL_neutral}

The prediction \cite{Landau-1957,Khalatnikov-1958,Abrikosov-1959,Lea-1973} of Fermi-surface shear collective modes and their detection in liquid $^3$He \cite{Roach-1976} represent milestones of Fermi-liquid theory. These milestones lay the foundations for the comparison with the transverse response of charged Fermi liquids in solid-state systems, as analyzed in subsequent sections. For these reasons, we begin by surveying the transverse (shear) response in the kinetic approach of Abrikosov and Khalatnikov \cite{Landau-1957,Abrikosov-1959, Nozieres-1999ql}. As nomenclature often differs in the liquid-helium and solid-state communities dealing with this subject, we describe such different terminologies, while setting univocal definitions to be consistently recalled throughout this paper. 

\subsection{Transverse collective mode with collisions}\label{FL_Tmode_coll}

The existence of shear collective modes in neutral Fermi liquids \cite{Landau-1957,Abrikosov-1959} is derived from the kinetic equation for Landau quasiparticles, the derivation of which is reported in Appendix \ref{kin_eq}. In essence, this kinetic equation describes the first-order deviation (or ``displacement'') $\epsilon_{\vec{k}}(\vec{q},\omega)$ of the quasiparticle distribution function at the Fermi surface with respect to global thermodynamic equilibrium. Such deviation is generated by quasiparticle interactions, collisions and external driving forces. We have 
\begin{multline}\label{eq:kin_k_L2}
\left( q v_{\vec{k},\sigma} \cos{\theta}  -\omega \right) \epsilon_{\vec{k}}(\vec{q},\omega) \\+ q v_{\vec{k},\sigma} \cos{\theta} \int \frac{d \Omega^{'}}{4 \pi} \sum_{l=0}^{+\infty} F_l^{S,A} \wp_l(\cos \theta^{'}) \epsilon_{\vec{k}^{'}} (\vec{q},\omega) \\={\mathscr{I}}_{coll}(\vec{q},\omega),
\end{multline}
where $\theta=\arccos \left(  \vec{q} \cdot \vec{v}_{\vec{k},\sigma}/\left|  \vec{q} \cdot \vec{v}_{\vec{k},\sigma}  \right|\right)$ is the angle between the wave vector $\vec{q}$ and the quasiparticle velocity $v_{\vec{k},\sigma}$. The interaction term $\sum_{l=0}^{+\infty} F_l^{S,A} \wp_l(\cos \theta^{'}) \epsilon_{\vec{k}^{'}} (\vec{q},\omega)$ is expanded in terms of Legendre polynomials $\wp_l(\cos \theta^{'})$ and Landau parameters $F_l^{S,A}$, in accordance with their respective standard definitions recalled in Appendix \ref{kin_eq}. The label $S$ and $A$ refer to the symmetric and antisymmetric channels, which correspond to density and spin excitations respectively \cite{Nozieres-1999ql,Dupuis-lect-2011}. Momentum-conserving scattering processes for quasiparticles are encoded by the collision integral ${\mathscr{I}}_{coll}(\vec{q},\omega)$.

We define the normalized velocity $s=\omega/(q v_{\vec{k},\sigma})$, $v_{\vec{k},\sigma}$ being the quasiparticle velocity for the state at wave vector $\vec{k}$ and spin $\sigma$ on the Fermi surface: this change of variables anticipates that soundlike collective modes of linear dispersion $\omega \propto q$ appear in some momentum and frequency regime, where $s$ becomes a constant. 
In terms of $s$, Eq.\@ (\ref{eq:kin_k_L2}) is 
\begin{multline}\label{eq:kin_k_L3}
\left( \cos{\theta}  -s \right) \epsilon_{\vec{k}}(\vec{q},\omega)\\ + \cos{\theta} \int \frac{d \Omega^{'}}{4 \pi} \sum_{l=0}^{+\infty} F_l^{S,A} \wp_l(\cos \theta^{'}) \epsilon_{\vec{k}^{'}} (\vec{q},\omega)\\=\frac{1}{q v_{\vec{k},\sigma}}{\mathscr{I}}_{coll}(\vec{q},\omega).
\end{multline}
We now expand the Fermi surface displacement $\epsilon_{\vec{k}}(\vec{q},\omega)$ similarly to what is done for the interactions. In three dimensions, the displacement is a function of the 3D solid angle and is expanded in spherical harmonics,
\begin{equation}\label{eq:exp_displ_FS}
 \epsilon_{\vec{k}}(\vec{q},\omega)=\sum_{l=0}^{+\infty} \sum_{m=-l}^{+l} \epsilon_{l,m}^{S,A} \mathscr{Y}_{l}^{m}(\theta,\phi),
\end{equation}
where $\hat{u}_{\vec{k}}$ is the unit vector along the direction of $\vec{k}$, and the definition of spherical harmonics $\mathscr{Y}_{l}^{m}(\theta,\phi)$ is recalled in Appendix \ref{kin_eq}.  

We consider the first transverse mode with $m=1$, which corresponds to transverse currents in the first spin-symmetric interaction channel. We truncate the sum over $l$ in Eqs.\@ (\ref{eq:kin_k_L3}) and (\ref{eq:exp_displ_FS}) to $l=1$, so that the interaction becomes $\sum_{l=0}^{+\infty} F_l^{S,A} \wp_l(\cos \alpha) \equiv F_0^S+F_1^S \cos{\alpha}$. Consistently, the Fermi surface displacement reads $\epsilon_{\vec{k}}(\vec{q},\omega) =\sum_{l=0}^{+\infty}\epsilon_{l,1}^{S}\mathscr{Y}_{l}^{1}(\theta,\phi) \equiv \epsilon^S(\theta) e^{i \phi}$, where $\epsilon^S(\theta)$ collects the $\theta$-dependent portion of the displacement. The kinetic equation becomes now 
\begin{multline}\label{eq:kin_k_m1}
\left( \cos{\theta}  -s \right) \epsilon^S(\theta) e^{i \phi} \\ + \cos{\theta} \int_0^{2\pi} \int_0^{\pi} \frac{d \phi' \sin{\theta'} d\theta'}{4 \pi} \left(F_0^S+F_1^S \cos{\alpha} \right)  \epsilon^S(\theta')e^{i \phi'}\\ ={\mathscr{I}}_{coll}(\vec{q},\omega),
\end{multline}
where we have defined ${\mathscr{I}}_{coll}(\epsilon^S(\theta))e^{i \phi}=\left(q v_{\vec{k},\sigma}\right)^{-1}{\mathscr{I}}_{coll}(\vec{q},\omega)$ as the expansion of the collision integral in spherical harmonics with $m=1$. By construction, the angle $\alpha$ is such that $\cos{\alpha}=\cos{\theta}\cos{\theta'}+\sin{\theta}\sin{\theta'}\cos{\left(\phi-\phi'\right)}$ \cite{Dupuis-lect-2011}. Inserting this expression for $\alpha$ into Eq.\@ (\ref{eq:kin_k_m1}) results in
\begin{multline}\label{eq:kin_k_int_coll2}
\left( \cos{\theta}  -s \right) \epsilon^S(\theta) e^{i \phi} \\ + \cos{\theta}  \int_0^{2\pi} \int_0^{\pi} \frac{d \phi' \sin{\theta'} d\theta'}{4 \pi} \left\{ \underbrace{F_0^S}_{\boxed{A}}+F_1^S \left[ \underbrace{\cos \theta \cos{\theta'} }_{\boxed{B}} \right. \right. \\ \left. \left. + \sin \theta \sin{\theta'} \cos \left(\phi-\phi'\right) \right] \right\}  \epsilon^S(\theta')e^{i \phi'} = 
\frac{1}{q v_F^*}{\mathscr{I}}_{coll}(\vec{q},\omega)
\end{multline}
The terms $\boxed{A}$ and $\boxed{B}$ give zero upon integration over the angles $\theta'$ and $\phi'$, and we are left with 
\begin{multline}\label{eq:kin_k_int_coll}
\left( \cos{\theta}  -s \right) \epsilon^S(\theta) e^{i \phi} \\ + \cos{\theta}  \int_0^{\pi} \frac{ \sin{\theta'} d\theta'}{4 \pi} F_1^S  \sin \theta \sin{\theta'} \pi e^{i \phi}  \epsilon^S(\theta') \\  =\frac{1}{q v_F^*}{\mathscr{I}}_{coll}(\vec{q},\omega).
\end{multline}
The conservation of particle number, energy and momentum in collisions imposes constraints on the form of the collision integral ${\mathscr{I}}_{coll}(\vec{q},\omega)$: the moments of the distribution function, obtained by phase-space integration of Eq.\@ (\ref{eq:kin_k_int_coll}), have to yield the continuity equation for particle density, as well as the conservation of energy and of momentum \cite{Abrikosov-1959}. Such conservation laws lead to the collision integral \cite{Lea-1973}
\begin{multline}\label{eq:coll_int_transv_2}
\frac{{\mathscr{I}}_{coll}(\vec{q},\omega)}{q v_F^*} \\ =-\frac{\epsilon^S(\theta)-\left[ \epsilon^S(\theta)\right]_{av}-3 \left[\epsilon^S(\theta) \sin \theta \right]_{av} \sin \theta}{i \omega \tau_c} s e^{i \phi},
\end{multline}
where the notation $\left[ \cdot \right]_{av}= \int_{0}^{\pi}d \theta (\sin \theta)^2/4$ denotes the angular average with respect to $\theta$. In this approach, the single collision time $\tau_c$ parametrizes the integral ${\mathscr{I}}_{coll}\left[\epsilon^S(\theta)\right]$: this allows one to model collisions independently from the microscopic origin of scattering \cite{Abrikosov-1959}, as done in the application to liquid $^3$He. In later sections, we will use a microscopic expression for $\tau_c$ stemming from the Pauli exclusion principle: see also Appendix \ref{app:FL_scat}. 

Contrarily to a longitudinal mode with $m=0$, the transverse mode with $m=1$ does not generate a net density flow, as it couples to transverse currents but not directly to density fluctuations, so that $\left[ \epsilon^S(\theta)\right]_{av}=0$ \cite{Lea-1973}. 
Using the parametrization (\ref{eq:coll_int_transv_2}), one can solve Eq.\@ (\ref{eq:kin_k_int_coll}) for the displacement $\epsilon^S(\theta)$ as detailed in Appendix \ref{transv_mode_coll}. In particular, a non-vanishing solution for $\epsilon^S(\theta)$ appears for specific combinations of momentum $\vec{q}$ and frequency $\omega$ in the absence of external driving forces. This solution is a collective mode with dispersion relation 
 \cite{Lea-1973,Roach-1976}
\begin{equation}\label{eq:Lea}
\left(\xi^2-1\right)\left[\frac{\xi}{2} \ln \left(\frac{\xi+1}{\xi-1}\right) -1 \right]= \frac{F_1^S-6-9 \beta}{3 F_1^S- 9 \beta},
\end{equation}
where 
\begin{subequations}\label{eq:Lea_var}
\begin{equation}\label{eq:Lea_var1}
\xi=s \left(1+\frac{i}{\omega \tau_c} \right),
\end{equation}
\begin{equation}\label{eq:Lea_var2}
s=\frac{\omega}{q v_F^{*}}, 
\end{equation}
\begin{equation}\label{eq:Lea_var3}
\beta=\frac{1}{i \omega \tau_c-1}.
\end{equation}
\end{subequations}
\begin{figure}[ht]
\includegraphics[width=1.0\columnwidth]{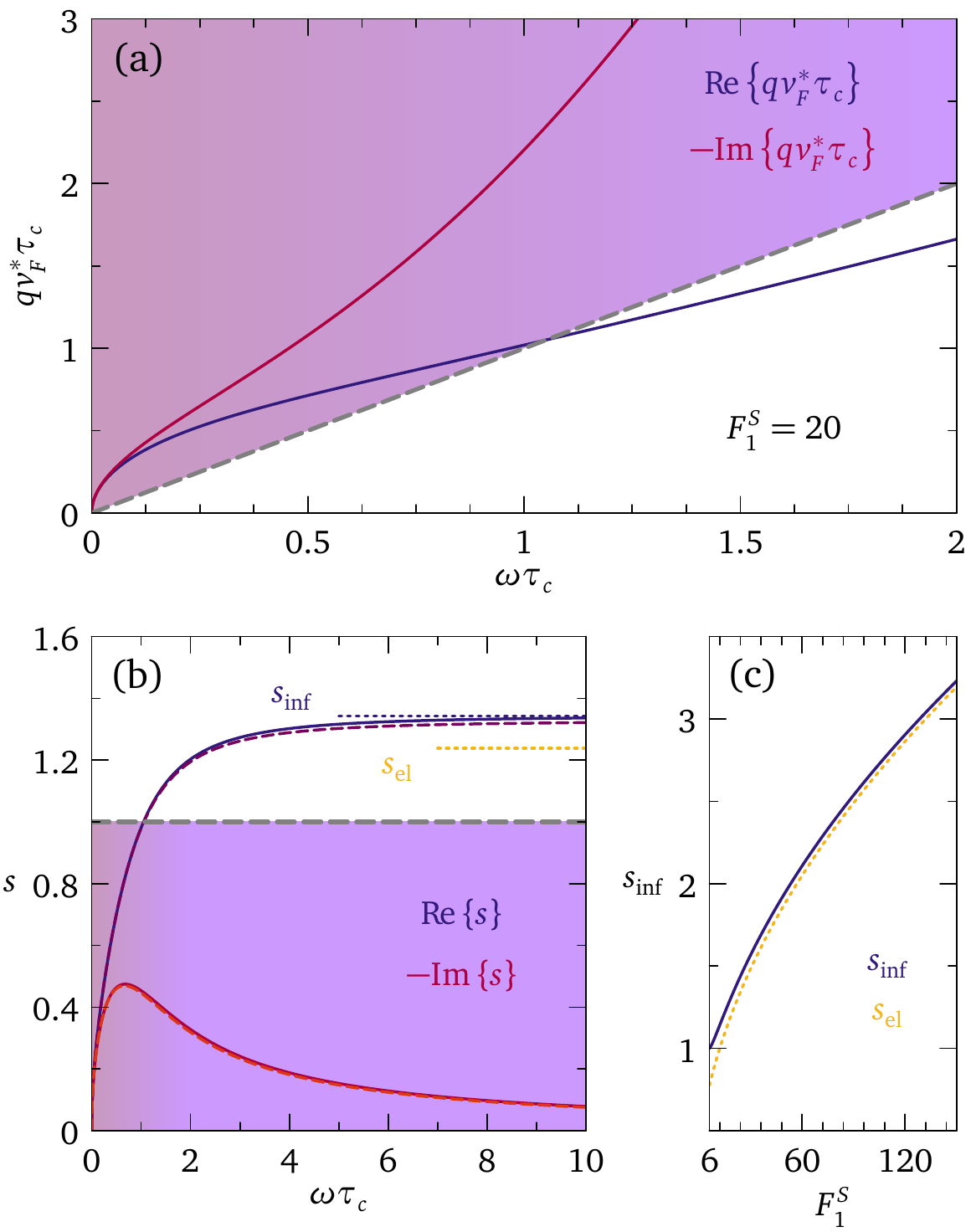}
\caption{\label{fig:unch_disp} Dispersion relation of the transverse shear collective mode from Eq.\@ (\ref{eq:Lea}). (a) Dimensionless wave vector $q v_F^* \tau_c$ as a function of dimensionless frequency $\omega \tau_c$ for first symmetric Landau parameter $F_1^S=20$. The blue and red curves show the real and imaginary part of the mode wave vector, respectively. The shaded purple area indicates the Lindhard electron-hole continuum $\omega<\left(v_F^* \mathrm{Re}\left\{q \right\}\right)$. (b) Dimensionless collective mode velocity (\ref{eq:Lea_var2}) as a function of $\omega \tau_c$ for $F_1^S=20$. The blue and red solid curves show the real and imaginary part of $s$, respectively. The purple (red) dashed curve shows the real (imaginary) part of the leading-order expansion (\ref{eq:vs_hydro}) in the hydrodynamic regime $\omega \tau_c \rightarrow 0$. The dashed blue horizontal line shows the asymptotic limit $s_{\mathrm{inf}}=\lim_{\omega \tau_c \rightarrow +\infty} s$ according to Eq.\@ (\ref{eq:vs_inf}). The dashed gold horizontal line indicates the elasticlike limit $s_{\mathrm{el}}$ from Eq.\@ (\ref{eq:vs_elast}). (c) Comparison between the collisionless asymptotic limit $s_{\mathrm{inf}}$ and the elasticlike expression $s_{\mathrm{el}}$ as a function of Landau parameter $F_1^S$: $s_{\mathrm{inf}}$ and $s_{\mathrm{el}}$ converge in the limit of large $F_1^S$. 
 }
\end{figure}
The mode described by the dispersion relation (\ref{eq:Lea}) has been labeled ``transverse sound'' in the liquid-helium literature, although physically it amounts to a shear oscillation of the Fermi surface with (complex-valued) dimensionless velocity (\ref{eq:Lea_var2}). It depends on Fermi-liquid interactions through the Landau parameter $F_1^S$, and on collisions through $\tau_c$. Besides, Landau damping strongly attenuates transverse sound in some regions of the $\vec{q}$-$\omega$ plane, due to the mode exchanging energy and momentum with incoherent electron-hole excitations \cite{Abrikosov-1959,Berthod-2018}: at small momentum, this happens when $\omega < v_F^* \left|q \right|$, i.e.\@, $\left|s \right|<1$. As $\mathrm{Re}\left\{s\right\} \gg \mathrm{Im}\left\{s\right\}$ where transverse sound emerges from the Lindhard continuum (see Fig.\@ \ref{fig:unch_disp}) an equivalent criterion for the propagation of the mode is $\mathrm{Re}\left\{s\right\}>1$.

Figure \ref{fig:unch_disp}(a) shows the the normalized wave vector $q v_F^{*} \tau_c$ of transverse sound from Eq.\@ (\ref{eq:Lea}) as a function of $\omega \tau_c$ for $F_1^S=20$. Blue and red lines refer to $\mathrm{Re}\left\{q v_F^{*} \tau_c \right\}$ and $\mathrm{Im}\left\{q v_F^{*} \tau_c \right\}$ respectively. The real part represents the dissipationless, i.e.\@ reactive, component of the collective mode, while the imaginary part encodes dissipative processes and is connected to the mode damping. The purple shadow highlights the Lindhard electron-hole continuum of quasiparticle excitations, from which transverse sound emerges at $\omega \tau_c \approx 1$. Figure \ref{fig:unch_disp}(b) shows the corresponding real and imaginary parts of the mode velocity (\ref{eq:Lea_var2}) as a function of $\omega \tau_c$, as blue and red solid lines respectively.  

In the hydrodynamic/collisional regime $\omega \tau_c \ll 1$: this condition establishes local thermodynamic equilibrium, so that the system behaves like a viscous fluid \cite{FZVM-2014,Nozieres-1999ql}. Hence, the collective mode is diffusive in nature and essentially governed by quasiparticle collisions. The relaxational mode \footnote{In the hydrodynamic/collisional regime $\omega \tau_c \ll 1$, the transverse collective mode is damped, i.e.\@, it has a complex sound velocity  $s=\omega/(q v_F^{*}) \in \mathbb{C}$ with equal real and imaginary parts \cite{Lea-1973}. In such conditions, the transverse mode is called \emph{relaxational mode} in part of the literature, distinguishing this regime from the one of propagating transverse sound. Other references employ the label \emph{damped transverse sound} in referring to damped transverse waves in the collisional regime. In this paper, we will employ the nomenclature \emph{relaxational mode} and \emph{damped transverse sound} as synonyms, meaning a transverse collective mode satisfying Eq.\@ (\ref{eq:Lea}) in the regime $\omega \tau_c \ll 1$. Generic solutions of Eq.\@ (\ref{eq:Lea}), without specifying whether we are in collisional/collisionless regime, will be named \emph{shear mode} or \emph{transverse sound} in this paper.} velocity in hydrodynamic regime at quadratic order in $\omega \tau_c \rightarrow 0$ is
\begin{multline}\label{eq:vs_hydro}
s=-\frac{i \left(3 +F_1^S\right)\sqrt{\omega \tau_c}}{\sqrt{-15 i\left(3+F_1^S\right)+15 F_1^S \omega \tau_c}}\\ = \frac{1}{v_F^*}\sqrt{ \omega \nu(0)} \frac{-i}{\sqrt{-i+F_1^S \omega \tau_c/\left(3+F_1^S\right)}},
\end{multline}
where $\nu(0)$ is the static viscosity coefficient (\ref{eq:nu_0}) of the isotropic Fermi liquid in three dimensions \cite{Conti-1999, Nozieres-1999ql, Abrikosov-1959}.
In the limit $\omega \tau_c \rightarrow 0$, (\ref{eq:vs_hydro}) implies
\begin{equation}\label{eq:vs_hydro_Re_Im}
\mathrm{Re}\left\{s\right\}=-\mathrm{Im}\left\{s\right\}=\frac{1}{v_F^*}\frac{\sqrt{\omega \nu(0)}}{\sqrt{2}},
\end{equation}
that is, the mode is critically damped in hydrodynamic regime, having equal and opposite real and imaginary parts. Figure \ref{fig:unch_disp}(b) shows the real and imaginary parts of Eq.\@ (\ref{eq:vs_hydro}) as a function of $\omega \tau_c$ as purple and red dashed lines, respectively. 
Notice that Eq.\@ (\ref{eq:vs_hydro}) implies the dispersion relation 
\begin{equation}\label{eq:q_hydro}
q=\sqrt{\frac{\omega}{\nu(0)}}\sqrt{\frac{i\left(3+F_1^S\right)-F_1^S \omega \tau_c}{3+F_1^S}}:
\end{equation}
this means $\omega \propto q^2$ in hydrodynamic regime for uncharged transverse sound, with the proportionality coefficient governed by the static shear viscosity (\ref{eq:nu_0}). The quadratic dispersion (\ref{eq:q_hydro}) in hydrodynamic regime is modified by the electric charge, as described in Sec.\@ \ref{Charged_FL_XT}.

At higher values of $\omega \tau_c$ the evolution of the shear response depends on the first Landau parameter $F_1^S$. If the interaction is sufficiently strong, such that the transverse mode persists in the absence of collisions, then the mode propagates and is essentially undamped for $\omega \tau_c \rightarrow +\infty$: this is the limit of transverse zero sound \footnote{In this paper we will refer to real solutions of Eq.\@ (\ref{eq:Lea}) in the collisionless limit $\omega \tau_c \rightarrow +\infty$ as \emph{transverse zero sound} or \emph{propagating shear}, consistently with the literature \cite{Abrikosov-1959,Lea-1973,Roach-1976}}, in which the system responds out of equilibrium in a reactive way, i.e.\@, without dissipation. Such reactive response is reminiscent of elasticity in solids, but in the Fermi liquid it occurs only at finite frequency \cite{Tokatly-2000,FZVM-2014}. The transverse zero sound velocity reaches the real constant determined by the numerical solution $s=s_{\mathrm{inf}}$ of \cite{Khalatnikov-1958}
\begin{equation}\label{eq:vs_inf}
\left(s^2-1\right)\left[\frac{s}{2} \ln \left(\frac{s+1}{s-1}\right) -1 \right]= \frac{F_1^S-6}{3 F_1^S}. 
\end{equation}
\begin{figure}[ht]
\includegraphics[width=0.8\columnwidth]{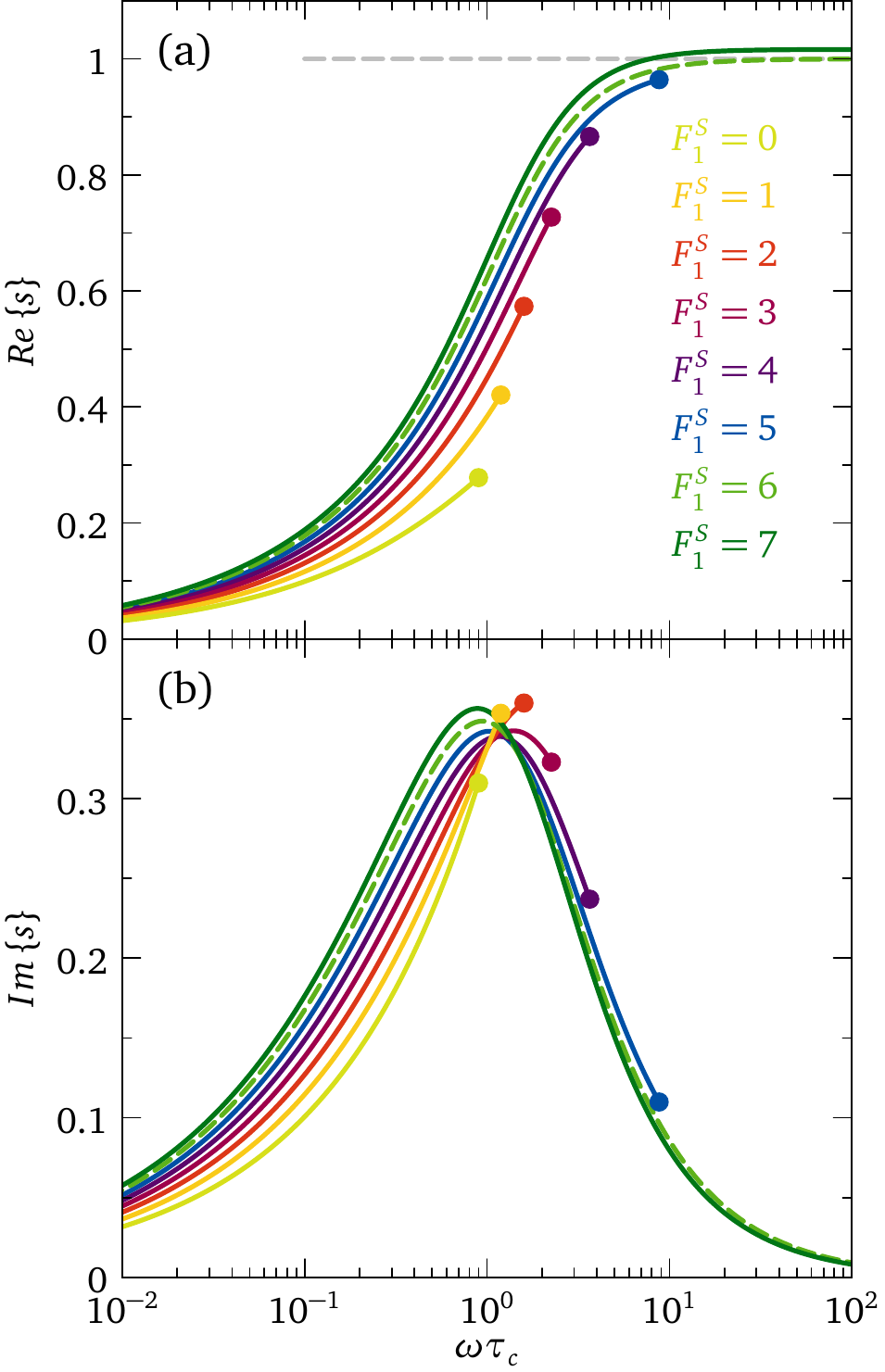}
\caption{\label{fig:vs_crit}
Transverse sound velocity, calculated from Eq.\@ (\ref{eq:Lea}), as a function of the product $\omega \tau_c$, for different values of the first Landau parameter ranging from $F_1^S=0$ to $F_1^S=7$. Panels (a) and (b) show the real part and the imaginary part of the transverse collective mode velocity, respectively. For $F_1^S<6$, a dot marks the critical value $(\omega \tau_c)_{crit}$ above which Eq.\@ (\ref{eq:Lea}) has no solution. The dashed gray line in panel (a) highlights the value of the renormalized Fermi velocity $v_F^*$, i.e. $\mathrm{Re}\left\{s \right\}=1$: for $F_1^S>6$, the asymptotic collisionless limit $\lim_{\omega \tau_c \rightarrow +\infty} \mathrm{Re}\left\{s \right\}>1$ exists and corresponds to the solution of Eq.\@ (\ref{eq:vs_inf}). 
}
\end{figure}
Strictly speaking, only in this limit is the labeling ``sound'' for the transverse mode fully justified, since $\omega \propto q$. 
For strong interactions $F_1^S \gg 1$, Eq.\@ (\ref{eq:vs_inf}) approaches the analytical result $s=s_{\mathrm{el}}$:
\begin{equation}\label{eq:vs_elast}
s_{\mathrm{el}}=\sqrt{\frac{1}{5} \left(1+\frac{F_1^S}{3}\right)}=\frac{\sqrt{\mu_s}}{v_F^{*}} \\ : \; F_1^S \gg 1
\end{equation}
In Eq.\@ (\ref{eq:vs_elast}), we have defined
\begin{equation}\label{eq:mu_FL}
\mu_s=\frac{1}{5} \frac{(v_F)^2}{1+ F_1^S/3},
\end{equation}
which is the reactive shear modulus of the Fermi liquid \cite{Conti-1999,Vignale-2005,Nozieres-1999ql}. 

For strong interaction $F_1^S \gg 1$, the reactive response of the Fermi surface in the limit $\omega \tau_c \rightarrow +\infty$ is analogous to the vibration of an elastic substance characterized by the shear modulus $\mu_s$ -- see Fig.\@ \ref{fig:unch_disp}(c). Notice that the Fermi-liquid shear modulus (\ref{eq:mu_FL}) satisfies $\mu_s=\nu(0)/\tau_c$ where $\nu(0)$ is the static viscosity (\ref{eq:nu_0}) \cite{Conti-1999, Vignale-2005}. Reducing the interaction to $F_1^S \rightarrow 6^+$, the exact mode velocity (\ref{eq:vs_inf}) approaches the renormalized Fermi velocity, i.e.\@, $s \rightarrow 1^+$, and it significantly differs from the elasticlike estimation (\ref{eq:vs_elast}). This is shown in Fig.\@ \ref{fig:unch_disp}(c) as a function of $F_1^S$, and by the dashed blue and gold horizontal lines in Fig.\@ \ref{fig:unch_disp}(b) for $s=s_{\mathrm{inf}}$ and $s=s_{\mathrm{el}}$, respectively. Such discrepancy is a consequence of Landau damping \cite{Dobbs-2000,Brooker-1967}, an effect not recounted by the elasticlike approximation (\ref{eq:vs_elast}). Notice that the reactive character of transverse sound emerges only in the high-frequency regime $\omega \tau_c \gg 1$ at fixed $\tau_c$, while in the static limit the system displays a dominant dissipative response (\ref{eq:vs_hydro_Re_Im}). Therefore, Fermi-surface rigidity is physically different from elasticity in classical solids, in that it is an inherently dynamical phenomenon. 

Figure \ref{fig:vs_crit} shows the shear mode velocity, as calculated from Eq.\@ (\ref{eq:Lea}), as a function of the product $\omega \tau_c$, for different values of the first Landau parameter ranging from $F_1^S=0$ to $F_1^S=7$. Figure \ref{fig:vs_crit}(a) shows the real part of the transverse sound velocity, while Fig.\@ \ref{fig:vs_crit}(b) displays the imaginary part of the velocity. For $F_1^S<6$, we see that the collective mode velocity $s$ resulting from Eq.\@ (\ref{eq:Lea}) is a continuous curve only up to a $F_1^S$-dependent critical value $(\omega \tau_c)_{crit}$, marked by a dot, above which the collective mode equation (\ref{eq:Lea}) has no solution. Physically, this means that interactions are not strong enough to sustain transverse zero sound without collisions: for $\omega \tau_c>(\omega \tau_c)_{crit}$ there are no transverse waves in the Fermi liquid due to strong Landau damping, and only incoherent electron-hole quasiparticles can be excited. Such a disappearance of solutions above a critical value $(\omega \tau_c)_{crit}$ was pointed out by Brooker \cite{Brooker-1967} for longitudinal sound in a Fermi liquid, and reported by Lea \textit{et al.}\@ \cite{Lea-1973} for transverse sound.

For $F_1^S>6$, Eq.\@ (\ref{eq:Lea}) admits a collective mode solution for any value of $\omega \tau_c$ with asymptotic value $\lim_{\omega \tau_c \rightarrow +\infty} s=s_{\mathrm{inf}}$ in accordance with Eq.\@ (\ref{eq:vs_inf}). The dashed green lines in Fig.\@ \ref{fig:vs_crit}(a) and \ref{fig:vs_crit}(b) identify the real and imaginary part of the transverse sound velocity for the critically Landau-damped case $F_1^S=6$, for which transverse sound exists for all values of $\omega \tau_c$ but the limit $\lim_{\omega \tau_c \rightarrow +\infty} \mathrm{Re} \left\{s\right\}=1$: the transverse mode is precisely at the edge of the electron-hole continuum.

\subsection{Interaction-dependent existence of transverse sound}\label{Exist_sound_T}

The  $F_1^S$-dependent quantity $(\omega \tau_c)_{crit}$ such that Eq.\@ (\ref{eq:Lea}) has no solution, either real or complex, for $\omega \tau_c > (\omega \tau_c)_{crit}$ can be calculated numerically from the dispersion relation (\ref{eq:Lea}), in analogy with Brooker's treatment \cite{Brooker-1967} for longitudinal sound. 
First, we notice that the real and imaginary parts of Eq.\@ (\ref{eq:Lea}) are discontinuous, because of the discontinuity inherent in the logarithmic term $\xi/2 \ln \left[\left(\xi+1\right)/\left(\xi-1\right)\right]$ in passing from the lower half of the complex plane to the upper half or viceversa. Such discontinuity occurs for $\xi \in \left(-1, \, 1 \right)$: if this happens at a real value of $(\omega \tau_c)_{crit}$, that means that the transverse wave solution existing for $\omega \tau_c <(\omega \tau_c)_{crit}$ is interrupted by the discontinuity. Therefore, in order to find the critical value $(\omega \tau_c)_{crit}$ at a given interaction $F_1^S$, we first impose that $\xi$ is real, which means $\mathrm{Im} \left\{ s \left[1+i/\left(\omega \tau_c\right) \right] \right\}=0 $. For $s \in \mathbb{C}$, i.e.\@, damped solutions, we have $\mathrm{Im} \left\{ s \left[1+i/\left(\omega \tau_c\right) \right] \right\}=\mathrm{Im}\left\{s \right\} + \mathrm{Re} \left\{ s \right\}/(\omega \tau_c)=0 $, so $\mathrm{Im} \left\{s \right\}=-\mathrm{Re} \left\{ s \right\}/(\omega \tau_c)$. 
\begin{figure}[ht]
\includegraphics[width=1.0\columnwidth]{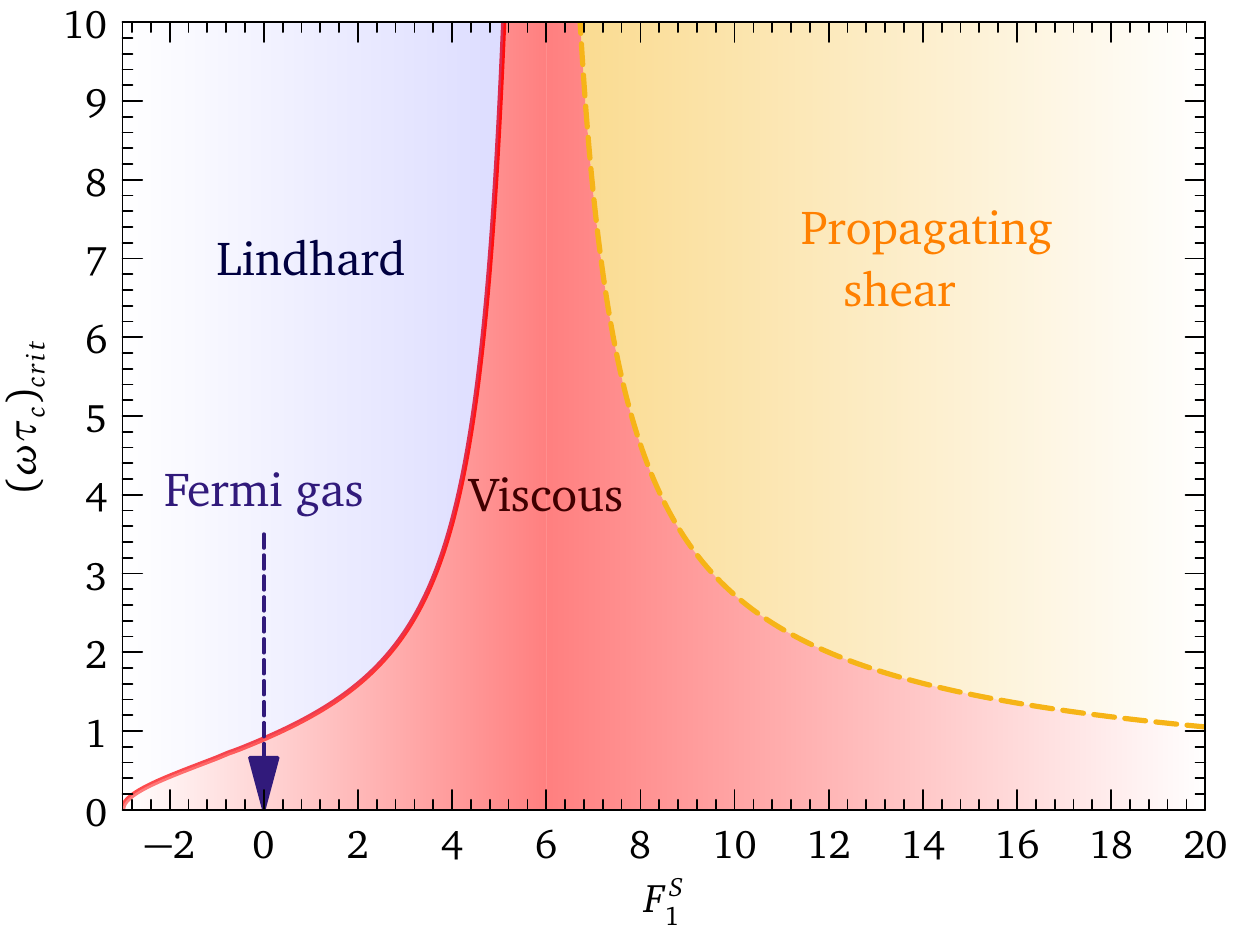}
\caption{\label{fig:omtau_crit} 
Transverse response ``phase diagram'' for a neutral Fermi liquid, as a function of product between frequency and collision time $\left(\omega \tau_c \right)_{crit}$ and of interaction strength quantified by the first Landau parameter $F_1^S$. The regions shaded in red, blue, and gold identify the viscous, Lindhard, and propagating regimes of the shear response, respectively. The red solid line represents the critical value $\left(\omega \tau_c \right)_{crit}$ above which no shear modes exist. The dashed orange curve marks the value of $\omega \tau_c$ above which the shear mode emerges from the electron-hole continuum and starts propagating according to Eq.\@ (\ref{eq:omega_cont}). 
}
\end{figure}
We impose the latter condition in Eq.\@ (\ref{eq:Lea}), which now depends on $\mathrm{Re} \left\{ s \right\}$ and $\omega \tau_c$. We then solve the imaginary part of the dispersion (\ref{eq:Lea}) numerically for $\mathrm{Re} \left\{ s \right\}$, and we insert the found value of $\mathrm{Re} \left\{ s \right\}$ in the real part of Eq.\@ (\ref{eq:Lea}). The latter is solved numerically on the real axis to give $\omega \tau_c \equiv (\omega \tau_c)_{crit}$ as shown in Fig.\@ \ref{fig:omtau_crit}. 

The red-shaded region in Fig.\@ \ref{fig:omtau_crit} identifies the viscous regime, defined as the regime where the relaxational mode (transverse damped sound)  exists for an interaction comprising only the first Landau parameter $F_1^S$. The blue-shaded area denotes the Lindhard regime, where no collective modes are allowed in the neutral Fermi liquid. The red solid line separating the viscous and Lindhard regions marks the critical value $\left(\omega \tau_c \right)_{crit}$ for $F_1^S<6$. Such line approaches zero at $F_1^S=-3$: this means that the Fermi-liquid ground state becomes unstable against collective modes, as the effective interaction $A_1^S=F_1^S/\left(F_1^S-3\right)$ between quasiparticles becomes negative for $F_1^S<-3$ \cite{vanderMarel-2011}. An analogous instability criterion for longitudinal sound reads $F_0^S<-1$ \cite{Nozieres-1999ql}. The blue arrow identifies the Fermi gas limit, meaning truly noninteracting quasiparticles, i.e.\@, $F_1^S=0$. We see that the noninteracting Fermi gas sustains a transverse relaxational mode for $\omega \tau_c < \left(\omega \tau_c \right)_{crit} \approx 0.87$. This Fermi-gas dynamics is qualitatively analogous in the longitudinal channel \cite{Brooker-1967,Nozieres-1999ql}. 
For $F_1^S>6$ a smooth crossover takes place between the viscous regime and the propagating shear regime, with the latter denoted by the gold-shaded area. In this area, the mode lies at frequencies higher than the electron-hole continuum, thus becoming propagating zero sound. The dashed orange curve marks the product $\omega \tau_c$ above which the collective mode emerges from the electron-hole continuum and starts propagating, i.e.\@, 
\begin{equation}\label{eq:omega_cont}
\mathrm{Re}\left\{s\right\}=\mathrm{Re}\left\{\frac{\omega_c}{v_F^* q}\right\}=1. 
\end{equation}

\section{Transverse response of charged Fermi liquids}\label{Charged_FL_XT}

Fermi-liquid theory must include the effect of electric charges to address electrons in metals. Each charged quasiparticle perceives finite electromagnetic potentials even in the absence of external fields, due to the electromagnetic fields generated by all other moving quasiparticles in the ensemble \cite{Coleman-2015mb}. In the kinetic equation approach summarized in Appendix \ref{kin_eq}, such electromagnetic potentials are the scalar $\phi_\sigma(\vec{r},t)$ and the vector $\vec{A}(\vec{r},t)$ --- see the quasiclassical force (\ref{eq:qc_force}). 
The modification of interactions among Landau quasiparticles due to the long-ranged Coulomb interaction $V_{\rm Coul}=e^2/\left(\epsilon_0 q^2\right)$ poses a challenge for the Fermi-liquid picture, since $V_{\rm Coul}(q)$ is divergent for $q \rightarrow 0$. By contrast, in neutral systems Landau theory assumes short-ranged quasiparticle interactions with a well-defined long-wavelength limit. The solution of this conundrum was first found by Silin \cite{Silin-1958a,Silin-1958b}: effects due to Coulomb interactions can be separated into a long-ranged part, representing the classical polarization field that provides dielectric screening, and a short-ranged quantum component, driven by the virtual creation of electron-hole pairs around a charged particle. The polarization field reduces the range of Coulomb interactions to a finite distance through dielectric screening, while electron-hole quantum fluctuations modify the short-ranged interactions with respect to the neutral case. The net consequence is that the long-ranged and spherically symmetric polarization screens quasiparticle interactions in the isotropic $l=0$ channel, thereby modifying the value of $F_0^{S}$ only. This alters Landau parameters according to $\tilde{F}_l^S(\vec{q})= \left[e^2 N_{el}^{*}(0)\right]/\left(\epsilon_0 q^2\right) \delta_{l0}+ F_l^S$ \cite{Coleman-2015mb}. Hence, electric charge does not introduce any momentum dependence of $F_1^S$, which is linked to transverse excitations in the $l=1$ channel as derived in Sec.\@ \ref{Kin_FL_neutral}. The only consequence of Coulomb interactions is that $F_1^S$ differs from the corresponding value in the neutral system, due to the electron-hole short-ranged quantum component. 
Furthermore, one can expect that the dispersion relations of collective modes are modified by the inclusion of electric charge. The transverse $l=1$ channel is particularly intriguing: such transverse excitations can couple to photons, which allows one to probe Fermi liquid collective modes in solids using electromagnetic fields. With this in mind, we now analyze the transverse susceptibility, linked to the current response of charged systems.

\subsection{Interacting transverse susceptibility in the absence of collisions}\label{FL_XT_nocoll}

In order to study the response of the Fermi liquid to transverse perturbations at any momentum and frequency, we have to calculate the full transverse susceptibility, which is the proportionality coefficient between the transverse current density $\vec{J}_T(\vec{q},\omega)$ and the respective Fourier component of the applied vector potential $\vec{A}_T(\vec{q},\omega)$. 
We first consider the case with no short-ranged interactions between quasiparticles and no collisions \cite{Dupuis-lect-2011}. The derivation is reported in Appendix \ref{X_T0} for the interested reader, while here we quote only the final result. It is 
\begin{multline}\label{eq:Xi_T0}
X_T^0(\vec{q},\omega)\equiv X_T^0(s)=3\frac{ n m^*}{m^2}\mathscr{I}(s) \\ =\frac{3}{4} \frac{n m^*}{m^2} \left[ -\frac{4}{3}+ 2 s^2 +s \left(1-s^2\right)\ln{\left(\frac{s+1}{s-1} \right)} \right]
\end{multline}
with the integral
\begin{multline}\label{eq:vs_int}
\mathscr{I}(s)=\frac{1}{4} \int_0^{\pi} \frac{(\sin \theta)^3 \cos{\theta}}{s-\cos \theta} d \theta \\ = -\frac{1}{3} + \frac{1}{2} s^2 + \frac{s}{4} \left( 1-s^2 \right) \ln \left(\frac{s+1}{s-1} \right).
\end{multline}
The noninteracting transverse susceptibility (\ref{eq:Xi_T0}) depends on frequency and momentum through the ratio $s=\omega/\left(v_F^{*} q\right)$. 
For $\omega=0$ (static limit), Eq.\@ (\ref{eq:Xi_T0}) gives $X_T^0(0)=-\frac{n}{m} \left(1+\frac{F_1^S}{3}\right)$. In the quasistatic nonlocal limit $\omega \ll q$, an expansion of the integral $\mathscr{I}(s)$ around $s=0$ gives $\mathscr{I}(s)=-\frac{1}{3}+i \pi \frac{s}{4}+o(s^2)$; this gives $X_T^0(s)=\frac{n}{m} \left[-\left(1+\frac{F_1^S}{3}\right)+ i \frac{3}{4} \pi s \right]$. In the local limit with $q=0$ and $\omega>0$, we have $s \rightarrow +\infty$ and $\lim_{s \rightarrow +\infty} \mathscr{I}(s)=0$, so that $\lim_{s\rightarrow +\infty}X_T^0(s)=0$.
Quasiparticle collisions modify the result (\ref{eq:Xi_T0}) as analyzed in the following section.  

\subsection{Interacting transverse paramagnetic susceptibility with collisions}\label{FL_XT_coll}

When we include short-ranged quasiparticle interactions and collisions, we have to employ the full kinetic equation (\ref{eq:kin_k_int_coll}). From there, we perform the same steps as in Sec.\@ \ref{FL_Tmode_coll}, but now explicitly including an external vector potential $\vec{A}(\vec{q},\omega)$. The result is 
\begin{multline}\label{eq:kin_k_int_compl}
\left( \cos{\theta}  -s \right) \epsilon^S(\theta) e^{i \phi}+ \cos{\theta}  \int_0^{2\pi} \int_0^{\pi} \frac{d \phi' \sin{\theta'} d\theta'}{4 \pi} \left\{ \underbrace{F_0^S}_{\boxed{A}} \right. \\ \left.+F_1^S \left[ \underbrace{\cos \theta \cos{\theta'} }_{\boxed{B}}+ \sin \theta \sin{\theta'} \cos \left(\phi-\phi'\right) \right] \right\}  \epsilon^S(\theta')e^{i \phi'} \\ +\frac{e}{m v_F^*} \cos \theta \vec{k} \cdot \vec{A}(\vec{q},\omega)  =\frac{1}{q v_F^*}{\mathscr{I}}_{coll}(\vec{q},\omega),
\end{multline}
where we assume an interaction of the form $F_0^S+F_1^S \cos{\alpha}$ as done in the neutral case. 
The integration along $\phi'$ gives zero for the terms $\boxed{A}$ and $\boxed{B}$ in Eq.\@ (\ref{eq:kin_k_int_compl}), and we are left with 
\begin{multline}\label{eq:kin_k_int_compl_2}
\left( \cos{\theta}  -s \right) \epsilon^S(\theta) e^{i \phi} \\ + \cos{\theta} \left[ \int_0^{\pi} \frac{ \sin{\theta'} d\theta'}{4 \pi} F_1^S  \sin \theta \sin{\theta'} \pi e^{i \phi}  \epsilon^S(\theta') \right. \\  \left. +\frac{e}{m v_F^*} \vec{k} \cdot \vec{A}(\vec{q},\omega) \right]  = \\ \frac{1}{q v_F^*}{\mathscr{I}}_{coll}(\vec{q},\omega)
\end{multline}
We now employ the parametrization (\ref{eq:coll_int_transv_2}) for the collision integral ${\mathscr{I}}_{coll}(\vec{q},\omega)$, as done for the neutral case of Sec.\@ \ref{FL_Tmode_coll}. Hence, we assume that the inclusion of electric charge does not qualitatively modify the collision integral.
In terms of the variables (\ref{eq:Lea_var}), the kinetic equation becomes 
\begin{multline}\label{eq:kin_k_int_compl_3}
\left( \cos{\theta}  -\xi \right) \epsilon^S(\theta) + \left[ 3  \epsilon^S(\theta) \sin \theta \right]_{av} \sin{\theta} \left(\frac{F_1^{S}}{3} \cos{\theta}- \beta \xi \right)\\ = -\cos{\theta} \frac{e}{m v_F^*} \vec{k} \cdot \vec{A}(\vec{q},\omega) e^{-i \phi}
\end{multline}
We now utilize the definition of the paramagnetic current density $\vec{J}_p(\vec{q},\omega)$ in a Fermi liquid, Eq.\@ (\ref{eq:FL_J_p}), and we define the paramagnetic susceptibility as the ratio between $\vec{J}_p(\vec{q},\omega)$ and the vector potential $\vec{A}(\vec{q},\omega)$. The details of this calculation are reported in Appendix \ref{Int_X_transv}. 
The final result for the interacting paramagnetic transverse response function is
\begin{multline}\label{eq:Xi_T}
X_T^P(\xi)=\frac{X_T^0(\xi)}{1-3\left(\frac{F_1^S}{3}-\beta \right)\frac{m^2}{3 n m^*} X_T^0(\xi) +\beta} \\ =\frac{n}{m} \frac{3\left(1+ \frac{F_1^S}{3} \right) \mathscr{I}(\xi)}{1-3 \left(\frac{F_1^S}{3}-\beta\right) \mathscr{I}(\xi)+\beta}
\end{multline}
where $X_T^0(\xi)$ is the noninteracting response function (\ref{eq:Xi_T0}) in terms of $\xi$. Arguments similar to the ones developed so far apply to the longitudinal susceptibility $X_L(\vec{q},\omega)$ \cite{Nozieres-1999ql, Dupuis-lect-2011}. The result (\ref{eq:Xi_T}) allows us to study the electromagnetic response of the charged system.

\section{Fermi-liquid optical conductivity and dielectric function}\label{FL_epsT_coll}

From the paramagnetic response function (\ref{eq:Xi_T}), we can calculate the optical properties of the charged Fermi liquid. In particular, the transverse dielectric function is obtained in two steps. First, we calculate the optical conductivity by the means of the Kubo formula for a translationally invariant system \cite{Mahan-2000,Bruus-2004mb,Berthod-2018}:
\begin{equation}\label{eq:Kubo_q}
\sigma_{T, \mu \nu}(\vec{q},\omega)=\frac{i e^2 }{\omega}\left[ \chi_{\vec{J}\vec{J}}^{\mu \nu}(\vec{q},\omega) + \delta_{\mu \nu} \frac{n}{m} \right],
\end{equation}
where $\chi_{\vec{J}\vec{J}}^{\mu \nu}(\vec{q},\omega)=\int_{-\infty}^{+\infty} dt e^{i \omega t} \left(-i/\hbar\right) \Theta(t) \times$ $\left\langle \left[J^{\mu}(\vec{q},t),J^{\nu}(-\vec{q},0)\right]\right\rangle$ is the retarded current-current correlation function, $\Theta(t)$ is the Heaviside step function, $\left[\hat{A}, \hat{B}\right]=\hat{A}\hat{B}-\hat{B}\hat{A}$ denotes the commutator between operators $\hat{A}$ and $\hat{B}$, and $J^{\mu}(\vec{q},t)$ is the component of the current density operator for momentum $\vec{q}$ and time $t$ along the spatial direction $\mu=\left\{x,y,z\right\}$.
We take the diagonal component $\chi_{\vec{J}\vec{J}}^{\mu \mu}(\vec{q},\omega)$ which corresponds to the transverse susceptibility $X_T^P(\vec{q},\omega)$, so that 
\begin{equation}\label{eq:Kubo_T}
\sigma_{T}(\vec{q},\omega)=\frac{i e^2 }{\omega}\left[ X_{T}^P(\vec{q},\omega) +\frac{n}{m} \right]
\end{equation}
The second term in square brackets in Eq.\@ (\ref{eq:Kubo_T}) is the diamagnetic susceptibility, which is connected to the diamagnetic part of the current response. Then, we can write
\begin{equation}\label{eq:sigma_T_mb}
\sigma_T(\vec{q},\omega)=\frac{i e^2 }{\omega}\left[ X_{T}^P(\vec{q},\omega) +\frac{n}{m} \right] =\frac{i e^2 }{\omega} X_{T}(\vec{q},\omega).
\end{equation}
In Eq.\@ (\ref{eq:sigma_T_mb}) we have defined the total current susceptibility $X_{T}(\vec{q},\omega)$, which considers both the paramagnetic term (\ref{eq:Xi_T}) and the diamagnetic term:
\begin{multline}\label{eq:Xi_T_tot}
X_{T}(\vec{q},\omega)= X_{T}^P(\vec{q},\omega) +\frac{n}{m} \\=\frac{n}{m} \frac{\left(1+ \beta  \right) \left[3\mathscr{I}(\xi) +1\right]}{1-3 \left(\frac{F_1^S}{3}-\beta\right) \mathscr{I}(\xi)+\beta}.
\end{multline}
In the collisionless limit $\tau_c\rightarrow +\infty$, from Eq.\@ (\ref{eq:Xi_T_tot}) we retrieve the known result \cite{Dupuis-lect-2011}
\begin{multline}\label{eq:Xi_T_colless} 
\lim_{\tau_c \rightarrow +\infty}X_T(\xi)\equiv X_T(s)=\frac{X_T^0(s)}{1-F_1^S\left[\frac{m^2}{3 n m^*} X_T^0(s) \right]} \\ =\frac{n}{m} \frac{3 \mathscr{I}(s) + 1 }{1-F_1^S \mathscr{I}(s)}
\end{multline}
Employing the general relation \cite{Dressel-2001}
\begin{equation}\label{eq:sigma_epsilon}
\sigma_{T}(\vec{q},\omega)=-i \epsilon_0 \omega \left[ \epsilon_T(\vec{q},\omega)-1\right],
\end{equation}
we arrive at the transverse dielectric function: 
\begin{equation}\label{eq:Kubo_T_eps}
\epsilon_{T}(\vec{q},\omega)=1-\frac{ e^2 }{\epsilon_0 \omega^2}\left[ X_{T}^P(\vec{q},\omega) +\frac{n}{m} \right]
\end{equation}
Using Eq.\@ (\ref{eq:Kubo_T_eps}) and (\ref{eq:sigma_T_mb}), we arrive at the explicit form
\begin{equation}\label{eq:epsilon_T_FL}
\epsilon_T(\vec{q},\omega)=1- \frac{(\omega_p)^2}{\omega^2}\frac{\left(1+ \beta  \right) \left[3\mathscr{I}(\xi)+1 \right]}{1-3 \left(\frac{F_1^S}{3}-\beta\right) \mathscr{I}(\xi)+\beta},
\end{equation}
where $\omega_p=\sqrt{n e^2/(m \epsilon_0)}$ is the electron plasma frequency. 

The pole of the dielectric function (\ref{eq:epsilon_T_FL}) corresponds exactly to the transverse sound dispersion relation (\ref{eq:Lea}). Indeed, in general the poles of $\epsilon_T(\vec{q},\omega)$ yield the collective modes existing in the \emph{matter sector} of the theory, i.e.\@, the modes which exist in the material in the absence of external electromagnetic fields. 
In the collisionless limit $\tau_c \rightarrow +\infty$, Eq.\@ (\ref{eq:epsilon_T_FL}) reduces to \cite{Dupuis-lect-2011}
\begin{equation}\label{eq:epsilon_T_FL_colless}
\lim_{\tau_c\rightarrow +\infty}\epsilon_T(\vec{q},\omega)=1- \frac{(\omega_p)^2}{\omega^2} \frac{3 \mathscr{I}(\xi) + 1 }{1-F_1^S \mathscr{I}(s)}
\end{equation}
The self-consistent solutions of Maxwell's equations inside the Fermi liquid give the collective modes of the system in the charged case. These modes are polaritons, whereby electromagnetic radiation couples to  Fermi-surface oscillations. Formally, these collective modes satisfy \cite{Nozieres-1999ql}
\begin{equation}\label{eq:coll_modes_charged}
\frac{q^2 c^2}{\omega^2}=\epsilon_T(\vec{q},\omega). 
\end{equation}
We can obtain analytical solutions by expanding Eqs.\@ (\ref{eq:coll_modes_charged}) and (\ref{eq:epsilon_T_FL}) in various physically important limits, as analyzed in the following sections. 

\subsection{Propagating shear regime}\label{Viscoel_q0}

We first analyze the dielectric function (\ref{eq:epsilon_T_FL}) in the high-frequency, long-wavelength regime $\omega \gg v_F^* \left|q \right|$, i.e.\@, for excitations outside of the electron-hole continuum. 
Then, $s \rightarrow +\infty$ according to Eq.\@ (\ref{eq:Lea_var}). In such limit, the leading-order series expansion of Eq.\@ (\ref{eq:epsilon_T_FL}) yields
\begin{equation}\label{eq:eps_T_series}
\epsilon_T(\vec{q},\omega)=1-\frac{(\omega_p)^2}{\omega^2}\left[1+\frac{1}{5}\left(1+\frac{F_1^S}{3}\right) \frac{\omega \tau_c}{(i + \omega \tau_c )s^2}\right]
\end{equation}
Since $s \rightarrow +\infty$ we can employ the Taylor series $(1+x)^{-1}=1-x+o(x^2)$ : $x\rightarrow 0$ on the term in square brackets in Eq.\@ (\ref{eq:eps_T_series}). Using the definition (\ref{eq:Lea_var2}) for $s$, we obtain
\begin{equation}\label{eq:eps_T_el_VE}
\epsilon_T(\vec{q},\omega)=1-\frac{(\omega_p)^2}{\omega^2+ i \omega \nu(\omega) q^2}
\end{equation}
where $\nu(\omega)$ is the generalized shear modulus (\ref{eq:nu_q0_vel}) of the isotropic Fermi liquid \cite{Conti-1999,Vignale-2005,Bedell-1982}.
Equations \@ (\ref{eq:eps_T_el_VE}) and (\ref{eq:nu_q0_vel}) are the fundamental result of this section. They show that the kinetic equation approach in the regime $\omega \gg v_F^{*} \left|q \right|$ predicts a spatially nonlocal dielectric response, where the effect of spatial nonlocality is encoded into the generalized shear modulus $\nu(\omega)$. The emerging macroscopic phenomenology is analogous to the one of a viscous charged fluid. In fact, the same dielectric function can be obtained from a macroscopic viewpoint, by combining the linearized Navier-Stokes equation for transverse currents -- which includes a frequency-dependent ``viscosity coefficient'' $\nu(\omega)$ -- with Maxwell's equations \cite{FZVM-2014}. However, such phenomenological arguments do not yield an explicit expression for the coefficient $\nu(\omega)$ in terms of microscopic parameters: such an expression requires a microscopic model, like Fermi-liquid theory yielding Eq.\@ (\ref{eq:nu_q0_vel}). 

The dielectric response is influenced by the dependence of $\nu(\omega)$ on frequency: the term $i \omega \nu(\omega)$ in Eq.\@ (\ref{eq:eps_T_el_VE}) crosses over from a viscous liquid/collisional regime $\omega \tau_c \ll 1$ to a reactive/collisionless regime $\omega \tau_c \gg 1$. In the viscous limit $i \nu(\omega) \approx i \nu(0)$ is predominantly imaginary, therefore it is equivalent to the dissipative viscosity coefficient (\ref{eq:nu_0}) \cite{FZVM-2014, Abrikosov-1959}. Such regime is adiabatically connected to the classical fluid formed by quasiparticles at high temperatures. 
\begin{figure}[t]
\includegraphics[width=0.75\columnwidth]{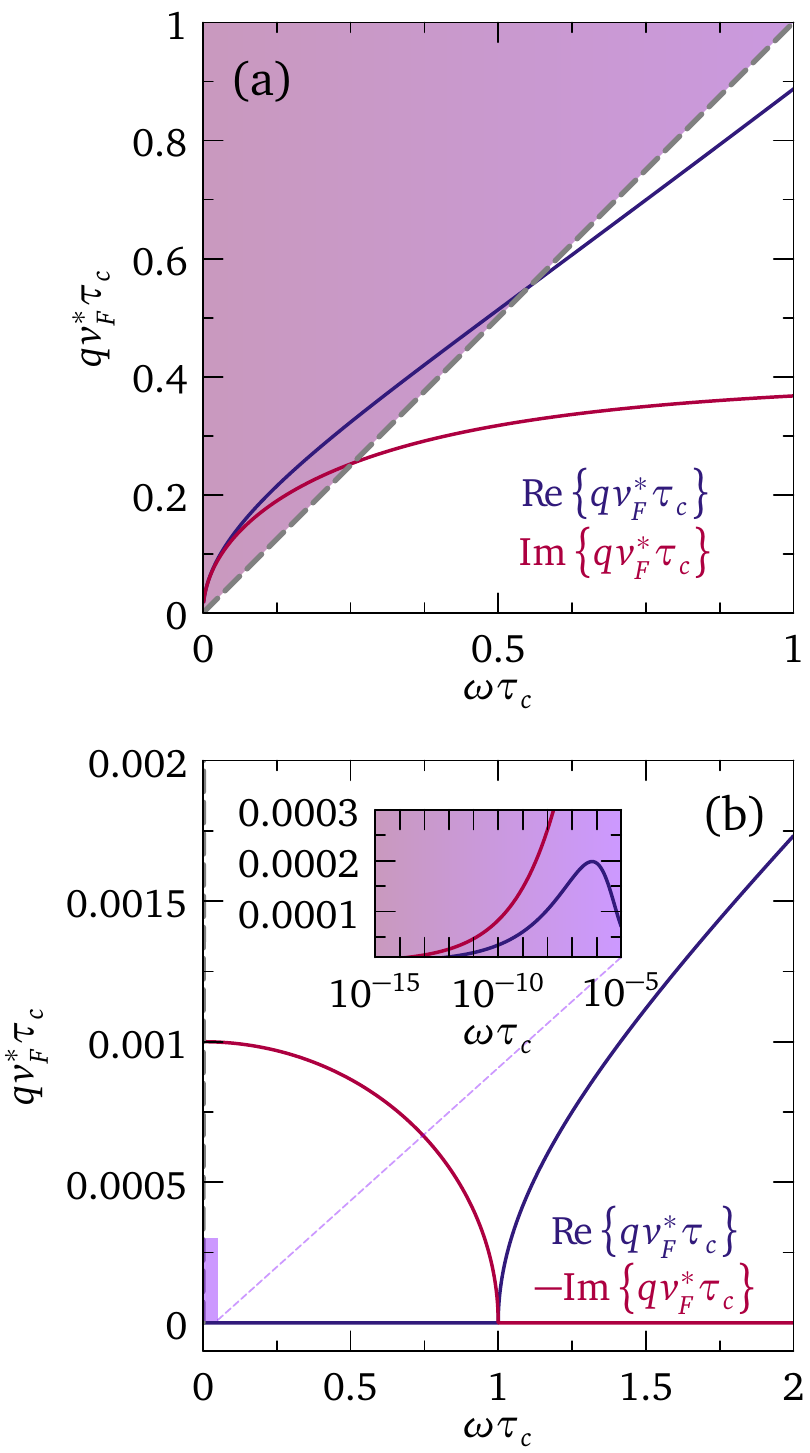}
\caption{\label{fig:VE_modes} Dispersion relation of the two frequency-degenerate polariton branches in propagating shear regime from Eq.\@ (\ref{eq:VE_modes}), for first symmetric Landau parameter $F_1^S=20$, renormalized Fermi velocity $v_F^*=10^{-3} c$, and plasma frequency fixed by $\omega_p \tau_c=1$. (a) Dimensionless wave vector $q v_F^* \tau_c$ of the shear-polariton as a function of dimensionless frequency $\omega \tau_c$. The blue and red curves show the real and imaginary part of the mode wave vector, respectively. The shaded purple area indicates the Lindhard continuum. (b) Dimensionless wave vector $q v_F^* \tau_c$ of the plasmon-polariton as a function of dimensionless frequency $\omega \tau_c$. The blue and red curves show the real and imaginary part of the mode wave vector, respectively. The inset zooms the same data at low frequencies, showing the quartic dispersion (\ref{eq:VE_modes_hydro}) valid for $\omega \rightarrow 0$. 
}
\end{figure}
In the collisionless regime, $i \nu(\omega)$ is essentially real-valued, i.e.\@, reactive, giving rise to a dissipationless nonlocal response. In such conditions the coefficient $\nu(\omega) \approx i \mu_s/\omega$ is related to the dynamic Fermi-liquid shear modulus (\ref{eq:mu_FL}) \cite{Conti-1999,Vignale-2005, Tokatly-2000}. The literature refers to $\nu(\omega)$ and its dissipative/reactive components with different terminologies: ``viscoelastic coefficients'' \cite{Conti-1999,Vignale-2005}, ``generalized hydrodynamics'' \cite{Tokatly-2000}, and ``generalized viscosity'' \cite{FZVM-2014}. Reference \onlinecite{Tokatly-2000} points out an intriguing connection with the phenomenology of highly viscous fluids: these systems respond as elastic solids on short timescales, but they behave as viscous liquids on long timescales. Indeed, the Fermi liquid is much more viscous than strongly interacting electron systems \cite{FZVM-2014}. In this paper $\nu(\omega)$ is labeled ``generalized shear modulus'', to stress that it refers to the shear response and that it has both viscous/dissipative and reactive/dissipationless components, each of which dominates in opposite frequency regimes. 

The polaritons satisfy Eq.\@ (\ref{eq:coll_modes_charged}). Using Eq.\@ (\ref{eq:eps_T_el_VE}), we obtain \emph{two complex-valued solutions}:
\begin{multline}\label{eq:VE_modes}
q^2=\frac{1}{2 c^2 \nu(\omega) }\left\{i c^2 \omega +\nu(\omega)  \omega^2 \right. \\ \left. \pm i \sqrt{\omega } \sqrt{\omega  \left[c^2+i \nu(\omega)  \omega \right]^2-4 i c^2 \nu(\omega)  (\omega_p)^2}\right\}
\end{multline}
Figure \ref{fig:VE_modes}(a) displays the dispersion relation of the polaritons (\ref{eq:VE_modes}). It shows the normalized wave vector $q v_F^{*} \tau_c$ as a function of $\omega \tau_c$ for $F_1^S=20$, with blue and red lines for $\mathrm{Re}\left\{q v_F^{*} \tau_c \right\}$ and $\mathrm{Im}\left\{q v_F^{*} \tau_c \right\}$, respectively. The purple shadow highlights the Lindhard electron-hole continuum. 
One of the two solutions becomes propagating, with an entirely real dispersion, above the plasma frequency $\omega_p$: for this reason, I label this solution as ``plasmon-polariton''. The other root lies much closer to the Lindhard continuum, and it emerges from the continuum at a $F_1^S$-dependent frequency, analogously to the uncharged chase of transverse sound. In the following, I refer to this solution as ``shear-polariton''. 
In the hydrodynamic/collisional regime $\omega \tau_c \ll 1$, where $\nu(\omega) \approx \nu(0)$ gives rise to viscous dissipation, Eq.\@ (\ref{eq:VE_modes}) admits the leading-order expansions for $\omega\rightarrow 0$
\begin{subequations}\label{eq:VE_modes_hydro}
\begin{equation}\label{eq:VE_modes_hydro1}
q=\sqrt{\frac{\omega_p}{c}}\left[\frac{\omega}{\nu(0)}\right]^{\frac{1}{4}} \left[\sin\left(\frac{\pi}{8}\right)-i \cos\left(\frac{\pi}{8}\right)\right]+o\left(\sqrt{\omega}\right),
\end{equation}
\begin{equation}\label{eq:VE_modes_hydro2}
q=\sqrt{\frac{\omega_p}{c}}\left[\frac{\omega}{\nu(0)}\right]^{\frac{1}{4}} \left[\cos\left(\frac{\pi}{8}\right)+i \sin\left(\frac{\pi}{8}\right)\right]+o\left(\sqrt{\omega}\right),
\end{equation}
\end{subequations}
for the plasmon-polariton and the shear-polariton, respectively. Equations (\ref{eq:VE_modes_hydro}) show that the dispersion is quartic, i.e.\@, $\omega \propto q^4$, in the collisional/hydrodynamic limit, as shown by Fig.\@ \ref{fig:VE_modes}(a) and by the inset of Fig.\@ \ref{fig:VE_modes}(b). The result (\ref{eq:VE_modes_hydro2}) for the shear-polariton is in contrast with the uncharged case, where the dispersion is quadratic according to Eq.\@ (\ref{eq:q_hydro}). Such difference is due to the combined effects of electric charge and of Landau damping. 
Firstly, radiation-matter interaction modifies the dispersion of the shear-polariton at momenta $q<\omega_p/c$ with respect to uncharged transverse sound, due to the latter being ``repelled'' by the nearby photon root $\omega=c q$ \cite{Nozieres-1999ql}: the charged shear-polariton, generated from the photon-matter mixing, acquires quadratic dispersion and is pushed down inside the electron-hole continuum -- see also Sec.\@ \ref{emerg_VE}. Such ``quadratic bending'' of the dispersion due to electromagnetic forces is also retrieved for reactive shear stresses in the isotropic version of the Wigner crystal formed by charged bosonic constituents \cite{Beekman-2017}. A similar mechanism is also at play in particular viscoelastic-like holographic models of strange metals \cite{Baggioli-2020}. 
At vanishing frequencies deep inside the continuum, Landau damping further modifies the quadratic shear-polariton dispersion into the quartic result (\ref{eq:VE_modes_hydro2}).

In the opposite collisionless regime $\omega \tau_c \gg 1$, $\nu(\omega)$ is predominantly imaginary, which leads to a dissipationless reactive contribution. Then, Eq.\@ (\ref{eq:VE_modes}) is expanded for $\omega\rightarrow +\infty$ as 
\begin{subequations}\label{eq:VE_modes_el}
\begin{equation}\label{eq:VE_modes_el1}
q=\frac{\omega}{c}+o\left(\omega^{-2}\right),
\end{equation}
\begin{equation}\label{eq:VE_modes_el2}
q=\frac{\omega}{\sqrt{\mu_s}},
\end{equation}
\end{subequations}
for the plasmon-polariton and shear-polariton, respectively, where $\mu_s$ is the Fermi-liquid shear modulus (\ref{eq:mu_FL}). Physically, Eq.\@ (\ref{eq:VE_modes_el1}) is the standard phenomenon whereby the plasmon dispersion asymptotically reaches the uncoupled photon root $\omega=c q$ at very large frequencies $\omega \gg \omega_p$, where radiation does not feel the coupling to fermionic quasiparticles. Furthermore, Eq.\@ (\ref{eq:VE_modes_el2}) is equivalent to the uncharged case (\ref{eq:vs_elast}): at high momenta $q \gg \omega_p/c$, the Fermi-surface oscillation is essentially uncoupled from radiation. Consequently, in this limit we recover the dispersion of uncharged transverse zero sound.

\subsection{Momentum dependence of the generalized shear modulus}\label{q4}

As mentioned in Sec.\@ \ref{FL_epsT_coll}, starting from the limit $\omega /(v_F^* \left|q\right|)\rightarrow+\infty$, an expansion in  $q \rightarrow 0$ of Eq.\@ (\ref{eq:epsilon_T_FL}) at order $q^{2n}, \, n\in \mathbb{N}$ gives $2n$ frequency-degenerate charge collective modes. Such procedure corresponds to a gradient expansion for the coordinate-dependent current density in real space \cite{FZVM-2014}. The first term of such expansion defines the generalized shear modulus $\nu(\omega)$. Therefore, $\epsilon_T(\vec{q},\omega)$ entails an increasing number of polaritons at higher orders $q^n$, $n>2$, up to the limit $\omega/(v_F^{*} \left| q\right|)\rightarrow 1$ where the expansion breaks down. Technically, terms of order higher than $q^2$ may be reformulated in terms of a momentum dependence of the generalized shear modulus. In this picture, $\nu(\omega)$ becomes a scale-dependent quantity in real space, as it occurs in 2D systems \cite{Krishna-Kumar-2017,Ledwith-2019}. Such an analysis for Fermi liquids in three dimensions is left for future work, while this paper focuses on the leading-order $q^2$ correction to the dielectric properties. 

\subsection{Emergence of the shear collective mode from the Lindhard continuum}\label{emerg_VE}

\begin{figure}[t]
\includegraphics[width=0.7\columnwidth]{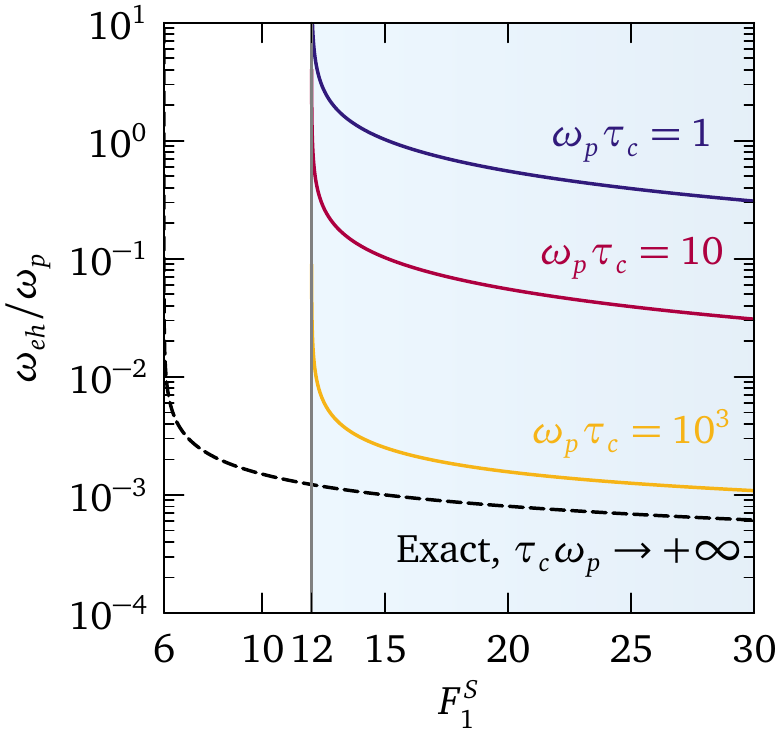}
\caption{\label{fig:eh_bound} Inferior limit on frequency $\omega_{eh}=v_F^{*} \mathrm{Re}\left\{q\right\}$ for the propagation of the shear-polariton above the Lindhard continuum, as a function of first Landau parameter $F_1^S$. The solid blue, red and gold curves show the estimation $\omega_{eh}=\omega_{eh}^{VE}$ from equation (\ref{eq:VE_modes}) valid in the propagating shear regime, for $\omega \tau_c=\left\{1,10,10^3\right\}$. The black dashed curve shows the exact calculation according to Eq.\@ (\ref{cont_bound}) in the collisionless limit $\omega \tau_c \rightarrow +\infty$. The continuum mechanics description (\ref{eq:VE_modes}) of the coupling between radiation and the Fermi liquid in terms of the generalized shear modulus $\nu(\omega)$ breaks down for $F_1^S \leq 12$, although the shear-polariton still propagates for $F_1^S>6$.
}
\end{figure}

As mentioned in Sec.\@ \ref{Viscoel_q0}, the shear-polariton no longer propagates when it submerges itself into the Lindhard electron-hole continuum, which occurs at frequencies $\omega \leq v_F^* \left|q\right|$. Since $\mathrm{Re}\left\{q\right\} \gg \mathrm{Im}\left\{q\right\}$, the threshold frequency can be estimated with $\omega_{eh}=v_F^{*} \mathrm{Re}\left\{q \right\}$. Such constraint can be deduced by the numerical solution of the dispersion (\ref{eq:VE_modes}) together with Eq.\@ (\ref{eq:omega_cont}). Figure \ref{fig:eh_bound} shows the frequency $\omega_{eh}^{VE}$ as a function of $F_1^S$. Within the assumptions of Sec.\@ \ref{Viscoel_q0}, the shear-polariton never exits from the continuum if $F_1^S<12$. 
More generally, we can derive a precise statement on the emergence of the shear-polariton from the continuum, by directly analyzing the uncharged dispersion relation (\ref{eq:Lea}) for transverse sound. In fact, since the upper bound on the renormalized Fermi velocity is $v_F^*/c \approx 0.001$ in standard metals, the condition $\omega=v_F^* \mathrm{Re}\left\{q\right\}$ also implies $\omega \ll c q$. In this regime we can let $c q /\omega \rightarrow +\infty$ at the left-hand side of Eq.\@ (\ref{eq:coll_modes_charged}), which means that the polariton solution reduces to the pole of the dielectric function (\ref{eq:epsilon_T_FL}), i.e.\@, to uncharged transverse sound: the collective mode is the same as in the uncharged system \cite{Nozieres-1999ql}. Furthermore, a microscopic analysis of the Fermi-liquid collision time $\tau_c(\omega, T)$ --- see Sec.\@ \ref{Scatter} --- implies that, for the first Landau parameter $F_1^S>6$, the condition (\ref{eq:omega_cont}) occurs for $\omega \tau_c(\omega, T) \gg 1$. The latter inequality implies that $\xi \approx s$ in Eq.\@ (\ref{eq:Lea}). One verifies that the shear collective mode is underdamped in such regime, i.e.\@ $\mathrm{Re}\left\{ q\right\} \gg \mathrm{Im}\left\{ q\right\}$. Therefore, to a high degree of accuracy, we can assume $s \in \mathbb{R}$ in Eqs.\@ (\ref{eq:Lea}) and (\ref{eq:omega_cont}), which give an analytical solution for $\omega=\omega_{eh}$:
\begin{equation}\label{cont_bound}
\omega_{eh}=\frac{\omega_p}{c}\frac{3 v_F^*}{\sqrt{(F_1^S-6) \left[1-\left(\frac{v_F^*}{c}\right)^2\right]}}.
\end{equation}
Equation (\ref{cont_bound}) confirms the early order-of-magnitude estimate by Nozi\`eres and Pines \cite{Nozieres-1999ql}: the shear-polariton plunges into the continuum at momenta lower than $\mathrm{Re}\left\{q\right\}=\omega/v_F^* \approx \omega_p/ c$. Consistently with Sec.\@ \ref{Exist_sound_T}, the shear-polariton never exits from the continuum if $F_1^S <6$, since Eq.\@ (\ref{cont_bound}) implies $\lim_{F_1^S\rightarrow 6^+} \omega_{eh}=+\infty$. 
The discrepancy between $\omega_{eh}^{VE}$ and $\omega_{eh}$ underlines the limits of the continuum-mechanics picture for shear modes: for $\omega_{eh}<\omega<\omega_{eh}^{VE}$ the shear-polariton still propagates in the system, however this propagation cannot be described within the $q^2$ expansion of the dielectric function performed in Sec.\@ \ref{Viscoel_q0}. As mentioned in section \ref{q4}, when $\omega/(v_F^* \mathrm{Re}\left\{q\right\}) \rightarrow 1^+$, higher-order momentum corrections with respect to the dielectric function (\ref{eq:eps_T_el_VE}) are non-negligible.

\subsection{Anomalous skin effect regime}\label{Anomal}

In the low-frequency and high-momentum regime $\left|q \right| v_F^* \gg \omega$ (quasistatic limit \cite{Nozieres-1999ql}) at finite $\omega \tau_c$, an expansion of the optical conductivity (\ref{eq:sigma_T_mb}) at leading order in $s\rightarrow 0$ gives
\begin{equation}\label{eq:sigma_anom_FL}
\sigma_T(\vec{q},\omega)=\epsilon_0 (\omega_p)^2 \frac{3 \pi}{4} \frac{1}{v_F q}.
\end{equation}
The real-valued conductivity (\ref{eq:sigma_anom_FL}) represents dissipative processes in which radiation transfers energy to electrons moving perpendicularly to $\vec{q}$ -- see Eq.\@ (3.122) in Ref.\@ \onlinecite{Nozieres-1999ql}. 
Using Eq.\@ (\ref{eq:sigma_epsilon}), we obtain the transverse dielectric function:
\begin{equation}\label{eq:epsilon_anom_FL}
\epsilon_T(\vec{q},\omega)=1+i\frac{(\omega_p)^2}{\omega} \frac{3 \pi}{4} \frac{1}{v_F q}. 
\end{equation} 
Corrections of higher order in $s$ with respect to Eq.\@ (\ref{eq:sigma_anom_FL}) depend on $\tau_c$. For instance, at order $s^2$ we have
\begin{equation}\label{eq:sigma_anom_FL_2}
\sigma_T(\vec{q},\omega)=\epsilon_0 (\omega_p)^2 \frac{3 \pi}{4} \frac{1}{v_F q}+\sigma_2(\vec{q},\omega)+o(s^3),
\end{equation}
where
\begin{multline}\label{eq:sigma2_corr_tauc}
\sigma_2(\vec{q},\omega)=\frac{(3 + F_1^S) (3 \pi^2-16)}{16 q^2 \tau_c (v_F)^2} \\ + 
 i \frac{ \omega \left[48 + F_1^S (16 - 3 \pi^2)\right]}{16 q^2 (v_F)^2}. 
\end{multline}
 Another convenient way to consider $\tau_c$-related corrections is to first expand the integral (\ref{eq:vs_int}) as $\mathscr{I}(a s)=-1/3 - 1/4 i \pi a  s+ o\left[ (a s)^2 \right]$. Using $s=\omega/(v_F^{*} q)$ and $a=1+i/(\omega \tau_c)$, we obtain 
\begin{multline}\label{eq:int_qstat}
\mathscr{I}(\xi)=-\frac{1}{3} -\frac{1}{4}\pi i \left(1+\frac{i}{\omega \tau_c} \right)s\\ =-\frac{1}{3} -\frac{1}{4}\pi i \left(\omega+\frac{i}{\tau_c}\right)\frac{1}{v_F^{*} q}.
\end{multline}
The term $i/\tau_c$ in Eq.\@ (\ref{eq:int_qstat}) is important in the regime $\omega \tau_c \ll 1$, i.e.\@ $\omega \rightarrow 0$ at finite $\tau_c$ (quasistatic limit) \cite{Nozieres-1999ql,Khoo-2021_preprint}. 
Inserting Eq.\@ (\ref{eq:int_qstat}) in the susceptibility (\ref{eq:Xi_T_tot}), and expanding the latter at leading order in $\omega\rightarrow 0^+$, we obtain
\begin{multline}\label{eq:XT_qstat}
X_{T}(\vec{q},\omega)=-\frac{n}{m}\frac{9 i \pi \omega \tau_c}{(3 + F_1^S) (-3 \pi + 4 q \tau_c v_F^{*})}+o(\omega^2) \\ = -\frac{n}{m}\frac{i\omega }{\frac{4}{3 \pi} v_F \left(-\frac{3 \pi}{4 v_F^{*} \tau_c} + q \right)}.
\end{multline}
The conductivity associated with Eq.\@ (\ref{eq:XT_qstat}) is finite in the static nonlocal limit. Explicitly, 
\begin{multline}\label{eq:XT_qstat2}
\sigma_T(\vec{q},0^+)=i e^2\frac{X_{T}(\vec{q},\omega)}{\omega} \\ =\frac{1}{4 \pi} \frac{e^2}{\hbar^3}\frac{m^2 (v_F)^3}{v_F q- \frac{3 \pi}{4} \left(1+\frac{F_1^S}{3}\right) (\tau_c)^{-1}}. 
\end{multline}
Equation (\ref{eq:XT_qstat2}) has the structure $\sigma_T(\vec{q},0^+) \sim m^{d-1} v_F^{d}/\left(-C_d/\tau_c+v_F q \right)$, with $d=3$ dimensionality and $C_3=3 \pi\left(1+F_1^S/3\right)/4$; such structure agrees with the leading-order series expansion of Eqs.\@ (22)-(24) in Ref.\@ \onlinecite{Khoo-2021_preprint} for $\tau_K \rightarrow +\infty$ and $v_F q \tau_c \gg 1$, in the 2D case ($d=2$). 
Notice that Eq.\@ (\ref{eq:XT_qstat2}) correctly reduces to Eq.\@ (\ref{eq:sigma_anom_FL}) when $q \gg (v_F^{*} \tau_c)^{-1}$. 

The collective modes of the charged system result from Eqs.\@ (\ref{eq:coll_modes_charged}) and (\ref{eq:epsilon_anom_FL}), giving
\begin{equation}\label{eq:coll_anom}
\frac{q^2 c^2}{\omega^2}=1+i\frac{ (\omega_p)^2}{\omega^2} \frac{3 \pi}{4} \frac{\omega}{v_F q}.
\end{equation}
Equation (\ref{eq:coll_anom}) has three polariton roots for $q$ in the complex plane. However, for $\omega/(v_F^{*} \left|q \right|) \rightarrow 0$ we have
\begin{equation}\label{eq:coll_anom2}
\left[1+i\frac{ (\omega_p)^2}{\omega^2} \frac{3 \pi}{4} \frac{\omega}{v_F q}\right]^{-1}=\frac{\omega^2}{q^2 (v_F^{*})^2}\frac{(v_F^{*})^2}{c^2}\approx 0.
\end{equation}
Employing the series expansion $\left(1+x\right)^{-1}=1-x+o(x^2)$ in Eq.\@ (\ref{eq:coll_anom2}) with $x=i\frac{ (\omega_p)^2}{\omega^2} \frac{3 \pi}{4} \frac{\omega}{v_F q} \rightarrow 0$, we obtain
\begin{equation}\label{eq:coll_anom3}
q=i \frac{3 \pi}{4} \frac{(\omega_p)^2}{\omega} \frac{1}{v_F},
\end{equation}
which is an entirely imaginary dispersion. Such dispersion again signals a dissipative coupling between radiation and Fermi-surface quasiparticles. Physically, this happens because photons do not excite shear modes but only incoherent electron-hole pairs in the Lindhard continuum for $\omega \ll v_F^{*} \left|q \right|$.

Notice that the real-valued optical conductivity (\ref{eq:sigma_anom_FL}) is independent of interactions and of frequency, but it depends only on momentum in the quasistatic limit \cite{Nozieres-1999ql,Dressel-2001}: it is exactly the momentum-dependent optical conductivity used in the theory of anomalous skin effect \cite{Dressel-2001,Sondheimer-2001}. Hence, in Sec.\@ \ref{Skin_anom} we will employ the formalism of anomalous skin effect to analyze the optical properties of charged Fermi liquids in the quasistatic limit. 

\subsection{Hydrodynamic regime}\label{Hydro}

When $\tau_c$ is the smallest timescale in the system, local thermodynamic equilibrium among quasiparticles allows for hydrodynamic flow \cite{Narozhny-2017,Narozhny-2019}. This hydrodynamic regime is realized by the dielectric function (\ref{eq:epsilon_T_FL}) when frequency and momentum both tend to zero, so that we can expand $\epsilon_T(\vec{q},\omega)$ at fixed finite $s$ and $\omega \tau_c \rightarrow 0$. 
For a clean Fermi liquid characterized by the collision rate $(\tau_c)^{-1} \propto (\hbar \omega)^2 +( \pi k_B T)^2$ -- see Sec.\@ \ref{Scatter} and Appendix \ref{app:FL_scat} -- $\omega \tau_c \rightarrow 0$ in two cases: $\omega \rightarrow +\infty$ at any temperature and $\omega \rightarrow 0$ at $T>0$. Therefore,  ``viscous'' hydrodynamic behavior develops at vanishing frequencies only in the finite-temperature case. Such dichotomy reflects that thermal effects at $T>0$ disrupt the $T=0$ Fermi-liquid order for $\hbar \omega \lesssim k_B T$ (thermal regime) \cite{Berthod-2013}. 

The leading-order expansion of the dielectric function (\ref{eq:epsilon_T_FL}) in $\omega \tau_c \rightarrow 0$ at fixed $s$ produces
\begin{equation}\label{eq:eps_T_hydro}
\epsilon_T(\vec{q},\omega)=1-\frac{(\omega_p)^2}{\omega^2}\left[1-i \frac{1}{5}\left(1+\frac{F_1^S}{3}\right) \frac{\omega \tau_c}{s^2}\right].
\end{equation}
As done in Sec.\@ \ref{Viscoel_q0}, we can resum Eq.\@ (\ref{eq:eps_T_hydro}) with $(1+x)^{-1}=1-x+o(x^2)$ for $x=\omega \tau_c/s^2\rightarrow 0$, so that
\begin{equation}\label{eq:eps_T_el_hydro}
\epsilon_T(\vec{q},\omega)=1-\frac{(\omega_p)^2}{\omega^2+ i \omega \nu(0) q^2}.
\end{equation}
The dielectric function (\ref{eq:eps_T_el_hydro}) describes the electromagnetic response of viscous charged fluids \cite{FZVM-2014}, written in terms of the static Fermi-liquid shear viscosity $\nu(0)$ from Eq.\@ (\ref{eq:nu_0}) \cite{Conti-1999,Vignale-2005, Abrikosov-1959}. It also corresponds to the $\omega \tau_c \rightarrow 0$ limit of Eq.\@ (\ref{eq:eps_T_el_VE}), consistently. There are still two polariton branches in hydrodynamic regime, which follow the dispersion (\ref{eq:VE_modes}) upon substituting $\nu(\omega)$ with $\nu(0)$. In particular, the shear-polariton is critically damped in hydrodynamic regime, i.e.\@, its dispersion has equal real and imaginary parts -- see Eq.\@ (\ref{eq:VE_modes_hydro2}) in Sec.\@ \ref{Viscoel_q0}.
\begin{figure}[ht]
\includegraphics[width=0.9\columnwidth]{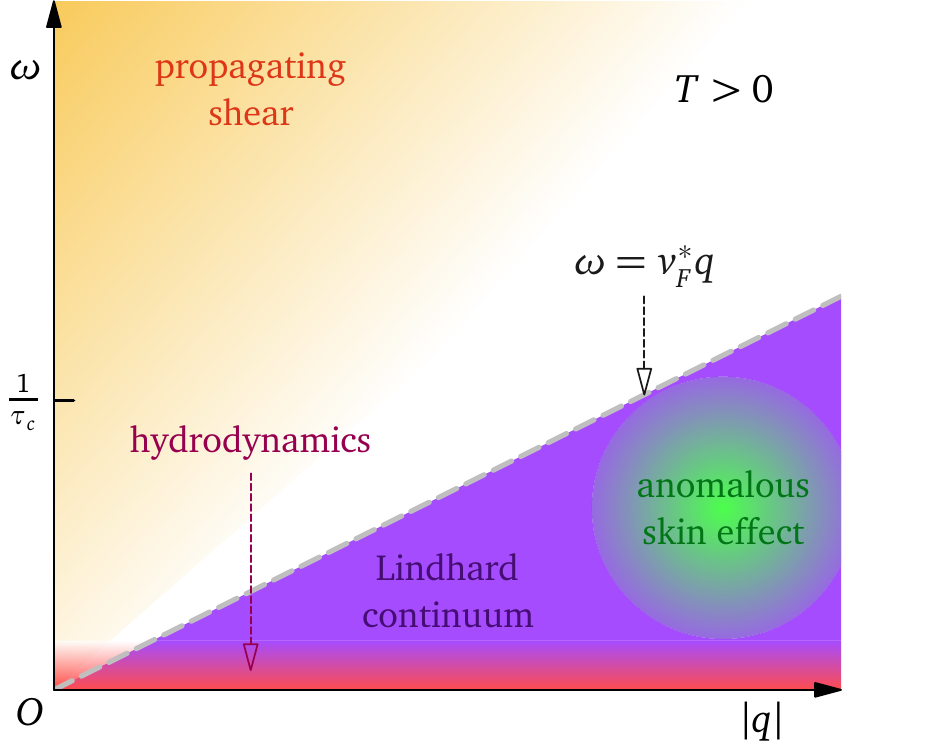}
\caption{\label{fig:regimes}
Sketch of the different regimes for the dielectric response of charged Fermi liquids at $T>0$ in the $\left(\left|q\right|, \omega \right)$ plane: the condition $\omega \gg v_F^{*} \left|q\right|$ corresponds to the propagating shear regime (Sec.\@ \ref{Viscoel_q0}); at both vanishing momentum and frequency and for $\omega \tau_c \rightarrow 0$ the response becomes hydrodynamic (Sec.\@ \ref{Hydro}); in the high-momentum, low-frequency regime $\omega \ll v_F^{*} \left|q\right|$ the Lindhard continuum dominates the response and produces anomalous skin effect (Sec.\@ \ref{Anomal}). At $T=0$ the hydrodynamic regime is absent; see discussion in Sec.\@ \ref{Hydro}.
}
\end{figure}
Figure \ref{fig:regimes} summarizes the different regimes for the Fermi-liquid dielectric response analyzed so far. In the next sections we analyze how such regimes are affected by momentum relaxation. 

\section{Optical conductivity with momentum relaxation}\label{VE_tauk}

In crystalline solids, the presence of the ionic lattice, of defects and impurities invariably breaks Galilean invariance, so that quasiparticle momentum is not a conserved quantity \footnote{For electrons interacting with an ideal ionic Bravais lattice, momentum is conserved only up to a reciprocal lattice vector \cite{Ashcroft-1976}: although the \emph{global} momentum of electrons and the lattice is conserved, the two individual momentum components for electrons and lattice vibrations are not \cite{Narozhny-2019}}. This induces momentum relaxation in the electron ensemble, which is essential in preventing the divergence of conductivities in metals and implies a finite mean-free path 
\begin{equation}\label{eq:l_MFP}
l_{MFP}=v_F \tau_K, 
\end{equation}
renormalized by interactions as $l_{MFP}^*=v_F^* \tau_{K}$. 
The interplay of momentum-conserving collisions and momentum-relaxing scattering depends on the specific lattice symmetry and quasiparticle dispersion. Its complexity has been carefully analyzed in 2D strongly interacting systems \cite{Pal-2012,Briskot-2015,Narozhny-2019,Maslov-2017}. While such detailed analysis is beyond the scope of this paper, a discussion is nevertheless in order, to qualitatively assess whether shear stress propagation can be observed in solid-state Fermi liquids for the 3D isotropic case here analyzed. 

To include damping of transverse currents due to momentum relaxation, we modify the kinetic equation (\ref{eq:kin_k_int_compl_2}) into 
\begin{multline}\label{eq:kin_k_int_compl_2r}
\left( \cos{\theta}  -s \right) \epsilon^S(\theta) e^{i \phi} \\ + \cos{\theta} \left[ \int_0^{\pi} \frac{ \sin{\theta'} d\theta'}{4 \pi} F_1^S  \sin \theta \sin{\theta'} \pi e^{i \phi}  \epsilon^S(\theta') \right. \\  \left. +\frac{e}{m v_F^*} \vec{k} \cdot \vec{A}(\vec{q},\omega) \right]  = \\ \frac{1}{q v_F^*}{\mathscr{I}}_{coll}(\vec{q},\omega)+\frac{1}{q v_F^*}{\mathscr{I}}_{r}(\vec{q},\omega)
\end{multline}
where ${\mathscr{I}}_{r}(\vec{q},\omega)$ describes momentum-relaxing scattering. We now follow Ref.\@ \onlinecite{Conti-1999} to relate the solution of Eq.\@ (\ref{eq:kin_k_int_compl_2r}) to the one without momentum relaxation: using a single-relaxation time approximation, we write  
\begin{equation}\label{eq:relax_int}
\frac{1}{q v_F^{*}}{\mathscr{I}}_{r}(\vec{q},\omega)=-\frac{\epsilon^S(\theta)-\epsilon^r(\theta)}{i \omega \tau_{K}}s e^{i \phi},
\end{equation}
so that momentum damping attempts to restore a ``locally relaxed'' equilibrium distribution function characterized by the displacement $\epsilon^r(\theta)$ \cite{Conti-1999}. Such displacement would be in equilibrium in the presence of collisions and of a vector potential $\vec{A}_r(\vec{q},\omega)$, but in the absence of relaxation. The scattering processes described by the relaxation time $\tau_{K}$ conserve particle number but not current. In this case, the transverse susceptibility with relaxation and collisions may be written in terms of the relaxationless susceptibility as outlined in Appendix \ref{XT_relax} \cite{Conti-1999}.
Explicitly
\begin{equation}\label{eq:Xi_T_r}
X_{T}^r(\vec{q},\omega)=\frac{n}{m} \frac{\omega}{\omega+i/\tau_K} \frac{\left(1+\tilde{\beta}\right)\left[1+3\mathscr{I}(\zeta)\right]}{1-3\left(\frac{F_1^S}{3}-\tilde{\beta}\right)\mathscr{I}(\zeta)+\tilde{\beta}},
\end{equation}
where
\begin{equation}\label{eq:zeta_var}
\zeta=\xi+s \frac{i}{\omega \tau_{K}} =s\left(1+\frac{i}{\omega \tau_m}\right),
\end{equation}
\begin{equation}\label{eq:tau_paral}
\tau_m=\left[\frac{1}{\tau_c}+\frac{1}{\tau_{K}}\right]^{-1},
\end{equation}
and $\tilde{\beta}=\beta \xi/\zeta$\footnote{I am grateful to F.\@ Pientka for pointing out the difference between the factors $\tilde{\beta}$ and $\beta$, with and without relaxation respectively.}. 
Physically, Eq.\@ (\ref{eq:zeta_var}) tells us that the smallest timescale in the system, related to either momentum relaxation or collisions, dominates the nonlocal part of the electrodynamic response.
Once inserted into the Kubo formula (\ref{eq:sigma_T_mb}), the susceptibility (\ref{eq:Xi_T_r}) yields the optical conductivity
\begin{multline}\label{eq:sigma_T_mb_r}
\sigma_T(\vec{q},\omega)=\frac{i e^2 }{\omega}X_{T}^r(\vec{q},\omega)=\\= \frac{i n e^2}{m\left(\omega+i/\tau_K\right)} \frac{\left(1+\tilde{\beta}\right)\left[1+3\mathscr{I}(\zeta)\right]}{1-3\left(\frac{F_1^S}{3}-\tilde{\beta}\right)\mathscr{I}(\zeta)+\tilde{\beta}}
\end{multline}
The dielectric function derived from Eqs.\@ (\ref{eq:sigma_T_mb_r}) and (\ref{eq:sigma_epsilon}) is
\begin{equation}\label{eq:epsilon_T_FL_r}
\epsilon_T(\vec{q},\omega)=1-\frac{(\omega_p)^2}{\omega\left(\omega+i/\tau_K\right)} \frac{\left(1+\tilde{\beta}\right)\left[1+3\mathscr{I}(\zeta)\right]}{1-3\left(\frac{F_1^S}{3}-\tilde{\beta}\right)\mathscr{I}(\zeta)+\tilde{\beta}}. 
\end{equation} 
Equations (\ref{eq:sigma_T_mb_r}) and (\ref{eq:epsilon_T_FL_r}) are the fundamental results of this section. In the following, we will specialize these results to all regimes previously analyzed in Sec.\@ \ref{FL_epsT_coll}, to assess the impact of momentum relaxation on the optical response.

\subsection{Propagating shear regime}\label{viscoel_q0_rel}

Momentum relaxation alters the results (\ref{eq:eps_T_el_VE}) for the dielectric function. In the regime $\omega \gg v_F^* \left|q\right|$, we perform an expansion of Eq.\@ (\ref{eq:epsilon_T_FL_r}) in $s \rightarrow +\infty$ analogously to Sec.\@ \ref{Viscoel_q0}. 
The leading-order expansion at finite $\omega \tau_{K}$ and $\omega \tau_c$ gives
\begin{multline}\label{eq:eps_T_series_rel_gen1}
\epsilon_T(\vec{q},\omega)=1-\frac{(\omega_p)^2}{\omega\left(\omega+i/\tau_K\right)} \times \\ \left[ 1-i \frac{1+F_1^S/3}{5 (i +\omega \tau_K)} \frac{(v_F^{*})^2 q^2 \tau_c (\tau_K)^2}{ \tau_c + \tau_K -i\omega \tau_c \tau_K} \right] +o(s^{-3}). 
\end{multline}
Resumming the $q^2$-dependent term in Eq.\@ (\ref{eq:eps_T_series_rel_gen1}) using $(1+x)^{-1}=1-x+o(x^2)$, we obtain
\begin{equation}\label{eq:eps_T_series_rel_gen1bis}
\epsilon_T(\vec{q},\omega) =1-\frac{(\omega_p)^2}{\omega^2+i \omega/\tau_K +i \omega \tilde{\nu}(\omega) q^2},
\end{equation}
with the generalized shear modulus $\tilde{\nu}(\omega)$ modified by momentum relaxation according to Eq.\@ (\ref{eq:nu_tilde}). 
The optical conductivity associated with the dielectric function (\ref{eq:eps_T_series_rel_gen1bis}) is 
\begin{equation}\label{eq:sigma_T_series_rel_gen1bis}
\sigma_T(\vec{q},\omega)=\frac{i n e^2/m}{\omega+i/\tau_{K} +i \tilde{\nu}(\omega) q^2}.
\end{equation}
Taking the limit $\zeta \rightarrow +\infty$, i.e.\@, $s \rightarrow +\infty$ and $q v_F^{*} \tau_m \rightarrow 0$, still produces Eq.\@ (\ref{eq:eps_T_series_rel_gen1}).
The limit $s \rightarrow +\infty$ with $\tau_c \ll \tau_{K} $ yields
\begin{equation}\label{eq:eps_T_series_rel_gen}
\epsilon_T(\vec{q},\omega)=1-\frac{(\omega_p)^2}{\omega \left(\omega+\frac{i}{\tau_{K}}\right)} \left[1+\frac{1}{5}\left(1+\frac{F_1^S}{3}\right) \frac{(v_F^{*})^2 \tau_c q^2}{\omega(i + \omega \tau_c )}\right]. 
\end{equation}
Resumming the $q$-dependent term in Eq.\@ (\ref{eq:eps_T_series_rel_gen}) at finite $\omega \tau_c$, as done for Eq.\@ (\ref{eq:eps_T_series_rel_gen1}), we achieve
\begin{equation}\label{eq:eps_T_el_VE_rel_m}
\epsilon_T(\vec{q},\omega)=1-\frac{(\omega_p)^2}{\omega^2+i \frac{\omega}{ \tau_{K} } + i \left( \omega+\frac{i}{\tau_{K}}\right) \nu(\omega) q^2},
\end{equation}
where $\nu(\omega)$ is the relaxationless Fermi-liquid generalized shear modulus (\ref{eq:nu_q0_vel}). Equation (\ref{eq:eps_T_el_VE_rel_m}) is the generalization of the dielectric function (\ref{eq:eps_T_el_VE}) to finite (weak) momentum relaxation for $ \tau_{K} \gg \tau_c$. 
Moreover, the term $i/\tau_{K} \nu(\omega) \propto \tau_c/\tau_{K} \rightarrow 0$ for $\tau_c \ll \tau_{K}$, so that
\begin{equation}\label{eq:eps_T_el_VE_rel}
\epsilon_T(\vec{q},\omega)\approx 1-\frac{(\omega_p)^2}{\omega^2+i \frac{\omega}{ \tau_{K} } + i \omega \nu(\omega) q^2},
\end{equation}
which is exactly the dielectric function stemming from the macroscopic phenomenology of viscous charged fluids \cite{FZVM-2014} in the presence of the momentum-relaxing term $i \omega / \tau_{K}$. Hence, the Fermi liquid responds akin to a viscous charged fluid for excitation energy far above the electron-hole continuum, and for a momentum relaxation rate lower than both collision rate and radiation frequency. 
The optical conductivity stemming from Eq.\@ (\ref{eq:eps_T_el_VE_rel}) is 
\begin{equation}\label{eq:sigma_T_el_VE_rel}
\sigma_T(\vec{q},\omega)=i \epsilon_0 \frac{(\omega_p)^2}{\omega+ i\left[ (\tau_{K})^{-1}+ \nu(\omega) q^2 \right]}.
\end{equation} 
Conversely, if $s \rightarrow +\infty$ and $\tau_{K} \ll \tau_c$ Eq.\@ (\ref{eq:eps_T_series_rel_gen1}) becomes 
\begin{multline}\label{eq:eps_T_series_rel_genk}
\epsilon_T(\vec{q},\omega)=1-\frac{(\omega_p)^2}{\omega \left(\omega+\frac{i}{\tau_{K}}\right)} \\ \times \left[1-\frac{1}{5}\left(1+\frac{F_1^S}{3}\right) (q v_F^{*} \tau_K)^2\right]. 
\end{multline}
At leading order in $q v_F^{*} \tau_K \rightarrow 0$, the $q$-dependent nonlocal term is negligible and we retrieve the Drude/Ohmic dielectric function \cite{Dressel-2001}
\begin{equation}\label{eq:eps_T_Drude}
\epsilon_T(\omega)=1-\frac{(\omega_p)^2}{\omega^2+i \omega / \tau_{K}}
\end{equation}
for $\left|\tilde{\nu}(\omega) q^2 \right| \ll \left|\omega +i/\tau_{K}\right|$. The competition between $\nu(\omega)$ and momentum relaxation also determines the DC conductivity as discussed in Sec.\@ \ref{FL_DC}. We defer a more detailed discussion on microscopic scattering rates in Fermi liquids to Sec.\@ \ref{Scatter}, while here we just comment on the qualitative impact of weak momentum relaxation on polaritons in the regime $\tau_{K} \gg \tau_c$. 

With momentum relaxation, two collective mode branches stem from Eqs.\@ (\ref{eq:coll_modes_charged}) and (\ref{eq:eps_T_el_VE_rel_m}): they still correspond to the plasmon-polariton and shear-polariton analyzed in Sec.\@ \ref{viscoel_q0_rel}, however their dispersion is affected by $\tau_{K}$. Formally, we have 
\begin{widetext}
\begin{equation}\label{eq:VE_modes_rel}
q^2=\frac{\omega^2 }{2 c^2}  +\frac{i \omega}{2 \nu(\omega)} \left\{1  \pm \frac{\sqrt{\tau_{K} \left\{\omega  \left[c^2+ i \nu(\omega) \omega \right]^2-4 i c^2 \nu(\omega) (\omega_p)^2\right\}+i \left[c^2+i \nu(\omega) \omega \right]^2}}{c^2 \sqrt{\omega  \tau_{K}+i}}\right\},
\end{equation}
\end{widetext}
which reduces to Eq.\@ (\ref{eq:VE_modes}) for $\tau_{K}\rightarrow +\infty$. Furthermore, for any ratio $\tau_c/\tau_{K}$ one can similarly solve Eqs.\@ (\ref{eq:coll_modes_charged}) and (\ref{eq:eps_T_series_rel_gen1bis}) for $q^2$, to find an expression which depends on $\tilde{\nu}(\omega)$ given by Eq.\@ (\ref{eq:nu_tilde}). 
The collective mode dispersion derived from Eqs.\@ (\ref{eq:coll_modes_charged}) and (\ref{eq:eps_T_el_VE_rel}) is formally identical to the one resulting from the semiclassical phenomenology of viscous charged fluids, stemming from the combination of Maxwell's and Navier-Stokes equations \cite{FZVM-2014}. The Fermi-liquid theory here developed allows us to assign the microscopic value (\ref{eq:nu_q0_vel}) to $\nu(\omega)$, which remains otherwise undetermined by the macroscopic phenomenology \footnote{In Ref.\@ \onlinecite{FZVM-2014} an expression for $\nu(\omega)$ as a function of frequency $\omega$ and collision time $\tau_c$ was obtained through an analytical fit of the numerical dispersion for transverse sound in a neutral Fermi liquid, i.e.\@ Eq.\@ (\ref{eq:Lea}). The fitted functional behavior of $\nu(\omega)$ qualitatively agrees with the generalized shear modulus (\ref{eq:nu_q0_vel}), which is derived directly from the Fermi-liquid kinetic equation in this paper.}.
Moreover, the expression for $q^2$ from Eqs.\@ (\ref{eq:coll_modes_charged}) and (\ref{eq:eps_T_el_VE_rel}) reduces to Eq. (3.43) in ref.\@ \onlinecite{Levchenko-2020} if one neglects displacement currents (considering only conduction currents) in Maxwell's equations --- see Eq.\@ (\ref{eq:wave_mat_q}) and associated discussion --- and if one approximates the generalized shear modulus (\ref{eq:nu_q0_vel}) by its zero-frequency value (\ref{eq:nu_0}), i.e.\@ the static viscosity $\nu(0)$. 
\begin{figure}[t]
\includegraphics[width=0.75\columnwidth]{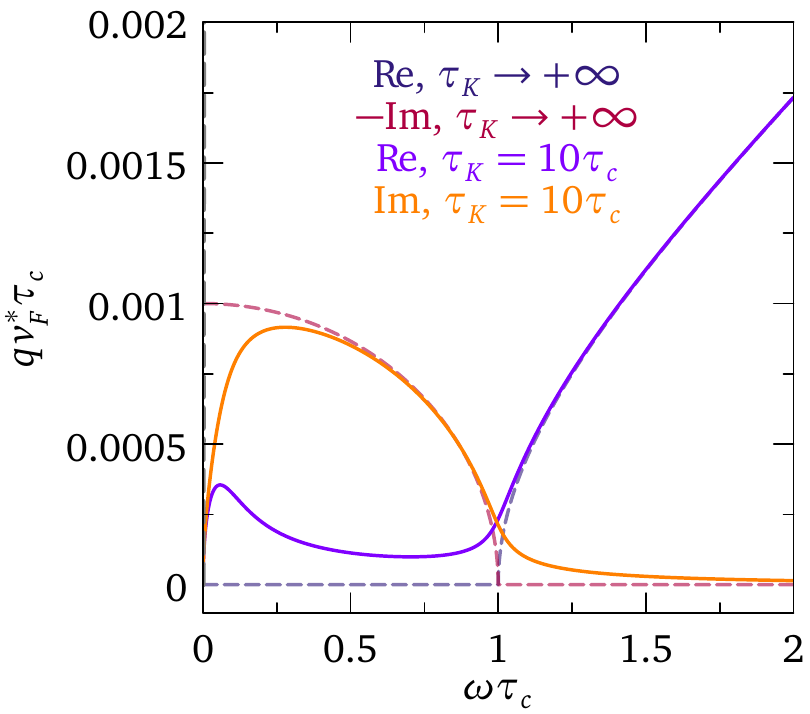}
\caption{\label{fig:VE_mode_tauk} Dispersion relation of the plasmon-polariton in the presence of momentum relaxation, from Eq.\@ (\ref{eq:VE_modes_rel}), for first symmetric Landau parameter $F_1^S=20$, renormalized Fermi velocity $v_F^*=10^{-3} c$, and plasma frequency fixed by $\omega_p \tau_c=1$. The purple and orange curves show the real and imaginary part of the mode wave vector, respectively. The dashed blue and red curves show the corresponding relaxation-free calculations in Fig.\@ \ref{fig:VE_modes}(b) for comparison. 
}
\end{figure}

Remarkably, the dispersion of the shear-polariton is robust against momentum relaxation, as it is negligibly affected by the value of $\tau_{K}$: it still looks as in Fig.\@ \ref{fig:VE_modes}(a), and it emerges from the continuum as in the relaxationless case described in Sec.\@ \ref{emerg_VE}. Significant differences with respect to the relaxationless case emerge only for $\tau_{K}\ll \tau_c$, where the calculation rapidly converges to the Drude/Ohmic limit in accordance with Eq.\@ (\ref{eq:eps_T_series_rel_genk}). On the other hand, $\tau_{K}$ significantly influences the plasmon-polariton dispersion, as illustrated in Fig.\@ \ref{fig:VE_mode_tauk}. Therefore, the shear- and plasmon-polariton dispersions are mainly governed by quasiparticle collisions and momentum relaxation, respectively. The character of the two modes is swapped at the bifurcation point $\omega \tau_{K}>1$ \cite{FZVM-2014}.

Two distinct refractive indexes $n_1(\omega)$ and $n_2(\omega)$ for radiation correspond to the two polaritons in the propagating shear regime, as found from the relation between the polariton dispersion (\ref{eq:VE_modes_rel}) and the definition of the refractive index $q=\omega n(\omega)/c$ \cite{Jackson-1962,Dressel-2001}. In the same way, the two refractive indexes satisfy
\begin{equation}\label{eq:n_charged}
\left[n_i(\omega)\right]^2=\epsilon_T\left[\frac{\omega}{c}n_i(\omega),\omega\right], \, i=\left\{1, \, 2\right\}
\end{equation}
where we use Eqs.\@ (\ref{eq:eps_T_series_rel_gen1bis}) or (\ref{eq:eps_T_el_VE}) with or without momentum relaxation respectively. Given the analogy of Eq.\@ (\ref{eq:n_charged}) with the formally equivalent result for viscous charged fluids \cite{FZVM-2014}, in the propagating shear regime we can calculate the optical properties of the charged Fermi liquid by utilizing the results of Ref.\@ \onlinecite{FZVM-2014}. Such task is performed in Sec.\@ \ref{Opt}, while in the next section we discuss the low-frequency, high-momentum regime. 

\subsection{Anomalous skin effect regime}\label{Anomal_tauk}

The results of Sec.\@ (\ref{Anomal}) for $v_F^* \left|q\right| \gg \omega$ are also sensitive to momentum relaxation. Expanding Eq.\@ (\ref{eq:sigma_T_mb_r}) at leading order in $s\rightarrow 0^+$ at finite $\omega \tau_{K}$ and $\omega \tau_c$, we achieve
\begin{equation}\label{eq:sigma_anom_FL_r1}
\sigma_T(\vec{q},\omega)=\frac{n e^2}{m}\left[\frac{3 \pi}{4} \frac{1}{v_F q}+ \sigma_2(\vec{q},\omega) \right] +o(s^3),
\end{equation}
where the second-order correction $\sigma_2(\vec{q},\omega)$ results
\begin{multline}\label{eq:sigma_anom_FL_corr}
\sigma_2(\vec{q},\omega)=\frac{(3+F_1^S)(3 \pi^2-16)}{16 \tau_c (v_F q)^2} \\ +\frac{\left[3F_1^S \pi^2-16(3+F_1^S)\right](1-i \omega \tau_K)}{16 \tau_K (v_F q)^2}.
\end{multline}
Hence, the leading-order term for $s\rightarrow 0$ depends on momentum but not on frequency or collision/relaxation time, and it still corresponds to the result (\ref{eq:sigma_anom_FL}), consistently with the theory of anomalous skin effect \cite{Nozieres-1999ql,Reuter-1948,Sondheimer-2001}. However, corrections due to $\tau_c$ and $\tau_K$ appear at order $s^2$, as shown by Eq.\@ (\ref{eq:sigma_anom_FL_corr}). 

Equations (\ref{eq:sigma_anom_FL_r1}) and (\ref{eq:sigma_epsilon}) lead to the dielectric function 
\begin{equation}\label{eq:epsilon_anom_FL_r}
\epsilon_T(\vec{q},\omega)= 1+i \frac{(\omega_p)^2}{\omega}\left[\frac{3 \pi}{4} \frac{1}{v_F q}+ \sigma_2(\vec{q},\omega) \right] +o(s^3),
\end{equation} 
which generates the dispersion of collective modes through Eq.\@ (\ref{eq:coll_modes_charged}). 
According to the microscopic models for scattering rates in Sec.\@ \ref{Scatter}, $\tau_{K} \gg \tau_c$ in Fermi liquids at low temperatures. In this case, Eq.\@ (\ref{eq:sigma_anom_FL_r1}) reduces to Eq.\@ (\ref{eq:sigma_anom_FL_2}). 
Conversely, in the high-temperature regime $\tau_c \gg \tau_{K}$, so that Eq.\@ (\ref{eq:sigma_anom_FL_corr}) becomes
\begin{equation}\label{eq:sigma_anom_FL_ter}
\sigma_2(\vec{q},\omega)=\left(\frac{1}{\tau_K}-i \omega\right)\frac{3 F_1^S \pi^2-16 \left(3+F_1^S\right)}{16 (v_F q)^2}. 
\end{equation} 
Moreover, in the quasistatic limit we can consider corrections to the leading-order result (\ref{eq:sigma_anom_FL_r1}) by expanding the integral (\ref{eq:vs_int}) as in the relaxationless case (\ref{eq:int_qstat}), but now substituting $\omega + i/\tau_c \mapsto \omega+ i/\tau_c+i/\tau_K$, and applying such expansion to the total susceptibility (\ref{eq:Xi_T_r}) with relaxation. Proceeding in the same way as in Eqs.\@ (\ref{eq:XT_qstat}) and (\ref{eq:XT_qstat2}), we arrive at the quasi-static conductivity: 
\begin{subequations}
\begin{equation}\label{eq:XT_qstat_rel}
\sigma_T(\vec{q},0^+) =\frac{e^2}{4 \pi \hbar^3}\frac{m^2 (v_F)^3}{Q(q)}, 
\end{equation}
\begin{equation}
Q(q)=v_F q-\frac{3 \pi}{4 \tau_c}-\frac{\pi}{4}F_1^S\left(\frac{1}{\tau_c}+\frac{1}{\tau_K}\right). 
\end{equation}
\end{subequations}
The conductivity (\ref{eq:XT_qstat_rel}), similarly to its relaxationless limit (\ref{eq:XT_qstat2}), has a structure analogous to the leading-order series expansion of Eqs.\@ (22)-(24) in Ref.\@ \onlinecite{Khoo-2021_preprint} for $v_F q \tau_m \gg 1$ and $(\tau_m)^{-1}=(\tau_c)^{-1}+(\tau_K)^{-1}$. 

The dielectric function (\ref{eq:epsilon_anom_FL_r}) at first order in $s \rightarrow 0$ entails three polariton branches, as in the relaxationless case of Sec.\@ \ref{Anomal}. Such three polaritons correspond to three frequency-degenerate refractive indexes for radiation. However, due to the connection of the conductivity (\ref{eq:sigma_anom_FL_r1}) with anomalous skin effect, it is convenient to directly deduce the refractive index from the surface impedance in anomalous skin effect regime, as we will do in Sec.\@ \ref{t_anomal}; see also the discussion in Sec.\@ \ref{ABC}.

\section{DC conductivity and transport}\label{FL_DC}

The zero-frequency limit of the Fermi-liquid optical conductivity (\ref{eq:sigma_T_mb_r}) gives the momentum-dependent DC conductivity $\sigma_{DC}(\vec{q})=\lim_{\omega \rightarrow 0} \sigma_T(\vec{q},\omega)$, measured in transport experiments \cite{Gurzhi-1968}. As discussed in Secs.\@ \ref{VE_tauk}-\ref{Anomal_tauk}, the result is different depending on the ratio $\omega/q=v_F^{*} s$ along which the zero-frequency limit is taken. 
Taking the latter at finite momentum $q$ and finite $\omega \tau_c$ implies $s\rightarrow 0$, therefore we can employ the results of Sec.\@ \ref{Anomal_tauk}: at linear order in $s$, we obtain the frequency-independent conductivity (\ref{eq:sigma_anom_FL}), linked to anomalous skin effect, while for both finite $\tau_c$ and $\tau_{K}$ we have the corrections $\sigma_2(\vec{q},\omega)$ at order $s^2$, in accordance with Eq.\@ (\ref{eq:sigma_anom_FL_corr}). As discussed in Sec.\@ \ref{Hydro}, the above limits are appropriate for Fermi liquids at $T=0$, where the vanishing frequency limit is equivalent to $s \rightarrow 0$. 

When both $\omega$ and $q$ are small, it is convenient to take the zero-frequency limit along radial lines in the $\left(\omega, \left|q\right|\right)$ plane at fixed $s$ \cite{Berthod-2018}, as done in Sec.\@ \ref{Hydro} in the relaxationless case. Such limit corresponds to setting $\omega=0$ in Eq.\@ (\ref{eq:sigma_T_el_VE_rel}) for $\tau_{K} \gg \tau_c$, such that
\begin{equation}\label{eq:sigma_T_el_VE_comp}
\sigma_{DC}(\vec{q})=\epsilon_0 \frac{(\omega_p)^2}{ (\tau_{K})^{-1}+ \nu(0) q^2}.
\end{equation}
Equation (\ref{eq:sigma_T_el_VE_comp}) shows that the DC transport properties depend on the competition between the momentum-relaxing term $i (\tau_{K})^{-1}$ and the spatially nonlocal, collision-dependent term $i \nu(0) q^2$. When the former dominates, we have
\begin{multline}\label{eq:VE_sigmaDrude}
\sigma_{DC}(\vec{q})\equiv \sigma_{DC}=\epsilon_0 (\omega_p)^2 \tau_{K} \\ =\frac{n e^2 \tau_{K} }{m}, \, \left| (v_F^{*})^2 \tau_c q^2 \right| \ll \frac{1}{\tau_{K}}, 
\end{multline}
which is the standard DC conductivity for a Drude/Ohmic conductor \cite{Dressel-2001}. On the contrary, for negligible momentum relaxation we obtain 
\begin{multline}\label{eq:VE_sigma_visc}
\sigma_{DC}(\vec{q})=\frac{\epsilon_0 (\omega_p)^2}{(v_F^{*})^2 \tau_c q^2} \\ =\frac{ m e^2  k_F}{ 3 \pi^2 \hbar^2 \tau_c q^2} \left(1+\frac{F_1^S}{3}\right), \, \left| (v_F^{*})^2 \tau_c q^2 \right| \gg \frac{1}{\tau_{K}},
\end{multline}
which depends on momentum as well as on the collision time $\tau_c$ through the Fermi-liquid viscosity $\nu(0)$. Once Fourier-transformed from momentum space into real-space coordinates $\vec{r}$, Eq.\@ (\ref{eq:VE_sigma_visc}) governs the diffusive transport properties in viscous electron fluids flowing through narrow channels, with consequences like a size-dependent resistivity controlled by viscosity \cite{Gurzhi-1963,Jaggi-1990,Alekseev-2016, Scaffidi-2017, Levchenko-2020}. 

\section{Collision time and momentum relaxation time from Mattheissen's rule}\label{Scatter}

In order to qualitatively estimate the interplay of momentum-conserving collisions and momentum-relaxing scattering, we have to consider the microscopic nature of the different available scattering channels for electrons in solids. How are the collision time $\tau_c$ and the relaxation time $\tau_{K}$ related in a realistic solid-state system? Magnetotransport experiments on the strongly interacting electrons in WP$_2$ suggest that $\tau_c \approx \tau_{K}$ at high temperatures, while $\tau_{K} \gg \tau_c$ in the low-temperature regime \cite{Gooth-2018}. In particular, for WP$_2$ the momentum-relaxing time saturates to a nearly constant value below $T \approx 10$ K, possibly due to residual impurity scattering, while $\tau_c \propto T^{-1}$ as expected for a quantum critical fluid \cite{Gooth-2018}. On the other hand, Ref.\@ \cite{FZVM-2014} assumed $\tau_c=\alpha_U \tau_{K}$ at all temperatures with ``Umklapp efficiency'' $\alpha_U \in \left(0, \, 1\right)$. Reference \onlinecite{Berthod-2013} compares the generalized Drude model, including a frequency-dependent optical scattering rate, with the microscopic optical conductivity derived from the Kubo formula in the assumption of a local Fermi-liquid self-energy \cite{Stricker-2014}. While such microscopic analyses transcend the scope of this paper, in the following I consider the example of acoustic phonons, impurities and Umklapp processes as independent relaxation channels.
\begin{figure}[ht]
\includegraphics[width=0.8\columnwidth]{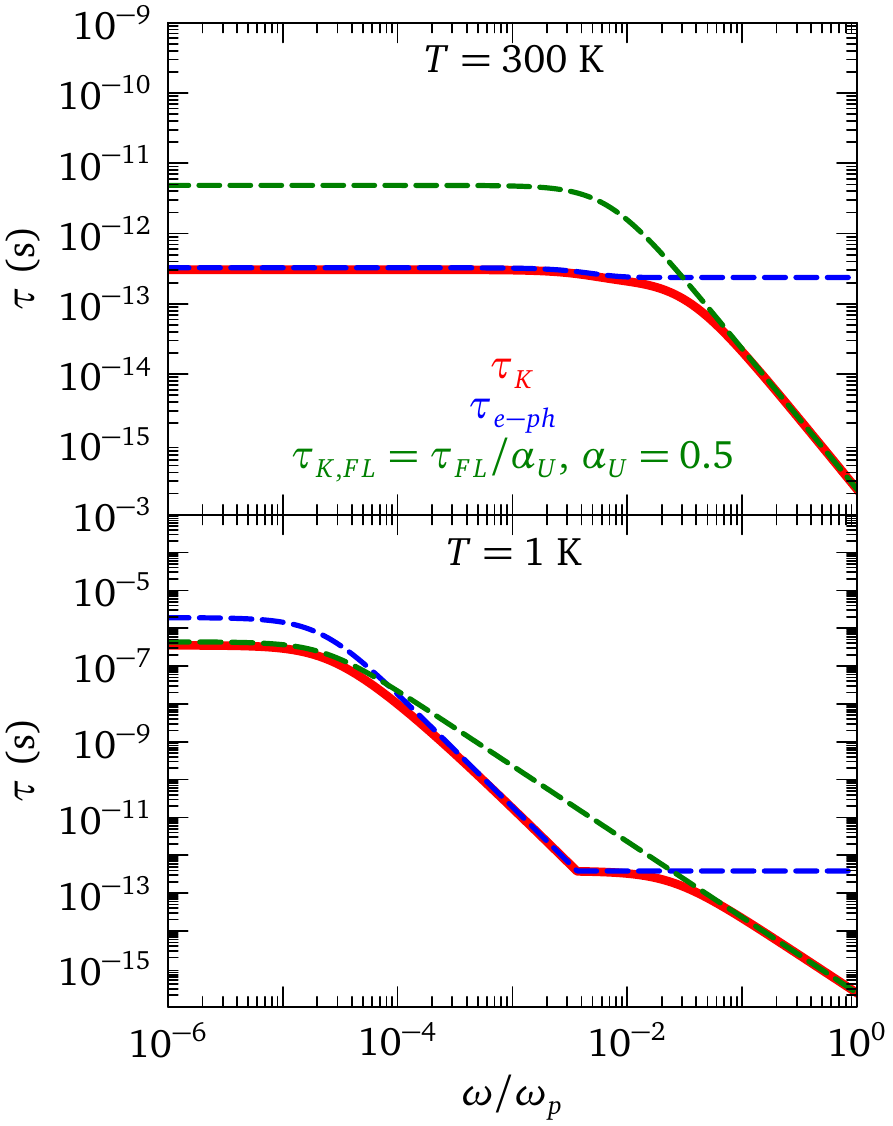}
\caption{\label{fig:tau_tot} 
Total momentum-relaxation time $\tau_{K}(\omega, T)$ $\left(s\right)$ as a function of frequency $\omega$ normalized to the plasma frequency $\omega_p$, according to Mattheissen's rule (\ref{eq:Mat_tau}) and without impurities, i.e.\ $\tau_i \rightarrow +\infty$. Solid curves show $\tau_{K}(\omega, T)$ at temperatures $T=300 \, K$ and $T=1 \, K$. Dashed curves display the individual contributions from Umklapp processes with efficiency $\alpha_U=0.5$ and from electron-phonon scattering, for which we use the same parameters as in Figs.\@ \ref{fig:tau_FL} and \ref{fig:tau_eph}, respectively. 
}
\end{figure} 
The electron-electron collision time $\tau_c$ is here assumed to follow the standard Fermi-liquid expression stemming from quasiparticle phase-spase restriction \cite{vanderMarel-2011}, the derivation of which is recalled in Appendix \ref{app:FL_scat}. This yields $\tau_c \propto E_F^{*} \left[ (\hbar \omega)^2 + (\pi k_B T)^2\right]^{-1}$, where $E_F^{*}=\hbar^2 k_F^2/(2 m^{*})$. 
The electron-phonon scattering time $\tau_{e-ph}(\omega,T)$ is calculated from the many-body self-energy in Sec.\@ \ref{app:e-ph_scat}. A constant impurity scattering rate $\tau_{i} \in \mathbb{R}^+$ is added in the first Born approximation \cite{Berthod-2018}. Following Ref.\@ \cite{FZVM-2014}, a proportionality constant is assumed between the Fermi-liquid collision time $\tau_c$ and the Umklapp relaxation time $\tau_{K,FL}$: $\tau_c=\alpha_U \tau_{K,FL}$, with $\alpha_U \in \left(0,1\right)$ Umklapp efficiency. Physically, at each collision a relative ratio $\alpha_U$ of momentum is transferred to the lattice, an assumption which works satisfactorily in transition metals \cite{Laurence-1973}.  
Following the empirical Mattheissen's rule, the total relaxation rate $\left[\tau_{K}(\omega,T)\right]^{-1}$ is the sum of the individual rates for Umklapp, electron-phonon and impurity channels: 
\begin{equation}\label{eq:Mat_tau}
\frac{1}{\tau_{K}(\omega,T)}=\frac{1}{\tau_{K,FL}(\omega,T)}+\frac{1}{\tau_{e-ph}(\omega,T)}+\frac{1}{\tau_{i}}.
\end{equation}
Equation (\ref{eq:Mat_tau}) assumes that all relaxation channels are independent from each other. We calculate the quasiparticle contribution from Eq.\@ (\ref{eq:tau_qp_fin2}) and the electron-phonon contribution from Eq.\@ (\ref{eq:tau_eph_expl}). Figure \ref{fig:tau_tot} shows $\tau_{K}(\omega,T)$ from Eq.\@ (\ref{eq:Mat_tau}) as a function of frequency $\omega$ normalized to the plasma frequency $\omega_p$. The impurity scattering time is $\tau_i \rightarrow +\infty$. Solid curves correspond to $\tau_K(\omega,T)$, at temperatures $T=300 \, K$ and $T=1 \, K$ for Fig.\@ \ref{fig:tau_tot}(a) and \ref{fig:tau_tot}(b). Dashed curves show the individual contributions from Umklapp and electron-phonon scattering, using the same parameters as in Figs.\@ \ref{fig:tau_FL} and \ref{fig:tau_eph} respectively. 
As seen in Fig.\@ \ref{fig:tau_tot}(a), at room temperature the phonon contribution to $\tau_{K}$ prevails at low frequency, while the Umklapp component takes over in the high-frequency regime. Conversely, the low-temperature result in Fig.\@ \ref{fig:tau_tot}(b) shows that the Umklapp contribution limits $\tau_{K}$ in the low- and high-frequency limits, separated by an intermediate frequency window dominated by phonon scattering. A finite impurity scattering time $\tau_i$ further constrains the static limit of $\tau_{K}$ when $\tau_i$ becomes the smallest timescale in Eq.\@ (\ref{eq:Mat_tau}), as further commented upon in Sec.\@ \ref{T_phon_imp}.

In principle, the insertion of the microscopic models (\ref{eq:tau_qp_fin2}) for $\tau_c$ and (\ref{eq:Mat_tau}) for $\tau_K$ into the Fermi-liquid dielectric function is subjected to constraints: Eq.\@ (\ref{eq:epsilon_T_FL_r}) is derived from the kinetic equation in relaxation-time approximation, which assumes that variations of the scattering rates occur at the same frequency as the variations of the induced Fermi-surface displacement. 

\section{Nonlocal optical properties of charged Fermi liquids}\label{Opt}

The results obtained in Secs.\@ \ref{Charged_FL_XT} - \ref{Scatter} equip us with ueful tools to investigate observable traces of Fermi-surface rigidity in optical spectroscopy experiments. Here we focus on the surface impedance and the transmission through a thin film. These setups require to consider how radiation is reflected, absorbed and transmitted at interfaces between different dielectric media. Such problem is well-defined in the standard Drude model through Maxwell's equations alone, however it becomes underdetermined when more than one optical mode is propagating in the material, as in the propagating shear case of section \ref{viscoel_q0_rel}. Multiple approaches are available to overcome this difficulty, and to those approaches the next section is dedicated for completeness. 

\subsection{Constitutive relations for electromagnetic fields at interfaces}\label{ABC}

Physically, reflection and transmission coefficients at interfaces between different dielectric media depend on how electrons at the boundary react to the incident electromagnetic radiation. When the electronic response is spatially local, it is sufficient to consider the boundary conditions stemming directly from Maxwell's equations at the boundary. At normal incidence, such conditions are the continuity of the electric field and of its derivative \cite{Dressel-2001}. However, when more than one polariton mode propagates into the material, the problem remains underdetermined by considering Maxwell's equations alone. The need for additional boundary conditions (ABCs) \cite{Venger-2004,Halevi-1984} originates from the fact that, rigorously, a dielectric function $\epsilon(\vec{q},\omega)$ depending on a single wave vector $\vec{q}$, or equivalently on the difference $\left|\vec{r}-\vec{r}'\right|$ between two coordinates $\vec{r}$ and $\vec{r}'$, is valid only for translationally invariant systems \cite{Berthod-2018}. A surface or interface breaks translation invariance, so that the nonlocal dielectric function $\epsilon(\vec{r},\vec{r}',\omega)$ should depend separately from $\vec{r}$ and $\vec{r}'$ at the boundary. To avoid complications implied by a more rigorous model of the surface \cite{Henneberger-1998, Chen-1993}, a common alternative is to retain the bulk expression $\epsilon(\vec{q},\omega)$ even at the boundary, at the cost of introducing ABCs deduced from the specific properties of the electrons that react to radiation \cite{Halevi-1992disp, Cocoletzi-2005}.  Notable examples of this approach are retrieved in the treatment of light-exciton coupling \cite{Halevi-1992disp, Cocoletzi-2005}, longitudinal plasmons \cite{Melnyk-1968,Melnyk-1970,Fuchs-1971}, and viscous charged liquids \cite{FZVM-2014}. The latter reference employed constitutive relations for the current density at boundaries stemming from fluid dynamics \cite{Bocquet-2007}, and expressed in terms of the slip length $\lambda_s \in \left(0,+\infty\right)$: $\lambda_s$ is the ratio between the shear dynamical viscosity of the moving fluid and the tangential friction per unit area exerted by the fluid on the boundary of the solid \cite{FZVM-2014}. Since in this model the friction force is proportional to the fluid velocity $\vec{v}$ at the boundary, a nonzero slip length implies a finite fluid velocity at the interface \cite{FZVM-2014}, with the models of Pekar \cite{Pekar-1958} and Ting-Frenkel-Birman \cite{Ting-1975} as limiting cases for $\lambda_s=0$ and $\lambda_s \rightarrow +\infty$ respectively. 
On the other hand, textbook treatments of anomalous skin effect in metals are derived in terms of microscopic models for the interface, in the form of specular of diffuse scattering of electrons at the boundary \cite{Reuter-1948,Sondheimer-2001}: in this approach, one assumes that a portion $p\in \left[0,1\right]$ of quasiparticles experiences specular scattering at the sample boundaries, while a portion $1-p$ undergoes diffusive scattering. 
The approaches in terms of $\lambda_s$ and $p$ give qualitatively consistent results, although differences appear at the quantitative level. In the following sections, we adopt both approaches and compare their results for the surface impedance. 

A potential reconciliation of the ABC and surface-modeling approaches is offered by Ref.\@ \onlinecite{Kiselev-2019}, which reports a microscopic calculation of $\lambda_s$ in 2D and 3D Fermi liquids for specular and diffusive scattering. In three dimensions and in terms of the generalized shear modulus (\ref{eq:nu_q0_vel}), we have
\begin{subequations}\label{eq:lambda_s}
\begin{multline}\label{eq:lambda_s_s}
\lambda_s^{s}=\frac{1}{h^2 (h')^2 (k_F)^4} \frac{45 \pi^2}{(k_F)^4 \hbar} n m \nu(0) \\ =\frac{3}{h^2 (h')^2 (k_F)^4} v_F^{*} \tau_c ,
\end{multline}
\begin{equation}\label{eq:lambda_s_d}
\lambda_s^{d}=\frac{8 \pi^2}{(k_F)^4 \hbar} n m \nu(0)=\frac{8}{15} v_F^{*} \tau_c,
\end{equation}
\end{subequations}
where $n$ is the 3D electron density, $k_F=(3 \pi^2 n)^{\frac{1}{3}}=m v_F/\hbar$ is the 3D Fermi wave vector, and $\left(\lambda_s^{s},\lambda_s^{d}\right)$ refer to specular and diffuse scattering, respectively. Notice that the slip length becomes itself a a function of frequency and temperature through $\tau_c \propto \left[(\hbar \omega)^2 + (\pi k_B T)^2 \right]^{-1}$. The ratio between $\lambda_s^{s}$ and $\lambda_s^{d}$ depends only on the ratio between the surface rugosity, parametrized by the height $h$ and characteristic length $h'$ of periodic corrugations \cite{Kiselev-2019}, and $k_F$: $\lambda_s^{d}/\lambda_s^{s}=\left[8 h^2 (h')^2 (k_F)^4\right]/45$. Therefore, as long as the surface corrugations occur on a length scale larger than the Fermi wavelength $(2 \pi)/k_F$, the slip length is larger in the diffusive case. 

\subsection{Surface impedance}\label{Zs}

The surface impedance $Z(\omega)$ is an ideal probe of spatial nonlocality in the electrodynamic response of metals: it is defined as the ratio of the electric field $\vec{E}$, normal to the metallic surface, to the total current density $\vec{J}$ induced in the bulk of the semi-infinite sample \cite{Jackson-1962,Dressel-2001}
\begin{equation}\label{eq:Z_gen}
Z(\omega)=\frac{E(z,\omega)_{z=0^+}}{\int_0^{+\infty} J(z,\omega) dz}. 
\end{equation}
Here, we defined the orthogonal coordinate $z$ with respect to the surface $z=0$, and $J(z,\omega)$ is the current density per unit area. Were the electrodynamic response local, the current density induced by the surface electric field $E(0^+,\omega)$ would be exclusively located at the surface $J(z,\omega) \equiv J(0^+,\omega)$. Any degree of spatial nonlocality generates a current response extending at $z>0$. A useful relation between $Z(\omega)$ and the reflection coefficient $r(\omega)$ at the boundary between vacuum and a semi-infinite dielectric medium is \cite{Jackson-1962,Dressel-2001}
\begin{equation}\label{eq:Z_r}
r(\omega)=\frac{Z(\omega)-Z_0}{Z(\omega)+Z_0},
\end{equation}
where $Z_0=\mu_0 c$ is the vacuum surface impedance. 
\begin{figure}[ht]
\includegraphics[width=0.8\columnwidth]{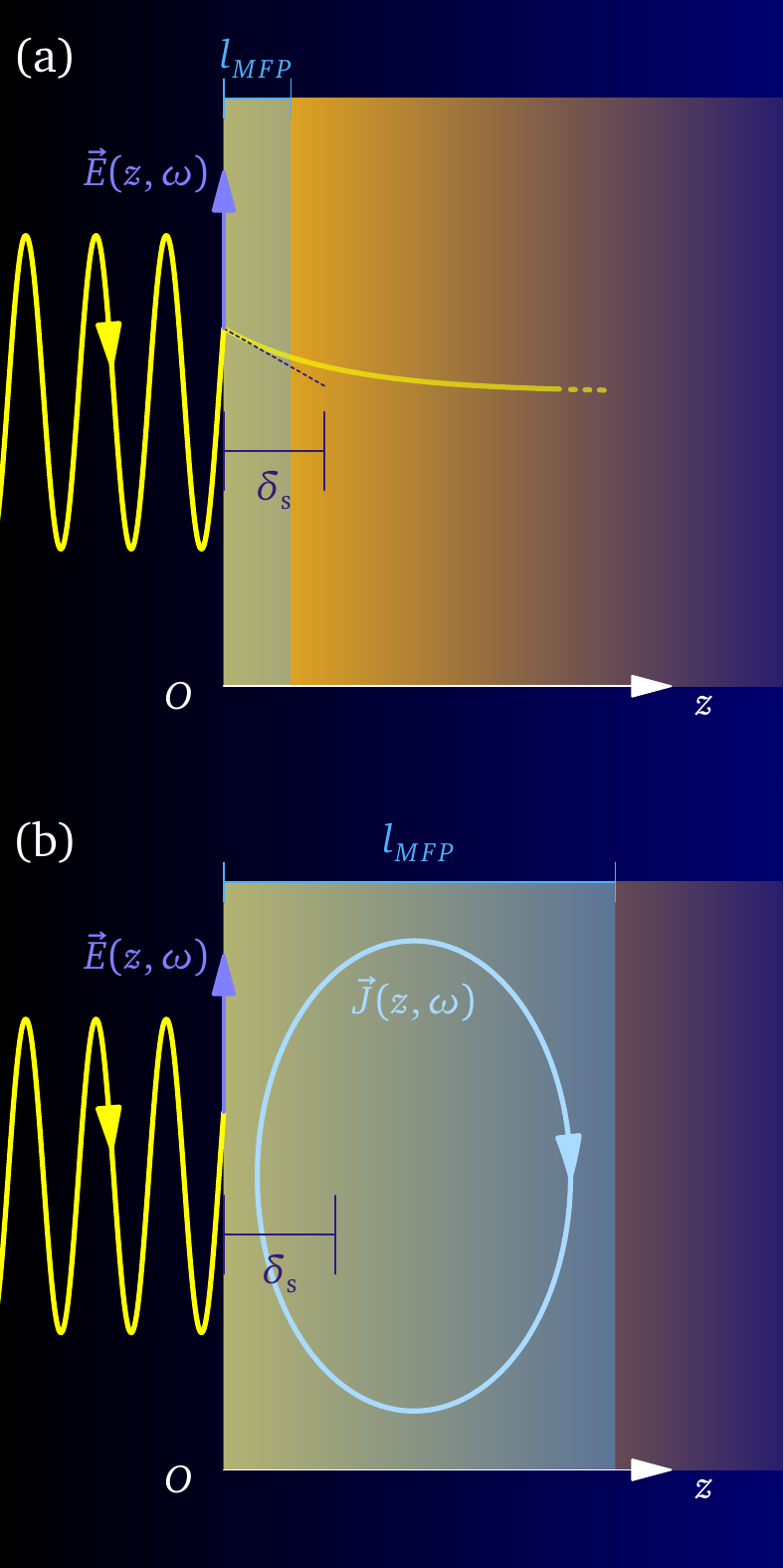}
\caption{\label{fig:RS_illustr} Schematic representation of the nonlocal current response generated inside a metal by an electric field at normal incidence. Panel (a) sketches the exponential damping of electric fields within a distance $\delta_s$ when $l_{MFP} \ll \delta_s$ (normal skin effect), while panel (b) highlights the formation of nonlocal currents $\vec{J}(z,\omega)$ in the regime $l_{MFP} \gg \delta_s$ (anomalous skin effect).
}
\end{figure}
Physically, to generate nonlocal currents in the bulk, electrons must be able to travel freely on a distance larger than the depth from $z=0$ within which electric fields are suppressed by dielectric screening. The latter is the Drude skin depth $\delta_s$ \cite{Jackson-1962,Dressel-2001} 
\begin{equation}\label{eq:delta_skin_Drude}
\delta_s=\sqrt{\frac{2}{\mu_0 \omega \sigma_{DC}}} \propto \frac{1}{\sqrt{\omega}},
\end{equation}
where $\sigma_{DC}$ is the Drude DC conductivity (\ref{eq:VE_sigmaDrude}). We will recall the derivation of Eq.\@ (\ref{eq:delta_skin_Drude}) in Sec.\@ \ref{Skin_Drude}. When $l_{MFP} \ll \delta_s$, electrons respond locally to the incident electromagnetic field, and the latter gets exponentially damped inside the metal within a characteristic length scale $\delta_s$, as illustrated in Fig.\@ \ref{fig:RS_illustr}(a). On the other hand, for $l_{MFP} \gg \delta_s$ the response becomes spatially nonlocal, and bulk transverse currents arise in the metal, as sketched in Fig.\@ \ref{fig:RS_illustr}(b). This allows for a much larger penetration depth of the electric field inside the metal, up to distances of the order of $l_{MFP}$, which in turn affects the optical properties. Hence, we can define the qualitative criterion $\delta_s \ll l_{MFP}$ for anomalous skin effect, while for $\delta_s \gtrsim l_{MFP}$ we retrieve the results of the Drude model, as described in Secs.\@ \ref{Skin_Drude} and \ref{Skin_anom}. \\ 
The ratio $\delta_s/l_{MFP} \propto (\tau_{K})^{-\frac{3}{2}}$ is governed by momentum relaxation. In a clean Fermi liquid with no relaxing processes, $\delta_s/l_{MFP} \rightarrow 0$ and spatial nonlocality dominates the dielectric response at all frequencies -- see Appendix \ref{Delta_skin_cons}. However, in crystalline solids phonons provide strong momentum damping at room temperature, so that $\delta_s \gg l_{MFP}$ and local electrodynamics occurs: this is why anomalous skin effect is observed in clean samples at low temperatures, such that the mean-free path increases due to freezing of phonon scattering channels \cite{Pippard-1947a,Pippard-1947b}. 

Textbook derivations of anomalous skin effect are usually performed in the regime $\omega \ll v_F^{*} \left|q\right|$ \cite{Dressel-2001, Sondheimer-2001} \footnote{An exception to this is the transverse impedance in a neutral Fermi liquid, which was analyzed in Ref.\@ \onlinecite{Shahzamanian-2006}.}. However, since the generalized shear modulus $\nu(\omega)$ entails spatial nonlocality, it produces a phenomenology analogous to anomalous skin effect in the propagating shear regime $\omega \gg v_F^{*} \left|q\right|$. Therefore, observing the characteristic phenomenology of anomalous skin effect in the regime $\omega \gg v_F^{*} \left|q\right|$ is direct evidence for a propagating shear mode in the solid-state Fermi liquid.  

Here we follow Dressel and Gr\"{u}ner \cite{Dressel-2001} in sketching the derivation of anomalous skin effect in the absence of an external magnetic field. The electromagnetic wave equation, for fields $E(z,\omega) \propto e^{i(q z-\omega t)}$ in the $xy$ plane and without free charge density, depends on the induced current density $J(z,\omega)$ inside the metal according to
\begin{multline}\label{eq:wave_mat}
\frac{\partial^2 E(z,\omega)}{\partial z^2} + \frac{\omega^2}{c^2} \int dz' \epsilon_T(z-z',\omega) E(z',\omega) \\ =-\mu_0 i \omega J(z,\omega)+\alpha_b,
\end{multline}
where $\alpha_b$ is a factor depending on the quality of boundary scattering. The integral on the left-hand side of Eq.\@ (\ref{eq:wave_mat}) takes into account spatial nonlocality in the dielectric function $\epsilon_T(z-z',\omega)$, which depends on the difference between the probe and reaction coordinates $z-z'$ in the translationally invariant case.  
In the following we assume specular scattering at the metal surface, which implies $\left. \partial E(z,\omega)/\partial z \right|_{z=0^+}=-\left. \partial E(z,\omega)/\partial z \right|_{z=0^-}$ and therefore $\alpha_b=2 \left. \partial E(z,\omega)/\partial z \right|_{z=0^-} \delta(z)$. Other types of boundary scattering do not qualitatively modify the results here outlined \cite{Dressel-2001}. Fourier-transforming Eq.\@ (\ref{eq:wave_mat}) to reciprocal space of momenta $q$ yields
\begin{multline}\label{eq:wave_mat_q}
-q^2 E(q,\omega) + \underbrace{\frac{\omega^2}{c^2} \epsilon_T(\omega) E(q,\omega)}_{\boxed{A}} \\ =-\underbrace{\mu_0 i \omega J(q,\omega)}_{\boxed{B}}+2 \left. \frac{\partial E(z,\omega)}{\partial z} \right|_{z=0^+}. 
\end{multline}
The term $\boxed{A}$ in Eq.\@ (\ref{eq:wave_mat_q}) stems from the displacement current, while conduction currents generate the term $\boxed{B}$. 
The general solution of Eq.\@ (\ref{eq:wave_mat_q}) satisfies Chambers' formula \cite{Chambers-1952,Dressel-2001}. If $\tau_c$ does not depend on $q$, one can also utilize the Boltzmann kinetic equation to calculate the current density $J(q,\omega)$, similarly to what we did in Sec.\@ \ref{FL_XT_coll}. In an even simpler approximation, one can utilize Ohm's law 
\begin{equation}\label{eq:Ohm_gen}
J(\vec{q},\omega)=\sigma_T(\vec{q},\omega)E(\vec{q},\omega)
\end{equation}
knowing the transverse conductivity $\sigma_T(\vec{q},\omega)$. 
We first recall the analysis of ``normal'' skin effect for the local Drude model, and we sketch the derivation of anomalous skin effect in the standard regime $\omega \ll v_F^{*} \left|q\right|$ \cite{Reuter-1948, Sondheimer-2001,Dressel-2001}. Then, we compare these results to the analogous phenomenon occurring when the shear polariton propagates. 

\subsubsection{``Normal'' skin effect in the Drude model}\label{Skin_Drude}

The standard derivation of the skin effect neglects the displacement current term $\boxed{A}$ in Eq.\@ (\ref{eq:wave_mat_q}) with respect to the conduction current $\boxed{B}$. This approximation holds for frequencies up to $\omega \approx \sqrt{\sigma_T(\vec{q},\omega)/\left(\epsilon_0 \tau_{K}\right)}$, which lie in the near ultraviolet region for most metals \cite{Reuter-1948}, i.e.\@, at $\omega \approx \omega_p$ where the metal becomes transparent. 
In this regime, the surface impedance results from the combination of Eq.\@ (\ref{eq:wave_mat_q}), the generalized Ohm's law (\ref{eq:Ohm_gen}), and the general definition (\ref{eq:Z_gen}):
\begin{equation}\label{eq:Z_gen_1D}
Z(\omega)=i \omega \mu_0 \frac{E(0,\omega)}{\left. \frac{\partial E(z,\omega)}{\partial z} \right|_{z=0^+}}. 
\end{equation}
To obtain the momentum-dependent electric field, we insert Eq.\@ (\ref{eq:Ohm_gen}) in the wave equation (\ref{eq:wave_mat_q}), which gives
\begin{equation}\label{eq:E_q_sigma}
E(q,\omega)=2 \left. \frac{\partial E(z,\omega)}{\partial z} \right|_{z=0^+} \frac{1}{\mu_0 i \omega \sigma_T(q,\omega) - q^2}. 
\end{equation}
We now employ the spatially local Drude conductivity with momentum relaxation from Eqs.\@ (\ref{eq:eps_T_Drude}) and (\ref{eq:sigma_epsilon}),
\begin{equation}\label{eq:Drude_sigma} 
\sigma_T(\omega)=i \epsilon_0 \omega \frac{(\omega_p)^2}{\omega^2+i \omega/\tau_{K}}. 
\end{equation}
which corresponds to the regime $\left|i \omega \nu(\omega) q^2\right| \ll \left|i \omega (\tau_{K})^{-1}\right|$ in Eq.\@ (\ref{eq:sigma_T_el_VE_rel}), i.e.\@ the regime in which momentum relaxation dominates over spatial nonlocality. 
After inserting Eq.\@ (\ref{eq:Drude_sigma}) into Eq.\@ (\ref{eq:E_q_sigma}), we transform back the latter to coordinate space, and at the interface $z=0$ we have
\begin{multline}\label{eq:E_int_Drude}
E(0,\omega)=\frac{1}{2 \pi} \left. \int_{-\infty}^{+\infty} dq E(q,\omega) e^{-i q z}\right|_{z=0^{-}} \\ =- \frac{1}{\pi} \left. \frac{\partial E(z,\omega)}{\partial z} \right|_{z=0^{-}} \int_{-\infty}^{+\infty} dq \left[-\frac{(\omega_p)^2}{c^2} \frac{\omega \tau_K}{i+\omega \tau_K}-q^2\right]^{-1} \\=  -\left. \frac{\partial E(z,\omega)}{\partial z} \right|_{z=0^+} \frac{c}{\omega_p} \sqrt{1+\frac{i}{\omega \tau_{K}}}, 
\end{multline}
where in the last step we have utilized the boundary condition on the electric field derivative at $z=0^{+}$. 
The Drude surface impedance follows from Eqs.\@ (\ref{eq:E_int_Drude}) and (\ref{eq:Z_gen_1D}): 
\begin{equation}\label{eq:Z_Drude}
Z(\omega)=-i \mu_0 \omega \frac{c}{\omega_p} \sqrt{1+\frac{i}{\omega \tau_{K}}}, \, \omega\ll \omega_p.
\end{equation}
Expanding Eq.\@ (\ref{eq:Z_Drude}) in the Hagen-Rubens regime $\omega \tau_{K} \ll 1$, we retrieve the well-known result for ``normal'' skin effect: 
\begin{equation}\label{eq:Z_Drude_HR}
Z(\omega)=\frac{\mu_0 c \sqrt{\omega}}{\omega_p \sqrt{\tau_{K}}} \frac{1-i}{\sqrt{2}}= \mu_0 \omega \delta_s \frac{1-i}{2},
\end{equation}
where $\delta_s$ is the classical skin depth (\ref{eq:delta_skin_Drude}), i.e.\@, the characteristic damping length of the electromagnetic field inside the Drude metal.
In the opposite regime $\omega \tau_{K} \gg 1$, Eq.\@ (\ref{eq:Z_Drude}) gives a purely imaginary result
\begin{equation}\label{eq:Z_Drude_highom}
Z(\omega)=-i \mu_0 c \frac{\omega}{\omega_p}. 
\end{equation}
Notice that one can also derive the above results from the relation (\ref{eq:Z_r}): the reflection coefficient between vacuum and a metal is 
\begin{equation}\label{eq:Drude_r}
r(\omega)=\frac{1-n(\omega)}{1+n(\omega)}
\end{equation}
where $n(\omega)=\sqrt{\epsilon_T(\omega)}$ is the refractive index. Using the Drude dielectric function (\ref{eq:eps_T_Drude}) and Eq.\@ (\ref{eq:Z_r}), we obtain
\begin{equation}\label{eq:Z_Drude_n}
Z(\omega)=\frac{\mu_0 c}{n(\omega)}=\mu_0 c \frac{\sqrt{1+\frac{i}{ \omega \tau_K}}}{\sqrt{1+\frac{i}{ \omega/\tau_K} -\frac{(\omega_p)^2}{\omega^2}}},
\end{equation}
which reduces to Eq.\@ (\ref{eq:Z_Drude}) in the Hagen-Rubens regime \cite{Dressel-2001}. 

\subsubsection{Anomalous skin effect regime}\label{Skin_anom}

\begin{figure}[ht]
\includegraphics[width=0.68\columnwidth]{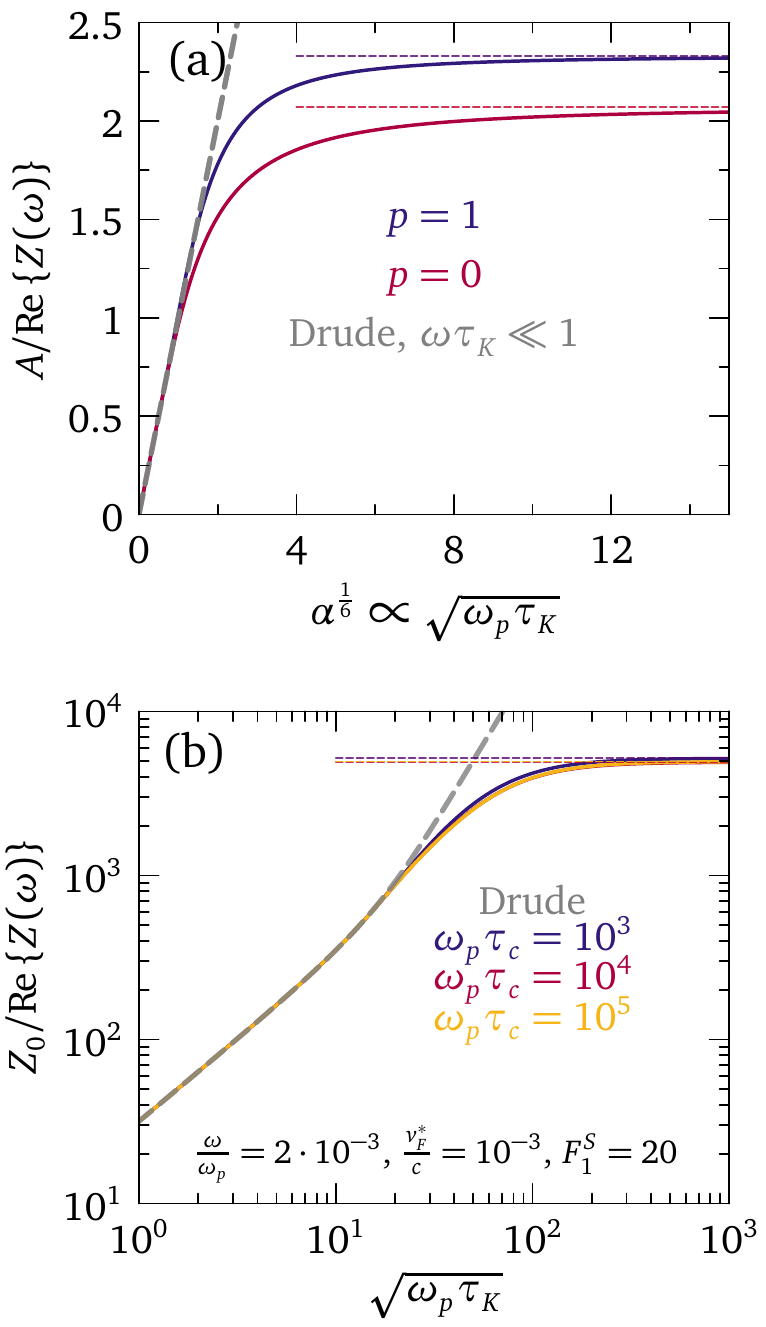}
\caption{\label{fig:Z_FL} Inverse surface resistance $\left[\mathrm{Re}\left\{Z(\omega)\right\}\right]^{-1}$ showing the characteristic saturation of anomalous skin effect. Panel (a) shows the Reuter-Sondheimer model according to Eqs.\@ (\ref{eq:Z_RandS}) as a function of $\alpha^{1/6} \propto \sqrt{\omega_p \tau_{K}}$, with blue and red curves for specular and diffuse scattering respectively. The dashed blue and red lines indicate the associated asymptotic limits: the specular-scattering case obeys Eq.\@ (\ref{eq:Z_anom}), which is $9/8$ of the diffusive-scattering result. The dashed gray curve shows the Drude result (\ref{eq:Z_Drude}). Panel (b) shows the Fermi-liquid calculation (\ref{eq:Z_VE_low}) in propagating shear regime as a function of $\sqrt{\omega_p \tau_{K}}$, at frequency $\omega=2 \cdot 10^{-3} \omega_p$, renormalized Fermi velocity $v_F^*=10^{-3}c$, and first Landau parameter $F_1^S=20$. The blue, red and gold solid curves indicate the results for $\omega_p \tau_c=\left\{10^3,10^4,10^5\right\}$ respectively. The dashed horizontal curves are the associated relaxationless limits (\ref{eq:Z_VE_taukinf}). The dashed gray curve is the Drude result (\ref{eq:Z_Drude_n}).
}
\end{figure}

The results of Sec.\@ \ref{Skin_Drude} are modified in the high-momentum regime $\omega \ll v_F^{*} \left|q\right|$, where the momentum dependence of the transverse conductivity $\sigma_T(\vec{q},\omega)$ leads to anomalous skin effect. To see this, we insert the conductivity (\ref{eq:sigma_anom_FL}), which we derived from the Fermi-liquid kinetic equation, into Eq.\@ (\ref{eq:E_q_sigma}) \cite{Dressel-2001}. Fourier-transforming back to real space implies
\begin{multline}\label{eq:E_int_anom}
E(0,\omega)=\frac{1}{2 \pi} \left. \int_{-\infty}^{+\infty} dq E(q,\omega) e^{-i q z}\right|_{z=0} \\ =- \frac{1}{\pi} \left. \frac{\partial E(z,\omega)}{\partial z} \right|_{z=0^-} \int_{-\infty}^{+\infty} dq \left[i \frac{\omega}{c^2}(\omega_p)^2 \frac{3 \pi}{4} \frac{1}{v_F q} - q^2 \right]^{-1} \\ =-\frac{1}{\pi} \left. \frac{\partial E(z,\omega)}{\partial z} \right|_{z=0^-} \int_{-\infty}^{+\infty} dq \left[i \mu_0 \omega \frac{3 \pi}{4} \frac{\sigma_{DC}}{l_{MFP} q} - q^2 \right]^{-1} \\ = \frac{1}{\pi} \left. \frac{\partial E(z,\omega)}{\partial z} \right|_{z=0^+}  \int_{-\infty}^{+\infty} dq \left[\frac{i}{(\delta_s^{\mathrm{an}})^3 q} - q^2 \right]^{-1},
\end{multline}
where we have defined the skin depth in anomalous regime: 
\begin{equation}\label{eq:skin_anom}
\delta_s^{\mathrm{an}}=\left[\frac{4}{3 \pi} \frac{c^2}{\omega (\omega_p)^2} v_F \right]^{\frac{1}{3}}=\left(\frac{4}{3 \pi \mu_0 \omega} \frac{l_{MFP}}{\sigma_{DC}} \right)^{\frac{1}{3}} \propto \omega^{-\frac{1}{3}}. 
\end{equation}
The $q$-integration in Eq.\@ (\ref{eq:E_int_anom}) gives
\begin{equation}\label{eq:E_int_anom2}
E(0,\omega)=-  \left. \frac{\partial E(z,\omega)}{\partial z} \right|_{z=0^+} \frac{2}{3} \delta_s^{\mathrm{an}} \left(1+\frac{i}{\sqrt{3}}\right). 
\end{equation}
Inserting Eq.\@ (\ref{eq:E_int_anom2}) in Eq.\@ (\ref{eq:Z_gen_1D}), we obtain the surface impedance
\begin{multline}\label{eq:Z_anom}
Z(\omega)=-i \mu_0 \omega \delta_s^{\mathrm{an}} \frac{2}{3}\left(1+\frac{i}{ \sqrt{3}}\right)\\ =\frac{2}{3}\left[\frac{4 (\mu_0)^2 \omega^2}{3 \pi} \frac{l_{MFP}}{\sigma_{DC}}\right]^{\frac{1}{3}} \left( \frac{1}{\sqrt{3}} - i \right). 
\end{multline}
Equation (\ref{eq:Z_anom}) coincides with the asymptotic value of the surface impedance for specular boundary scattering, found for $l_{MFP} \gg \delta_s$ from the more general theory by Reuter and Sondheimer \footnote{In Refs.\@ \onlinecite{Reuter-1948, Sondheimer-2001} a positive imaginary part of $Z(\omega)$ is obtained, since the convention $E(z,\omega) \propto e^{i \omega t}$ is adopted.} \cite{Reuter-1948,Sondheimer-2001}. The value for diffuse boundary scattering is obtained by multiplying Eq.\@ (\ref{eq:Z_anom}) by $9/8$ \cite{Reuter-1948,Dressel-2001}. Figure \ref{fig:Z_FL}(a) shows the asymptotic specular limit (\ref{eq:Z_anom}) and its diffusive counterpart as horizontal blue and red dashed lines, respectively. 

In anomalous skin effect regime, the electric field penetrating into the metal is exponentially damped as a function of coordinate $z$ [Eq.\@ (\ref{eq:E_int_anom2})] as for normal skin effect --- cf.\@ equation (\ref{eq:E_int_Drude}): the majority of the field is confined within a length scale of the order of the skin depth (\ref{eq:skin_anom}). However, the frequency dependence $\delta_s^{\mathrm{an}} \propto \omega^{-\frac{1}{3}}$ found in anomalous regime is different from the Hagen-Rubens expression $\delta_s \propto \omega^{-\frac{1}{2}}$, and electrons respond coherently at distances up to $l_{MFP}$ from the sample surface due to spatial nonlocality. 
An accurate calculation of the surface impedance at finite $\tau_{K}$ requires to go beyond the leading-order high-momentum expansion (\ref{eq:sigma_anom_FL}). This task has been performed by Reuter and Sondheimer, by solving the inhomogeneous Boltzmann equation for free electrons in relaxation-time approximation \cite{Reuter-1948,Sondheimer-2001}. Such solution provides a quantitatively precise description of anomalous skin effect for any value of $\tau_{K}$, including the crossover to the Drude model, while it neglects quasiparticle interactions and momentum-conserving collisions. 
In principle, one could also calculate the surface impedance for any ratio $\tau_c/\tau_{K}$ from the general transverse susceptibility (\ref{eq:Xi_T_r}) complemented by boundary conditions at the vacuum-Fermi liquid interface, and compare the results with Reuter-Sondheimer theory in the regime $\tau_c/\tau_{K} \rightarrow +\infty$. Such detailed comparison is left for future work, while here we quote the Reuter-Sondheimer surface impedance in the microwave spectral region \cite{Reuter-1948, Sondheimer-2001}:
\begin{subequations}\label{eq:Z_RandS}
\begin{equation}
Z(\omega)=-i \sqrt{\frac{8}{3}} A \alpha^{1/3} \frac{1}{l_{MFP}}\frac{E(0,\omega)}{\left. \frac{\partial E(z,\omega)}{\partial z} \right|_{z=0^+}},
\end{equation}
\begin{equation}
A=\frac{\sqrt{6}}{2^{\frac{1}{3}}} \left( \frac{m v_F}{3 n e^2}\right)^{\frac{1}{3}} (\mu_0 \omega)^{\frac{2}{3}},
\end{equation}
\begin{equation}
\alpha=\frac{3}{2} \left(\frac{l_{MFP}}{\delta_s}\right)^2=\frac{3}{4} \left(\frac{v_F^*}{c}\right)^2 \omega \tau_{K} (\tau_{K})^2 (\omega_p)^2,
\end{equation}
\end{subequations}
where $\delta_s$ is the classical skin depth (\ref{eq:delta_skin_Drude}). The parameter $\alpha$ is precisely the ratio between mean-free path and skin depth, which determines the crossover between the Drude/local and anomalous/nonlocal regimes, as mentioned at the beginning of Sec.\@ \ref{Zs}. \\
The solid blue (red) curve in Fig.\@ \ref{fig:Z_FL}(a) shows the inverse surface resistance from Eqs.\@ (\ref{eq:Z_RandS}) in the specular (diffuse) scattering case, as a function of $\alpha^{\frac{1}{6}} \propto \sqrt{\tau_{K}}$ as commonly done after Pippard \cite{Pippard-1947a,Pippard-1947b,Pippard-1947c}: this highlights the transition from the Drude result (dashed gray curve) to the asymptotic saturation (\ref{eq:Z_anom}) as the momentum-relaxation time $\tau_{K}$ increases. Appendix \ref{Delta_skin_cons} considers the ratio $\delta_s/l_{MFP}$ for a Fermi liquid, employing the total relaxation time from Sec.\@ \ref{Scatter}. 

\subsubsection{Propagating shear regime at low frequencies}\label{Skin_VE_low}

We now analyze the surface impedance in the presence of the generalized shear modulus (\ref{eq:nu_q0_vel}). Let us first determine the general expression for $Z(\omega)$, and then study its different physical limits in relation to the skin depth.

Our starting point is again Eq.\@ (\ref{eq:E_q_sigma}), assuming a Ohmic relation (\ref{eq:Ohm_gen}) between current density and electric field. Using the dielectric function (\ref{eq:eps_T_series_rel_gen1bis}) and the relation (\ref{eq:sigma_epsilon}), we achieve
\begin{multline}\label{eq:E_int_VE_low}
E(0,\omega)=\frac{1}{2 \pi} \left. \int_{-\infty}^{+\infty} dq E(q,\omega) e^{-i q z}\right|_{z=0^{-}}\\ =\frac{1}{\pi} \left. \frac{\partial E(z,\omega)}{\partial z} \right|_{z=0^{-}} \lambda_L \int_{-\infty}^{+\infty} d z \left[\frac{1}{A+ B z^2}+z^2\right]^{-1},
\end{multline}
where 
\begin{equation}\label{eq:q_z}
z=q \lambda_L
\end{equation}
and we have identified a characteristic length scale 
\begin{equation}\label{eq:London_depth}
\lambda_L=\frac{c}{\omega_p}=\sqrt{\frac{m}{\mu_0 n e^2}},
\end{equation}
which is the London penetration depth \cite{Tinkham-1996int,Levchenko-2020}. The terms $A$ and $B$ are 
\begin{subequations}\label{eq:E_int_VE_low_2}
\begin{equation}\label{eq:A_c}
A=1+\frac{i}{ \omega \tau_{K}},
\end{equation}
\begin{equation}\label{eq:B_nu} 
B= i \frac{(\omega_p)^2}{c^2} \frac{\tilde{\nu}(\omega)}{\omega}=i \left( \frac{\tilde{\delta}_{\nu}}{\lambda_L}\right)^2,
\end{equation}
\end{subequations} 
where we have identified another (complex-valued) length scale of the problem:
\begin{equation}\label{eq:delta_nu_tilde}
\tilde{\delta}_{\nu}=\sqrt{\frac{\tilde{\nu}(\omega)}{\omega}}.
\end{equation}
The variable $\tilde{\delta}_{\nu}$ depends on the generalized shear modulus (\ref{eq:nu_tilde}) valid for any ratio $\tau_c/\tau_K$ with $\omega \gg v_F^{*} \left|q \right|$. 

We perform the momentum integration in Eq.\@ (\ref{eq:E_int_VE_low}) and we write the result in terms of the variables (\ref{eq:E_int_VE_low_2}): 
\begin{subequations}
\begin{equation}
E(0,\omega)=- \left. \frac{\partial E(z,\omega)}{\partial z} \right|_{z=0^+} \lambda_L \frac{\sqrt{2} \left(A+\sqrt{B}\right)}{\sqrt{C}+2 \sqrt{B/C}},
\end{equation}
\begin{equation}
C=\sqrt{A^2-4 B}+A.
\end{equation}
\end{subequations} 
One verifies that Eqs.\@ (\ref{eq:E_int_VE_low_2}) reduce to Eq.\@ (\ref{eq:E_int_Drude}) for $\nu(\omega)=0$. 
The surface impedance then is derived from Eq.\@ (\ref{eq:Z_gen_1D}), giving 
\begin{equation}\label{eq:Z_VE_low}
\frac{Z(\omega)}{c \mu_0}=- i \frac{\omega}{\omega_p} \frac{\sqrt{2} \left(A+\sqrt{B}\right)}{\sqrt{C}+2 \sqrt{B/C}}.
\end{equation} 
\begin{figure}[ht]
\includegraphics[width=0.75\columnwidth]{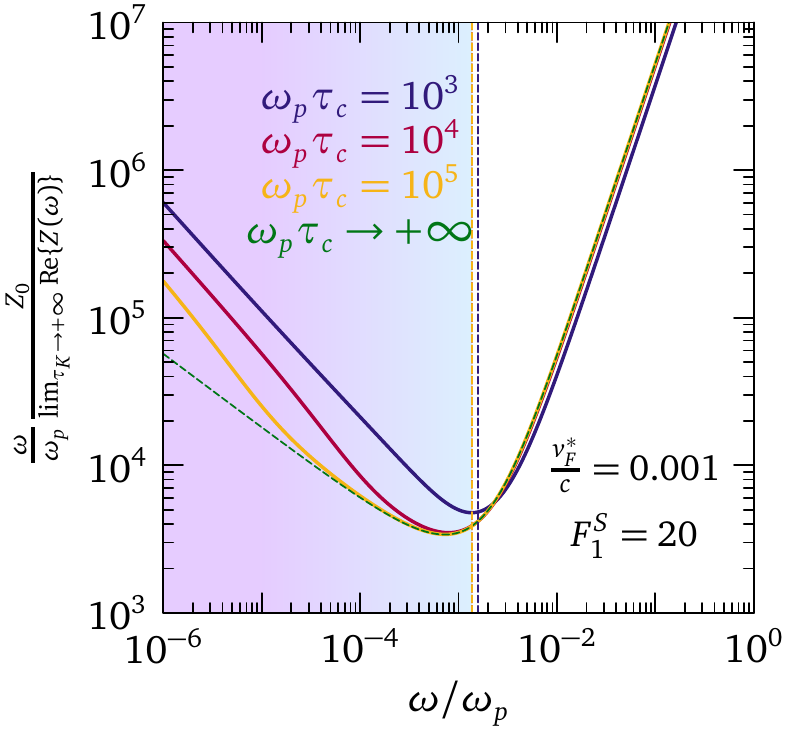}
\caption{\label{fig:RS_inf} Relaxationless limit of the inverse surface resistance for a 3D Fermi liquid in the propagating shear regime as a function of normalized frequency $\omega/\omega_p$, according to Eq.\@ (\ref{eq:Z_VE_taukinf}), for renormalized Fermi velocity $v_F^*=10^{-3}c$ and first Landau parameter $F_1^S=20$. The collision times $\omega_p \tau_c=\left\{10^3,10^4,10^5\right\}$ give the blue, red and gold curves respectively. The green dashed curve shows the purely reactive limit $\omega \tau_c \rightarrow +\infty$. Vertical dashed lines mark the frequency below which the shear-polariton is in the Lindhard continuum for each value of $\tau_c$. 
}
\end{figure}
Remarkably, the propagating-shear surface impedance (\ref{eq:Z_VE_low}) displays the typical phenomenology of anomalous skin effect, i.e.\@, the saturation of $\mathrm{Re}\left\{Z(\omega)\right\}^{-1}$ in the relaxationless limit $\tau_K \rightarrow +\infty$. This is evident from Fig.\@ \ref{fig:Z_FL}(b), which displays $\left[\mathrm{Re}\left\{Z(\omega)\right\}\right]^{-1}$ as a function of $\sqrt{\omega_p \tau_{K}}$. Here I use a renormalized Fermi velocity $v_F^*/c=0.001$, first Landau parameter $F_1^S=20$, and frequency $\omega/\omega_p=0.002$. The blue, red and golden curves show the result for $\omega_p \tau_c=\left\{10^3,\, 10^4, \, 10^5\right\}$, respectively. For consistency, the frequency $\omega/\omega_p=0.002$ is chosen such that the shear-polariton mode is outside of the electron-hole continuum for $\omega_p \tau_c \geq 10^3$, in accordance with Sec.\@ \ref{emerg_VE}; see also Fig.\@ \ref{fig:eh_bound}. 
In the limit $\tau_{K}\rightarrow +\infty$, Eq.\@ (\ref{eq:Z_VE_low}) reaches the asymptotic value  
\begin{multline}\label{eq:Z_VE_taukinf} 
\lim_{\tau_{K} \rightarrow +\infty} \frac{ Z(\omega)}{c \mu_0} \\ =-i \frac{\omega}{\omega_p} \frac{\sqrt{2} \left(\sqrt{B'}+1\right)}{\sqrt{\sqrt{1-4 B'}+1}+2 \sqrt{B'/\left(\sqrt{1-4 B'}+1\right)}},
\end{multline}
where 
\begin{equation}\label{eq:B_nu_inf}
B'=i \left(\frac{\omega_p}{c}\right)^2 \frac{ \nu(\omega) }{\omega}=i \left(\frac{\delta_{\nu}}{\lambda_L}\right)^2. 
\end{equation}
Notice that the term (\ref{eq:B_nu_inf}) depends only on the ratio between the London penetration depth (\ref{eq:London_depth}) and the complex-valued length scale
\begin{equation}\label{eq:delta_nu}
\delta_{\nu}=\sqrt{\frac{\nu(\omega)}{\omega}},
\end{equation}
governed by the generalized shear modulus (\ref{eq:nu_q0_vel}). Therefore, the asymptotic limit (\ref{eq:Z_VE_taukinf}) is controlled by $\nu(\omega)$ through Eq.\@ (\ref{eq:B_nu_inf}). Such limit is marked by dashed blue, red and gold lines in Fig.\@ \ref{fig:Z_FL}(b) for $\omega_p \tau_c=\left\{10^3,\, 10^4, \, 10^5\right\}$, respectively.

It is also interesting to see how the saturation limit (\ref{eq:Z_VE_taukinf}) depends on frequency $\omega$ and collision time $\tau_c$ through $\nu(\omega)$. Figure \ref{fig:RS_inf} shows the asymptotic value $\lim_{\tau_{K} \rightarrow +\infty} \mathrm{Re}\left\{Z(\omega)\right\}$ as a function of normalized frequency $\omega/\omega_p$, for different values of collision time. The purely reactive/elastic limit $\omega \tau_c \rightarrow +\infty$ is depicted by the green curve. The vertical dashed lines mark the frequency at which the shear-polariton emerges from the continuum, at the corresponding value of $\omega_p \tau_c$ (see also Fig.\@ \ref{fig:eh_bound}): the calculation is valid for frequencies greater than the threshold marked by the dashed line. We see that, in the propagating/reactive regime for the shear polariton, the asymptotic value (\ref{eq:Z_VE_taukinf}) is a nearly universal function of $\omega/\omega_p$, at fixed renormalized Fermi velocity $v_F^*/c$ and first Landau parameter $F_1^S$. In particular, in this regime the curves overlap with the one for the purely reactive/elasticlike limit $\omega \tau_c \rightarrow +\infty$: hence, the saturation limit of the inverse surface resistance is determined by the reactive/elasticlike component of the generalized hear modulus $\nu(\omega)$. In other words, the phenomenology of anomalous skin effect for $\omega \gg v_F^* \left|q\right|$ reflects the reactive character of the Fermi-surface shear response, enforced by the incident electromagnetic field. 

We now discuss how the anomalous behavior of the surface impedance in the propagating shear regime results from the competition between the different length scales, such as (\ref{eq:delta_skin_Drude}), (\ref{eq:London_depth}), and (\ref{eq:delta_nu}). For simplicity we focus on the regime $\tau_K \gg \tau_c$, which implies values of $\omega \tau_K$ close to the saturation limit of $\mathrm{Re}\left\{Z(\omega)\right\}^{-1}$ in Fig.\@ \ref{fig:Z_FL}. In such regime, the interface electric field (\ref{eq:E_int_VE_low}) becomes
\begin{multline}\label{eq:E_0_dimful}
E(0,\omega)=-\frac{1}{\pi}\left. \frac{\partial E(z,\omega)}{\partial z} \right|_{z=0^-} \\ \times \int_{-\infty}^{+\infty} d q \left\{\frac{-(\omega_p)^2}{c^2}  \left[1+\frac{i}{\omega \tau_K}+i \frac{\nu(\omega)}{\omega} q^2 \right]^{-1}-q^2\right\}^{-1} \\= \frac{1}{\pi}\left. \frac{\partial E(z,\omega)}{\partial z} \right|_{z=0^-} \\ \times \int_{-\infty}^{+\infty} d q\left\{\frac{-i}{(\lambda_L)^2 \left[ \frac{1-i \omega \tau_K}{\omega \tau_K}+(\delta_{\nu})^2 q^2\right] }+q^2\right\}^{-1}.
\end{multline}
Equation (\ref{eq:E_0_dimful}) can also be recast in a form that highlights another length scale of the problem: 
\begin{multline}\label{eq:E_0_dimful_lG}
E(0,\omega)=-\frac{1}{\pi}\left. \frac{\partial E(z,\omega)}{\partial z} \right|_{z=0^-} \int_{-\infty}^{+\infty} d q \left\{i/\left\{(\lambda_L)^2(\delta_\nu)^2 \times \right.\right. \\  \left.\left. \left[(1-i \omega \tau_K)(1-i\omega \tau_c) (l_G)^{-2}  + q^2\right] \right\} -q^2\right\}^{-1},
\end{multline}
where 
\begin{equation}\label{eq:l_G}
l_G=\sqrt{\nu(0) \tau_K}
\end{equation}
is the Gurzhi length \cite{Gurzhi-1968,Levchenko-2020}, which is often invoked in studies of size effects on the DC conductivity of viscous liquids. 

When $\left| (\omega \tau_K)^{-1}-i \right| \gg \left| \delta_\nu \right|$, we can neglect the nonlocal $\delta_\nu$-dependent term in Eq.\@ (\ref{eq:E_0_dimful}) and we directly recover the standard result (\ref{eq:E_q_sigma}) of the Drude model. Further specializing to the limit $\omega \tau_K \rightarrow 0$ gives the associated skin depth $\delta_s \equiv \lambda_L/\sqrt{\omega \tau_K}$, i.e.\@, Eq.\@ (\ref{eq:delta_skin_Drude}). 
On the other hand, if $\omega \tau_K \rightarrow +\infty$ we retrieve the asymptotic limit (\ref{eq:Z_VE_taukinf}), which implies
\begin{multline}\label{eq:E_0_dimful_tauKinf}
E(0,\omega)= \frac{1}{\pi}\left. \frac{\partial E(z,\omega)}{\partial z} \right|_{z=0^-} \\ \times  \int_{-\infty}^{+\infty} d q \left[\frac{1}{(\lambda_L)^2 +i (\lambda_L)^2(\delta_{\nu})^2 q^2 }+q^2\right]^{-1}.
\end{multline}
From Eq.\@ (\ref{eq:E_0_dimful_tauKinf}) we see that the interface electric field, and the surface impedance through Eq.\@ (\ref{eq:Z_gen_1D}), is determined by the competition between $\lambda_L$ and $\delta_{\nu}$ in the relaxationless limit. In the dissipative (viscous) regime $\omega \tau_c \ll 1$ (still $\omega \gg v_F^{*} \left|q \right|$), we have $\nu(\omega)\approx \nu(0)$, where $\nu(0)$ is the DC viscosity coefficient (\ref{eq:nu_0}), so that
\begin{multline}\label{eq:E_0_dimful_tauKinf2}
E(0,\omega)= \frac{1}{\pi}\left. \frac{\partial E(z,\omega)}{\partial z} \right|_{z=0^-}  \\ \times \int_{-\infty}^{+\infty} d q \left\{\left[(\lambda_L)^2 +i (\lambda_L)^2(\delta_{\nu}^0)^2 q^2\right]^{-1} +q^2\right\}^{-1}, 
\end{multline}
where we have identified the real-valued length scale \cite{Levchenko-2020}
\begin{equation}\label{eq:delta_nu_0}
\delta_\nu^0=\sqrt{\frac{\nu(0)}{\omega}}=\lim_{\omega\rightarrow 0} \delta_\nu. 
\end{equation}
In the ``viscous'' low-frequency limit $1\ll \left| (\delta_\nu^0 q)^2 \right|$, which means $\omega \ll \nu(0)\left|q^2 \right|$, the $q$-integration in Eq.\@ (\ref{eq:E_0_dimful_tauKinf2}) gives the surface impedance directly from Eq.\@ (\ref{eq:Z_gen_1D}):
\begin{multline}\label{eq:Z_viscous}
Z(\omega)=\omega \mu_0 \frac{\delta_s^{\rm hd}}{\sqrt{2}} \left[\sin\left(\frac{\pi}{8}\right)+i \cos\left(\frac{\pi}{8}\right)\right], \\ \omega \ll \nu(0)\left|q^2 \right|.
\end{multline}
In Eq.\@ (\ref{eq:Z_viscous}) we recognize a skin depth in viscous regime,
\begin{equation}\label{eq:skin_depth_viscous}
\delta_s^{\rm hd}=\sqrt{\lambda_L \delta_{\nu}^0} \propto \omega^{-\frac{1}{4}},
\end{equation}
which is the geometric average between the London penetration depth $\lambda_L$ and the viscosity-related length scale $\delta_\nu^0$. 
Thus, with Eqs.\@ (\ref{eq:Z_viscous}) and (\ref{eq:skin_depth_viscous}) we retrieve the results (3.41) and (3.48) in Ref.\@ \onlinecite{Levchenko-2020} as a special case of the general formulation (\ref{eq:Z_VE_taukinf}). From a macroscopic viewpoint, these equations can be obtained directly through Eqs.\@ (\ref{eq:wave_mat_q}) (neglecting the displacement current $\boxed{A}$) and (\ref{eq:Z_gen_1D}), using the Navier-Stokes equation to relate the current density $\vec{J}(\vec{q},\omega)$ to $\vec{E}(\vec{q},\omega)$, and neglecting the frequency dependence of the generalized shear modulus $\nu(\omega)$. The formal equivalence between the result (\ref{eq:Z_viscous}) and Ref.\@ \onlinecite{Levchenko-2020} highlights the universality of the skin effect phenomenology in ``viscous'' (hydrodynamic \cite{Gurzhi-1968}) regime $\omega \tau_c \rightarrow 0$: coupled electron-phonon fluids \cite{Levchenko-2020} and Fermi-liquid interactions (this paper) lead to the same expression for $Z(\omega)$, while differences between the models appear in the respective microscopic expressions for the viscosity coefficient $\nu(0)$. This also means that Fermi-liquid results of this paper in the regime $\omega \tau_c \rightarrow 0$, like Eq.\@ (\ref{eq:E_0_dimful_tauKinf2}), may be readily extended to take into account the influence of electron-phonon coupling on electron viscosity \cite{Gurzhi-1959,Levchenko-2020}. 

Turning to the collisionless regime $\omega \tau_c \rightarrow +\infty$, we realize that the interface electric field (\ref{eq:E_0_dimful_tauKinf}) becomes
\begin{multline}\label{eq:E_0_dimful_shear}
E(0,\omega)= \frac{1}{\pi}\left. \frac{\partial E(z,\omega)}{\partial z} \right|_{z=0^-} \\ \times \int_{-\infty}^{+\infty} d q \left\{\left[(\lambda_L)^2 - (\lambda_L)^2(\delta_{\rm sh})^2 q^2\right]^{-1} +q^2\right\}^{-1}, 
\end{multline}
where we have identified a length scale
\begin{equation}\label{eq:skin_VE_low}
\delta_s^{\mathrm{sh}}=\frac{\sqrt{\mu_s}}{\omega}= \frac{v_F^{*}}{\omega} \sqrt{\frac{1}{5} \left(1+\frac{F_1^S}{3}\right)} \propto \omega^{-1},
\end{equation}
linked to the Fermi-liquid reactive shear modulus $\mu_s$; see Eq.\@ (\ref{eq:mu_FL}). If $1 \ll \left|\delta_s^{\rm sh} q^2 \right|$, which means $\omega \ll \sqrt{\mu_s} \left|q \right|^2$, the $q$-dependent term in square brackets in Eq.\@ (\ref{eq:E_0_dimful_shear}) dominates over $(\lambda_L)^2$. Then, the $q$-integration in Eq.\@ (\ref{eq:E_0_dimful_shear}) yields the surface impedance from Eq.\@ (\ref{eq:Z_gen_1D}):
\begin{equation}\label{eq:Z_shear2}
Z(\omega)=\omega \mu_0 \delta_s^{\rm el} \frac{1-i}{2}, \omega \ll \sqrt{\mu_s} \left|q \right|^2,
\end{equation}
controlled by the skin depth
\begin{equation}\label{eq:skin_depth_reactive}
\delta_s^{\rm el}=\sqrt{\lambda_L \delta_s^{\mathrm{sh}}}.
\end{equation}
Hence the skin depth (\ref{eq:skin_depth_reactive}) is determined by the geometric average between $\lambda_L$ and $\delta_s^{\mathrm{sh}}$, and therefore it is controlled by the reactive shear modulus (\ref{eq:mu_FL}). Table \ref{tab:skin_depth} summarizes the values assumed by the skin depth in all analyzed regimes. 

\begin{table}
	\centering
		\begin{tabular}{| c | c |} 
 \hline
 Drude/Ohmic & $\delta_s(\omega)=\frac{\lambda_L}{\sqrt{\omega \tau_K}}=\sqrt{\frac{2}{\mu_0 \omega \sigma_{DC}}} \propto \omega^{-\frac{1}{2}}$ (\ref{eq:delta_skin_Drude}) \\
$$ & $\lambda_L=\frac{c}{\omega_p}=\sqrt{\frac{m}{\mu_0 n e^2}}$ (\ref{eq:London_depth}) \\
 \hline
 Anomalous & $\delta_s^{\mathrm{an}}=\left[\frac{4}{3 \pi} \frac{c^2}{\omega (\omega_p)^2} v_F \right]^{\frac{1}{3}} \propto \omega^{-\frac{1}{3}}$ (\ref{eq:skin_anom}) \\ 
 \hline
Hydrodynamic & $\delta_s^{\rm hd}=\sqrt{\lambda_L \delta_{\nu}^0} \propto \omega^{-\frac{1}{4}}$ (\ref{eq:skin_depth_viscous}) \\ 
$$ & $\delta_\nu^0=\sqrt{\left(1+\frac{F_1^S}{3}\right)\frac{(v_F^{*})^2 \tau_c}{5 \omega}} \propto \sqrt{\frac{\tau_c}{\omega}}$ (\ref{eq:delta_nu_0}) \\
 \hline
Collisionless & $\delta_s^{\rm el}=\sqrt{\lambda_L \delta_s^{\mathrm{sh}}}\propto \omega^{-\frac{1}{2}}$ (\ref{eq:skin_depth_reactive}) \\
$$ & $\delta_s^{\mathrm{sh}}=\frac{v_F^{*}}{\omega} \sqrt{\frac{1}{5} \left(1+\frac{F_1^S}{3}\right)} \propto \omega^{-1}$ (\ref{eq:skin_VE_low}) \\ 
 \hline
\end{tabular} \caption{\label{tab:skin_depth} Skin depth in various regimes for the charged Fermi liquid. The Drude value (\ref{eq:delta_skin_Drude}) is valid for finite $\tau_{K}$, while all other quoted results hold in the relaxationless regime $\tau_{K} \rightarrow +\infty$.}
\end{table}

Crossovers among the different regimes characterized by the skin depths \ref{tab:skin_depth} may be envisaged on the basis of qualitative criteria analogous to the normal/anomalous condition $\delta_s(\omega)=l_{MFP}$ in Section \ref{Skin_anom}. In particular, in relaxationless Fermi liquids at finite temperatures, a high-frequency crossover between the normal and propagating-shear skin effects may be qualitatively set at
\begin{equation}\label{eq:cross_skin_ps}
\delta_s<l_c=v_F^{*} \tau_c,
\end{equation}
where the skin depth assumes the shear-propagation value (\ref{eq:skin_VE_low}), consistently with the saturation of the surface resistance in Fig.\@ \ref{fig:Z_FL}. In the same way, one expects a crossover between anomalous and hydrodynamic regimes for 
\begin{equation}\label{eq:cross_skin_hydro}
\delta_s^{\mathrm{an}}<l_c,
\end{equation}
whereby the skin depth becomes hydrodynamic, i.e.\@, (\ref{eq:skin_depth_viscous}) \cite{FZVM-new}. Using the Fermi-liquid collision time (\ref{eq:tau_qp_fin2}) in Eq.\@ (\ref{eq:cross_skin_hydro}), one realizes that such hydrodynamic regime occurs at frequencies close to the DC limit, $\omega \lesssim \left(10^{-19} \div 10^{-15}\right) \omega_p$, for $F_1^S=\left(6 \div 50\right)$ and a representative electron density $n=10^{23}$ cm$^{-3}$ for standard metals. As such, the hydrodynamic skin effect is practically undetectable by optical means, its consequences being mainly visible in transport experiments on clean samples; see Sec.\@ \ref{FL_DC}.

\subsubsection{High-frequency regime in the Drude model}\label{Local_high}

To contrast the results in Sec.\@ \ref{Skin_Drude}, it is instructive to analyze what happens to the Drude dielectric response when $\omega > \sqrt{\sigma_T(\vec{q},\omega)/(\epsilon_0 \tau_{K})}$, that is $\omega>\omega_p$: this amounts to neglecting the conduction term $\boxed{B}$ in Eq.\@ (\ref{eq:wave_mat_q}) in favor of the displacement current term $\boxed{A}$. In this case, we have
\begin{equation}\label{eq:E_Drude_high}
E(0,\omega)=2 \left. \frac{\partial E(z,\omega)}{\partial z} \right|_{z=0^-} \frac{1}{\frac{\omega^2}{c^2}\epsilon_0 \epsilon_T(\vec{q},\omega)-q^2},
\end{equation} 
which corresponds to Eq.\@ (\ref{eq:E_q_sigma}) if we substitute $\epsilon_T(\vec{q},\omega)-1$ in the latter with $\epsilon_0 \epsilon_T(\vec{q},\omega)$. Using the Drude dielectric function and Fourier-transforming back to real space, we achieve 
\begin{multline}\label{eq:E_int_Drude_high}
E(0,\omega)=\frac{1}{2 \pi} \left. \int_{-\infty}^{+\infty} dq E(q,\omega) e^{-i q z}\right|_{z=0} \\ =- \left. \frac{\partial E(z,\omega)}{\partial z} \right|_{z=0^-} \frac{c}{\omega_p} \sqrt{\frac{i + \omega \tau_{K}}{\omega \tau_{K} \left[1-\left(\frac{\omega}{\omega_p}\right)^2\right]-i \left(\frac{\omega}{\omega_p}\right)^2}}. 
\end{multline}
From Eqs.\@ (\ref{eq:Z_gen_1D}) and (\ref{eq:E_int_Drude_high}), we arrive at the surface impedance: 
\begin{equation}\label{eq:Z_Drude_high}
Z(\omega)= \mu_0 \omega \frac{c}{\omega_p}  \sqrt{\frac{i+ \omega \tau_{K}}{
  i \left(\frac{\omega}{\omega_p}\right)^2 - \omega \tau_{K} + \left(\frac{\omega}{\omega_p}\right)^3  \tau_{K}}},
\end{equation}
which is equivalent to the expression (\ref{eq:Z_Drude_n}) obtained from the Drude refractive index. For $\omega >\omega_p$ the surface impedance quickly drops to the constant real value $\lim_{\omega\rightarrow +\infty} Z(\omega)=\frac{c}{\omega_p}$ due to the displacement current. 
These results are modified by the presence of a generalized shear modulus (\ref{eq:nu_q0_vel}). 

\subsubsection{Propagating shear regime at high frequencies}\label{Skin_VE_high}

In principle, the spatial nonlocality of the dielectric function (\ref{eq:eps_T_series_rel_gen1bis}) could affect the surface impedance even for $\omega>\omega_p$. Hence, it is worth checking whether quantitative differences with respect to the Drude calculation \ref{Local_high} emerge in such regime. In this case, Eq.\@ (\ref{eq:wave_mat_q}), with the displacement current term $\boxed{A}$ and neglecting the conduction term $\boxed{B}$, gives
\begin{multline}\label{eq:E_VE_high}
E(q,\omega)=2 \left. \frac{\partial E(z,\omega)}{\partial z} \right|_{z=0^-}\\ \cdot  \left[\mu_0 i \omega \sigma_T(\vec{q},\omega)+\frac{\omega^2}{c^2} \epsilon_T(\vec{q},\omega)-q^2\right]^{-1}= \\ 2 \left. \frac{\partial E(z,\omega)}{\partial z} \right|_{z=0^+} \left[\frac{\omega^2}{c^2} \epsilon_T(\vec{q},\omega)-q^2\right]^{-1}. 
\end{multline}
However, it turns out that the differences between the Drude surface impedance and the one stemming from Eq.\@ (\ref{eq:E_VE_high}) are negligible, as shown in Appendix \ref{Shear_Z_high}. Therefore, the best regime to seek signatures of Fermi-surface shear reactance in the surface impedance is the one of section \ref{Skin_VE_low}, i.e.\@, $\omega< \omega_p$. 

\subsubsection{Surface impedance from the refractive indexes and slip length}\label{eq:Z_VE_n}

The derivations in Secs.\@ \ref{Skin_VE_low} and \ref{Skin_VE_high} illustrate how the phenomenology of anomalous skin effect is affected by the propagation of the shear-polariton in the Fermi liquid. However, so far we have considered the boundary conditions of the problem in terms of specular/diffusive scattering at the vacuum-metal interface \cite{Sondheimer-2001, Kiselev-2019}. 
An alternative way to calculate the Fermi-liquid surface impedance in propagating shear regime is to exploit the relation between $Z(\omega)$, the reflection coefficient $r(\omega)$, and the refractive index, similarly to Eq.\@ (\ref{eq:Z_r}) for the Drude model. 
As argued in section \ref{viscoel_q0_rel}, Fermi-liquid theory gives two polariton modes (\ref{eq:VE_modes}), hence two refractive indexes $n_1(\omega)$ and $n_2(\omega)$, for each $\omega$ in the regime $\omega \gg v_F^* \left|q\right|$ and $\tau_c \ll \tau_{K}$, where the dielectric function is (\ref{eq:eps_T_el_VE_rel}). In such regime, the vibration of the Fermi surface macroscopically resembles the response of a viscous charged liquid, so that the constitutive relations at the interface with vacuum may be derived from fluid dynamics and expressed in terms of the slip length $\lambda_s$. This allows us to utilize the results of Ref.\@ \onlinecite{FZVM-2014} in the calculation of the surface impedance. 
\begin{figure}[ht]
\includegraphics[width=0.65\columnwidth]{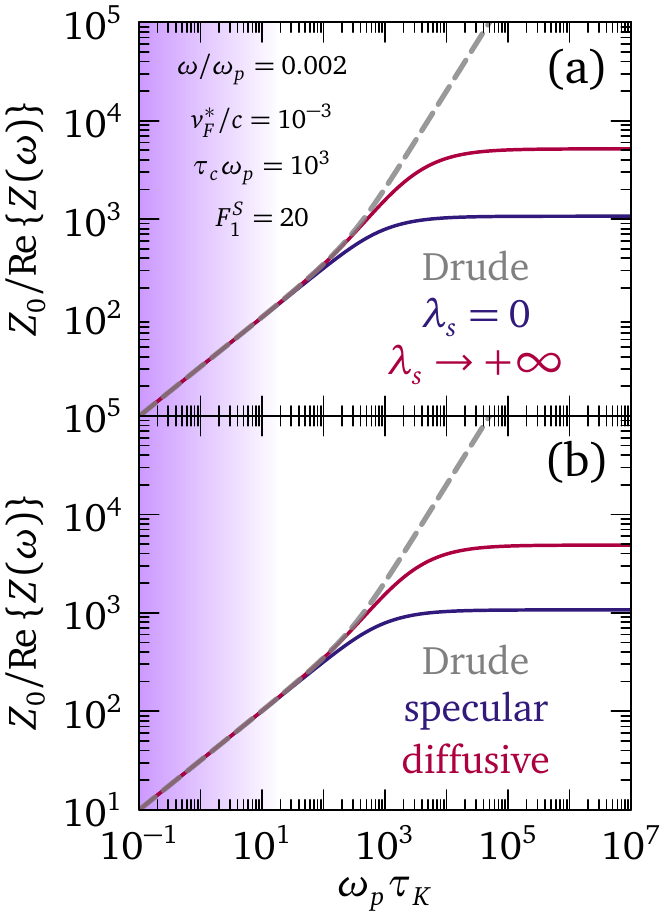}
\caption{\label{fig:Z_FL_n}  Inverse surface resistance $\mathrm{Re}\left\{Z(\omega)\right\}^{-1}$ in propagating shear regime as a function of $\omega \tau_K$, using the slip-length constitutive relations \cite{FZVM-2014} in accordance with Eq.\@ (\ref{eq:Z_FZVM}), at frequency $\omega=2 \cdot 10^{-3}\omega_p$, with renormalized Fermi velocity $v_F^*=10^{-3}c$, collision time $\tau_c=10^3 \omega_p$, and first Landau parameter $F_1^S=20$. In the purple-shaded area $\tau_c > \tau_{K}$ and the dielectric function (\ref{eq:eps_T_el_VE_rel}) is no longer accurate. Panel (a) shows the opposite cases $\lambda_s=0$ and $\lambda_s\rightarrow +\infty$ as blue and red solid curves, respectively. The dashed gray curve is the Drude result (\ref{eq:Z_Drude_n}). Panel (b) shows the results for specular and diffusive boundary scattering, stemming from Eqs.\@ (\ref{eq:lambda_s}), as blue and red solid curves, respectively. In the specular case, we assume a surface roughness $h=h'=100/k_F$.  
}
\end{figure}

In terms of the refractive indexes, the reflection coefficient at normal incidence for a viscous electron liquid reads
\begin{subequations}
\begin{equation}\label{eq:r_VE}
r(\omega)=t_1+t_2-1,
\end{equation}
\begin{multline}\label{eq:t_VE}
t_{j}=\frac{2 (n_{\bar{j}}-1)}{(n_j+1)(n_{\bar{j}}-n_j)}\frac{1-i \omega/c n_{\bar{j}} \lambda_s}{1+(1-n_{\bar{j}}-n_j) i \omega/c \lambda_s}, \\ \left\{j, \bar{j}\right\}=\left\{1,2\right\}. 
\end{multline}
\end{subequations}
where $\lambda_s \in \left(0,+\infty\right)$ is the slip length, introduced in Sec.\@ \ref{ABC}.
With the expression (\ref{eq:r_VE}), we can again employ the relation (\ref{eq:Z_r}) for the surface impedance, which yields
\begin{equation}\label{eq:Z_FZVM}
\frac{Z(\omega)}{Z_0}=\frac{n_1+n_2+i\frac{\lambda_s}{c} \omega \left[ 1-(n_1)^2-(n_2)^2-n_1 n_2 \right]}{1+ n_1 n_2 \left[ 1- i \frac{\lambda_s}{c} \omega (n_1+n_2) \right]}
\end{equation}
One verifies that the surface impedance (\ref{eq:Z_FZVM}) reduces to the Drude expression (\ref{eq:Drude_r}) consistently for $\nu(\omega) \rightarrow 0$: in this case, only one optical mode is propagating in the metal, that is $n_1(\omega) \rightarrow +\infty$ or $n_2(\omega) \rightarrow + \infty$. The expression (\ref{eq:Z_FZVM}) reduces to Eqs.\@ (3.44) and (3.46) of Ref.\@ \onlinecite{Levchenko-2020}, for $\lambda_s=0$ and $\lambda_s \rightarrow +\infty$ respectively, if one neglects the displacement current, approximates $\nu(\omega) \approx \nu(0)$, and assumes no reflection at the vacuum-metal interface.

Figure \ref{fig:Z_FL_n}(a) shows the inverse surface resistance $Z_0/\mathrm{Re}\left\{Z(\omega)\right\}$ as a function of momentum-relaxation time $\omega_p \tau_{K}$, for renormalized Fermi velocity $v_F^*/c=10^{-3}$, first Landau parameter $F_1^S=20$, collision time $\tau_c \omega_p=1000$, and at frequency $\omega/\omega_p=2 \cdot 10^{-3}$. Blue and red curves show the results for $\lambda_s=0$ and $\lambda_s \rightarrow +\infty$, respectively. We see that, even for constitutive relations in terms of the slip length, we retrieve the phenomenology of anomalous skin effect for $\omega \gg v_F^* \left|q\right|$: in particular, with increasing relaxation time $\tau_{K}$ we approach the regime $\tau_c \ll \tau_{K}$ of Sec.\@ \ref{viscoel_q0_rel}.  
The slip-length parametrization may be converted into specular or diffuse interface scattering by the means of Eqs.\@ (\ref{eq:lambda_s}). Using the latter into Eq.\@ (\ref{eq:Z_FZVM}), we obtain the surface impedance displayed in Fig.\@ \ref{fig:Z_FL_n}(b). In the specular case we assume $h=h'=100/k_F$, which corresponds to $h \approx 5$ nm for the present parameters. Since the slip length for diffusive scattering is larger than for specular scattering, the associated red and blue curves in Fig.\@ \ref{fig:Z_FL_n}(b) are close to the infinite- and zero-slip length results in Fig.\@ \ref{fig:Z_FL_n}(a), respectively. 
Overall, quantitative differences with respect to the specular-scattering results in Fig.\@ \ref{fig:Z_FL} emerge in the asymptotic limit $\tau_{K}\rightarrow +\infty$, however the qualitative outcome remains the same using the slip-length approach. Hence, the saturation of the inverse surface resistance for $\tau_{K} \rightarrow +\infty$ is a robust feature of the propagating shear regime, with an asymptotic limit governed by the generalized shear modulus $\nu(\omega)$. The numerical value of such limit depends parametrically on the boundary conditions assumed for the motion of the electron fluid at the sample interface \cite{Sondheimer-2001}. 

\subsection{Thin-film transmission}\label{Film_T}

Another notable spectroscopic probe of spatial nonlocality in Fermi liquids is the optical transmission through a thin film of thickness $d$. In fact, the nonlocal response of the quasiparticles to the incident electromagnetic wave modulates the transmission coefficient $t_{\rm film}(\omega)$ as a function of frequency $\omega$ in a different way with respect to the Drude model \cite{FZVM-2014}. 
In the following, we analyze the thin-film transmission for the two opposite cases of Sec.\@ \ref{Anomal_tauk}, i.e.\@, in the anomalous skin effect regime, and of Sec.\@ \ref{viscoel_q0_rel}, i.e.\@, in the propagating shear regime. 
\begin{figure}[ht]
\includegraphics[width=0.8\columnwidth]{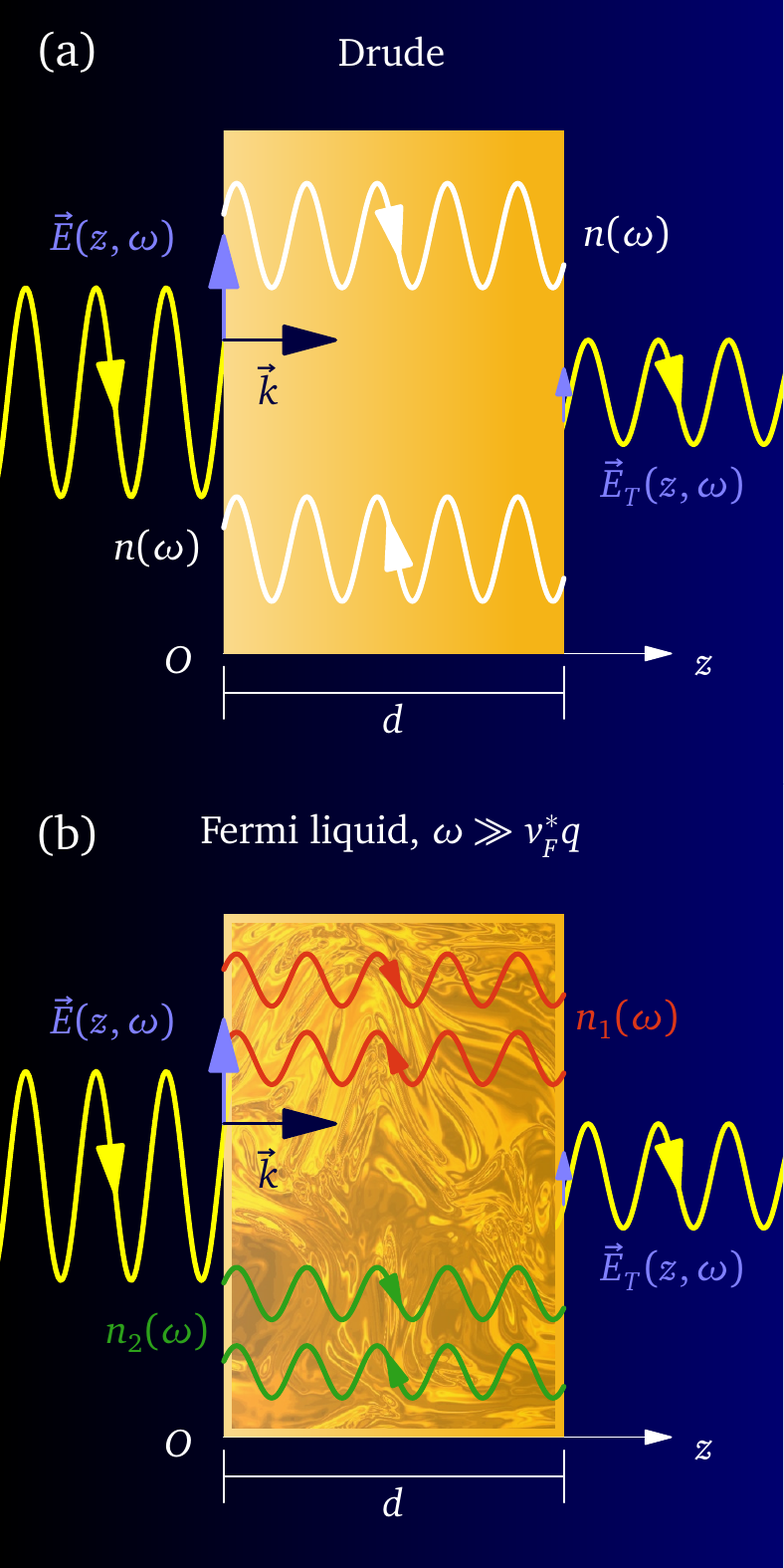}
\caption{\label{fig:slab} Schematic representation of the optical transmission through a Fermi-liquid thin film of thickness $d$. $\vec{E}_T(z,\omega)$ is the transmitted electric field at frequency $\omega$, while $\vec{E}(z,\omega)$ is the electric field at normal incidence with wave vector $\vec{k}$. Oriented wavy lines depict electromagnetic waves. In the Drude/Ohmic case, $\vec{E}_T(z,\omega)$ results from the superposition of two counterpropagating rays with refractive index $n(\omega)$, as depicted in panel (a). In the propagating shear regime of a Fermi liquid, the two different refractive indexes $n_1(\omega)$ ad $n_2(\omega)$ give rise to a total of four rays inside the slab, as illustrated in panel (b).
}
\end{figure}

\subsubsection{Drude regime}\label{t_Drude}

Whenever a single polariton mode propagates into the system, Fresnel equations are sufficient to determine the optical transmission and reflection coefficients at interfaces \cite{Jackson-1962,Dressel-2001}. This is true for the Drude regime --- cf.\@ Eqs.\@ (\ref{eq:eps_T_series_rel_genk}) and (\ref{eq:eps_T_Drude}), and Sec.\@ \ref{Skin_Drude} --- where the only optical mode is the plasmon-polariton. In this case, to obtain the thin-film transmission coefficient $t_{\rm film}(\omega)$ including Fabry-Perot internal reflections inside the slab, one can represent the electric field inside the film as the superposition of two counterpropagating rays with refractive index $n(\omega)=\sqrt{\epsilon_T(\omega)}$, as depicted in Fig.\@ \ref{fig:slab}(a). The continuity of the electric field and of its derivative at the slab boundaries give rise to a system of four equations, from which the slab transmission amplitude results: 
\begin{subequations}\label{eq:t_film_Drude}
\begin{equation}
t_{\rm film}(\omega)= \theta_1 e^{i \frac{\omega}{c} n(\omega) d}+ \theta_2 e^{-i \frac{\omega}{c} n(\omega) d},
\end{equation}
\begin{equation}\label{eq:theta_1}
\theta_1=-\frac{2 \left[ n(\omega)+1\right]}{e^{2 i \frac{\omega}{c} n(\omega) d} \left[ n(\omega)-1\right]^2 - \left[ n(\omega)+1\right]^2 },
\end{equation}
\begin{equation}\label{eq:theta_2}
\theta_2=\frac{2 \left[ n(\omega)-1 \right]}{e^{-2 i \frac{\omega}{c} n(\omega) d} \left[ n(\omega)+1\right]^2 - \left[ n(\omega)-1\right]^2 }.
\end{equation}
\end{subequations}
Equations (\ref{eq:t_film_Drude}) produce the dashed gray lines in Figs.\@ \ref{fig:t_VE} and \ref{fig:tVE_eph}, for the parameters there specified. 

\subsubsection{Anomalous skin effect regime}\label{t_anomal}

In anomalous skin regime, at leading order in $\omega/(v_F^{*} \left|q\right|) \rightarrow 0$ there are three polariton branches, each associated to a frequency-degenerate refractive index, as mentioned in Secs.\@ \ref{Anomal} and \ref{Anomal_tauk}. Hence, in principle one could adopt the approach of the previous Sec.\@ \ref{t_Drude} to calculate the thin-film transmission, i.e.\@, considering constitutive relations at the slab boundaries. However, as mentioned in Sec.\@ \ref{ABC}, Fresnel equations leave the system underdetermined, and one would require ABCs to form a closed problem. The ABCs approach is convenient in the regime $\omega \gg v_F^{*} \left|q\right|$, as an ABC required to close the problem can be constructed in analogy with fluid dynamics as detailed in Sec.\@ \ref{t_viscoel}. Instead, in the regime $\omega/(v_F^{*} \left|q\right|) \rightarrow 0$ an alternative convenient approach is to deduce a single refractive index $n(\omega)$ from the surface impedance $Z(\omega)$, written in terms of the specularity coefficient $p$ for boundary scattering. This builds directly upon the results of Sec.\@ \ref{Skin_anom}, as one can still utilize standard optical formulas like (\ref{eq:t_film_Drude}) in anomalous regime, provided that the refractive index is deduced from $Z(\omega)$ \cite{Dingle-1953a,Dingle-1953d}:
\begin{equation}\label{eq:n_Z}
n(\omega)=\left[Z(\omega)\right]^{-1},
\end{equation}
as results from Eqs.\@ (\ref{eq:Z_r}) and (\ref{eq:Drude_r}). In the absence of momentum relaxation, $Z(\omega)$ is given by Eq.\@ (\ref{eq:Z_anom}).  The latter, together with Eqs.\@ (\ref{eq:n_Z}) and (\ref{eq:t_film_Drude}), determines $t_{\rm film}$. At finite momentum relaxation and for $\tau_c \rightarrow+\infty$ one retrieves Reuter-Sondheimer theory of anomalous skin effect, expressed by Eqs.\@ (\ref{eq:Z_RandS}) in the microwave regime. 

Based on Eqs.\@ (\ref{eq:Z_RandS}), the ratio (\ref{eq:deltas_mfp}) between the mean-free path and the Drude skin depth determines the crossover from the Drude regime $l_{MFP} \lesssim \delta_s$ to anomalous skin effect regime $l_{MFP} \gg \delta_s$. $\tau_{K}$ from Umklapp and acoustic-phonons scattering, as determined by the total relaxation time (\ref{eq:Mat_tau}), gives the following qualitative picture, substantiated by Appendix \ref{Delta_skin_cons}: at room temperature, the skin depth is larger than, or comparable to, the mean-free path at all frequencies, so that the results from the Drude model are expected to hold. At cryogenic temperatures, there is an extended low-frequency window in which $\delta_s/l_{MFP}\ll 1$, so that anomalous skin effect develops. 
The case including both finite $\tau_{K}$ and $\tau_c$ as well as quasiparticle interactions requires a generalization of Reuter-Sondheimer theory, which in principle can be obtained from the Fermi-liquid optical conductivity (\ref{eq:sigma_T_mb_r}) and deserves future work. In this respect, the low-temperature regime in which $\tau_c >\tau_{K}$ is the most interesting, since a finite collision rate may modify the optical conductivity and the surface impedance with respect to Reuter-Sondheimer theory, as exemplified by Eq.\@ (\ref{eq:sigma_anom_FL_r1}). 
The low-frequency blue (red) solid curves in the green-shaded areas of Fig.\@ \ref{fig:t_VE} show the thin-film transmission in anomalous skin regime for specular (diffusive) boundary scattering, for particle density $n=10^{23}$ cm$^{-3}$, first Landau parameter $F_1^S=20$ and $\tau_{K} \rightarrow +\infty$. In this case, the optical conductivity is (\ref{eq:sigma_anom_FL}). Notice that this conductivity is independent of $\tau_c$, which makes $t_{\rm film}(\omega)$ temperature-independent in the absence of relaxation: this is why Figs.\@ \ref{fig:t_VE}(a) and \ref{fig:t_VE}(b) report identical results in anomalous skin regime. 

A qualitative constraint on the frequency domain where Eq.\@ (\ref{eq:sigma_anom_FL}) holds can be derived from the condition $\omega \ll v_F^{*} \left|q\right|$. In such regime, $\mathrm{Re}\left\{q\right\}\ll \mathrm{Im}\left\{q\right\}$, so that an equivalent criterion is $\left|\mathrm{Im}\left\{q\right\}\right|=\frac{\omega}{c}\left|\mathrm{Im}\left\{n(\omega)\right\}\right|> \frac{\omega}{v_F^{*}}$, or
\begin{equation}\label{eq:n_EHbound}
\left|\mathrm{Im}\left\{n(\omega)\right\}\right|> n_{\rm min}=\frac{c}{v_F^{*}}. 
\end{equation}
Using Eqs.\@ (\ref{eq:n_Z}) and (\ref{eq:n_EHbound}), we obtain the frequency $\omega_{\rm max}$ such that $\left|\mathrm{Im}\left\{n(\omega_{\rm max})\right\}\right|=n_{\rm min}$: this frequency corresponds to the blue (red) vertical dashed line in Fig.\@ \ref{fig:t_VE} for specular (diffuse) boundary scattering. 

Finally, hydrodynamic skin effect in finite-temperature charged Fermi liquids with $\tau_{K} \gg \tau_c$ develops for frequencies close to the DC limit as indicated by the criterion (\ref{eq:cross_skin_hydro}). However, as discussed in Sec.\@ \ref{Skin_VE_low}, this regime is practically beyond the detective power of standard optical techniques and it mainly shows in transport experiments.

\subsubsection{Propagating shear regime}\label{t_viscoel}

\begin{figure}[t]
\includegraphics[width=0.9\columnwidth]{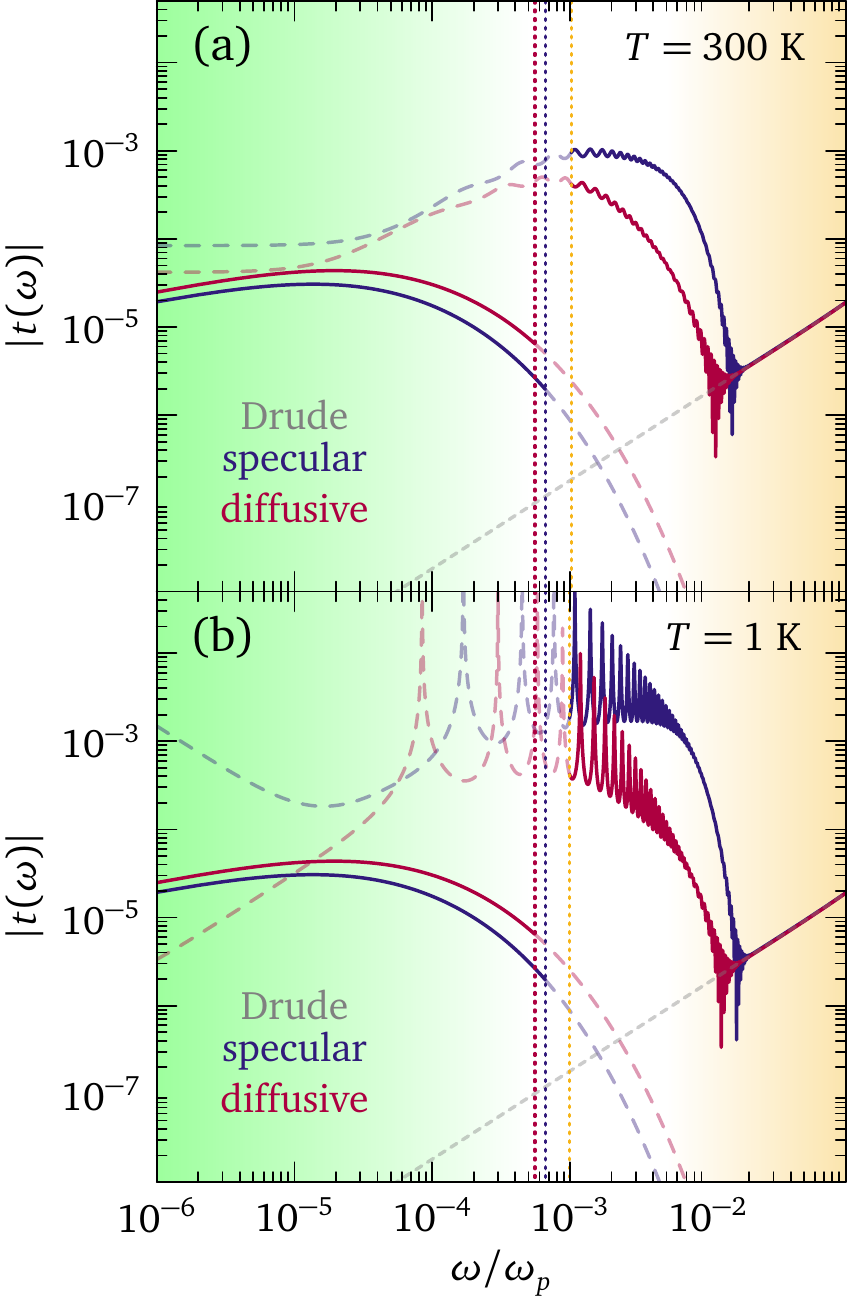}
\caption{\label{fig:t_VE} Modulus of the thin-film transmission coefficient $\left|t_{\rm film}(\omega)\right|$ for a 3D Fermi liquid as a function of normalized frequency $\omega/\omega_p$, for electron density $n=10^{23}$ cm$^{-3}$, Landau parameters $F_0^S=1$, $F_1^S=20$, in the absence momentum relaxation. Blue and red curves refer to specular and diffuse boundary scattering respectively. The curves for frequencies up to the green dashed vertical line result from the refractive index (\ref{eq:n_Z}) and the surface impedance (\ref{eq:Z_anom}) in anomalous skin effect regime, in accordance with the constraint (\ref{eq:n_EHbound}). Curves for frequencies above the orange dashed line stem from the propagating-shear model of Eq.\@ (\ref{eq:t_viscoel}) with the two polariton branches (\ref{eq:VE_modes}), consistently with the constraint (\ref{eq:n_EHbound2}). Panels (a) and (b) show the result at $T=300$ K and $T=1$ K, respectively.  The dashed gray curve shows the result of the Drude model, according to Eq.\@ (\ref{eq:t_film_Drude}). 
 }
\end{figure}
In propagating shear regime $\omega \gg v_F^{*} \left|q\right| \rightarrow +\infty$, the two frequency-degenerate polariton modes (\ref{eq:VE_modes_rel}) give rise to the two refractive indexes (\ref{eq:n_charged}). In this case, we are able to identify appropriate ABCs from the analogy between Fermi-liquid electrodynamics and the electromagnetic response of viscous charged fluids \cite{FZVM-2014}, which makes the computation of the thin-film transmission a closed problem. Such approach follows Sec.\@ IIB of Ref.\@ \onlinecite{FZVM-2014}: in essence, the continuity of the electric field and of its derivative, together with the constitutive relation stemming from the linearized Navier-Stokes equation, generates a system of six linear equations. Here we quote the final result: 
\begin{multline}\label{eq:t_viscoel}
t(\omega)=\frac{t_{\rm film}(\omega)}{t_v}=e^{-i \frac{\omega}{c} d} \left[t_1e^{i \frac{\omega}{c} n_1(\omega) d}+\theta_1 e^{-i \frac{\omega}{c} n_1(\omega) d} \right. \\ \left. + t_2e^{i \frac{\omega}{c} n_2(\omega) d}+\theta_2 e^{-i \frac{\omega}{c} n_2(\omega) d} \right].
\end{multline}
In Eq.\@ (\ref{eq:t_viscoel}), the thin-film transmission coefficient is calibrated against a slice of vacuum of the same thickness $d$, as done in experimental practice.
The coefficients $\left(t_1,\theta_1,t_2,\theta_2\right)$ in Eq.\@ (\ref{eq:t_viscoel}), which describe the amplitude of two counter-propagating waves inside the slab for each optical mode, are determined numerically through Eq.\@ (9) of Ref.\@ \onlinecite{FZVM-2014}. \\

In the following, we use the relations (\ref{eq:lambda_s}) between the slip length $\lambda_s$ and the generalized shear modulus $\nu(\omega)$, as done in Sec.\@ \ref{eq:Z_VE_n}, to convert the slip-length parametrization into diffuse/specular boundary scattering. The blue and red solid curves in the orange-shaded areas of Fig.\@ \ref{fig:t_VE} show the modulus of the thin-film transmission coefficient $\left|t_{\rm film}(\omega)\right|$ as a function of frequency $\omega$ in the relaxationless limit $\tau_{K}\rightarrow +\infty$, for specular and diffuse boundary scattering respectively. Particle density is fixed to $n=10^{23}$ cm$^{-3}$ and the first Landau parameter is $F_1^S=20$. The Drude result for the same parameters is shown by the dashed gray line for comparison. 

The vertical dashed orange line shows the frequency $\omega_{VE}$ corresponding to the frequency limit $\omega \gg \omega_{eh}^{VE}$ from Sec.\@ \ref{q4}, as the phenomenology of Eq.\@ (\ref{eq:eps_T_el_VE}) applies for $\omega \gg \omega_{eh}^{VE}$. Such constraint translates into
\begin{equation}\label{eq:n_EHbound2}
\mathrm{Re}\left\{n_i(\omega)\right\}<\frac{c}{v_F^{*}}
\end{equation}
for the refractive index, where $n_i(\omega)$ is the refractive index of the shear-polariton ($n_1(\omega)$ and $n_2(\omega)$ swap character at $\omega \tau_{K}=1$). 

Figure \ref{fig:t_VE}(a) suggests a higher absolute value of the thin-film transmission coefficient in propagating shear regime even at $T=300$ K without momentum relaxation, with respect to the Drude model. The room-temperature wiggles as a function of $\omega$ in Fig.\@ \ref{fig:t_VE}(a) are amplified into sizable oscillations in the low-temperature case of Fig.\@ \ref{fig:t_VE}(b), and stem from the mutual interference of the two polaritons (\ref{eq:VE_modes_rel}). The amplification of the oscillations with decreasing temperature originates from the increase in $\tau_c$, which brings the system towards the collisionless regime $\omega \tau_c \gg 1$. In the latter condition, the Fermi surface resonates with a dominant dissipationless reactive response, as analyzed in Sec.\@ \ref{Viscoel_q0}, and radiation couples to such reactive resonance by exchanging energy and momentum: the less dissipative the Fermi-surface resonance is, the more efficient is its coupling to photons, and the more its traces in the optical transmission are visible. 
However, as noted in Secs.\@ \ref{VE_tauk} and \ref{Scatter}, in a realistic solid-state system Galilean invariance is inevitably broken by the crystalline lattice and impurities, which cause the relaxation of quasiparticle momentum. Hence, we have to address the question of whether some traces of spatial nonlocality persist in the transmission spectrum at finite $\tau_{K}$. The next section deals with such question. 

\subsubsection{Role of impurity and phonon scattering}\label{T_phon_imp}
\begin{figure}[t]
\includegraphics[width=0.77\columnwidth]{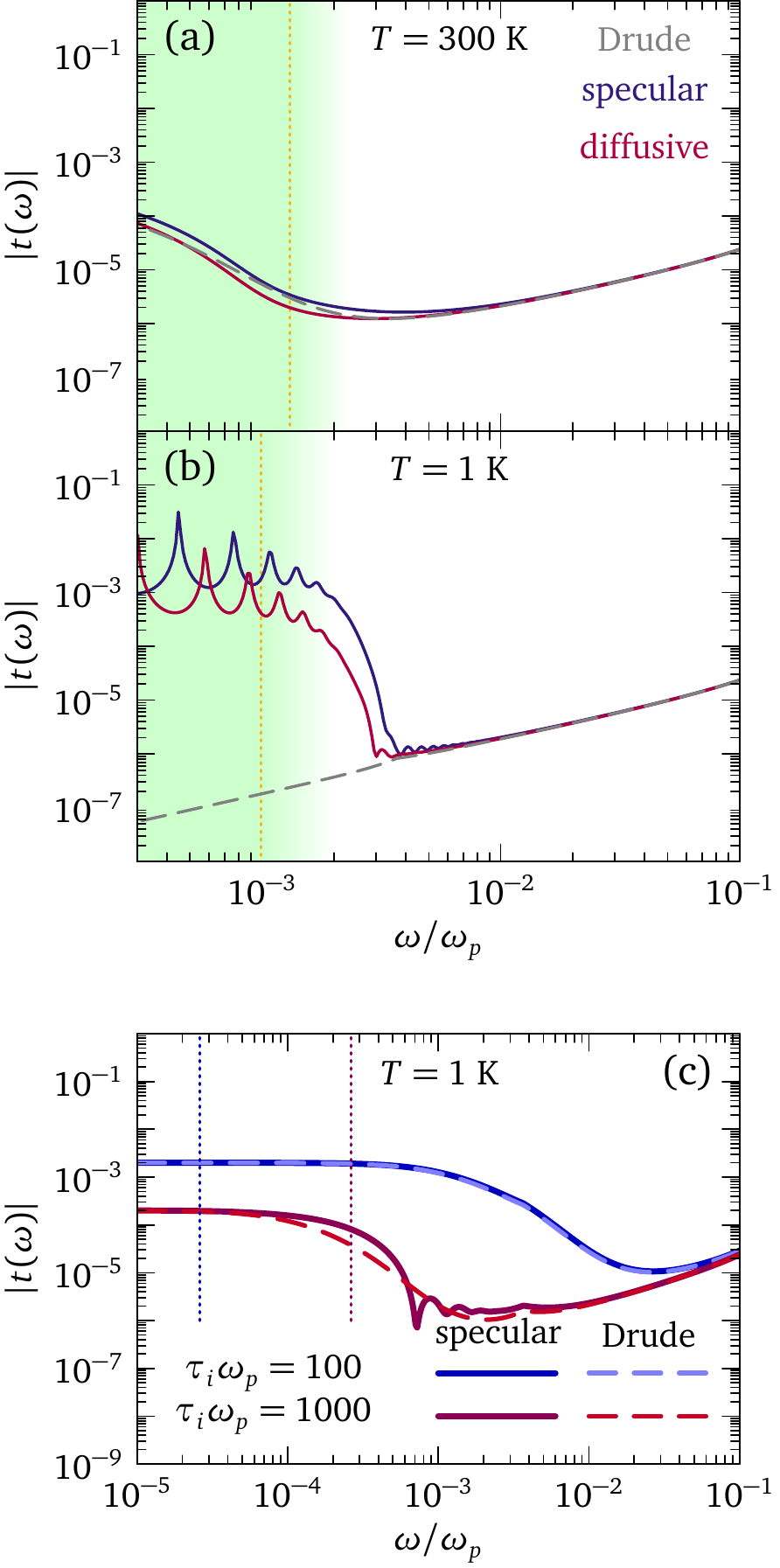}
\caption{\label{fig:tVE_eph} 
Transmission modulus $\left|t_{\rm film}(\omega)\right|$ from Eq.\@ (\ref{eq:t_viscoel}) including impurity and electron-phonon scattering through Eq.\@ (\ref{eq:Mat_tau}). The blue (red) curve refers to specular (diffuse) boundary scattering. Panels (a) and (b) show the results with acoustic phonons and no impurities, at $T=300$ K and $T=1$ K respectively. The blue and red solid curves in panel (c) additionally include impurity scattering using $\tau_i \omega_p=\left\{100, 1000\right\}$, respectively, assuming specular boundary scattering. Dashed curves show the Drude calculation for the same $\tau_{K}$. All other parameters are fixed as in Fig.\@ \ref{fig:t_VE}. 
}
\end{figure}
In the presence of momentum relaxation, the dielectric function in the propagating shear regime becomes (\ref{eq:eps_T_el_VE_rel}). Here I assume a momentum-relaxation time given by Mathiessen's rule (\ref{eq:Mat_tau}), with independent contributions stemming from acoustic phonons, impurities and Umklapp scattering as described in Sec.\@ \ref{Scatter}. As in the momentum-conserving case, the two roots of Eq.\@ (\ref{eq:eps_T_el_VE_rel}) give the frequency-degenerate refractive indexes to use in Eq.\@ (\ref{eq:t_viscoel}) \cite{FZVM-2014}. The resulting absolute value of the transmission coefficient $\left|t(\omega)\right|$ is displayed in Fig.\@ \ref{fig:tVE_eph}: Fig.\@ \ref{fig:tVE_eph}(a) shows the case  at $T=300$ K with acoustic phonon and Umklapp scattering but without impurities, Fig.\@ \ref{fig:tVE_eph}(b) illustrates the case at $T=1$ K with acoustic phonons and Umklapp scattering but no impurities, and Fig.\@ \ref{fig:tVE_eph}(c) displays the calculation at $T=1$ K with phonons, Umklapp scattering and different values of impurity scattering time $\tau_{i}$. In all panels, dashed lines show the result from the standard Drude model, using the same relaxation time $\tau_{K}$ as in the propagating-shear calculation. An orange vertical dashed line in Fig.\@ \ref{fig:tVE_eph}(a) and Fig.\@ \ref{fig:tVE_eph}(b) marks the frequency below which the shear-polariton enters into the Lindhard continuum, so that the viscoelastic model is no longer valid in accordance with the constraint (\ref{eq:n_EHbound2}). The same constraint gives the vertical dashed red and blue lines in Fig.\@ \ref{fig:tVE_eph}(c), for $\tau_{i}\omega_p=\left\{100, 1000\right\}$ respectively. 

The calculations in Fig.\@ \ref{fig:tVE_eph}(a) confirm the educated guess that momentum relaxation conceals any traces of shear-mode propagation at room temperature: the curves for both specular and diffuse scattering are very close to the Drude transmission, even for a Landau parameter as high as $F_1^S=20$. This is because the relaxation time $\tau_{K} \ll \tau_c$ is severely limited by phonons at room temperature, such that momentum-relaxing processes dominate over the nonlocal term $i \omega \nu(\omega) q^2$ in Eq.\@ (\ref{eq:eps_T_el_VE_rel}). A lower $F_1^S>12$ increases $\tau_c$, which in principle favors shear propagation; however, it also rapidly reduces the frequency window in which the model (\ref{eq:eps_T_el_VE_rel}) holds, because it pushes the emergence of the shear-polariton from the continuum to higher frequency, until the shear-polariton never exits from the continuum for $F_1^S\rightarrow 12$ --- see also discussion in Sec.\@ \ref{emerg_VE}. 
However, qualitative differences between the Drude and propagating-shear models emerge in Fig.\@ \ref{fig:tVE_eph}(b), even in the presence of acoustic phonons and Umklapp scattering. At low temperature, $\tau_{K}$ increases due to the progressive freezing of phonon scattering, and oscillations induced by $\nu(\omega)$ emerge for $\omega/\omega_p=\left( 10^{-4} \div 10^{-2}\right)$. These are the same oscillations of Fig.\@ \ref{fig:t_VE}, which partially persist at low temperature even in the presence of relaxation. 

A constraint on sample purity results from Fig.\@ \ref{fig:tVE_eph}(c): decreasing the impurity scattering time $\tau_{i}$ effectively abates nonlocal effects by decreasing $\tau_K$, until the calculation from the propagating-shear model becomes practically indistinguishable from the Drude result at $\tau_{i} \omega_p=100$. For a typical electron density $n \approx 10^{23}$ cm$^{-3}$, this means $\tau_{i}\approx 5.6 \cdot 10^{-15}$ s, while typical Drude scattering rates of metals lie in the range $\tau_{K}\approx \left( 10^{-15} \div 10^{-13}\right)$ s at $T=77$ K \cite{Ashcroft-1976}. Therefore, a sufficiently low impurity concentration is essential to unveil phenomena connected to the Fermi-surface shear rigidity in low-temperature optical spectra. 

\section{Extracting the generalized shear modulus from optics}\label{Opt_exp_nu}

The results of Secs.\@ \ref{Zs} and \ref{Film_T} show that the shear rigidity of the Fermi surface, encoded by the generalized shear modulus $\nu(\omega)$, qualitatively modifies the optical properties of strongly interacting Fermi liquids at low temperature. Conversely, observing such shear-rigidity effects in optical spectroscopy may allow one to extract $\nu(\omega)$ from experiments on a given solid-state Fermi liquid at cryogenic temperatures. This section outlines some strategies to deduce values for $\nu(\omega)$ in the propagating shear regime analyzed in earlier sections. 

The analysis below requires some previous characterization of Fermi-liquid microscopic parameters, namely: the first Landau parameter $F_1^S$, for instance deducible from the effective mass $m^{*}$ extracted from the electronic specific heat; the Fermi velocity $v_F^{*}$, which depends on $m^{*}$ and on electron density $n$ as measured by Hall transport data at high magnetic fields. Complementary experimental knowledge of the relaxation time $\tau_{K}$ and of the collision time $\tau_c$ is also desirable, to compare with the theoretical models (\ref{eq:Mat_tau}) and (\ref{eq:tau_qp_fin2}) and to better identify the regime in which the observed shear response takes place. The latter identification makes use of the theoretical refractive indexes $n_1(\omega)$ and $n_2(\omega)$ derived in Secs.\@ \ref{Viscoel_q0} (for $\tau_K=0$) and \ref{viscoel_q0_rel} (for $\tau_K\neq0$).

A first method to deduce $\nu(\omega)$ utilizes the saturation limit (\ref{eq:Z_VE_taukinf}) of the surface impedance. As discussed in Sec.\@ \ref{Skin_VE_low}, in propagating shear regime $\omega \gg v_F^{*} \left|q\right|$ a phenomenology analogous to anomalous skin effect occurs, namely the saturation of the (inverse) surface resistance $\mathrm{Re}\left\{Z(\omega)\right\}^{-1}$ controlled by $\nu(\omega)$ in the relaxationless limit $\omega \tau_K \rightarrow +\infty$. In the absence of shear rigidity, e.g.\@ for $\nu(\omega)\rightarrow 0$, such saturation would be absent and the standard Drude (ohmic) result (\ref{Skin_Drude}) would be observed. Although this propagating-shear phenomenon implies a phenomenology analogous to anomalous skin effect, the latter is normally retrieved for $\omega \ll v_F^{*} \left|q\right|$ (deep in the Lindhard continuum). Hence, observing the surface resistance saturation in the propagating shear regime is itself a qualitative demonstration of a non-negligible $\nu(\omega)$. The generalized shear modulus can then be deduced, at given $\omega$ and $\tau_K \gg \tau_c$, through Eqs.\@ (\ref{eq:Z_VE_taukinf}) and (\ref{eq:B_nu_inf}). Depending on $\omega \tau_c$ -- with $\tau_c(\omega,T)$ depending on $\omega$, $T$ and $F_1^S$ through Eq.\@ (\ref{eq:tau_qp_fin2}) -- the skin depth may then be extracted from $Z(\omega)$, and compared with the expressions collected in Table \ref{tab:skin_depth} for the viscous/dissipative or the reactive/collisionless cases. This comparison also provides a consistency check of the assumed response regime: the theoretical and experimental values of $\omega/\left(v_F^{*} \left|q\right|\right)$, $\omega \tau_K$ and $\omega \tau_c$ have to be consistent with one another. 
A further check to distinguish the propagating shear and anomalous skin effect regimes stems from the dependence of the impedance saturation limit on $\omega$: in anomalous skin regime $Z(\omega) \propto \delta_s^{\rm an} \propto \omega^{-1/3}$, while in propagating shear regime with $\left|\delta_\nu q \right| \gg 1$ the frequency dependence ranges from $Z(\omega) \propto \delta_s^{\rm hd} \propto \omega^{-1/4}$ (hydrodynamic/dissipative case) to $Z(\omega) \propto \delta_s^{\rm el} \propto \omega^{-1/2}$ (reactive/dissipationless case) if $\omega \tau_c \ll 1$ or $\omega \tau_c \gg 1$, respectively. The general case for any $\omega \tau_c$, i.e.\@ Eq.\@ (\ref{eq:Z_VE_taukinf}), depends on two length scales $\lambda_L$ and $\delta_\nu$ identified in Eqs.\@ (\ref{eq:B_nu_inf}), (\ref{eq:London_depth}), and (\ref{eq:delta_nu}). 
Furthermore, in the hydrodynamic case the value of the shear viscosity $\nu(0)$ extracted from $Z(\omega)$ in the limit $\omega \tau_K \rightarrow +\infty$ might be compared with size-dependent transport measurements on the same material; see Sec.\@ \ref{FL_DC}. 

\begin{figure}[t]
\includegraphics[width=0.9\columnwidth]{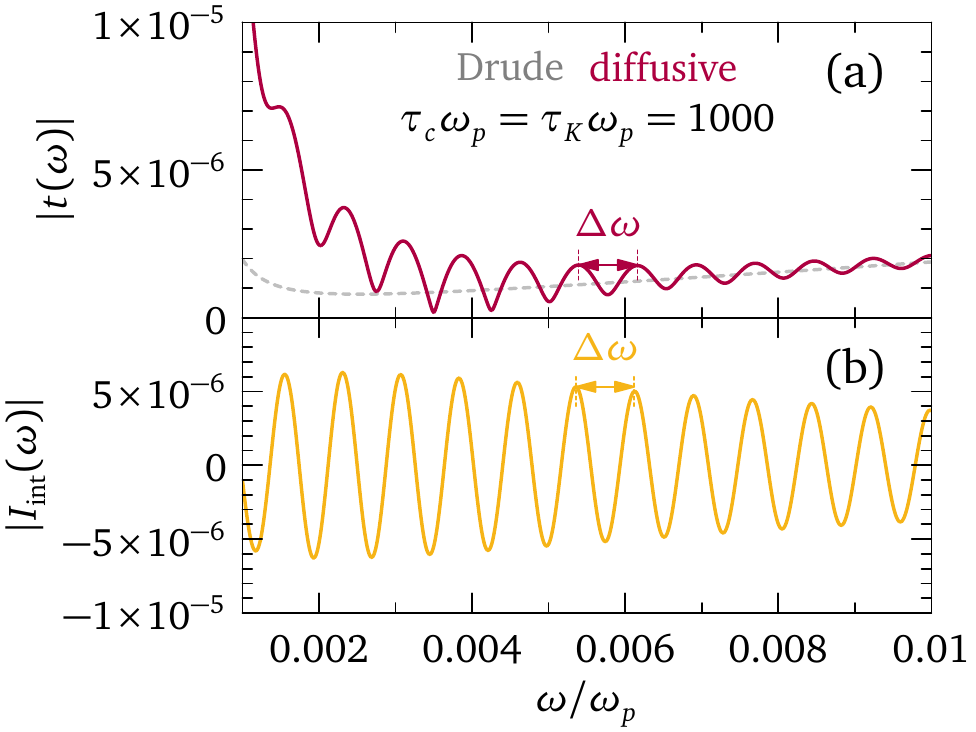}
\caption{\label{fig:t_oscill} 
(a) Transmission modulus $\left|t(\omega)\right|$ from Eq.\@ (\ref{eq:t_viscoel}) for fixed scattering rates $\tau_c=\tau_K=1000/\omega_p$ and thickness $d=10 c/\omega_p$. The red curve assumes diffuse boundary scattering, while the gray dashed curve is the Drude result. $\Delta \omega$ denotes the frequency interval between successive transmission oscillation. (b) Interference part of the transmitted intensity $\left|I_{\rm int}\right|$ across the vacuum-film interface assuming no Fabry-Perot interference in the slab, according to Eq.\@ (\ref{eq:t_thick_int}), for the same parameters as in panel (a). }

\end{figure}

Another strategy to deduce information on $\nu(\omega)$ from optics is to analyze the oscillations in frequency of the thin-film transmission modulus $\left|t_{\rm film}(\omega)\right|$ in propagating shear regime. If such oscillations are experimentally discerned, the data may be fitted to Eq.\@ (\ref{eq:t_viscoel}) as a function of frequency to extract the evolution of $\nu(\omega)$ with $\omega$ (knowing $\tau_K$). To better understand the effect of $\nu(\omega)$, let us qualitatively analyze the expected signal for sufficiently thick films, i.e.\@, $d \gg c/\left[\omega \left|n_i(\omega) \right|\right], \, i=\left\{1,2\right\}$. In this case, we can neglect Fabry-Perot interferences stemming from multiple reflections at $z=d$, so that the terms involving $\theta_1$ and $\theta_2$ may be neglected in Eq.\@ (\ref{eq:t_viscoel}). Then, the total transmission can be written as \cite{Valentinis-thesis,FZVM-2014}
\begin{subequations}\label{eq:t_thick}
\begin{equation}\label{eq:t_thick_tot}
\left|t_{\rm film}(\omega)\right|^2=I_1(\omega)+I_2(\omega)+I_{\rm int}(\omega)
\end{equation}
\begin{multline}\label{eq:t_thick_j}
I_j(\omega)=\left|t_je^{i\frac{\omega}{c}n_j(\omega) d}\right|^2 \\ =\left|t_j e^{i\frac{\omega}{c}\mathrm{Re}\left\{n_j(\omega)\right\} d}\right|^2 e^{-2\frac{\omega}{c}\mathrm{Im}\left\{n_j(\omega)\right\} d}, \\ \, j=\left\{1,2\right\},
\end{multline}
\begin{multline}\label{eq:t_thick_int}
I_{\rm int}(\omega)=2 \mathrm{Re}\left\{t_1e^{i\frac{\omega}{c}n_1(\omega) d} \left[ t_2 e^{i\frac{\omega}{c}n_2(\omega) d}\right]^{\ast} \right\} \\ =2\left|t_1\right| \left|t_2\right|e^{-\frac{\omega}{c}\left[\mathrm{Im}\left\{n_1(\omega)\right\}+\mathrm{Im}\left\{n_2(\omega)\right\}\right] z} \\ \times \cos\left[\mathrm{Arg}\left\{t_1\right\}-\mathrm{Arg}\left\{t_2 \right\}+\Delta \omega z \right]. 
\end{multline}
\end{subequations}
Therefore, the total transmission comprises two exponentially damped terms (\ref{eq:t_thick_j}), with damping linked to the imaginary part of the refractive index, and an interference term between the two polaritons. Such interference term depends on the arguments $\mathrm{Arg}\left\{t_1\right\}$ and $\mathrm{Arg}\left\{t_2\right\}$, but also on the difference between the real parts of the polariton refractive indexes through
\begin{equation}\label{eq:Delta_omega}
\Delta \omega= \frac{\omega}{c} \left( \mathrm{Re}\left\{n_1(\omega)\right\}-\mathrm{Re}\left\{n_2(\omega)\right\} \right).
\end{equation}
The term (\ref{eq:t_thick_int}) generates the transmission oscillations in propagating shear regime. In turn, since $n_1(\omega)$ and $n_2(\omega)$ are governed by the generalized shear modulus through Eqs.\@ (\ref{eq:VE_modes_rel}) and (\ref{eq:n_charged}), $\nu(\omega)$ ultimately controls the exponential damping of the terms (\ref{eq:t_thick_j}) as well as the interference oscillatory term (\ref{eq:t_thick_int}). 

In thick films of strongly interacting Fermi liquids we expect the interference term (\ref{eq:t_thick_int}) to be visible when the amplitudes of the two polariton modes are comparable, that is when $I_1(\omega) \approx I_2(\omega)$ \cite{Valentinis-thesis}. The characteristic wavelenght (in frequency) between successive oscillations $\Delta \omega$ from Eq.\@ (\ref{eq:t_thick_int}) may then be compared with the experimental oscillations (if visible) to extract $\nu(\omega)$ through the frequency dependence of the refractive indexes. Figure \ref{fig:t_oscill} shows a comparison between the full numerical $\left|t_{\rm film}(\omega)\right|$ from Eq.\@ (\ref{eq:t_viscoel}) and the interference term (\ref{eq:t_thick_int}) for thick films, in Fig.\@ \ref{fig:t_oscill}(a) and Fig.\@ \ref{fig:t_oscill}(b) respectively. For simplicity and visual clarity, in Fig.\@ \ref{fig:t_oscill} I fix the scattering rates to $\tau_c=\tau_K=1000/\omega_p$, although the dependence of these rates on frequency and temperature does not affect the comparison. Other parameters of the calculations are: thickness $d=10 c/\omega_p$, Landau parameter $F_1^S=20$, renormalized Fermi velocity $v_F^{*}=10^{-3}c$, and diffusive interface scattering according to Eq.\@ (\ref{eq:lambda_s_d}). The Drude result from Eq.\@ (\ref{eq:t_film_Drude}) for the same thickness is also shown for reference as a gray dashed curve: it corresponds to the limit $\nu(\omega)=0$ in the transmission (\ref{eq:t_viscoel}). We see that, for a thickness $d=10 c/\omega_p$, the thick-film analysis in Fig.\@ \ref{fig:t_oscill}(b) essentially captures the oscillatory pattern and the associated period $\Delta \omega$ in the thin-film transmission. 
In the more general case of arbitrary thickness $d$, instead of two interfering terms like in Eq.\@ (\ref{eq:t_thick}) we would have four interfering terms corresponding to $t_j$ and $\theta_j$, $j=\left\{1,2\right\}$ [Eq.\@ (\ref{eq:t_viscoel})] but the mechanism for the appearance of transmission oscillations due to $\nu(\omega)$ is analogous to the thick-film case. 

The procedures here sketched to deduce the generalized shear modulus from optical measurements may be best performed together for consistency, in order to check the values of $\nu(\omega)$ extracted from different methods. Moreover, other optically accessible quantities, like the reflection coefficient \cite{FZVM-2014}, are affected by the propagation of the shear polariton and may be used for complementary tests.

\section{Conclusions and perspectives}\label{Concl}

In summary, the transverse shear response of an electrically charged Fermi liquid is spatially nonlocal, due to correlations among quasiparticles entailed by interactions and collisions. The nature of this response depends on the ratio between frequency $\omega$ and momentum $\vec{q}$, as realized by calculating the transverse susceptibility in three dimensions in the kinetic equation approach of Landau, Abrikosov and Khalatnikov (Sec.\@ \ref{Charged_FL_XT}). The regime $\omega \ll v_F^{*} \left|q\right|$ corresponds to incoherent dissipative electron-hole excitations in the Lindhard continuum, and the associated $q$-dependent conductivity gives anomalous skin effect (Sec.\@ \ref{Anomal}). In the regime $\omega \gg v_F^{*}\left|q\right|$ the collective response of Fermi-surface quasielectrons is analogous to the response of a ``visco-elastic'' substance endowed with a frequency-dependent generalized shear modulus $\nu(\omega)$ \cite{Conti-1999,Tokatly-2000,Vignale-2005} (Sec.\@ \ref{Viscoel_q0}). Two polaritons for each frequency exist: the plasmon-polariton, reminiscent of the usual optical plasmon, and the shear-polariton, resulting from the mixing of photons with a Fermi-surface shear resonance, i.e.\@, transverse sound in the charged Fermi liquid.  
In the collisional/hydrodynamic regime $\omega \tau_c \ll 1$, at any finite $\omega /\left( v_F^{*} \left|q\right|\right)$ ratio, the response is predominantly dissipative as in a highly viscous liquid \cite{Tokatly-2000, FZVM-2014}, since the collision time $\tau_c$ is the smallest timescale in the system and allows for local thermodynamic equilibrium (Sec.\@ \ref{Hydro}). In the opposite collisionless/elasticlike regime $\omega \tau_c \gg 1$, with $\omega \gg v_F^{*} \left|q\right|$ and for sufficiently strong interactions $F_1^S>6$, the Fermi surface responds with a predominantly dissipationless (reactive) contribution: over short timescales, the Fermi liquid reacts similarly to an elastic solid. 

In solid-state systems, momentum relaxation imposed by the breaking of Galilean invariance modifies the Fermi-liquid dielectric function through the relaxation time $\tau_K$ (Sec.\@ \ref{VE_tauk}). Specifically, in the propagating shear regime $\omega \gg v_F^*\left|q\right|$ the leading-order nonlocal correction to the dielectric function is encoded by a generalized shear modulus $\tilde{\nu}(\omega)$ (Sec.\@ \ref{viscoel_q0_rel}) modified by momentum relaxation: $\tilde{\nu}(\omega)$ depends on the collision time $\tau_c$ as well as on $\tau_K$. If $\tau_c \ll \tau_K$ (weak momentum relaxation), the spatially nonlocal part of $\epsilon_T(\vec{q},\omega)$ again depends on $\nu(\omega)$. When $\tau_K \ll \tau_c$, the Fermi-liquid dielectric function reduces to the Drude case with relaxation time $\tau_K$. 
The competition between relaxation and spatial nonlocality affects the response at low frequencies (close to the DC limit) and finite temperature, such that hydrodynamic behavior develops when momentum relaxation is negligible (Sec.\@ \ref{FL_DC}). 

Optical experiments represent ideal instruments to probe Fermi-surface rigidity in solid-state Fermi liquids, since electromagnetic waves are transversely polarized and couple to the electric charge of quasiparticles, thus exciting a shear response. 
The surface impedance of semi-infinite samples (Sec.\@ \ref{Zs}) resulting from the Fermi-liquid optical conductivity retrieves the phenomenology of anomalous skin effect, including the saturation of the surface resistance in the relaxationless limit. Specializing to the regime $\omega \ll v_F^{*}\left|q\right|$, we retrieve the relaxationless limit predicted by the theory of Reuter and Sondheimer \cite{Reuter-1948,Sondheimer-2001}. Such limit is modified by spatial nonlocality and controlled by $\nu(\omega)$ in the propagating shear regime $\omega \gg v_F^{*}\left|q\right|$. In this regime, an analytical expression for the surface impedance is written in terms of the refractive indexes for the plasmon-polariton and shear-polariton. The skin depth, that is the characteristic penetration depth of electromagnetic fields in the Fermi liquid, is also affected by spatial nonlocality through $\nu(\omega)$ (Sec.\@ \ref{Skin_VE_low}): in particular, in hydrodynamic/collisional regime it is $\delta_s^{\rm hd} \propto \sqrt{\nu(0)/\omega} \propto \omega^{-1/4}$, while in reactive/collisionless regime it is $\delta_s^{\rm el} \propto \left[\sqrt{\mu_s}/\omega\right]^{1/2} \propto \omega^{-1/2}$, where $\nu(0)$ is the viscosity coefficient and $\mu_s$ is the reactive shear modulus (see Table \ref{tab:skin_depth}), and it exhibits qualitatively distinct frequency dependences in different regimes. 

Fermi-liquid shear modes can be also probed by the optical transmission of thin films (Sec.\@ \ref{Film_T}). In the propagating shear regime (Sec.\@ \ref{t_viscoel}), the thin-film transmission modulus increases with respect to its Drude counterpart (Sec.\@ \ref{t_Drude}), with pronounced oscillations due to the interference of the two coherent polaritons. Such oscillations are amplified in the collisionless regime $\omega \tau_c \gg 1$, reflecting the coupling of radiation to the reactive shear mode of the Fermi surface, with the latter resonating like an elastic membrane. Momentum relaxation, considered in a simple model with acoustic phonons, Umklapp scattering and impurities as independent relaxation channels, suppresses the effects of shear viscoelasticity at room temperature, while residual traces of spatial nonlocality persist at cryogenic temperatures in strongly interacting Fermi liquids, with a high effective mass $m^*=m\left(1+F_1^S/3\right) \gg 1$ and a low impurity concentration (Sec.\@ \ref{T_phon_imp}). 

The various assumptions considered in this paper may call into question the feasibility of observing shear-propagation effects in 3D solids. For instance, I assumed a single parabolic band, while non-parabolicity destroys the proportionality between current density $\vec{J}$ and quasiparticle momentum $\vec{k}$: in the latter situation, the low-momentum transverse response will not simply be connected to a generalized shear modulus $\nu(\omega)$, as other dissipative and/or nondissipative transport coefficient arise \cite{Bradlyn-2012,Gorbar-2018}. Likewise, I considered nearly isotropic Fermi liquids, in which translational invariance is broken only on the microscopic scale of the crystalline lattice. This allows for a simplification of the viscosity tensor \cite{Bradlyn-2012}, with the bulk and shear viscosity coefficients as the only independent components. The number of components increases with lower symmetry of the system \cite{Cook-2019}, as in the case of graphene which is neither Galilean-, nor Lorentz-invariant \cite{Briskot-2015, Link-2018, Narozhny-2019}. Moreover, I considered the simplest parametrization of the transverse shear response in terms of the first Landau parameter $F_1^S$. Considering more Landau parameters modifies quantitatively the results: for instance, transverse zero sound propagates even for $F_1^S<6$ if the second Landau parameter is $F_2^S>0$ \cite{Lea-1973}. The quantitative impact of momentum relaxation on the calculations depends on the microscopic scattering sources considered in the model. In particular, Matthiessen's rule for adding scattering rates can be justified for independent relaxation channels. A more microscopic understanding of the interplay between spatial nonlocality and relaxation deserves future work, even for 3D Fermi liquids. 

Within the assumptions of this paper, the main requirements which a candidate 3D material must satisfy to observe Fermi-liquid shear propagation are dictated by the frequency-dependent coefficient $\nu(\omega)$: one needs a heavy effective mass $m^* \gg 1$, or equivalently a high first Landau parameter $F_1^S \gg 1$, and very clean samples at cryogenic temperatures. Fermi liquids formed by heavy-fermion systems, or the low-density doped perovskyte SrTiO$_3$ \cite{Cook-2019} might provide appropriate platforms. A more detailed discussion of suitable materials will be reported in a companion paper \cite{FZVM-new}.
More generally, the macroscopic approach of continuum mechanics applied to electron ensembles, together with appropriate microscopic models for transport and electrodynamic coefficients, may open new avenues in the investigation of low-temperature correlations in metals: the reactive coefficients describing low-momentum electrodynamics persist even in the presence of pairing correlations \cite{Gochan-2019}, while the dissipative components disappear in the superconducting state. The evolution of the reactive shear coefficients across a superconducting transition, experimentally probed by optical spectroscopy, could offer valuable insight into the finite-momentum dynamics of Cooper pairing in unconventional superconductors \cite{Gochan-2019, Liao-2019}. 

\section{Acknowledgments}

The author is grateful to J.\@ Zaanen, D.\@ van der Marel, J.\@ Schmalian, A.\@ Morpurgo, B.\@ Narozhny, I.\@ Gornyi, C.\@ Berthod, S.\@ Hartnoll, E.\@ Kiselev, A.\@ Kuzmenko, G.\@ A.\@ Inkof, M.\@ Filippone, P.\@ Abbamonte, J.\@ Y.\@ Khoo, F.\@ Pientka, and I.\@ Sodemann for insightful discussions. This work was supported by the Swiss National Science Foundation (SNSF) through project 200021 – 162628 and through the SNSF Early Postdoc.Mobility Grant P2GEP2$\_$181450. 

\appendix

\section{Derivation of the Fermi liquid kinetic equation}\label{kin_eq}
This appendix contains a detailed derivation of the kinetic equation for the nonequilibrium distribution function of a Fermi liquid. Our main references for this section will be the works of Nozières and Pines \cite{Nozieres-1999ql} and Abrikosov and Khalatnikov \cite{Abrikosov-1959}. 
The essence of the kinetic equation approach is to consider how the nonequilibrium distribution function $\texttt{N}_{\vec{p}}(\vec{r},t)$ for electron-hole quasiparticles evolves in space $\vec{r}$ and time $t$, in response to perturbations. As such, the kinetic equation is equivalent to a quantum Boltzmann equation for the distribution function. 
Following Nozières and Pines \cite{Nozieres-1999ql}, given a total energy of the Fermi liquid $E_{FL}(n,t)$ which is a function of electron density $n$, the local excitation energy of a quasiparticle is 
\begin{multline}
\tilde{E}_n\left[\vec{p}(\vec{r},t), \sigma \right]=\frac{\partial E_{FL}(n,t)}{\partial \delta \texttt{N}_{\vec{p}',\sigma'}(\vec{r},t)}=E_n\left[\vec{p}(\vec{r},t), \sigma \right] \\ +\sum_{\vec{p}',\sigma'}\mathrm{f}_{\vec{p},\sigma,\vec{p}',\sigma'}\delta \texttt{N}_{\vec{p}',\sigma'}(\vec{r},t).
\end{multline}
This means that the local excitation energy is the sum of the energy of the quasiparticle $E_n\left[\vec{p}(\vec{r},t), \sigma \right]$, with momentum $\vec{p}$ and spin $\sigma$, and the local interaction energy of that quasiparticle interacting with other quasiparticles in its surroundings; the latter term is similar to an effective mean field of the Fermi liquid, given by the reaction of all other quasiparticles when we perturb the energy of a given state with momentum $\vec{p}$. The short-range interaction matrix elements between quasiparticles of momenta $\left\{  \vec{p},\vec{p}' \right\}$ and spin $\left\{ \sigma, \sigma' \right\}$ is $\mathrm{f}_{\vec{p},\sigma,\vec{p}',\sigma'}$.
Landau considered the excitation energy $\tilde{E}_n\left[\vec{p}(\vec{r},t), \sigma \right]$ as an effective Hamiltonian for the Fermi liquid, to describe the transport properties of the system. For electrically charged quasiparticles, we write the microscopic Hamiltonian as
\begin{multline}\label{eq:H_FL_gen}
\hat{H}_{qp}(\vec{r},\vec{p},\sigma)=E_n\left[\vec{p}+e\vec{A}(\vec{r},t), \sigma \right] \\  +\sum_{\vec{p}',\sigma'}\mathrm{f}_{\vec{p},\sigma,\vec{p}',\sigma'}\delta \texttt{N}_{\vec{p}',\sigma'}(\vec{r},t)-e \phi_{\sigma}(\vec{r},t).
\end{multline}
In Eq.\@ (\ref{eq:H_FL_gen}), we recognize the quasiparticle energy eigenvalues $E_n\left[\vec{p}+e\vec{A}(\vec{r},t), \sigma \right]$, modified by the presence of the vector potential $\vec{A}(\vec{r},t)$, as well as by the short-range interaction matrix elements $\mathrm{f}_{\vec{p},\sigma,\vec{p}',\sigma'}$. The quasiparticle velocities are defined from the crystalline momentum $ \vec{v}_p = \nabla_p E_n(\vec{p})=\vec{p}/m^*$. 
The nonequilibrium distribution function $N_{\vec{p}}(\vec{r},t)=\delta \texttt{N}_{\vec{p},\sigma}(\vec{r},t)+N_0\left(E_n\left[\vec{p}+e\vec{A}(\vec{r},t), \sigma \right]\right)$ contains the departure of the quasiparticle statistics from local equilibrium $N_0\left(E_n\left[\vec{p}+e\vec{A}(\vec{r},t), \sigma \right]\right)$, and can be calculated from particle number conservation and gauge invariance \cite{Abrikosov-1959, Conti-1999}. We also include a scalar potential $\phi_{\sigma}(\vec{r},t)$. 

Equation (\ref{eq:H_FL_gen}) is valid in the regime $k_B T \ll E_F$ and $\hbar \omega \ll E_F$, with $E_F=\hbar^2 (k_F)^2/(2 m)$ Fermi energy, under the assumptions of large quasiparticle lifetime $\tau_{K} \gg (E_F)^{-1}$ and a well-defined Fermi surface. Physically, these conditions signify that thermal agitation and the external perturbation are not so strong to drive quasielectrons completely out of equilibrium, thus destroying the Fermi surface. From the Hamiltonian (\ref{eq:H_FL_gen}), one derives a linearized kinetic equation for quasiparticles of velocity $v_{\vec{k},\sigma}$ on the Fermi surface. The response to the perturbation depends on the interactions among quasiparticles and on the angle $\theta$ between the excitation wave vector $\vec{q}$ and the quasiparticle wave vector $\vec{k}$. 

Starting from the Hamiltonian (\ref{eq:H_FL_gen}), we linearize the changes in the quasiparticle energies $E_n\left[\vec{p}+e\vec{A}(\vec{r},t), \sigma \right]$ and in the distribution function $\delta \texttt{N}_{\vec{p},\sigma}(\vec{r},t)$, with respect to the vector potential $\vec{A}(\vec{r},t)$, as in 
\begin{multline}\label{eq:ev_distr}
\delta \texttt{N}_{\vec{p},\sigma}(\vec{r},t)=\texttt{N}_{\vec{p}}(\vec{r},t)-\texttt{N}_0\left(E_n\left[\vec{p}+e\vec{A}(\vec{r},t), \sigma \right]\right) \\ \approx \texttt{N}_{\vec{p}}(\vec{r},t)-\texttt{N}_0\left[E_n\left(\vec{p} \right)\right]-\frac{d \texttt{N}_0\left[E_n\left(\vec{p} \right)\right]}{d E_n\left(\vec{p},\sigma \right)} \frac{\vec{p} \cdot \vec{A}(\vec{r},t)}{m^*}.
\end{multline}
The nonequilibrium quasiparticle distribution function $\texttt{N}_{\vec{p}}(\vec{r},t)$ is derived directly from Liouville's equation, for a classical flow characterized by volume-conserving evolution in phase space. Therefore, we can write a quasiclassical mean-field kinetic equation for $\texttt{N}_{\vec{p}}(\vec{r},t)$, which is
\begin{multline}\label{eq:ev_distr2}
\frac{\partial \texttt{N}_{\vec{p},\sigma}(\vec{r},t)}{\partial t} +\frac{1}{\hbar} \frac{\partial \hat{H}_{qp}(\vec{r},\vec{p},\sigma)}{\partial \vec{p}} \frac{\partial\texttt{N}_{\vec{p},\sigma}(\vec{r},t)}{\partial \vec{r}} \\ -\frac{1}{\hbar} \frac{\partial \hat{H}_{qp}(\vec{r},\vec{p},\sigma)}{\partial \vec{r}} \frac{\partial\texttt{N}_{\vec{p},\sigma}(\vec{r},t)}{\partial \vec{p}}={\mathscr{I}}_{coll}(\vec{r},t).
\end{multline}
In Eq.\@ (\ref{eq:ev_distr}), the collisional term ${\mathscr{I}}_{coll}(\vec{r},t)={\left[ \partial \texttt{N}_{\vec{p},\sigma}(\vec{r},t)/\partial t \right]}_{coll}$ describes damping by collision processes that are not included into the mean-field Hamiltonian $\hat{H}_{qp}(\vec{r},\vec{p},\sigma)$. In fact, the Fermi liquid phenomenology relies on the existence of low-energy nearly independent quasiparticles, which do not collide, and additional collisions must be introduced by an external damping term ${\mathscr{I}}_{coll}(\vec{r},t)$. This collisional integral makes quasielectrons and quasiholes acquire a finite lifetime, $\tau_{qel}<+\infty$ and $\tau_{qh}<+\infty$ respectively: as common in quantum mechanics, we are treating a time-dependent problem with the eigenstates of a stationary configuration, in this case the one for free fermions, but including time-dependent scattering as a finite lifetime for the stationary states. Collisions redistribute momentum and energy among the quasiparticles, but the total momentum and the total energy of the ensemble is conserved. 
For the \emph{true} thermodynamic equilibrium, we know that the distribution function is the Fermi-Dirac statistics $f_{FD}\left[E_n\left(\vec{p},\sigma \right) \right]$\cite{Ashcroft-1976}: since this is the configuration for global thermodynamic equilibrium, it is a stationary state and we must have a vanishing collision term ${\left[ \partial  f_{FD}\left[E_n\left(\vec{p},\sigma \right) \right]/\partial t \right]}_{coll}=0$, which specifies the relation between the lifetimes 
$\tau_{qel}$ and $\tau_{qh}$. The nonequilibrium distribution function $\texttt{N}_{\vec{p},\sigma}(\vec{r},t)$ can be written as a deviation with respect to the true Fermi-Dirac global equilibrium, as $\texttt{N}_{\vec{p},\sigma}(\vec{r},t)=f_{FD}\left[E_n\left(\vec{p},\sigma \right) \right]+ \Xi_{\vec{k},\sigma}(\vec{r},t)$. Inserting the latter into Eq.\@ (\ref{eq:ev_distr}), we arrive at
\begin{multline}\label{eq:pert_distr}
\frac{\partial \Xi_{\vec{k},\sigma}(\vec{r},t)}{\partial t} +  \vec{v}_{\vec{k},\sigma} \cdot \frac{\partial \Xi_{\vec{k},\sigma}(\vec{r},t)}{\partial \vec{r}} \\ + \vec{v}_{\vec{k},\sigma} \cdot \vec{F}_{\sigma} (\vec{r},t) \delta \left[ E_n\left(\vec{k},\sigma \right)- \mu \right] =  {\mathscr{I}}_{coll}(\vec{r},t).
\end{multline}
In Eq.\@ (\ref{eq:pert_distr}), we have defined the total quasiclassical force acting on the individual quasiparticles
\begin{multline}\label{eq:qc_force}
\vec{F}_{\sigma} (\vec{r},t)=-\nabla_{\vec{r}}  \left[ -e \phi_{\sigma}(\vec{r},t) \right. \\ \left. +\sum_{\vec{k}',\sigma'} \mathrm{f}_{\vec{k},\sigma,\vec{k}',\sigma'} \Xi_{\vec{k}',\sigma'}(\vec{r},t) + \frac{ \hbar \vec{k}}{m} \cdot e \vec{A}(\vec{r},t) \right].
\end{multline}
The linearized kinetic equation (\ref{eq:pert_distr}) is manifestly gauge-invariant, and it satisfies the continuity equation for the flow of quasiparticles \cite{Abrikosov-1959}: this is the starting point for many applications of Landau phenomenology of Fermi liquids, for example the study of collective modes in the absence of perturbations. 

Now, we concentrate on density collective modes in the Fermi liquid. Collective modes propagate even in the absence of perturbations, so we set the electromagnetic potentials to zero as $\phi_{\sigma}(\vec{r},t)=0, \; \vec{A}(\vec{r},t)=0$. 
Furthermore, collective modes have wavelike evolution, therefore the deviation of the distribution function with respect to global equilibrium will be periodic in space and time, i.e.\@, $\Xi_{\vec{k},\sigma}(\vec{r},t) \propto e^{i \left( \vec{q} \cdot \vec{r}- \omega t \right)}$, with $\vec{q}$ transferred wave vector for the excitation. 
Inserting this wave-like form in Eq.\@ (\ref{eq:pert_distr}) without electromagnetic potentials and collisions leads to
\begin{multline}\label{eq:kin_q_gen}
\left(\vec{q} \cdot v_{\vec{k},\sigma} -\omega \right) \Xi_{\vec{k},\sigma}(\vec{r},t)+  \vec{q} \cdot v_{\vec{k},\sigma} \delta \left[ E_n\left(\vec{k},\sigma \right)- \mu \right] \\ \cdot \sum_{\vec{k}',\sigma'} \mathrm{f}_{\vec{k},\sigma, \vec{k}', \sigma'} \Xi_{\vec{k}',\sigma'}(\vec{r},t)={\mathscr{I}}_{coll}(\vec{r},t).
\end{multline}
In principle, excitations can alter the quasiparticle energies $E_n\left(\vec{k},\sigma \right)$ either by changing the wave vector $\vec{k}$ or by flipping the spin $\sigma$; in the following, having in mind soundlike collective modes, we assume that alterations of $E_n\left(\vec{k},\sigma \right)$ stem from variations in the quasiparticle momentum $\vec{k}$. 
Expanding the distribution deviations to first order in the energies $E_n\left(\vec{q},\sigma \right)$, we have $\Xi_{\vec{k},\sigma}(\vec{q},\omega)=-\epsilon_{\vec{k}}(\vec{q},\omega) \left. \partial f_{FD}\left[ E_n\left(\vec{k},\sigma \right) \right]/\partial E_n\left(\vec{k},\sigma \right) \right|_{\mu}$, with $\left. \partial f_{FD}\left[ E_n\left(\vec{k},\sigma \right) \right]/\partial E_n\left(\vec{k},\sigma \right) \right|_{\mu} \equiv -\delta \left[ E_n\left(\vec{k},\sigma \right)- \mu \right]$ for the Fermi-Dirac distribution \cite{Berthod-2018, Ashcroft-1976}. 
If we had wanted to analyze magnetic modes in the Fermi liquid, like spin waves, we could also have assumed a distribution function change of the kind $\Xi_{\vec{k},\sigma}(\vec{q},\omega)=-s_{\sigma}(\vec{q},\omega) \sigma \left. \partial f_{FD}\left[ E_n\left(\vec{k},\sigma \right) \right]/\partial E_n\left(\vec{k},\sigma \right) \right|_{\mu}=s_{\sigma}(\vec{q},\omega) \sigma\delta \left[ E_n\left(\vec{k},\sigma \right)- \mu \right]$, with the quasiparticle energies modified by spin flipping \cite{Abrikosov-1959}. 
A Taylor series expansion at first order of the change in the distribution function selects quasiparticles at the chemical potential $\mu$: only excitations around $\mu$ contribute to collective modes, since states at $E \ll \mu$ deep down into the Fermi sea are occupied and Pauli-blocked. 

The kinetic equation (\ref{eq:kin_q_gen}) in local equilibrium, written in reciprocal space of momenta $\vec{q}$ and frequency $\omega$, becomes
\begin{multline}\label{eq:kin_k_vec}
\left(\vec{q} \cdot \vec{v}_{\vec{k},\sigma}-\omega \right) \epsilon_{\vec{k}}(\vec{q},\omega)+ \vec{q} \cdot \vec{v}_{\vec{k},\sigma} \\ \cdot \left\{ \sum_{\vec{k}',\sigma'} \mathrm{f}_{\vec{k},\sigma, \vec{k}', \sigma'} \epsilon_{\vec{k}'} (\vec{q},\omega) \delta\left[ E_n(\vec{k}',\sigma')-\mu \right]  \right\} \\ ={\mathscr{I}}_{coll}(\vec{q},\omega),
\end{multline}
where ${\mathscr{I}}_{coll}(\vec{q},\omega)$ is the Fourier transform of the collision term ${\mathscr{I}}_{coll}(\vec{r},t)$ with respect to space and time. 
The delta functions $\delta\left[ E_n(\vec{k},\sigma)-\mu \right]$ and $\delta\left[ E_n(\vec{k}',\sigma')-\mu \right]$ fix the value of $k$ and $k'$, respectively: therefore, the character of the collective mode, including its polarization direction, will be determined by the respective orientation of the excitation wave vector $\vec{q}$ with respect to $\vec{k}$ and $\vec{k}'$. 

Anticipating the emergence of sound waves with acoustic dispersion $\omega_{\lambda}(\vec{q})= v_{s} q$, with $v_{s}$ sound velocity for the Fermi liquid, we write Eq.\@ (\ref{eq:kin_k_vec}) in terms of $v_s$ and obtain
\begin{multline}\label{eq:kin_k}
\left(v_{\vec{k},\sigma} \cos{\theta}-v_{s} \right) \epsilon_{\vec{k}}(\vec{q},\omega)+  v_{\vec{k},\sigma} \cos{\theta} \\ \cdot \left\{ \sum_{\vec{k}',\sigma'} \mathrm{f}_{\vec{k},\sigma, \vec{k}', \sigma'} \epsilon_{\vec{k}'} (\vec{q},\omega) \delta\left[ E_n(\vec{k}',\sigma')-\mu \right]  \right\} \\ ={\mathscr{I}}_{coll}(\vec{q},\omega),
\end{multline}
where $\theta= \arccos \left( \vec{q} \cdot \vec{v}_{\vec{k},\sigma}/\left|  \vec{q} \cdot \vec{v}_{\vec{k},\sigma}   \right|\right)$.  

To continue, we need to expand the angular dependence of the interaction matrix elements $\mathrm{f}_{\vec{k},\sigma, \vec{k}', \sigma'}$, which determines the spatial polarization of the perturbation. Being an angular distribution in 3D space, a suitable expansion basis for $\mathrm{f}_{\vec{k},\sigma, \vec{k}', \sigma'}$ is given by spherical harmonics, in terms of Legendre polynomials \cite{Landau-2013stat, Abrikosov-1959}. We recall that Legendre polynomials are $\wp_n(x)=(2^n n!)^{-1} d^n\left[(x^2-1)^n\right]/dx^n$ with associated Legendre polynomials $P_l^m(x)=(-1)^m (1-x^2)^{m/2} d^m\left[\wp_l(x)\right]/dx^m$. The latter are used in the definition of spherical harmonics $Y_{l,m}(\theta,\phi)=\sqrt{\left[(2l+1)(l-m)!\right]/\left[4 \pi(l+m)!\right]} P_l^m(\cos{\theta}) e^{i m \phi}, \; -l \leq m \leq l$.
The Fermi-liquid expansion of the interaction matrix elements is usually done in terms of Landau parameters $F_n^\alpha, \; n\in\mathbb{N}, \; \alpha=\left\{S,A \right\}$ \cite{Landau-2013stat, Abrikosov-1959}, which represent the respective magnitudes of quasiparticle interactions for each possible angular pattern of the collective mode at the Fermi surface, i.e.\@, for all spherical harmonics. Having selected quasiparticle energies at the Fermi surface, the interaction matrix elements will be at the Fermi surface $FS$: $\mathrm{f}_{\vec{k},\sigma, \vec{k}', \sigma'} \equiv \textsl{f}_{\sigma,\sigma'}^{FS}\delta\left[ E_n(\vec{k},\sigma)-\mu \right]\delta\left[ E_n(\vec{k}',\sigma')-\mu \right]$. 
The Fermi-liquid Landau parameters in three dimensions are defined as
\begin{equation}\label{eq:Landau_par3D}
F_l^{S,A}=\frac{N_0^{*}(0)}{2} \int \frac{d \Omega}{4 \pi} \left[ \textsl{f}_{\uparrow,\uparrow}^{FS}(\cos \theta) \pm \textsl{f}_{\uparrow,\downarrow}^{FS}(\cos \theta) \right] \frac{\wp_l(\cos \theta)}{2 l +1}.
\end{equation}
The superscript $S$ or $A$ in Eq.\@ (\ref{eq:Landau_par3D}) refers to the additive or subtractive combination of $\textsl{f}_{\uparrow,\uparrow}^{FS}(\cos \theta) \pm \textsl{f}_{\uparrow,\downarrow}^{FS}(\cos \theta)$, which distinguishes between symmetric and antisymmetric Landau parameters. The former kind is associated to density (charge) perturbations, while the latter deals with the spin (magnetic) response. The density of states per unit volume $N_0^{*}(0)$ includes renormalization effects due to quasiparticle interactions, and therefore differs from the free fermions one $N_0^{el}(0)$. 
We concentrate on the problem in three dimensions, for which the inverse relation of (\ref{eq:Landau_par3D}) is
\begin{equation}\label{eq:Landau_matr_el_3D}
\textsl{f}_{\uparrow,\uparrow}^{FS}(\cos \theta) \pm \textsl{f}_{\uparrow,\downarrow}^{FS}(\cos \theta)= \frac{2}{N_0^{*}(0)} \sum_{l=0}^{+\infty} F_l^{S,A} \wp_l(\cos \theta).
\end{equation}
With these definitions, we can act on the term between curly brackets in Eq.\@ (\ref{eq:kin_k_vec}). The sums over $\vec{k}'$ and $\sigma'$ of the delta function $\delta\left[ E_n(\vec{k}',\sigma^{'})-\mu \right]$ give the density of states $N_0^{*}(0)$ by definition. All remains of the sum is an integration over angular variables, which depends on the angles between the wave vectors $\vec{k}$ and $\vec{k}'$. With a factor $1/2$ to avoid double counting of interactions for states $\vec{k}$ and $\vec{k}'$, we have
\begin{multline}\nonumber
\sum_{\vec{k}',\sigma'} \textsl{f}_{\vec{k},\sigma, \vec{k}', \sigma'} \epsilon_{\vec{k}'} (\vec{q},\omega) \delta\left[ E_n(\vec{k}',\sigma')-\mu \right] \\ \equiv \frac{1}{2} \sum_{\vec{k}',\sigma'}\delta\left[ E_n(\vec{k}',\sigma')-\mu \right] \int \frac{d \Omega'}{4 \pi} \sum_{\sigma'} \textsl{f}_{\sigma, \sigma'}^{FS} \epsilon_{\vec{k}'} (\vec{q},\omega) \\
=\frac{N_0^{*}(0)}{2} \int \frac{d \Omega'}{4 \pi} \sum_{\sigma'} \textsl{f}_{\sigma, \sigma'}^{FS} \epsilon_{\vec{k}'} (\vec{q},\omega).
\end{multline}
Equation (\ref{eq:kin_k_vec}) thus becomes
\begin{multline}\label{eq:kin_k_int}
\left(\vec{q} \cdot \vec{v}_{\vec{k},\sigma}-\omega \right) \epsilon_{\vec{k}}(\vec{q},\omega) \\ + \vec{q} \cdot \vec{v}_{\vec{k},\sigma}  \frac{N_0^{*}(0)}{2} \int \frac{d \Omega'}{4 \pi} \sum_{\sigma'} \textsl{f}_{\sigma, \sigma'}^{FS} \epsilon_{\vec{k}'} (\vec{q},\omega) \\ ={\mathscr{I}}_{coll}(\vec{q},\omega).
\end{multline}
Employing the Landau parameter expansion (\ref{eq:Landau_matr_el_3D}) of the interaction matrix elements leads to Eq.\@ (\ref{eq:kin_k_L2}).

\section{Derivation of the dispersion relation for transverse sound with collisions} \label{transv_mode_coll}

To find the dispersion relation of the first Fermi-surface transverse mode with $m=1$ in the presence of collisions, we have to study the following kinetic equation, first obtained by Lea \textit{et al.\@} \cite{Lea-1973,Roach-1976}: 
\begin{multline}\label{eq:kin_k_coll_T}
\left( \cos{\theta}  -\xi \right) \epsilon^S(\theta) \\ + \left[3 \epsilon^S(\theta) \sin \theta \right]_{av} \sin \theta \left(\frac{F_1^S}{3} \cos \theta - \beta \xi \right)=0,
\end{multline}
where we have defined $\xi$ and $\beta$ as in Eqs.\@ (\ref{eq:Lea_var}), and we have used the parametrization (\ref{eq:coll_int_transv_2}) for the collision integral in terms of a single collision time $\tau_c$.
From Eq.\@ (\ref{eq:kin_k_coll_T}), we immediately deduce the Fermi surface displacement $\epsilon^S(\theta)=\left\{\left[3 \epsilon^S(\theta) \sin \theta \right]_{av} \sin \theta \left(\frac{F_1^S}{3} \cos \theta -\beta \xi\right)\right\}/(\xi-\cos \theta)$, which means $\epsilon^S(\theta)\propto \left[ \sin \theta \left(\frac{F_1^S}{3} \cos \theta -\beta \xi\right)\right]/(\xi-\cos \theta)$. Therefore, the average $\left[3 \epsilon^S(\theta) \sin \theta \right]_{av}$ results in
\begin{multline}\label{eq:angle_av_term}
\left[3 \epsilon^S(\theta) \sin \theta \right]_{av}=3 \int_0^\pi \frac{(\sin \theta)^2 d \theta}{4} \frac{F_1^S}{3} \frac{\sin \theta \cos \theta}{\xi-\cos \theta} \\  - \beta \int_0^\pi \frac{(\sin \theta)^2 d \theta}{4} \ \frac{\xi \sin \theta }{\xi-\cos \theta}.
\end{multline}
The first term at the right-hand side of Eq.\@ (\ref{eq:angle_av_term}) stems from short-ranged quasiparticle interactions, and it is present also in the absence of collisions, while the second term arises only at finite $\tau_c$. Integrating over the angle $\theta$ in Eq.\@ (\ref{eq:angle_av_term}), we obtain 
\begin{multline}\label{eq:av_term_Lea}
\left[3 \epsilon^S(\theta) \sin \theta \right]_{av} \\ =3 \frac{F_1^S}{3} \left[-\frac{1}{3} + \frac{\xi^2}{2} -\frac{\xi}{4} (\xi^2-1) \ln \left(\frac{\xi+1}{\xi-1}\right) \right] \\ - 3 \beta \left[ \frac{\xi^2}{2} -\frac{\xi}{4} (\xi^2-1) \ln \left(\frac{\xi+1}{\xi-1}\right) \right]\\ =3 \left(\frac{F_1^S}{3}- \beta \right) \left[-\frac{1}{3} + \frac{\xi^2}{2} -\frac{\xi}{4} (\xi^2-1) \ln \left(\frac{\xi+1}{\xi-1}\right) \right]-\beta \\ =3 \left(\frac{F_1^S}{3}-\beta \right) \left[-\frac{m^2}{3 n m^*} X_T^0(\xi) \right]  - \beta.
\end{multline}
Using $\epsilon^S(\theta)\propto  \sin \theta \left(\frac{F_1^S}{3} \cos \theta -\beta \xi \right)/(\xi-\cos \theta)$ \cite{Lea-1973}, the collective mode kinetic equation becomes
\begin{multline}\nonumber
\left( \cos{\theta}  -\xi \right) \frac{ \sin \theta \left(\frac{F_1^S}{3} \cos \theta -\beta \xi \right)}{\xi-\cos \theta} \\  + \left[3 \epsilon^S(\theta) \sin \theta \right]_{av} \sin \theta \left(\frac{F_1^S}{3} \cos \theta -\beta \xi \right)=0,
\end{multline}
which means 
\begin{equation}\label{eq:Lea_coll_eq_impl}
-1 + \left[3 \epsilon^S(\theta) \sin \theta \right]_{av}=0.
\end{equation}
Inserting the explicit expression of $\left[3 \epsilon^S(\theta) \sin \theta \right]_{av}$ from Eq.\@ (\ref{eq:av_term_Lea}), we have 
\begin{multline}\label{eq:Lea_coll_eq}
3 \left(\frac{F_1^S}{3}-\beta \right) \left[-\frac{1}{3} +\frac{\xi^2}{2}-\frac{\xi}{4} \left( \xi^2-1 \right) \ln \left(\frac{\xi+1}{\xi-1} \right) \right] \\ -\beta=1.
\end{multline}
Rearranging Eq.\@ (\ref{eq:Lea_coll_eq}), we finally achieve 
\begin{equation}\nonumber
-\frac{1}{3} +\frac{\xi^2}{2}-\frac{\xi}{4} \left( \xi^2-1 \right) \ln \left(\frac{\xi+1}{\xi-1} \right) =\frac{1+\beta}{F_1^S-3 \beta},
\end{equation}
or in other words
\begin{multline}\nonumber
\xi^2-\frac{\xi}{2} \left( \xi^2-1 \right) \ln \left(\frac{\xi+1}{\xi-1} \right) -1 \\ =2\left[\frac{3+ 3\beta}{3F_1^S-9 \beta}+\frac{1}{3} \right]-1 \\ =\frac{-F_1^S+6+9 \beta}{3 F_1^S-9 \beta},
\end{multline}
which gives Eq.\@ (\ref{eq:Lea}). 

\section{Derivation of the noninteracting transverse susceptibility without collisions}\label{X_T0}

We start from the kinetic equation (\ref{eq:pert_distr}) with the total quasiparticle force (\ref{eq:qc_force}). This includes the effect of the transverse vector potential $A(\vec{q},\omega)=A_T(\vec{q},\omega)$. Linearizing the change in the quasiparticle distribution function with respect to global equilibrium, similarly to Appendix \ref{kin_eq}, we obtain 
\begin{multline}\label{eq:kin_k_pot}
\left(\vec{q} \cdot \vec{v}_{\vec{k},\sigma}-\omega \right) \epsilon_{\vec{k}}(\vec{q},\omega)  +\vec{q} \cdot \vec{v}_{\vec{k},\sigma}  \\ \cdot \left\{ \sum_{\vec{k}',\sigma'} \textsl{f}_{\vec{k},\sigma, \vec{k}', \sigma'} \epsilon_{\vec{k}'} (\vec{q},\omega) \delta\left[ E_n(\vec{k}',\sigma')-\mu \right]  \right\}   ={\mathscr{I}}_{coll}(\vec{q},\omega) \\ -\frac{e}{m v_F^*} \left(\vec{q} \cdot \vec{v}_{\vec{k},\sigma}\right) \left[\vec{k} \cdot \vec{A}(\vec{q},\omega)\right].
\end{multline}
We transform the sum over $\vec{k}'$ and $\sigma'$ in an integral over angular coordinates at the Fermi wave vector $\vec{k}_F$, similarly to the procedure in Appendix \ref{kin_eq}. This fixes the quasiparticle velocity to the renormalized Fermi velocity, $\vec{v}_{\vec{k},\sigma} \equiv v_F^*$. Then, Eq.\@ (\ref{eq:kin_k_pot}) becomes 
\begin{multline}\label{eq:kin_k_int_coll_A}
\left( \cos{\theta}  -s \right) \epsilon_{\vec{k}}(\vec{q},\omega)+ \cos{\theta} \\ \cdot \left[ \int \frac{d \Omega'}{4 \pi} \sum_{l=0}^{+\infty} F_l^{S,A} \wp_l(\cos \theta') \epsilon_{\vec{k}'} (\vec{q},\omega) +\frac{e}{m v_F^*} \vec{k} \cdot \vec{A}(\vec{q},\omega) \right] \\ = \frac{1}{q v_F^*}{\mathscr{I}}_{coll}(\vec{q},\omega),
\end{multline}
which is just Eq.\@ (\ref{eq:kin_k_L3}) with the addition of the driving term that depends on the applied vector potential $\vec{A}(\vec{q},\omega)$. 
From this point, we perform the same operations on the kinetic equation (\ref{eq:kin_k_int_coll_A}) as we have done for Eq.\@ (\ref{eq:kin_k_L3}): we expand the Fermi surface displacement function $\epsilon_{\vec{k}}(\vec{q},\omega)$ in spherical harmonics using Eq.\@ (\ref{eq:exp_displ_FS}), we consider the first transverse mode with $m=1$ in the density, i.e.\@, symmetric $S$, interaction channel, and we truncate the infinite sum over $l$ to $l=1$, so that the interaction becomes $\sum_{l=0}^{+\infty} F_l^{S,A} \wp_l(\cos \alpha) \equiv F_0^S+F_1^S \cos{\alpha}$, and the Fermi surface displacement can be written as $\epsilon_{\vec{k}}(\vec{q},\omega) =\sum_{l=0}^{+\infty}\epsilon_{l,1}^{S}\mathscr{Y}_{l}^{1}(\theta,\phi) \equiv \epsilon^S(\theta) e^{i \phi}$, where $\epsilon^S(\theta)$ collects the $\theta$-dependent portion of the displacement. We also neglect collisions in this section, i.e.\@, ${\mathscr{I}}_{coll}(\vec{q},\omega)=0$. The resulting kinetic equation is
\begin{multline}\label{eq:kin_k_int_2}
\left( \cos{\theta}  -s \right) \epsilon^S(\theta) e^{i \phi}+ \\ \cos{\theta}  \int_0^{2\pi} \int_0^{\pi} \frac{d \phi' \sin{\theta'} d\theta'}{4 \pi} \left(F_0^S+F_1^S \cos{\alpha} \right)  \epsilon^S(\theta')e^{i \phi'}\\ +\frac{e}{m v_F^*}\cos{\theta} \vec{k} \cdot \vec{A}(\vec{q},\omega)  =0
\end{multline}
By kinetics the angle $\alpha$ is such that $\cos{\alpha}=\cos{\theta}\cos{\theta'}+\sin{\theta}\sin{\theta'}\cos{\left(\phi-\phi'\right)}$ \cite{Dupuis-lect-2011}. 
If we assume no interactions --- i.e.\@, $F_0^S=0$ and $F_1^S=0$ --- the Fermi surface displacement $\epsilon^S(\theta) e^{i \phi}$ follows immediately: 
\begin{equation}\label{eq:nonint_displ}
\epsilon^S(\theta) e^{i \phi}=\frac{e}{m v_F^*} \frac{\cos{\theta} \vec{k} \cdot \vec{A}(\vec{q},\omega)}{s-\cos{\theta}}
\end{equation}
The paramagnetic current density is \cite{Abrikosov-1959,Conti-1999}
\begin{multline}\label{eq:J_para_0}
\vec{J}(\vec{q},\omega)=\frac{1}{\mathscr{V}} \sum_{\vec{k},\sigma} \frac{\vec{k}}{m} v_F^* \delta(\xi_{\vec{k}}) \epsilon_{\vec{k}}(\vec{q},\omega) \\ =\frac{2}{\mathscr{V}} \frac{e}{m^2} \sum_{\vec{k}} \delta(\xi_{\vec{k}}) \frac{\cos{\theta}}{s-\cos{\theta}} \vec{k} \cdot \vec{A}(\vec{q},\omega)= \\ \sum_\nu X_{\mu \nu}^0 (\vec{q},\omega) e A_{\nu} (\vec{q},\omega)
\end{multline}
where the last line defines the noninteracting paramagnetic susceptibility tensor in Landau theory:
\begin{equation}\label{eq:X_0_def}
X_{\mu \nu}^0 (\vec{q},\omega)=\frac{2}{\mathscr{V}} \frac{1}{m^2} \sum_{\vec{k}} \delta(\xi_{\vec{k}}) \frac{\cos{\theta}}{s-\cos{\theta}} k_{\mu} k_{\nu}
\end{equation}
In the transverse case, for polarization along $x$ we take $\mu=\nu=x$ and we obtain
\begin{multline}\nonumber
X_T^0 (\vec{q},\omega)=X_{x x}^0 (\vec{q},\omega) =\frac{2}{\mathscr{V}} \frac{1}{m^2} \sum_{\vec{k}} \delta(\xi_{\vec{k}}) \frac{\cos{\theta}}{s-\cos{\theta}} (k_x)^2 \\ = \frac{2}{m^2} N_{el}^*(E_F) \int_0^{2 \pi} d \phi \int_0^{\pi} \frac{\sin{\theta} d\theta}{4 \pi}  \frac{\cos{\theta}}{s-\cos{\theta}} \left(\frac{k_F^* \sin{\theta}}{2} \right)^2  \\ = 2 \frac{1}{m^2} N_{el}^*(E_F) (k_F^*)^2 \frac{1}{8} \int_0^{\pi} \frac{(\sin{\theta})^3 \cos{\theta}}{s-\cos{\theta}} d \theta  \\ = 3/4 \frac{ n m^*}{m^2} \left[ -\frac{4}{3} + 2 s^2 + s \left( 1-s^2 \right) \ln \left(\frac{s+1}{s-1} \right)\right]  \\ = 3\frac{ n m^*}{m^2}\mathscr{I}(s)
\end{multline}
where we have defined the integral (\ref{eq:vs_int}).
In conclusion we have \cite{Dupuis-lect-2011}
\begin{multline}\label{eq:Xi_T02}
X_T^0(\vec{q},\omega)\equiv X_T^0(s)=3\frac{ n m^*}{m^2}\mathscr{I}(s) \\ =\frac{3}{4} \frac{n m^*}{m^2} \left[ -\frac{4}{3}+ 2 s^2 +s \left(1-s^2\right)\ln{\left(\frac{s+1}{s-1} \right)} \right]
\end{multline}
where the effective mass of Landau quasiparticles is $m^*=m \left(1+ F_1^S/3\right)$. Equation (\ref{eq:Xi_T02}) coincides with Eq.\@ (\ref{eq:Xi_T0}).

\section{Derivation of the interacting transverse susceptibility with collisions}\label{Int_X_transv}
In this section, we detail the derivation of the interacting transverse susceptibility of a Fermi liquid in the presence of collisions. 
Starting from the kinetic equation (\ref{eq:kin_k_int_compl_3}), we calculate the paramagnetic current density in a Fermi liquid using the definition
\begin{multline}\label{eq:FL_J_p}
\vec{J}_p(\vec{q},\omega)=\frac{1}{\mathscr{V}} \sum_{\vec{k},\sigma} \frac{\vec{k}}{m} v_F^* \delta(\xi_{\vec{k}}) \epsilon_{\vec{k},\sigma}(\vec{q},\omega) \\ =\frac{2}{\mathscr{V}} \sum_{\vec{k}} \frac{\vec{k}}{m} v_F^* \delta(\xi_{\vec{k}}) \epsilon_{\vec{k}}(\vec{q},\omega).
\end{multline}
The Fermi surface displacement $\epsilon_{\vec{k}}(\vec{q},\omega)$ can be deduced directly from Eq.\@ (\ref{eq:kin_k_int_compl_3}) if we solve the latter equation for $\epsilon^S(\theta)$: 
\begin{multline}\label{eq:FL_displ_T}
\epsilon^S(\theta) e^{i \phi}= \frac{\left[3 \epsilon^S(\theta) \sin \theta \right]_{av} \sin \theta  e^{i \phi} \left(\frac{F_1^S}{3} \cos \theta - \beta \xi \right)}{\xi-\cos \theta} \\ + \frac{ \frac{e}{m v_F^*} \vec{k} \cdot \vec{A}(\vec{q},\omega) \cos \theta}{\xi-\cos \theta}.
\end{multline} 
Inserting Eq.\@ (\ref{eq:FL_displ_T}) in Eq.\@ (\ref{eq:FL_J_p}), we have
\begin{multline}\label{eq:FL_J_p_2}
\vec{J}^P(\vec{q},\omega)=\frac{2}{\mathscr{V}} \sum_{\vec{k}}\delta(\xi_{\vec{k}}) \frac{\vec{k}}{m} \\ \cdot \frac{\left[3 \epsilon^S(\theta) \sin \theta \right]_{av} \sin \theta  e^{i \phi} \left(\frac{F_1^S}{3} \cos \theta - \beta \xi \right)}{\xi-\cos \theta}  + \vec{J}_0^P(\vec{q},\omega) \\ \equiv \sum_\nu X_{T, \mu \nu}^P(\vec{q},\omega) e A_\nu(\vec{q},\omega)
\end{multline} 
where $\vec{J}_0^P(\vec{q},\omega)=e/(m v_F^*) \vec{k} \cdot \vec{A}(\vec{q},\omega) \cos \theta /\left(\xi-\cos \theta\right)$ is the noninteracting paramagnetic current density, in accordance with the results of Sec.\@ \ref{FL_XT_nocoll}. In the last step of Eq.\@ (\ref{eq:FL_J_p_2}), we have defined the paramagnetic interacting transverse susceptibility $X_{T, \mu \nu}^P(\vec{q},\omega)$ as the ratio between the total paramagnetic current density component $J_{\mu}^P(\vec{q},\omega)$ and the vector potential component $A_\nu(\vec{q},\omega)$. 
We take the transverse component $\mu=\nu=x$ of the transverse current density (\ref{eq:FL_J_p_2}), and we find
\begin{multline}\label{eq:FL_susc_T}
X_T^{P}(\vec{q},\omega)=X_{T,xx}^{P}(\vec{q},\omega)= \\ \frac{2}{\mathscr{V}} \sum_{\vec{k}}\delta(\xi_{\vec{k}}) \frac{k_x}{m} \frac{e^{i \phi}}{e A_x(\vec{q},\omega)} \\ \cdot \frac{\left[3 \epsilon^S(\theta) \sin \theta \right]_{av} \sin \theta \left(\frac{F_1^S}{3} \cos \theta - \beta \xi \right)}{\xi-\cos \theta}  + X_T^0(\xi),
\end{multline} 
where $X_T^0(\vec{q},\omega) \equiv X_T^0(\xi)$ is the noninteracting paramagnetic transverse susceptibility, in accordance with the result (\ref{eq:Xi_T0}) in Sec.\@ \ref{FL_XT_nocoll}. 
Now, from the kinetic equation, not counting the linear term in the vector potential which is already included in $X_T^0(\xi)$, we expect a Fermi surface displacement $\epsilon^S(\theta) \propto \left[ \sin \theta \left( \frac{F_1^S}{3} \cos \theta- \beta \xi \right)\right]/\left( \xi- \cos \theta\right)$ as found previously. Therefore, we can write the term multiplying $\left[3 \epsilon^S(\theta) \sin \theta \right]_{av}$ in Eq.\@ (\ref{eq:FL_susc_T}) as 
\begin{multline}\nonumber
\frac{2}{\mathscr{V}} \sum_{\vec{k}}\delta(\xi_{\vec{k}}) \frac{k_x}{m} \frac{e^{i \phi}}{e A_x(\vec{q},\omega)} \\ \times \frac{\left[3 \epsilon^S(\theta) \sin \theta \right]_{av} \sin \theta \left(\frac{F_1^S}{3} \cos \theta - \beta \xi \right)}{\xi-\cos \theta} \\ \equiv \frac{2}{\mathscr{V} e A_x(\vec{q},\omega)} \sum_{\vec{k}} \delta(\xi_{\vec{k}}) \frac{k_x}{m} \epsilon^S(\theta) e^{i \phi}= X_T^P(\vec{q},\omega).
\end{multline} 
Hence, Eq.\@ (\ref{eq:FL_susc_T}) becomes
\begin{equation}\label{eq:FL_susc_T_2}
X_T^{P}(\vec{q},\omega)=X_T^{P}(\vec{q},\omega)\left[3 \epsilon^S(\theta) \sin \theta \right]_{av} + X_T^0(\vec{q},\omega).
\end{equation}
We are left with the calculation of $\left[3 \epsilon^S(\theta) \sin \theta \right]_{av}$, which gives Eq.\@ (\ref{eq:av_term_Lea}). Inserting the latter into Eq.\@ (\ref{eq:FL_susc_T_2}), we finally achieve
\begin{multline}\label{eq:FL_susc_T_3}
X_T^{P}(\vec{q},\omega)\left\{ 1- 3 \left(\frac{F_1^S}{3}-\beta \right) \left[-\frac{m^2}{3 n m^*} X_T^0(\xi) \right]  + \beta \right\} \\ =X_T^0(\xi).
\end{multline}
Solving Eq.\@ (\ref{eq:FL_susc_T_3}) for $X_T^{P}(\vec{q},\omega)$, we obtain Eq.\@ (\ref{eq:Xi_T}). 

\section{Derivation of the transverse susceptibility with collisions and momentum relaxation}\label{XT_relax}

The transverse susceptibility with momentum relaxation stemming from the kinetic equation (\ref{eq:kin_k_int_compl_2r}) can be related to the momentum-conserving solution (\ref{eq:Xi_T}) as follows \cite{Conti-1999}. If we parametrize the relaxation integral $\mathscr{I}_r(\vec{q},\omega)$ in the relaxation-time approximation (\ref{eq:relax_int}), it is convenient to define a ``dynamic correction to the displacement function'' $\epsilon^S(\theta)$ \cite{Conti-1999}:
\begin{equation}\label{eq:dynam_corr_displ}
\epsilon^D(\theta)=\epsilon^S(\theta)-\frac{i/\tau_{K}}{\omega+i/\tau_{K}}\epsilon^r(\theta),
\end{equation}
where $\epsilon^r(\theta)$ is the ``locally relaxed'' equilibrium displacement function for the relaxing processes (\ref{eq:relax_int}).
We also define the associated modified vector potential $\vec{A}_D(\vec{q},\omega)$,
\begin{multline}\label{eq:A_D}
\vec{A}_D(\vec{q},\omega)=\vec{A}(\vec{q},\omega)-\frac{i/\tau_{K}}{\omega+i/\tau_{K}}\vec{A}_r(\vec{q},\omega) \\ \equiv \frac{\omega}{\omega+i/\tau_{K}}\vec{A}(\vec{q},\omega),
\end{multline}
where $\vec{A}_r(\vec{q},\omega) \equiv \vec{A}(\vec{q},\omega)$ for processes that do not conserve current \cite{Conti-1999}. Inserting Eqs.\@ (\ref{eq:dynam_corr_displ}) and (\ref{eq:A_D}) into the kinetic equation (\ref{eq:kin_k_int_compl_2r}), and using the equilibrium condition for $\epsilon_r(\theta)$ \cite{Conti-1999} 
\begin{multline}\label{eq:equil_rel}
q v_F^{*} \cos{\theta} \left\{ \epsilon^r(\theta) e^{i \phi}  +   \int_0^{\pi} \frac{ \sin{\theta'} d\theta'}{4} F_1^S  \sin \theta \sin{\theta'}  \epsilon^r(\theta') e^{i \phi} \right. \\ \left.  +\frac{e}{m v_F^*} \vec{k} \cdot \vec{A}_r(\vec{q},\omega) \right\}= {\mathscr{I}}_{r}(\vec{q},\omega)=\\= i\frac{\epsilon^r(\theta)-3 \left[\epsilon^r(\theta) \sin \theta \right]_{av} \sin \theta}{\tau_c} e^{i \phi},
\end{multline}

we see that the momentum-relaxing kinetic equation (\ref{eq:kin_k_int_compl_2r}) for $\epsilon^S(\theta)$ is equivalent to the kinetic equation without relaxation for $\epsilon^D(\theta)$, in the presence of the vector potential $\vec{A}_D(\vec{q},\omega)$: 
\begin{multline}\label{eq:kin_k_int_D}
\left( \cos{\theta}  -\zeta \right) \epsilon^D(\theta) e^{i \phi} \\  + \cos{\theta} \left[ \int_0^{\pi} \frac{ \sin{\theta'} d\theta'}{4} F_1^S  \sin \theta \sin{\theta'}  \epsilon^D(\theta') e^{i \phi} \right. \\ \left.  +\frac{e}{m v_F^*} \vec{k} \cdot \vec{A}_D(\vec{q},\omega) \right] = \frac{3 \left[\epsilon^D(\theta) \sin \theta \right]_{av} \sin \theta}{i \omega \tau_c} s e^{i \phi}, 
\end{multline}
where we have defined the variable (\ref{eq:zeta_var}). 
Therefore, from this point on we can follow Appendices \ref{X_T0} and \ref{Int_X_transv}, to calculate the paramagnetic current density from the definition (\ref{eq:FL_J_p}) and the associated paramagnetic susceptibility $X_T^{P}(\vec{q},\omega)$. Specifically, Eq.\@ (\ref{eq:kin_k_int_D}) can be rewritten as
\begin{multline}\label{eq:kin_k_int_eps_D_2}
\left( \cos{\theta}  -\zeta \right) \epsilon^D(\theta) e^{i \phi}\\  + \left[ 3 \epsilon^D(\theta) \sin \theta\right]_{av} \sin \theta e^{i \phi} \left( \frac{F_1^S}{3} \cos \theta-\beta \xi \right)  \\=-\cos \theta \frac{e}{m v_F^*} \vec{k} \cdot \vec{A}_D(\vec{q},\omega). 
\end{multline}
The Fermi-surface displacement $\epsilon_{\vec{k}}(\vec{q},\omega) \equiv \epsilon^D(\theta)$ from Eq.\@ (\ref{eq:kin_k_int_eps_D_2}) is 
\begin{multline}\label{eq:eps_D_solve}
\epsilon^D(\theta)e^{i \phi}=\frac{\left[ 3 \epsilon^D(\theta) \sin \theta\right]_{av} \sin \theta e^{i \phi} \left( \frac{F_1^S}{3} \cos \theta-\beta \xi \right)}{\zeta- \cos \theta}\\ +\frac{e}{m v_F^*}\frac{\cos \theta \vec{k} \cdot \vec{A}_D(\vec{q},\omega)}{\zeta- \cos \theta}.
\end{multline}
The paramagnetic current density is then
\begin{multline}\label{eq:J_P_eps_D}
\vec{J}_P(\vec{q},\omega)=\frac{2}{\mathscr{V}} \sum_{\vec{k}} \frac{\vec{k}}{m} v_F^{*} \delta(\xi_{\vec{k}}) \\ \times \frac{\left[ 3 \epsilon^D(\theta) \sin \theta\right]_{av} \sin \theta e^{i \phi} \left( \frac{F_1^S}{3} \cos \theta-\beta \xi \right)}{\zeta- \cos \theta} +\vec{J}_P^0(\zeta),
\end{multline}
where we have identified the noninteracting paramagnetic current density
\begin{multline}\label{eq:J_P_0}
\vec{J}_P^0(\vec{q},\omega)=\vec{J}_P^0(\zeta) \\ =\frac{2}{\mathscr{V}} \sum_{\vec{k},\sigma} \frac{\vec{k}}{m} \delta(\xi_{\vec{k}}) \frac{e}{m} \vec{k} \cdot \vec{A}_D(\vec{q},\omega) \frac{\cos \theta}{\zeta- \cos \theta}.
\end{multline}
Since in linear-response theory the paramagnetic transverse susceptibility tensor $\tilde{X}_{T, \mu \nu}^P$ satisfies
\begin{equation}\label{eq:J_P_X_P}
\vec{J}_P(\vec{q},\omega)=\sum_\nu \tilde{X}_{T, \mu \nu}^P(\vec{q},\omega) e A_{D,\mu}(\vec{q},\omega),
\end{equation}
in our case choosing $\nu=\mu=x$ we have
\begin{multline}\label{eq:tilde_X_T_P}
\tilde{X}_{T, x x}^P(\vec{q},\omega)\equiv \tilde{X}_{T}^P(\vec{q},\omega) \\ =\frac{2}{\mathscr{V}} \sum_{\vec{k}} \frac{k_x}{e m A_{D,x}(\vec{q},\omega)} v_F^{*} \delta(\xi_{\vec{k}}) \\ \times \frac{\left[ 3 \epsilon^D(\theta) \sin \theta\right]_{av} \sin \theta e^{i \phi} \left( \frac{F_1^S}{3} \cos \theta-\beta \xi \right)}{\zeta- \cos \theta}+\tilde{X}_T^0(\zeta),
\end{multline}
where the noninteracting transverse susceptibility $\tilde{X}_T^0(\vec{q},\omega)\equiv\tilde{X}_T^0(\zeta)$ is 
\begin{equation}\label{eq:X_T^0_zeta}
\tilde{X}_T^0(\vec{q},\omega)\equiv\tilde{X}_T^0(\zeta)=\frac{J_P^0(\vec{q},\omega)}{e A_{D,x}(\vec{q},\omega)}.
\end{equation}
$\tilde{X}_T^0(\zeta)$ formally satisfies Eq.\@ (\ref{eq:Xi_T0}) upon substitution $s \mapsto \zeta$.
We notice that, not counting the term depending on the vector potential which is already contained in $\tilde{X}_T^0(\zeta)$, 
\begin{multline}\nonumber
\epsilon^D(\theta) \propto \frac{2}{\mathscr{V}} \sum_{\vec{k},\sigma} \frac{k_x}{e m A_{D,x}(\vec{q},\omega)} v_F^{*} \delta(\xi_{\vec{k}}) \\ \times \frac{ \sin \theta \left( \frac{F_1^S}{3} \cos \theta-\beta \xi \right)}{\zeta- \cos \theta},
\end{multline}
so that
\begin{multline}\label{eq:displ_D_equiv}
\frac{2}{\mathscr{V}} \sum_{\vec{k},\sigma} \frac{k_x}{e m A_{D,x}(\vec{q},\omega)} v_F^{*} \delta(\xi_{\vec{k}}) e^{i \phi} \\ \times \frac{\left[ 3 \epsilon^D(\theta) \sin \theta\right]_{av} \sin \theta \left( \frac{F_1^S}{3} \cos \theta-\beta \xi \right)}{\zeta- \cos \theta} \\ \equiv \tilde{X}_{T}^P(\vec{q},\omega)
\end{multline}
Inserting Eq.\@ (\ref{eq:displ_D_equiv}) into Eq.\@ (\ref{eq:tilde_X_T_P}), we obtain the self-consistency condition
\begin{equation}\label{eq:X_T_P_expl}
\tilde{X}_{T}^P(\vec{q},\omega)=\frac{\tilde{X}_T^0(\vec{q},\omega)}{1-\left[ 3 \epsilon^D(\theta) \sin \theta\right]_{av}}.
\end{equation}
We finally evaluate $\left[ 3 \epsilon^D(\theta) \sin \theta\right]_{av}$ as follows:
\begin{multline}\label{eq:eps_D_av}
\left[ 3 \epsilon^D(\theta) \sin \theta\right]_{av}=3 \int_0^\pi \frac{ (\sin \theta)^2 d \theta}{4} \frac{F_1^S}{3} \frac{\sin \theta \cos \theta}{\zeta-\cos \theta} \\ - \beta \int_0^\pi \frac{ (\sin \theta)^2 d \theta}{4}  \frac{ \xi \sin \theta }{\zeta-\cos \theta} \\ =3 \frac{F_1^S}{3} \left[-\frac{1}{3} + \frac{\xi^2}{2} -\frac{\xi}{4} (\xi^2-1) \ln \left(\frac{\xi+1}{\xi-1}\right) \right] \\  - 3 \beta \xi \left[ \frac{\zeta}{2} -\frac{1}{4} (\zeta^2-1) \ln \left(\frac{\zeta+1}{\zeta-1}\right) \right]\\ = 3 \frac{F_1^S}{3} \left[-\frac{1}{3} + \frac{\zeta^2}{2} -\frac{\zeta}{4} (\zeta^2-1) \ln \left(\frac{\zeta+1}{\zeta-1}\right) \right] \\  - 3 \beta \frac{\xi}{\zeta} \left[-\frac{1}{3}+ \frac{\zeta^2}{2} -\frac{\zeta}{4} (\zeta^2-1) \ln \left(\frac{\zeta+1}{\zeta-1}\right) \right] -\beta \frac{\xi}{\zeta} \\= 3 \left(\frac{F_1^S}{3}-\beta \frac{\xi}{\zeta}\right) \mathscr{I}(\zeta)-\beta \frac{\xi}{\zeta},
\end{multline}
where $\mathscr{I}(\zeta)$ satisfies Eq.\@ (\ref{eq:vs_int}). 

Using Eqs.\@ (\ref{eq:A_D}) and (\ref{eq:Xi_T0}) and $m^{*}=m\left(1+F_1^S/3\right)$, Eq.\@ (\ref{eq:X_T_P_expl}) becomes
\begin{multline}\label{eq:X_T_P_expl2}
\tilde{X}_{T}^P(\vec{q},\omega)=\frac{J_P(\vec{q},\omega)}{e A_{D,x}(\vec{q},\omega)} \\ =\frac{n}{m}\frac{3 \left(1+\frac{F_1^S}{3}\right)\mathscr{I}(\zeta)}{1-3\left(\frac{F_1^S}{3}-\tilde{\beta}\right)\mathscr{I}(\zeta)+\tilde{\beta}},
\end{multline}
where we have defined
\begin{equation}\label{eq:beta_tilde}
\tilde{\beta}= \frac{\xi}{\zeta} \beta =\frac{1+i/(\omega \tau_c)}{1+i/(\omega \tau_c)+i/(\omega \tau_K)} \frac{1}{i \omega \tau_c-1}.
\end{equation}
The total transverse susceptibility, including the paramagnetic and diamagnetic terms, is then 
\begin{multline}\label{eq:X_T_expl_coll}
\tilde{X}_{T}(\vec{q},\omega)=\tilde{X}_{T}^P(\vec{q},\omega)+\frac{n}{m}\\ =\frac{n}{m}\frac{\left(1+\tilde{\beta}\right)\left[1+3\mathscr{I}(\zeta)\right]}{1-3\left(\frac{F_1^S}{3}-\tilde{\beta}\right)\mathscr{I}(\zeta)+\tilde{\beta}}.
\end{multline}
Now, remembering that $\tilde{X}_{T}(\vec{q},\omega)$ represents the linear response to the vector potential $\vec{A}_D(\vec{q},\omega)$ which satisfies Eq.\@ (\ref{eq:A_D}) \cite{Conti-1999}, the response function $X_T(\vec{q},\omega)$ corresponding to the physical vector potential $\vec{A}(\vec{q},\omega)$ is then $X_T^r(\vec{q},\omega)=\omega/\left(\omega+i/\tau_K\right) \tilde{X}_T (\vec{q},\omega)$. This yields Eq.\@ (\ref{eq:Xi_T_r}). 

\section{Derivation of Landau quasiparticle scattering time}\label{app:FL_scat}
In a Fermi liquid, short-ranged quasiparticle residual interactions at the Fermi surface are usually expressed in terms of the symmetric and antisymmetric Landau parameters (\ref{eq:Landau_par3D}) \cite{Abrikosov-1959}. The many-body polarization of the medium renormalizes these interactions, producing renormalized Landau parameters \cite{vanderMarel-2011, Coleman-2015mb}
\begin{equation}\label{eq:Landau_A}
A_l^j=\frac{F_l^j}{1+F_l^j/(2l+1)}.
\end{equation}
The quantities (\ref{eq:Landau_A}) represent the scattering amplitudes between quasiparticles, and they can mediate Cooper pairing in superconducting materials \cite{vanderMarel-2011}. They also enter into the Fermi-liquid collision time, due to phase-space limitation of quasiparticle scattering \cite{vanderMarel-2011, Mahan-2000,Coleman-2015mb}, in accordance with
\begin{equation}\label{eq:tau_qp}
\frac{1}{\tau_{qp}(\omega,T)}=\frac{(m^{*})^3}{12 \pi^2 \hbar^3} \left[(\hbar \omega)^2+( \pi k_B T)^2 \right] \left\langle \frac{W(\theta, \phi)}{\cos{\left(\frac{\theta}{2}\right)}} \right\rangle,
\end{equation}
where $W(\theta, \phi)$ is the transition probability governing inelastic scattering at the Fermi surface, and $\left\{\theta, \phi\right\}$ are the angles between the Fermi momentum $\vec{k}_F$ and the excitation momentum $\vec{q}$ in 3D space. The brackets $\left\langle \cdot \right\rangle$ represent an average over the solid angle in momentum space. 
In s-p approximation \cite{vanderMarel-2011,Mahan-2000}, we consider short-range quasiparticle interactions only in the angular channels $l=\left\{0,1\right\}$, comprising both symmetric and antisymmetric parts. Then, the angular average in Eq.\@ (\ref{eq:tau_qp}) is performed on the transition probability
\begin{equation}\label{eq:ang_qp}
W(\theta, \phi)=\frac{\pi}{4 \hbar} \left[ A_S(\theta,\phi) + A_t(\theta,\phi) \right]^2+\frac{\pi}{2 \hbar} \left[ A_t(\theta,\phi) \right]^2,
\end{equation}
where 
\begin{equation}\label{eq:A_S}
A_S(\theta,\phi)=\frac{1}{N_0^{el}(0)} \left[(A_0^S-3 A_0^A)+(A_1^S-3 A_1^A \cos{\theta}) \right]
\end{equation}
and 
\begin{equation}\label{eq:A_T}
A_t(\theta,\phi)=\frac{1}{N_0^{el}(0)} \left[(A_0^S+ A_0^A)+(A_1^S+ A_1^A \cos{\theta}) \right]\cos{\phi}
\end{equation}
are the scattering amplitudes for singlet and triplet state, respectively \cite{vanderMarel-2011, Mahan-2000}. 
Averaging over the angular coordinates $\theta$ and $\phi$ gives
\begin{multline}\label{eq:ang_qp_av}
\left\langle \frac{W(\theta, \phi)}{\cos{\frac{\theta}{2}}} \right\rangle=\int \frac{\sin{\theta} d\theta d\phi}{4 \pi} \frac{W(\theta, \phi)}{\cos{\frac{\theta}{2}}}=\\=12 (\lambda_t)^2 \frac{\pi^5 \hbar^5}{(m^*)^3 E_F^*},
\end{multline}
where $12 (\lambda_t)^2$ contains contributions from the Landau interaction parameters $\left\{ A_0^S, A_0^A, A_1^S, A_1^A \right\}$ with numerical coefficients, in accordance with Eq.\@ (B.7) of Ref.\@ \cite{vanderMarel-2011}. 
\begin{figure}[t]
\includegraphics[width=0.9\columnwidth]{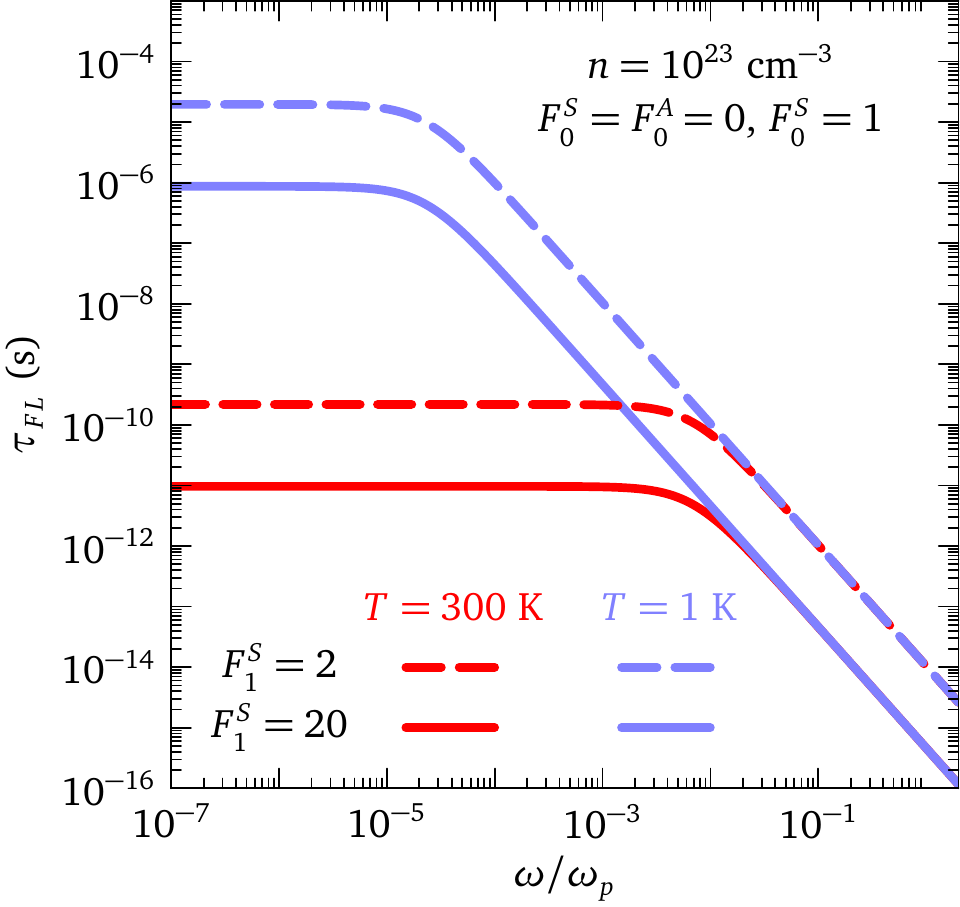}
\caption{\label{fig:tau_FL} 
Fermi-liquid quasiparticle collision time $\tau_{FL}(\omega, T)$ as a function of frequency $\omega$ normalized to the plasma frequency $\omega_p$, according to Eq.\@ (\ref{eq:tau_qp_fin2}). The red curve is for $T=300 \, K$ and the purple curve is for $T=1 \, K$. Dashed curves are for $F_1^S=2$ and solid curves for $F_1^S=20$. We assume the following material parameters: renormalized Fermi velocity $v_F^*/c=10^{-3}$; Landau parameters $F_0^S=F_0^A=F_1^A=0$. The resulting plasma frequency is $\hbar \omega_p=19.09 \, eV$. 
}
\end{figure}
In evaluating the collision time (\ref{eq:tau_qp}) we have to consider that Fermi-liquid quasiparticles cannot scatter on an energy scale larger than the Fermi energy $E_F$. This sets a lower bound for the scattering time
\begin{equation}\label{eq:tau_FL_min}
\tau_{min}=\frac{\hbar}{E_F}.
\end{equation}
Inserting Eq.\@ (\ref{eq:ang_qp_av}) into Eq.\@ (\ref{eq:tau_qp}) finally yields a quadratic energy-temperature dependence of the collision time \cite{vanderMarel-2011, Berthod-2013}, 
\begin{multline}\label{eq:tau_qp_fin2}
\tau_{FL}(\omega,T)=\tau_{qp}(\omega,T) \\ =\left\{\frac{\pi}{\hbar E_F^{*}} (\lambda_t)^2  \left[(\hbar \omega)^2+( \pi k_B T)^2 \right]\right\}^{-1},
\end{multline}
where $\hbar \omega$ is the excitation frequency, $k_B T$ is the thermal energy at temperature $T$, $E_F^{*}=\left[\hbar^2 (k_F)^2\right]/(2 m^*)$ is the renormalized Fermi level, and $(\lambda_t)^2$ stems from the angular integration of the scattering probability and depends on the Landau parameters $\left\{F_l^S,F_l^A\right\}$. When we retain only $F_0^S$ and $F_1^S$ in the expansion of the quasiparticle interactions in terms of spherical harmonics, $(\lambda_t)^2$ becomes
\begin{equation}\label{eq:lambda_t_01}
(\lambda_t)^2=\frac{1}{12} \left[ \frac{7}{24} (A_1^S)^2+\frac{5}{8}(A_0^S)^2-\frac{5}{12} A_0^S A_1^S \right].
\end{equation}
From Eqs.\@ (\ref{eq:tau_qp_fin2}) and (\ref{eq:lambda_t_01}), we see that the collisionless limit $\tau_{FL} \rightarrow +\infty$ is reached only when \emph{both} $\omega=0$ and $T=0$, or $(\lambda_t)^2=0$ which is the noninteracting Fermi-gas limit. 
Figure \ref{fig:tau_FL} shows the Fermi-liquid scattering time (\ref{eq:tau_qp_fin2}) in s-p approximation as a function of frequency $\omega$, at temperatures $T=300 \, K$ and $T=1 \, K$ and for the following material parameters: renormalized Fermi velocity $v_F^{*}/c=10^{-3}$, first Landau parameters $F_1^S=\left\{2, \, 20\right\}$, and other Landau parameters $F_0^S=F_0^A=F_1^A=0$. 

\section{Electron-phonon collision time for Debye phonon spectrum}\label{app:e-ph_scat}

A relaxation-time approximation for the Boltzmann equation is justified for acoustic phonon scattering \cite{Lavasani-2019}. In the following, we derive simple expressions for the electronic relaxation rate due to acoustic phonons from many-body theory. 

The self-energy of electrons due to one-phonon exchange processes satisfies
\begin{equation}\label{eq:Sigma_eph}
\Sigma(\vec{k},i \omega_n)=-\frac{k_B T}{\mathscr{V}} \sum_{\vec{q}, i \Omega_n} V_{e-ph} (\vec{q},i \Omega_n) \mathscr{G}(\vec{k}-\vec{q},i \omega_n-i\Omega_n),
\end{equation}
where $\mathscr{G}(\vec{k},i \omega_n)$ is the Matsubara form of the electron propagator $\mathscr{G}(\vec{k},\tau)=-\left\langle \mathscr{T}_\tau \left[c_{\vec{k}}(\tau)c^{\dagger}_{\vec{k}}(0)\right]\right\rangle$ with $c_{\vec{k}}, \, c^\dagger_{\vec{k}}$ electronic field operators \cite{Mahan-2000,Bruus-2004mb,Berthod-2018}. The electron-phonon interaction potential $V_{e-ph} (\vec{q},i \Omega_n)$ is
\begin{equation}\label{eq:V_eph}
V_{e-ph} (\vec{q},i \Omega_n) = \left|g_\lambda(\vec{q})\right|^2 \mathscr{D}_\lambda(\vec{q},i \Omega_n).
\end{equation}
In Eq.\@ (\ref{eq:V_eph}), $\mathscr{D}_\lambda(\vec{q},i \Omega_n)$ is the Matsubara form of the phonon propagator $\mathscr{D}_\lambda(\vec{q},\tau)=-\left\langle \mathscr{T}_\tau \left[B_{\vec{q},\lambda}(\tau)B^{\dagger}_{\vec{q},\lambda}(0)\right]\right\rangle$ for branch $\lambda$, and $g_\lambda(\vec{q})$ is the electron-phonon vertex. We also defined the linear combination $B^\dagger_{\vec{q},\lambda}=b^\dagger_\lambda(\vec{q})+b_\lambda(-\vec{q})$ of the phonon field operators $b_{\lambda}(\vec{q}), \, b_{\lambda}(\vec{q})$. 

We assume the free-fermion propagator 
\begin{equation}\label{eq:G_0}
\mathscr{G}^0(\vec{k},i \omega_n)=\frac{1}{i \omega_n-\xi_{\vec{k}}}
\end{equation}
and the free phonon propagator
\begin{equation}\label{eq:D_0}
\mathscr{D}^0_\lambda(\vec{q},i \Omega_n)=\frac{1}{i \Omega_n-\omega_\lambda(\vec{q})}-\frac{1}{i \Omega_n+\omega_\lambda(\vec{q})}. 
\end{equation}
Inserting Eqs.\@ (\ref{eq:G_0}) and (\ref{eq:D_0}) into Eq.\@ (\ref{eq:Sigma_eph}), we obtain 
\begin{multline}\label{eq:Sigma_eph2}
\Sigma(\vec{k},i \omega_n)=-\frac{k_B T}{\mathscr{V}} \sum_{\vec{q}, \lambda}\left|g_\lambda(\vec{q})\right|^2 \\ \times \sum_{i \Omega_n} \left[ \frac{1}{i \Omega_n-\omega_\lambda(\vec{q})}-\frac{1}{i \Omega_n+\omega_\lambda(\vec{q})} \right] \frac{1}{i \omega_n-i\Omega_n-\xi_{\vec{k}-\vec{q}}}. 
\end{multline}
We use the property of the Matsubara sum whereby $k_B T \sum_{i \nu_m} F(i \nu_m)$, with $F(i \nu_m)$ function of Matsubara frequencies, gives the sum of the residues of $f_{BE}(z) F(z)$ for bosons, and of $-f_{FD}(z) F(z)$ for fermions, at the poles of $F(z)$ : $z \in \mathbb{C}$, if $\lim_{\left|z\right|\rightarrow +\infty} f_{BE}(z) F(z) =0$, and $\lim_{\left|z\right|\rightarrow +\infty} -f_{FD}(z) F(z) =0$, respectively. This gives
\begin{multline}\label{eq:Sigma_eph3}
\Sigma(\vec{k},i \omega_n)=-\frac{1}{\mathscr{V}} \sum_{\vec{q}, \lambda}\left|g_\lambda(\vec{q})\right|^2 \\ \times \left\{ \frac{-f_{BE}\left[\omega_\lambda(\vec{q})\right]}{i \omega_n-\omega_\lambda(\vec{q})-\xi_{\vec{k}-\vec{q}}}-\frac{-f_{BE}\left[-\omega_\lambda(\vec{q})\right]}{i \omega_n+\omega_\lambda(\vec{q})-\xi_{\vec{k}-\vec{q}}} \right. \\+ \left. f_{BE}(i \omega_n-\xi_{\vec{k}-\vec{q}}) \right. \\ \left. \times \left[\frac{1}{i \omega_n-\xi_{\vec{k}-\vec{q}}-\omega_\lambda(\vec{q})}-\frac{1}{i \omega_n-\xi_{\vec{k}-\vec{q}}+\omega_\lambda(\vec{q})}\right] \right\}. 
\end{multline}
Using $f_{BE}(-\omega)=1-f_{BE}(\omega)$, $f_{BE}(i\omega_n-\xi)=-f_{FD}(-\xi)$, and $f_{FD}(-\xi)=1-f_{FD}(\xi)$, we have
\begin{multline}\label{eq:Sigma_eph4}
\Sigma(\vec{k},i \omega_n)=-\frac{1}{\mathscr{V}} \sum_{\vec{q}, \lambda}\left|g_\lambda(\vec{q})\right|^2 \\ \times \left\{ \frac{f_{FD}(-\xi_{\vec{k}-\vec{q}})+f_{BE}\left[\omega_\lambda(\vec{q})\right]}{i \omega_n-\omega_\lambda(\vec{q})-\xi_{\vec{k}-\vec{q}}} \right. \\ \left. + \frac{f_{FD}(\xi_{\vec{k}-\vec{q}})+f_{BE}\left[\omega_\lambda(\vec{q})\right]}{i \omega_n+\omega_\lambda(\vec{q})-\xi_{\vec{k}-\vec{q}}} \right\}.
\end{multline}
We now overlook any anisotropy in the electron-phonon matrix element $g_\lambda(\vec{q}) \equiv g\left[\omega_\lambda(\vec{q})\right]$, which allows us to transform the sum over momenta $\vec{q}$ in Eq.\@ (\ref{eq:Sigma_eph4}) into an integral over energy $\Omega$, making use of the phonon density of states (PDOS) per unit volume $N_{ph}(\Omega)=\mathscr{V}^{-1} \sum_{\vec{q},\lambda} \delta\left[\Omega-\omega_\lambda(\vec{q})\right]$. We identify the electron-phonon spectral function with the product of the electron-phonon matrix element times the PDOS, 
\begin{equation}\label{eq:alpha2F}
\alpha^2F(\Omega)=\left|g(\Omega)\right|^2 N_{ph}(\Omega).
\end{equation}
With the definition (\ref{eq:alpha2F}), Eq.\@ (\ref{eq:Sigma_eph4}) becomes 
\begin{multline}\label{eq:Sigma_eph5}
\Sigma(\vec{k},i \omega_n)=-\int_{-\infty}^{+\infty} d \Omega \alpha^2F(\Omega) \\ \times \left\{ \frac{f_{FD}(-\xi_{\vec{k}-\vec{q}})+f_{BE}\left[\omega_\lambda(\vec{q})\right]}{i \omega_n-\omega_\lambda(\vec{q})-\xi_{\vec{k}-\vec{q}}} \right. \\ \left. + \frac{f_{FD}(\xi_{\vec{k}-\vec{q}})+f_{BE}\left[\omega_\lambda(\vec{q})\right]}{i \omega_n+\omega_\lambda(\vec{q})-\xi_{\vec{k}-\vec{q}}} \right\}.
\end{multline}
We can achieve a momentum-independent self-energy with an appropriate average of Eq.\@ (\ref{eq:Sigma_eph4}) over momenta $\vec{k}$:
\begin{multline}\label{eq:Sigma_eph6}
\left\langle \Sigma(i \omega_n)\right\rangle_{\vec{k}}=\int_{-\infty}^{+\infty} d \xi N_0^{el}(\xi_{\vec{k}}) \Sigma(\vec{k},i \omega_n)= \\ -\int_{-\infty}^{+\infty} d \Omega \alpha^2F(\Omega) \int_{-\infty}^{+\infty} d \xi N_0^{el}(\xi) \\ \times \left\{ \frac{f_{FD}(-\xi)+f_{BE}\left[\omega_\lambda(\vec{q})\right]}{i \omega_n-\omega_\lambda(\vec{q})-\xi} + \frac{f_{FD}(\xi)+f_{BE}\left[\omega_\lambda(\vec{q})\right]}{i \omega_n+\omega_\lambda(\vec{q})-\xi} \right\},
\end{multline}
where we have used the electron density of states $N_0^{el}(E)=\mathscr{V}^{-1} \sum_{\vec{k}} \delta \left(E-\xi_{\vec{k}}\right)$. In the following, we assume a constant density of states within a bandwidth $\mathscr{W}$ much larger than any other energy scale of the problem. Formally,
\begin{equation}\label{eq:el_DOS}
N_0^{el}(\xi)=\begin{cases} N_0^{el}(0), \, \left|\xi \right| < \mathscr{W}, \\ 0, \, \left|\xi \right| > \mathscr{W} \end{cases}
\end{equation}
If we are interested in the energy and temperature dependence of the scattering rate due to electron-boson interactions, we can use the identity $\mathrm{Im}\left\{ (x+i 0^+)^{-1} \right\}=-\pi \delta(x)$ to find the imaginary part of the momentum-averaged self-energy (\ref{eq:Sigma_eph6}). The integral over the quasiparticle energies $\xi$ is convergent in the limit $\mathscr{W} \rightarrow +\infty$. Employing the symmetries $\alpha^2F(-\Omega)=-\alpha^2F(\Omega)$, $f_{FD}(-\epsilon)=1-f_{FD}(\epsilon)$, and $f_{BE}(-\Omega)=-1-f_{BE}(\Omega)$, we obtain 
\begin{multline}\label{eq:Im_Sigma_eph}
-\mathrm{Im}\left\langle \Sigma(\omega)\right\rangle_{\vec{k}}=-\pi N_0^{el}(0) \int_0^{+\infty} d\Omega \alpha^2F(\Omega) \\ \times \left[f_{FD}(\Omega-\omega)+f_{FD}(\Omega+\omega)+2 f_{BE}(\Omega) \right]. 
\end{multline}
It is customary to define the strength of the electron-phonon coupling by a coupling constant, which is derived from Eq.\@ (\ref{eq:Sigma_eph6}):
\begin{equation}\label{eq:lambda_eph}
\lambda=\left. \frac{d \mathrm{Re}\left\langle \Sigma(i \omega_n)\right\rangle_{\vec{k}}}{d \omega} \right|_{\omega=0, T=0}=2 \int_0^{+\infty} \frac{d\Omega}{\Omega} \alpha^2 F(\Omega),
\end{equation}
where in the last step we employed the property $\alpha^2F(-\Omega)=-\alpha^2F(\Omega)$. 

In the following, we consider the simple case of an acoustic phonon spectrum, characterized by the dispersion relation $\omega_\lambda(\vec{q}) = v_a q$ where $v_a$ is the acoustic phonon sound velocity. Given a density of lattice ions $n_{\rm ions}$, the Debye energy is then $\hbar \omega_D=\hbar \left( 6 \pi^2 n_{\rm ions} \right)^{\frac{1}{3}} v_a$ \cite{Ashcroft-1976}. The density of states for such acoustic phonon spectrum is then
\begin{equation}\label{eq:Debye_DOS}
N_D(E)=\frac{3 E^2}{2 (v_a)^3 \pi^3} \Theta(\hbar \omega_D-E)
\end{equation}
with $\Theta(x)$ the Heaviside step function. Assuming a constant electron-phonon matrix element $g(\Omega)\equiv g_a \, \forall \Omega$, we can write the electron-phonon spectral function (\ref{eq:alpha2F}) of the acoustic spectrum as 
\begin{equation}\label{eq:alpha2F_ac}
\alpha^2F_a(\Omega)=\lambda \frac{\Omega^2}{(\hbar \omega_D)^2} \Theta(\hbar \omega_D-\Omega),
\end{equation}
where $\lambda=3\left|g_a\right|^2 6 \pi^2 n_{ions}(2 \hbar \omega_D)$. Inserting the Debye-spectrum electron-phonon spectral function (\ref{eq:alpha2F_ac}) into Eq.\@ (\ref{eq:Im_Sigma_eph}), we obtain the imaginary part of the self-energy due to exchange of acoustic phonons 
\begin{equation}\label{eq:el-ph_tau_D}
-\mathrm{Im}\left\langle \Sigma(i \omega_n)\right\rangle_{\vec{k}}=\pi \lambda \hbar \omega_D \mathscr{S} \left(\frac{\omega}{\hbar \omega_D}, \frac{k_B T}{\hbar \omega_D}\right),
\end{equation}
where we have defined the function 
\begin{multline}\label{eq:el-ph_S}
\mathscr{S} \left(x,t\right)=t^3 \int_0^{\frac{1}{t}} du u^2 \left(\frac{1}{e^{u-\frac{x}{t}}+1}+\frac{1}{e^{u+\frac{x}{t}}+1} \right. \\ \left. +\frac{2}{e^u-1}\right),
\end{multline}
which possesses an analytical solution in terms of polylogarithms. For $t \gg 1$, we have $t^{-1} \rightarrow 0$ and we can expand the integrand in Eq.\@ (\ref{eq:el-ph_S}) in powers of $u$: this gives $2u + o(u^3)$, from which $\lim_{t\rightarrow +\infty} \mathscr{S}(x,t)=t$. Hence we can write
\begin{equation}\label{eq:el-ph_ht}
-\mathrm{Im}\left\langle \Sigma(\omega)\right\rangle_{\vec{k}}=\pi \lambda k_B T \, : \, \frac{k_B T}{\hbar \omega_D}\gg 1.
\end{equation}
\begin{figure}[t]
\includegraphics[width=0.65\columnwidth]{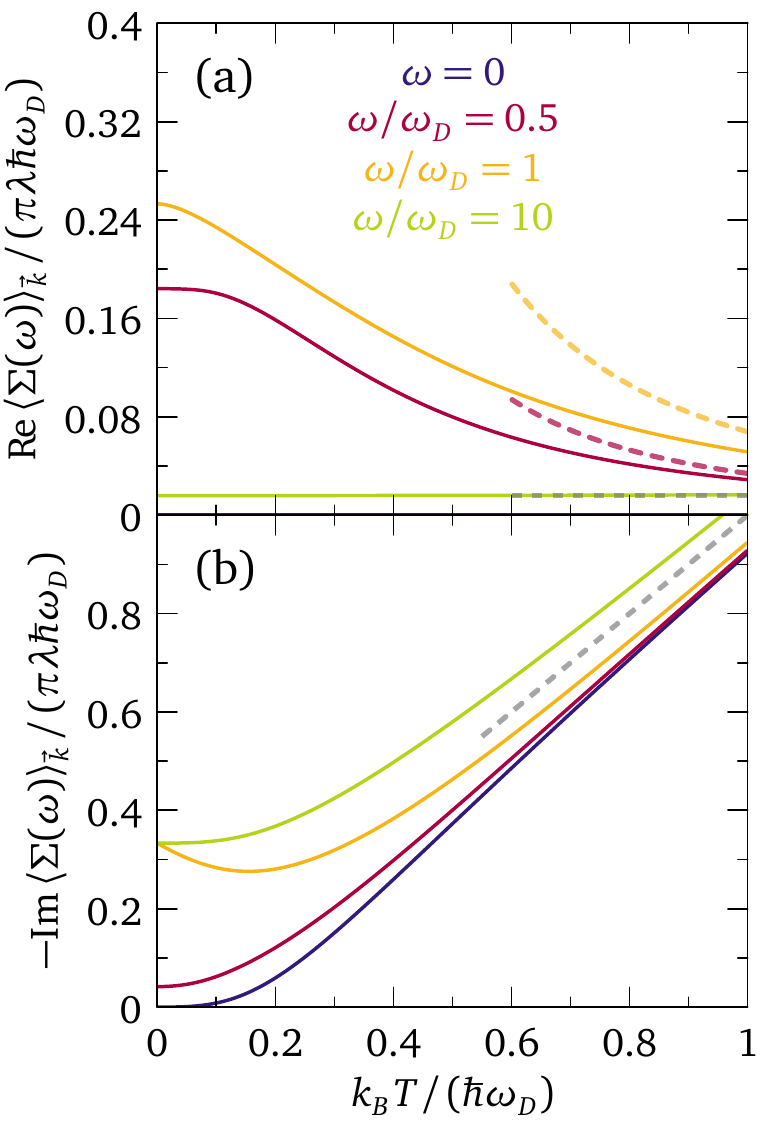}
\caption{\label{fig:sigma_eph} 
Normalized momentum-averaged self-energy $\left\langle \Sigma(\omega)\right\rangle_{\vec{k}}$ as a function of normalized temperature $k_B T/(\hbar \omega_D)$ for scattering off acoustic phonons. The blue, red, gold and green curves are calculated at frequency $\omega/\omega_D=\left\{0,0.5,1,10\right\}$ respectively. Solid lines in panel (a) show the real part of the self-energy according to Eq.\@ (\ref{eq:Re_Sigma_ac}), with dashed lines indicating the corresponding high-temperature expansion (\ref{eq:Re_Sigma_ac_ht}). Solid lines in panel (b) show the imaginary part of the self-energy from Eq.\@ (\ref{eq:el-ph_tau_D}). The gray dashed line is $k_B T/(\hbar \omega_D)$. 
}
\end{figure}
Consider now the opposite limit $t\rightarrow 0$. We can focus on $x>0$ since the function $\mathscr{S} \left(-x,t\right)=\mathscr{S} \left(x,t\right) \, \forall x$. The first term in brackets in Eq.\@ (\ref{eq:el-ph_S}) becomes the step function 
\begin{equation}\nonumber
\lim_{t\rightarrow 0} e^{u-\frac{x}{t}}+1=\begin{cases} 1, \, u<\frac{x}{t} \\ 0, \, u>\frac{x}{t} \end{cases},
\end{equation}
which cuts the superior limit of the integral in Eq.\@ (\ref{eq:el-ph_S}) to $\min{\left(t^{-1}, x/t\right)}$. The second term in the integrand vanishes and the third one is negligible in the limit $t\rightarrow 0$ due to the $t^3$ prefactor in Eq.\@ (\ref{eq:el-ph_S}). All in all, we obtain $\lim_{t\rightarrow 0} \mathscr{S}(x,t)=\min{\left(\left|x\right|^3, 1 \right)}$, so that
\begin{multline}\label{eq:el-ph_0t}
-\mathrm{Im}\left\langle \Sigma(\omega)\right\rangle_{\vec{k}}=\pi \lambda \hbar \omega_D \min{\left(\left|\frac{\omega}{\hbar \omega_D}\right|^3, 1 \right)} \\ : \, \frac{k_B T}{\hbar \omega_D} \rightarrow 0. 
\end{multline}
To calculate the real part of the momentum-averaged self-energy $\left\langle \Sigma(\omega)\right\rangle_{\vec{k}}$, we can employ Kramers-Kronig transformations on Eq.\@ (\ref{eq:Sigma_eph6}) with the electronic DOS (\ref{eq:el_DOS}). We must keep $\mathscr{W}<+\infty$ until the very end of the calculation to ensure the convergence of the $\xi$ integral. We have 
\begin{multline}\label{eq:Re_Sigma}
\mathrm{Re}\left\langle \Sigma(\omega)\right\rangle_{\vec{k}}=-\int_0^{+\infty} d\Omega \alpha^2F(\Omega) \\ \times \Pint_{-\mathscr{W}}^{\mathscr{W}} d\xi \left[ \frac{f_{FD}(-\xi)+f_{BE}(\Omega)}{\omega-\Omega-\xi} \right. \\ \left. +\frac{f_{FD}(\xi)+f_{BE}(\Omega)}{\omega+\Omega-\xi} \right]
\end{multline}
where the symbol $\Pint$ denotes the Cauchy principal value of the integral. Equation (\ref{eq:Re_Sigma}) splits into $\mathrm{Re}\left\langle \Sigma(\omega)\right\rangle_{\vec{k}}=A+B$, depending on $f_{BE}(\Omega)$ and $f_{FD}(\xi)$ respectively. Integrating the former term gives
\begin{multline}\label{eq:Re_Sigma_BE}
A=-\int_0^{+\infty} d\Omega \alpha^2F(\Omega) \\ \times \Pint_{-\mathscr{W}}^{\mathscr{W}} d\xi \left[\frac{f_{BE}(\Omega)}{\omega-\Omega-\xi}+\frac{f_{BE}(\Omega)}{\omega+\Omega-\xi} \right] \\ =\int_0^{+\infty} d\Omega \alpha^2F(\Omega) f_{BE}(\Omega) \ln\left| \frac{(\mathscr{W}+\omega)^2-\Omega^2}{(\mathscr{W}-\omega)^2-\Omega^2} \right|
\end{multline}
so that $\lim_{\mathscr{W} \rightarrow +\infty}A=0$. In that limit, we are left with term $B$, which is
\begin{multline}\label{eq:Re_Sigma_FD}
\mathrm{Re}\left\langle \Sigma(\omega)\right\rangle_{\vec{k}}=-\int_0^{+\infty} d\Omega \alpha^2F(\Omega) \\ \times \Pint_{-\mathscr{W}}^{\mathscr{W}} d\xi \left[ \frac{f_{FD}(-\xi)}{\omega-\Omega-\xi}+\frac{f_{FD}(\xi)}{\omega+\Omega-\xi} \right] \\ = -\int_0^{+\infty} d\Omega \alpha^2 F(\Omega) \left[ \Pint_{-\mathscr{W}+\omega}^{\mathscr{W}+\omega} d\epsilon \frac{f_{FD}(\epsilon-\omega)}{\epsilon-\Omega}\right. \\ \left. + \Pint_{-\mathscr{W}-\omega}^{\mathscr{W}-\omega} d\epsilon \frac{f_{FD}(\epsilon+\omega)}{-\epsilon+\Omega} \right]
\end{multline}
In the limit $ \mathscr{W} \gg \left|\omega\right| \, \forall \, \omega$, up to errors of order $\left|\omega\right|/\mathscr{W}$ we have 
\begin{multline}\label{eq:Re_Sigma_FD2}
\mathrm{Re}\left\langle \Sigma(\omega)\right\rangle_{\vec{k}}=-\int_0^{+\infty} d\Omega \alpha^2F(\Omega) \\ \times \Pint_{-\mathscr{W}}^{\mathscr{W}} d\epsilon \frac{f_{FD}(\epsilon-\omega)-f_{FD}(\epsilon+\omega)}{\epsilon-\Omega} +o\left(\frac{\left|\omega \right|}{\mathscr{W}}\right). 
\end{multline}
Now the integrand over $\epsilon$ in Eq.\@ (\ref{eq:Re_Sigma_FD2}) is explicitly cut by the Fermi-Dirac distributions to the range of $\pm \left(\omega+ \alpha \right)$, where $\alpha$ is a few times the thermal energy $k_B T$. This allows us to safely take the limit $\mathscr{W} \rightarrow +\infty$. Further using the Matsubara representation of the Fermi-Dirac distribution, we have 
\begin{multline}\label{eq:Re_Sigma_FD3}
\mathrm{Re}\left\langle \Sigma(\omega)\right\rangle_{\vec{k}}=-\int_0^{+\infty} d\Omega \alpha^2F(\Omega) \\ \times \Pint_{-\infty}^{\infty} d\epsilon \frac{f_{FD}(\epsilon-\omega)-f_{FD}(\epsilon+\omega)}{\epsilon-\Omega+i 0^+} \\ =-\int_0^{+\infty} d\Omega \alpha^2F(\Omega) \\ \times \mathrm{Re} \left\{ k_B T \sum_{i \omega_n} \int_{-\infty}^{+\infty} d\epsilon \left( \frac{1}{i \omega_n-\epsilon +\omega} \right. \right. \\ \left. \left. -\frac{1}{i \omega_n-\epsilon -\omega} \right) \frac{1}{\epsilon-\Omega+i 0^+} \right\}.
\end{multline}
The integral over $\epsilon$ in Eq.\@ (\ref{eq:Re_Sigma_FD3}) can be solved by the residue theorem, by closing the contour on the lower half of the complex plane, thus enclosing the pole at $\epsilon=\Omega-i0^+$. This yields
\begin{multline}\label{eq:Int_FD}
\int_{-\infty}^{+\infty} d\epsilon \left( \frac{1}{i \omega_n-\epsilon +\omega}-\frac{1}{i \omega_n-\epsilon -\omega} \right) \frac{1}{\epsilon-\Omega+i 0^+} \\ =2 \pi i \Theta(\omega_n) \frac{2 \omega}{(i \omega_n-\epsilon)^2-\omega^2}
\end{multline}
\begin{figure}[t]
\includegraphics[width=0.75\columnwidth]{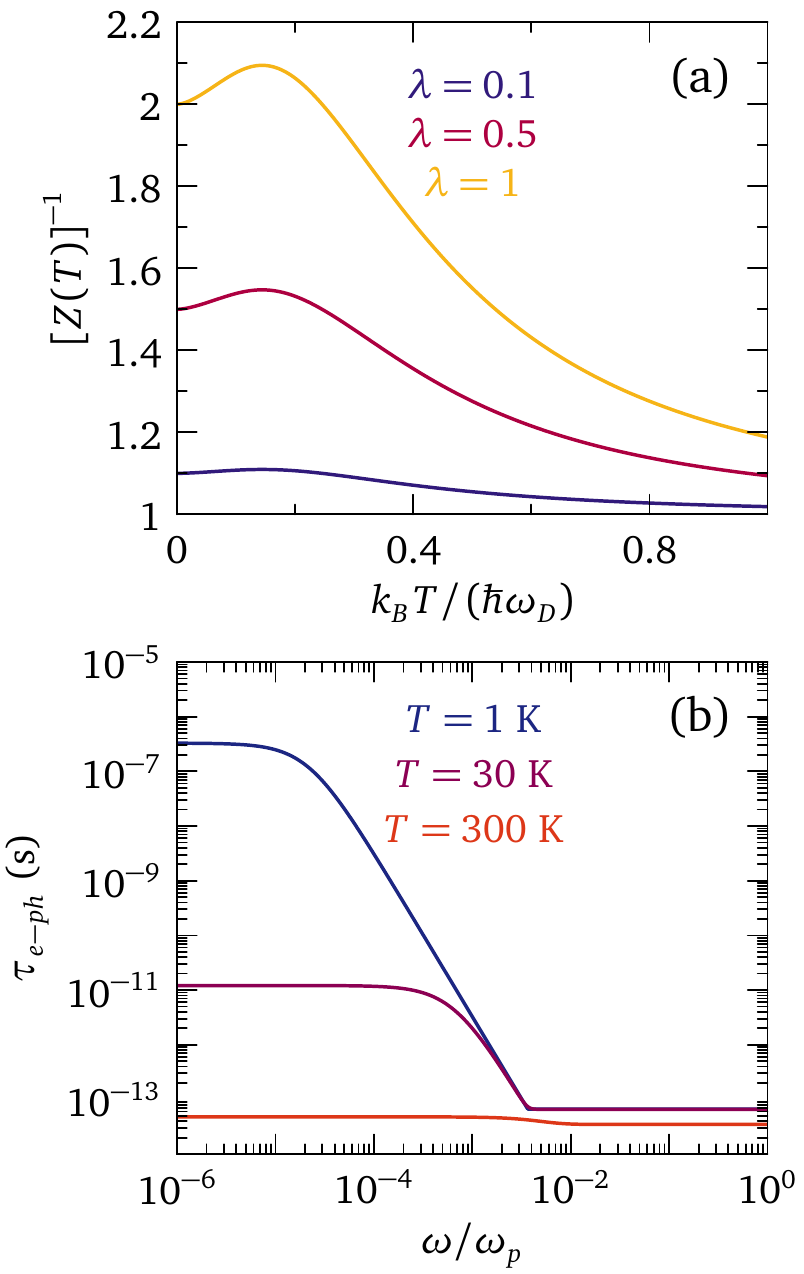}
\caption{\label{fig:tau_eph} 
(a) Quasiparticle weight for acoustic-phonon scattering as a function of normalized temperature $k_B T/(\hbar \omega_D)$, according to Eq.\@ (\ref{eq:Z_ac}). Blue, red and gold curves show the results for electron-phonon coupling constants $\lambda=\left\{0.1,0.5,1\right\}$, respectively. (b) Electron-phonon collision time $\tau_{e-ph}(\omega, T) \, \left(s\right)$ as a function of frequency $\omega$ normalized to the plasma frequency $\omega_p$, stemming from Eq.\@ (\ref{eq:tau_eph_expl}), for an electron-phonon coupling constant $\lambda=0.1$. The blue, purple and red lines correspond to the results at temperatures $T=\left\{1,30,300\right\}$ K, respectively. 
}
\end{figure}
The remaining Matsubara sum in Eq.\@ (\ref{eq:Re_Sigma_FD3}) is convergent and can be written in terms of the Digamma function, $\psi(z)=\lim_{M\rightarrow +\infty} \left(\ln M -\sum_{n=0}^{M} (n+z)^{-1}\right)$. The final result is
\begin{multline}\label{eq:Re_Sigma_gen}
\mathrm{Re}\left\langle \Sigma(\omega)\right\rangle_{\vec{k}}=-\int_0^{+\infty} d\Omega \alpha^2F(\Omega) \\ \times \mathrm{Re} \left[ \psi\left(\frac{1}{2}-i \frac{\omega-\Omega}{2 \pi k_B T} \right) -\psi\left(\frac{1}{2}+i \frac{\omega+\Omega}{2 \pi k_B T} \right) \right]. 
\end{multline}
The result (\ref{eq:Re_Sigma_gen}) applies to any isotropic phonon spectrum $\alpha^2F(\Omega)$. Specializing the calculation to the acoustic phonon spectrum (\ref{eq:alpha2F_ac}) yields 
\begin{equation}\label{eq:Re_Sigma_ac}
\mathrm{Re}\left\langle \Sigma(\omega)\right\rangle_{\vec{k}}=-\lambda \hbar \omega_D \mathscr{R}\left(\frac{\omega}{\hbar \omega_D}, \frac{k_B T}{\hbar \omega_D} \right)
\end{equation}
where we defined the function 
\begin{multline}\label{eq:R_ac}
\mathscr{R}(x,t)=\int_0^1 du u^2 \\ \times \mathrm{Re} \left[ \psi\left(\frac{1}{2}-i \frac{x-u}{2 \pi t} \right) -\psi\left(\frac{1}{2}+i \frac{x+u}{2 \pi t} \right) \right]
\end{multline}
The numerical solution of Eqs.\@ (\ref{eq:Re_Sigma_ac}) and (\ref{eq:R_ac}) gives the desired real part of the self-energy as a function of energy and temperature. In the high-temperature limit $T \rightarrow +\infty$, Eqs.\@ (\ref{eq:Re_Sigma_ac}) and (\ref{eq:R_ac}) simplify to 
\begin{multline}\label{eq:Re_Sigma_ac_ht}
\mathrm{Re}\left\langle \Sigma(\omega)\right\rangle_{\vec{k}}=-\lambda \hbar \omega  \left(\frac{\hbar \omega_D} {k_B T}\right)^2 \psi''\left(\frac{1}{2}\right)  \\ \approx 16.83 \lambda \hbar \omega \left(\frac{\hbar \omega_D} {k_B T}\right)^2. 
\end{multline}
Figure \ref{fig:sigma_eph} shows the real and imaginary parts of the acoustic-phonons self-energy as a function of normalized temperature $k_B T/(\hbar \omega_D)$, calculated at different frequencies according to Eqs.\@ (\ref{eq:el-ph_ht}) and (\ref{eq:Re_Sigma_ac}), with dashed lines indicating the high-temperature expansion (\ref{eq:Re_Sigma_ac_ht}) for $\mathrm{Re}\left\langle \Sigma(\omega)\right\rangle_{\vec{k}}$.

The scattering rate of electrons due to the exchange of acoustic phonons satisfies
\begin{equation}\label{eq:tau_eph}
\left[\tau_{e-ph}(\omega,T)\right]^{-1}=-2 Z(T) \mathrm{Im} \left\langle \Sigma(\omega)\right\rangle_{\vec{k}}
\end{equation}
where the quasiparticle spectral weight is defined as
\begin{equation}\label{eq:Z_def}
\left[ Z(T) \right]^{-1}= 1-\left. \frac{\partial \mathrm{Re} \left\langle \Sigma(\omega)\right\rangle_{\vec{k}}}{\partial \omega} \right|_{\omega=0}. 
\end{equation}
For the acoustic phonon spectrum, utilizing Eq.\@ (\ref{eq:Re_Sigma_ac}) and calculating the limit
\begin{multline}\label{eq:Der_Re_ac}
\lim_{x\rightarrow 0} \frac{1}{\hbar \omega_D} \mathrm{Re} \left\{ \psi\left(\frac{1}{2}-i \frac{x-u}{k_B T}\right)-\psi\left(\frac{1}{2}+i \frac{x+u}{k_B T}\right) \right\} \\ =\frac{1}{\hbar \omega_D} \frac{\mathrm{Im} \left\{ \psi'\left(\frac{1}{2}+i\frac{u}{2 \pi t}\right)\right\}}{\pi t},
\end{multline}
where $\psi'(z)$ is the first derivative of the Digamma function, we have
\begin{equation}\label{eq:Z_ac}
\left[ Z(T) \right]^{-1}= 1-\frac{\lambda \hbar \omega_D}{\pi k_B T} \mathscr{Q}\left( \frac{k_B T}{\hbar \omega_D} \right), 
\end{equation}
where we have defined the function
\begin{equation}\label{eq:Q_ac}
\mathscr{Q}(t)=\int_0^1 du u^2 \mathrm{Im}\left\{ \psi'\left(\frac{1}{2}+i \frac{u}{2 \pi t} \right)\right\}. 
\end{equation}
At zero temperature, we have $\lim_{t\rightarrow 0} \mathscr{Q}(t)=-1$, so that we retrieve
\begin{equation}\label{eq:Z_ac_0}
\left[ Z(0) \right]^{-1}= 1+\lambda, 
\end{equation}
which holds at $T=0$ for any spectrum of bosons exchanged by electrons. 
Inserting Eqs.\@ (\ref{eq:Z_def}), (\ref{eq:Der_Re_ac}), and (\ref{eq:el-ph_tau_D}) into Eq.\@ (\ref{eq:tau_eph}), we obtain an explicit expression for the scattering rate due to acoustic phonon exchange:
\begin{multline}\label{eq:tau_eph_expl}
\frac{1}{\tau_{e-ph}(\omega,T)}=\\=\frac{2}{1-\frac{\lambda \hbar \omega_D}{\pi k_B T} \mathscr{Q}\left( \frac{k_B T}{\hbar \omega_D} \right)} \pi \lambda \hbar \omega_D \mathscr{S} \left(\frac{\omega}{\hbar \omega_D}, \frac{k_B T}{\hbar \omega_D}\right).
\end{multline} 
Figure \ref{fig:tau_eph}(a) illustrates the temperature evolution of the quasiparticle weight (\ref{eq:Z_ac}): blue, red and gold curves show the results for electron-phonon coupling constants $\lambda=\left\{0.1,0.5,1\right\}$, respectively. Figure \ref{fig:tau_eph}(b) shows the electron-phonon collision time $\tau_{e-ph}(\omega, T) \, \left(s\right)$ as a function of frequency $\omega$ normalized to the plasma frequency $\omega_p$, resulting from Eq.\@ (\ref{eq:tau_eph_expl}), for an electron-phonon coupling constant $\lambda=0.1$ and at temperatures $T=\left\{1,30,300\right\}$ K. 
\begin{figure}[t]
\includegraphics[width=0.9\columnwidth]{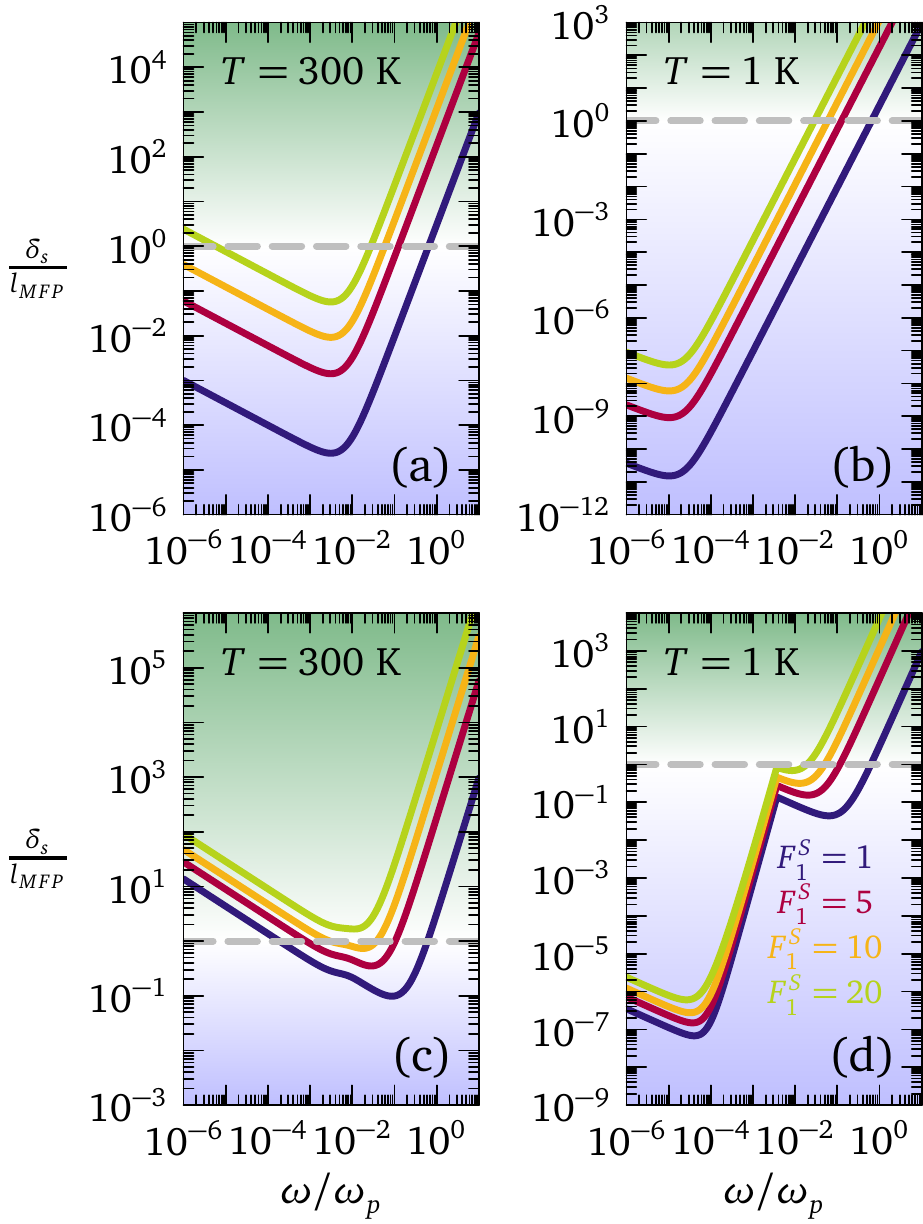}
\caption{\label{fig:skin_depth} Ratio between the Ohmic skin depth $\delta_s$ and the relaxational mean-free path $l_{MFP}$ as a function of frequency $\omega$ normalized to the plasma frequency $\omega_p$. Common parameters to all panels are: electron density $n=10^{29}$ cm$^{-3}$, and zero-order Landau parameter $F_0^S=1$. Different curves in all panels refer to first Landau parameter $F_1^S=\left\{1,5,10,20 \right\}$. Panels (a) and (b) consider only Umklapp scattering giving a momentum-relaxation time $\tau_{K}=\tau_c/\alpha_U$ with Umklapp efficiency $\alpha_U=0.5$ (see Appendix \ref{app:FL_scat}), at $T=300$ K and $T=1$ K respectively. Panels (c) and (d) include the additional contribution of acoustic phonons (see Appendix \ref{app:e-ph_scat}) with Debye temperature $T_D=500$ K and electron-phonon coupling constant $\lambda=0.1$, at $T=300$ K and $T=1$ K respectively.
}
\end{figure}

\section{Momentum-relaxation dependence of the skin depth for the Fermi liquid}\label{Delta_skin_cons}

The ratio between the Drude skin depth $\delta_s$ and the mean-free path $l_{MFP}$ is
\begin{equation}\label{eq:deltas_mfp}
\frac{\delta_s}{l_{MFP}}=\frac{c}{v_F^{*}} \sqrt{\frac{2}{\omega (\omega_p)^2 (\tau_{K})^3}},
\end{equation}
from Eqs.\@ (\ref{eq:delta_skin_Drude}) and (\ref{eq:l_MFP}). Therefore, such ratio depends on the momentum-relaxation time $\tau_{K}$ at a given frequency $\omega$. According to Reuter-Sondheimer theory, such ratio determines the crossover from local electrodynamics to anomalous skin effect; see Eq.\@ (\ref{eq:Z_RandS}). 

Figure \ref{fig:skin_depth} shows the ratio $\delta_s/l_{MFP}$ as a function of frequency $\omega$ normalized to the plasma frequency $\omega_p$, for a momentum-relaxation time $\tau_{K}=\tau_c/\alpha_U$ resulting from Umklapp processes, according to Sec.\@ \ref{app:FL_scat}, and acoustic phonons, in accordance with Sec.\@ \ref{app:e-ph_scat}. The parameters used are: electron density $n=10^{29}$ cm$^{-3}$, Umklapp efficiency $\alpha_U=0.5$, Debye temperature $T_D=500$ K, electron-phonon coupling constant $\lambda=0.1$, and zero-order Landau parameter $F_0^S=1$ (its inclusion does not qualitatively modify the results). The different curves in all panels refer to first symmetric Landau parameter $F_1^S=\left\{1,5,10,20 \right\}$. In the blue region $\delta_s<l_{MFP}$, so the electrodynamic response is nonlocal giving anomalous skin effect, while in the green region $\delta_s>l_{MFP}$ and local electrodynamics is at play, yielding a normal skin effect. The gray dashed line marks the condition $\delta_s=l_{MFP}$. Figures \ref{fig:skin_depth}(a) and \ref{fig:skin_depth}(b) neglect the phonon contribution, and consider only Umklapp processes, at $T=300$ K and $T=1$ K respectively: in both cases, there is an extended frequency window in which $\delta_s\ll l_{MFP}$. Upon adding the contribution from acoustic phonons, we obtain Figs.\@ \ref{fig:skin_depth}(c) and \ref{fig:skin_depth}(d), at $T=300$ K and $T=1$ K respectively: we see that typically $\delta_s/l_{MFP}\gtrapprox 1$ at all frequencies for the room-temperature case, so that electrodynamics is local and normal skin effect occurs; in the low-temperature calculation, there is a low-frequency range in which $\delta_s/l_{MFP}\ll 1$, where anomalous skin effect develops. The analysis here outlined is further modified by the inclusion of the collision time $\tau_c$: different regimes arise for the skin depth, as discussed in Sec.\@ \ref{Skin_VE_low}.
\begin{figure}[ht]
\includegraphics[width=0.7\columnwidth]{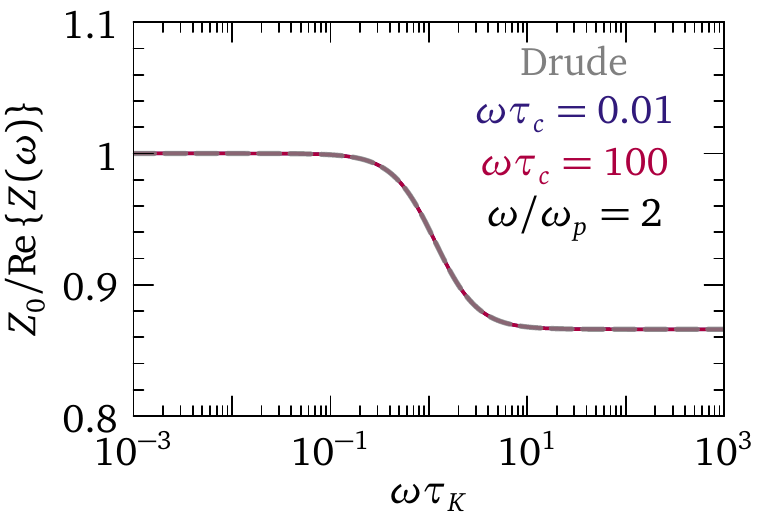}
\caption{\label{fig:RS_high} Inverse surface resistance $\mathrm{Re}\left\{Z(\omega)\right\}^{-1}$ as a function of $\omega \tau_{K}$ for a 3D Fermi liquid in the high-frequency regime, according to Eqs.\@ (\ref{eq:E_VE_high_2}) and (\ref{eq:Z_gen_1D}), at frequency $\omega=2 \omega_p$, renormalized Fermi velocity $v_F^*=10^{-3}c$, and first Landau parameter $F_1^S=20$. The blue and red solid curves refer to $\omega_p \tau_c=\left\{0.01,100\right\}$ respectively. The dashed gray curve is the Drude calculation from Eq.\@ (\ref{eq:Z_Drude_n}).
}
\end{figure}

\section{Surface impedance in the high-frequency propagating shear regime}\label{Shear_Z_high}

The high-frequency surface impedance for the propagating shear regime stems from the real-space electric field profile $E(z,\omega)$ that results from inserting Eq.\@ (\ref{eq:eps_T_series_rel_gen1bis}) into Eq.\@ (\ref{eq:E_VE_high}), and Fourier-transforming back to real space according to
\begin{multline}\label{eq:E_VE_high_2} 
E(z,\omega)=\frac{1}{\pi} \left. \frac{\partial E(z,\omega)}{\partial z} \right|_{z=0^+} \\ \times \int_{-\infty}^{+\infty} dq \left\{\frac{\omega^2}{c^2} \left[1-\frac{(\omega_p)^2}{\omega^2 +i \frac{\omega}{\tau_{K}} +i \omega \tilde{\nu}(\omega) q^2} \right]-q^2 \right\}^{-1} \\ = \frac{1}{\pi} \left. \frac{\partial E(z,\omega)}{\partial z} \right|_{z=0^+} \frac{c}{\omega_p} \\ \times \int_{-\infty}^{+\infty} dz \left\{\frac{\omega^2}{(\omega_p)^2} -\frac{1}{A +B z^2}-z^2 \right\}^{-1},
\end{multline}
where the variables $A$ and $B$ are given by Eq.\@ (\ref{eq:E_int_VE_low_2}) and the integration variable is (\ref{eq:q_z}). 
Performing the momentum integration in Eq.\@ (\ref{eq:E_VE_high_2}), one obtains a lengthy analytical expression for $E(z,\omega)$ in terms of $A$ and $B$, which translates into the surface impedance by the means of Eq.\@ (\ref{eq:Z_gen_1D}).
Figure \ref{fig:RS_high} illustrates the comparison of the inverse surface resistance stemming from Eq.\@ (\ref{eq:E_VE_high_2}) with the Drude surface resistance (\ref{eq:Z_Drude_high}): the two curves for $\mathrm{Re}\left\{Z(\omega)\right\}^{-1}$ are visually indistinguishable, which shows that spatial nonlocality due to shear-mode propagation gives negligible corrections to the surface reactance for $\omega>\omega_p$. Hence, we conclude that suitable conditions to observe the propagation of the shear-polariton in $Z(\omega)$ are $\omega \ll \omega_p$ and $\omega>v_F^* \left|q\right|$, as described in Sec.\@ \ref{Skin_VE_low}.


\begin{thebibliography}{117}%
\makeatletter
\providecommand \@ifxundefined [1]{%
 \@ifx{#1\undefined}
}%
\providecommand \@ifnum [1]{%
 \ifnum #1\expandafter \@firstoftwo
 \else \expandafter \@secondoftwo
 \fi
}%
\providecommand \@ifx [1]{%
 \ifx #1\expandafter \@firstoftwo
 \else \expandafter \@secondoftwo
 \fi
}%
\providecommand \natexlab [1]{#1}%
\providecommand \enquote  [1]{``#1''}%
\providecommand \bibnamefont  [1]{#1}%
\providecommand \bibfnamefont [1]{#1}%
\providecommand \citenamefont [1]{#1}%
\providecommand \href@noop [0]{\@secondoftwo}%
\providecommand \href [0]{\begingroup \@sanitize@url \@href}%
\providecommand \@href[1]{\@@startlink{#1}\@@href}%
\providecommand \@@href[1]{\endgroup#1\@@endlink}%
\providecommand \@sanitize@url [0]{\catcode `\\12\catcode `\$12\catcode
  `\&12\catcode `\#12\catcode `\^12\catcode `\_12\catcode `\%12\relax}%
\providecommand \@@startlink[1]{}%
\providecommand \@@endlink[0]{}%
\providecommand \url  [0]{\begingroup\@sanitize@url \@url }%
\providecommand \@url [1]{\endgroup\@href {#1}{\urlprefix }}%
\providecommand \urlprefix  [0]{URL }%
\providecommand \Eprint [0]{\href }%
\providecommand \doibase [0]{https://doi.org/}%
\providecommand \selectlanguage [0]{\@gobble}%
\providecommand \bibinfo  [0]{\@secondoftwo}%
\providecommand \bibfield  [0]{\@secondoftwo}%
\providecommand \translation [1]{[#1]}%
\providecommand \BibitemOpen [0]{}%
\providecommand \bibitemStop [0]{}%
\providecommand \bibitemNoStop [0]{.\EOS\space}%
\providecommand \EOS [0]{\spacefactor3000\relax}%
\providecommand \BibitemShut  [1]{\csname bibitem#1\endcsname}%
\let\auto@bib@innerbib\@empty
\bibitem [{\citenamefont {Nozieres}\ and\ \citenamefont
  {Pines}(1989)}]{Nozieres-1999ql}%
  \BibitemOpen
  \bibfield  {author} {\bibinfo {author} {\bibfnamefont {P.}~\bibnamefont
  {Nozieres}}\ and\ \bibinfo {author} {\bibfnamefont {D.}~\bibnamefont
  {Pines}},\ }\href@noop {} {\emph {\bibinfo {title} {Theory Of Quantum
  Liquids}}},\ Advanced Books Classics Series\ (\bibinfo  {publisher} {Westview
  Press},\ \bibinfo {address} {Boulder},\ \bibinfo {year} {1989})\BibitemShut
  {NoStop}%
\bibitem [{\citenamefont {Abrikosov}\ and\ \citenamefont
  {Khalatnikov}(1959)}]{Abrikosov-1959}%
  \BibitemOpen
  \bibfield  {author} {\bibinfo {author} {\bibfnamefont {A.~A.}\ \bibnamefont
  {Abrikosov}}\ and\ \bibinfo {author} {\bibfnamefont {I.~M.}\ \bibnamefont
  {Khalatnikov}},\ }\bibfield  {title} {\bibinfo {title} {The theory of a
  {Fermi} liquid (the properties of liquid $^3${He} at low temperatures)},\
  }\href {http://stacks.iop.org/0034-4885/22/i=1/a=310} {\bibfield  {journal}
  {\bibinfo  {journal} {Rep. Prog. Phys.}\ }\textbf {\bibinfo {volume} {22}},\
  \bibinfo {pages} {329} (\bibinfo {year} {1959})}\BibitemShut {NoStop}%
\bibitem [{\citenamefont {Tokatly}\ and\ \citenamefont
  {Pankratov}(2000)}]{Tokatly-2000}%
  \BibitemOpen
  \bibfield  {author} {\bibinfo {author} {\bibfnamefont {I.~V.}\ \bibnamefont
  {Tokatly}}\ and\ \bibinfo {author} {\bibfnamefont {O.}~\bibnamefont
  {Pankratov}},\ }\bibfield  {title} {\bibinfo {title} {Hydrodynamics beyond
  local equilibrium: Application to electron gas},\ }\href
  {https://doi.org/10.1103/PhysRevB.62.2759} {\bibfield  {journal} {\bibinfo
  {journal} {Phys. Rev. B}\ }\textbf {\bibinfo {volume} {62}},\ \bibinfo
  {pages} {2759} (\bibinfo {year} {2000})}\BibitemShut {NoStop}%
\bibitem [{\citenamefont {Landau}\ and\ \citenamefont
  {Lifshitz}(2013)}]{Landau-2013stat}%
  \BibitemOpen
  \bibfield  {author} {\bibinfo {author} {\bibfnamefont {L.}~\bibnamefont
  {Landau}}\ and\ \bibinfo {author} {\bibfnamefont {E.}~\bibnamefont
  {Lifshitz}},\ }\href@noop {} {\emph {\bibinfo {title} {Statistical
  Physics}}},\ \bibinfo {series} {Course of theoretical physics}\ No.~\bibinfo
  {number} {5}\ (\bibinfo  {publisher} {Elsevier Science},\ \bibinfo {address}
  {Oxford},\ \bibinfo {year} {2013})\BibitemShut {NoStop}%
\bibitem [{\citenamefont {Maeno}\ \emph {et~al.}(1997)\citenamefont {Maeno},
  \citenamefont {Yoshida}, \citenamefont {Hashimoto}, \citenamefont
  {Nishizaki}, \citenamefont {Ikeda}, \citenamefont {Nohara}, \citenamefont
  {Fujita}, \citenamefont {Mackenzie}, \citenamefont {Hussey}, \citenamefont
  {Bednorz},\ and\ \citenamefont {Lichtenberg}}]{Maeno-1997}%
  \BibitemOpen
  \bibfield  {author} {\bibinfo {author} {\bibfnamefont {Y.}~\bibnamefont
  {Maeno}}, \bibinfo {author} {\bibfnamefont {K.}~\bibnamefont {Yoshida}},
  \bibinfo {author} {\bibfnamefont {H.}~\bibnamefont {Hashimoto}}, \bibinfo
  {author} {\bibfnamefont {S.}~\bibnamefont {Nishizaki}}, \bibinfo {author}
  {\bibfnamefont {S.-i.}\ \bibnamefont {Ikeda}}, \bibinfo {author}
  {\bibfnamefont {M.}~\bibnamefont {Nohara}}, \bibinfo {author} {\bibfnamefont
  {T.}~\bibnamefont {Fujita}}, \bibinfo {author} {\bibfnamefont
  {A.}~\bibnamefont {Mackenzie}}, \bibinfo {author} {\bibfnamefont
  {N.}~\bibnamefont {Hussey}}, \bibinfo {author} {\bibfnamefont
  {J.}~\bibnamefont {Bednorz}},\ and\ \bibinfo {author} {\bibfnamefont
  {F.}~\bibnamefont {Lichtenberg}},\ }\bibfield  {title} {\bibinfo {title}
  {Two-dimensional {Fermi} liquid behavior of the superconductor
  {Sr$_2$RuO$_4$}},\ }\href {https://doi.org/10.1143/JPSJ.66.1405} {\bibfield
  {journal} {\bibinfo  {journal} {J. Phys. Soc. Jpn.}\ }\textbf {\bibinfo
  {volume} {66}},\ \bibinfo {pages} {1405} (\bibinfo {year}
  {1997})}\BibitemShut {NoStop}%
\bibitem [{\citenamefont {Bergemann}\ \emph {et~al.}(2003)\citenamefont
  {Bergemann}, \citenamefont {Mackenzie}, \citenamefont {Julian}, \citenamefont
  {Forsythe},\ and\ \citenamefont {Ohmichi}}]{Bergemann-2003}%
  \BibitemOpen
  \bibfield  {author} {\bibinfo {author} {\bibfnamefont {C.}~\bibnamefont
  {Bergemann}}, \bibinfo {author} {\bibfnamefont {A.~P.}\ \bibnamefont
  {Mackenzie}}, \bibinfo {author} {\bibfnamefont {S.~R.}\ \bibnamefont
  {Julian}}, \bibinfo {author} {\bibfnamefont {D.}~\bibnamefont {Forsythe}},\
  and\ \bibinfo {author} {\bibfnamefont {E.}~\bibnamefont {Ohmichi}},\
  }\bibfield  {title} {\bibinfo {title} {Quasi-two-dimensional {Fermi} liquid
  properties of the unconventional superconductor {Sr$_2$RuO$_4$}},\ }\href
  {https://doi.org/10.1080/00018730310001621737} {\bibfield  {journal}
  {\bibinfo  {journal} {Adv. Phys.}\ }\textbf {\bibinfo {volume} {52}},\
  \bibinfo {pages} {639} (\bibinfo {year} {2003})}\BibitemShut {NoStop}%
\bibitem [{\citenamefont {Stricker}\ \emph {et~al.}(2014)\citenamefont
  {Stricker}, \citenamefont {Mravlje}, \citenamefont {Berthod}, \citenamefont
  {Fittipaldi}, \citenamefont {Vecchione}, \citenamefont {Georges},\ and\
  \citenamefont {van~der Marel}}]{Stricker-2014}%
  \BibitemOpen
  \bibfield  {author} {\bibinfo {author} {\bibfnamefont {D.}~\bibnamefont
  {Stricker}}, \bibinfo {author} {\bibfnamefont {J.}~\bibnamefont {Mravlje}},
  \bibinfo {author} {\bibfnamefont {C.}~\bibnamefont {Berthod}}, \bibinfo
  {author} {\bibfnamefont {R.}~\bibnamefont {Fittipaldi}}, \bibinfo {author}
  {\bibfnamefont {A.}~\bibnamefont {Vecchione}}, \bibinfo {author}
  {\bibfnamefont {A.}~\bibnamefont {Georges}},\ and\ \bibinfo {author}
  {\bibfnamefont {D.}~\bibnamefont {van~der Marel}},\ }\bibfield  {title}
  {\bibinfo {title} {Optical response of {Sr}$_{2}${RuO}$_{4}$ reveals
  universal {Fermi}-liquid scaling and quasiparticles beyond {Landau} theory},\
  }\href {https://doi.org/10.1103/PhysRevLett.113.087404} {\bibfield  {journal}
  {\bibinfo  {journal} {Phys. Rev. Lett.}\ }\textbf {\bibinfo {volume} {113}},\
  \bibinfo {pages} {087404} (\bibinfo {year} {2014})}\BibitemShut {NoStop}%
\bibitem [{\citenamefont {Tytarenko}\ \emph {et~al.}(2015)\citenamefont
  {Tytarenko}, \citenamefont {Huang}, \citenamefont {De~Visser}, \citenamefont
  {Johnston},\ and\ \citenamefont {Van~Heumen}}]{Tytarenko-2015}%
  \BibitemOpen
  \bibfield  {author} {\bibinfo {author} {\bibfnamefont {A.}~\bibnamefont
  {Tytarenko}}, \bibinfo {author} {\bibfnamefont {Y.}~\bibnamefont {Huang}},
  \bibinfo {author} {\bibfnamefont {A.}~\bibnamefont {De~Visser}}, \bibinfo
  {author} {\bibfnamefont {S.}~\bibnamefont {Johnston}},\ and\ \bibinfo
  {author} {\bibfnamefont {E.}~\bibnamefont {Van~Heumen}},\ }\bibfield  {title}
  {\bibinfo {title} {Direct observation of a {Fermi} liquid-like normal state
  in an iron-pnictide superconductor},\ }\href
  {https://www.nature.com/articles/srep12421#article-info} {\bibfield
  {journal} {\bibinfo  {journal} {Sci. Rep.}\ }\textbf {\bibinfo {volume}
  {5}},\ \bibinfo {pages} {12421} (\bibinfo {year} {2015})}\BibitemShut
  {NoStop}%
\bibitem [{\citenamefont {Conti}\ and\ \citenamefont
  {Vignale}(1999)}]{Conti-1999}%
  \BibitemOpen
  \bibfield  {author} {\bibinfo {author} {\bibfnamefont {S.}~\bibnamefont
  {Conti}}\ and\ \bibinfo {author} {\bibfnamefont {G.}~\bibnamefont
  {Vignale}},\ }\bibfield  {title} {\bibinfo {title} {Elasticity of an electron
  liquid},\ }\href {https://doi.org/10.1103/PhysRevB.60.7966} {\bibfield
  {journal} {\bibinfo  {journal} {Phys. Rev. B}\ }\textbf {\bibinfo {volume}
  {60}},\ \bibinfo {pages} {7966} (\bibinfo {year} {1999})}\BibitemShut
  {NoStop}%
\bibitem [{\citenamefont {Principi}\ \emph {et~al.}(2016)\citenamefont
  {Principi}, \citenamefont {Vignale}, \citenamefont {Carrega},\ and\
  \citenamefont {Polini}}]{Principi-2016}%
  \BibitemOpen
  \bibfield  {author} {\bibinfo {author} {\bibfnamefont {A.}~\bibnamefont
  {Principi}}, \bibinfo {author} {\bibfnamefont {G.}~\bibnamefont {Vignale}},
  \bibinfo {author} {\bibfnamefont {M.}~\bibnamefont {Carrega}},\ and\ \bibinfo
  {author} {\bibfnamefont {M.}~\bibnamefont {Polini}},\ }\bibfield  {title}
  {\bibinfo {title} {Bulk and shear viscosities of the two-dimensional electron
  liquid in a doped graphene sheet},\ }\href
  {https://doi.org/10.1103/PhysRevB.93.125410} {\bibfield  {journal} {\bibinfo
  {journal} {Phys. Rev. B}\ }\textbf {\bibinfo {volume} {93}},\ \bibinfo
  {pages} {125410} (\bibinfo {year} {2016})}\BibitemShut {NoStop}%
\bibitem [{\citenamefont {Bradlyn}\ \emph {et~al.}(2012)\citenamefont
  {Bradlyn}, \citenamefont {Goldstein},\ and\ \citenamefont
  {Read}}]{Bradlyn-2012}%
  \BibitemOpen
  \bibfield  {author} {\bibinfo {author} {\bibfnamefont {B.}~\bibnamefont
  {Bradlyn}}, \bibinfo {author} {\bibfnamefont {M.}~\bibnamefont {Goldstein}},\
  and\ \bibinfo {author} {\bibfnamefont {N.}~\bibnamefont {Read}},\ }\bibfield
  {title} {\bibinfo {title} {{Kubo} formulas for viscosity: {Hall} viscosity,
  {Ward} identities, and the relation with conductivity},\ }\href
  {https://doi.org/10.1103/PhysRevB.86.245309} {\bibfield  {journal} {\bibinfo
  {journal} {Phys. Rev. B}\ }\textbf {\bibinfo {volume} {86}},\ \bibinfo
  {pages} {245309} (\bibinfo {year} {2012})}\BibitemShut {NoStop}%
\bibitem [{\citenamefont {Landau}(1957)}]{Landau-1957}%
  \BibitemOpen
  \bibfield  {author} {\bibinfo {author} {\bibfnamefont {L.~D.}\ \bibnamefont
  {Landau}},\ }\bibfield  {title} {\bibinfo {title} {Oscillations in a {Fermi}
  liquid},\ }\href@noop {} {\bibfield  {journal} {\bibinfo  {journal} {J.
  Exptl. Theoret. Phys. (U.S.S.R.)}\ }\textbf {\bibinfo {volume} {32}},\
  \bibinfo {pages} {59} (\bibinfo {year} {1957})},\ \bibinfo {note}
  {[\href{http://www.jetp.ac.ru/files/Landau2en.pdf}{{Soviet Phys. JETP}}
  \textbf{5}, 101 (1957)]}\BibitemShut {NoStop}%
\bibitem [{\citenamefont {Landau}\ \emph {et~al.}(2012)\citenamefont {Landau},
  \citenamefont {Lifshitz}, \citenamefont {Kosevich},\ and\ \citenamefont
  {Pitaevski{\u\i}}}]{Landau-1986el}%
  \BibitemOpen
  \bibfield  {author} {\bibinfo {author} {\bibfnamefont {L.}~\bibnamefont
  {Landau}}, \bibinfo {author} {\bibfnamefont {E.}~\bibnamefont {Lifshitz}},
  \bibinfo {author} {\bibfnamefont {A.}~\bibnamefont {Kosevich}},\ and\
  \bibinfo {author} {\bibfnamefont {L.}~\bibnamefont {Pitaevski{\u\i}}},\
  }\href@noop {} {\emph {\bibinfo {title} {Theory of Elasticity}}},\ \bibinfo
  {series} {Course of theoretical physics}\ No.~\bibinfo {number} {7}\
  (\bibinfo  {publisher} {Elsevier},\ \bibinfo {address} {Oxford},\ \bibinfo
  {year} {2012})\BibitemShut {NoStop}%
\bibitem [{\citenamefont {Link}\ \emph {et~al.}(2018)\citenamefont {Link},
  \citenamefont {Sheehy}, \citenamefont {Narozhny},\ and\ \citenamefont
  {Schmalian}}]{Link-2018}%
  \BibitemOpen
  \bibfield  {author} {\bibinfo {author} {\bibfnamefont {J.~M.}\ \bibnamefont
  {Link}}, \bibinfo {author} {\bibfnamefont {D.~E.}\ \bibnamefont {Sheehy}},
  \bibinfo {author} {\bibfnamefont {B.~N.}\ \bibnamefont {Narozhny}},\ and\
  \bibinfo {author} {\bibfnamefont {J.}~\bibnamefont {Schmalian}},\ }\bibfield
  {title} {\bibinfo {title} {Elastic response of the electron fluid in
  intrinsic graphene: The collisionless regime},\ }\href
  {https://doi.org/10.1103/PhysRevB.98.195103} {\bibfield  {journal} {\bibinfo
  {journal} {Phys. Rev. B}\ }\textbf {\bibinfo {volume} {98}},\ \bibinfo
  {pages} {195103} (\bibinfo {year} {2018})}\BibitemShut {NoStop}%
\bibitem [{\citenamefont {Giuliani}\ and\ \citenamefont
  {Vignale}(2005)}]{Vignale-2005}%
  \BibitemOpen
  \bibfield  {author} {\bibinfo {author} {\bibfnamefont {G.}~\bibnamefont
  {Giuliani}}\ and\ \bibinfo {author} {\bibfnamefont {G.}~\bibnamefont
  {Vignale}},\ }\href@noop {} {\emph {\bibinfo {title} {Quantum Theory of the
  Electron Liquid}}}\ (\bibinfo  {publisher} {Cambridge University Press},\
  \bibinfo {address} {Cambridge},\ \bibinfo {year} {2005})\BibitemShut
  {NoStop}%
\bibitem [{\citenamefont {Lucas}\ and\ \citenamefont
  {Das~Sarma}(2018)}]{Lucas-2018b}%
  \BibitemOpen
  \bibfield  {author} {\bibinfo {author} {\bibfnamefont {A.}~\bibnamefont
  {Lucas}}\ and\ \bibinfo {author} {\bibfnamefont {S.}~\bibnamefont
  {Das~Sarma}},\ }\bibfield  {title} {\bibinfo {title} {Electronic
  hydrodynamics and the breakdown of the {Wiedemann}-{Franz} and {Mott} laws in
  interacting metals},\ }\href {https://doi.org/10.1103/PhysRevB.97.245128}
  {\bibfield  {journal} {\bibinfo  {journal} {Phys. Rev. B}\ }\textbf {\bibinfo
  {volume} {97}},\ \bibinfo {pages} {245128} (\bibinfo {year}
  {2018})}\BibitemShut {NoStop}%
\bibitem [{\citenamefont {Andreev}\ \emph {et~al.}(2011)\citenamefont
  {Andreev}, \citenamefont {Kivelson},\ and\ \citenamefont
  {Spivak}}]{Andreev-2011}%
  \BibitemOpen
  \bibfield  {author} {\bibinfo {author} {\bibfnamefont {A.~V.}\ \bibnamefont
  {Andreev}}, \bibinfo {author} {\bibfnamefont {S.~A.}\ \bibnamefont
  {Kivelson}},\ and\ \bibinfo {author} {\bibfnamefont {B.}~\bibnamefont
  {Spivak}},\ }\bibfield  {title} {\bibinfo {title} {Hydrodynamic description
  of transport in strongly correlated electron systems},\ }\href
  {https://doi.org/10.1103/PhysRevLett.106.256804} {\bibfield  {journal}
  {\bibinfo  {journal} {Phys. Rev. Lett.}\ }\textbf {\bibinfo {volume} {106}},\
  \bibinfo {pages} {256804} (\bibinfo {year} {2011})}\BibitemShut {NoStop}%
\bibitem [{\citenamefont {Briskot}\ \emph {et~al.}(2015)\citenamefont
  {Briskot}, \citenamefont {Sch\"utt}, \citenamefont {Gornyi}, \citenamefont
  {Titov}, \citenamefont {Narozhny},\ and\ \citenamefont
  {Mirlin}}]{Briskot-2015}%
  \BibitemOpen
  \bibfield  {author} {\bibinfo {author} {\bibfnamefont {U.}~\bibnamefont
  {Briskot}}, \bibinfo {author} {\bibfnamefont {M.}~\bibnamefont {Sch\"utt}},
  \bibinfo {author} {\bibfnamefont {I.~V.}\ \bibnamefont {Gornyi}}, \bibinfo
  {author} {\bibfnamefont {M.}~\bibnamefont {Titov}}, \bibinfo {author}
  {\bibfnamefont {B.~N.}\ \bibnamefont {Narozhny}},\ and\ \bibinfo {author}
  {\bibfnamefont {A.~D.}\ \bibnamefont {Mirlin}},\ }\bibfield  {title}
  {\bibinfo {title} {Collision-dominated nonlinear hydrodynamics in graphene},\
  }\href {https://doi.org/10.1103/PhysRevB.92.115426} {\bibfield  {journal}
  {\bibinfo  {journal} {Phys. Rev. B}\ }\textbf {\bibinfo {volume} {92}},\
  \bibinfo {pages} {115426} (\bibinfo {year} {2015})}\BibitemShut {NoStop}%
\bibitem [{\citenamefont {Levitov}\ and\ \citenamefont
  {Falkovich}(2016)}]{Levitov-2016}%
  \BibitemOpen
  \bibfield  {author} {\bibinfo {author} {\bibfnamefont {L.}~\bibnamefont
  {Levitov}}\ and\ \bibinfo {author} {\bibfnamefont {G.}~\bibnamefont
  {Falkovich}},\ }\bibfield  {title} {\bibinfo {title} {Electron viscosity,
  current vortices and negative nonlocal resistance in graphene},\ }\href
  {http://www.nature.com/nphys/journal/vaop/ncurrent/full/nphys3667.html?WT.feed_name=subjects_fluid-dynamics}
  {\bibfield  {journal} {\bibinfo  {journal} {Nat. Phys.}\ }\textbf {\bibinfo
  {volume} {12}},\ \bibinfo {pages} {672} (\bibinfo {year} {2016})}\BibitemShut
  {NoStop}%
\bibitem [{\citenamefont {Narozhny}\ \emph {et~al.}(2017)\citenamefont
  {Narozhny}, \citenamefont {Gornyi}, \citenamefont {Mirlin},\ and\
  \citenamefont {Schmalian}}]{Narozhny-2017}%
  \BibitemOpen
  \bibfield  {author} {\bibinfo {author} {\bibfnamefont {B.~N.}\ \bibnamefont
  {Narozhny}}, \bibinfo {author} {\bibfnamefont {I.~V.}\ \bibnamefont
  {Gornyi}}, \bibinfo {author} {\bibfnamefont {A.~D.}\ \bibnamefont {Mirlin}},\
  and\ \bibinfo {author} {\bibfnamefont {J.}~\bibnamefont {Schmalian}},\
  }\bibfield  {title} {\bibinfo {title} {Hydrodynamic approach to electronic
  transport in graphene},\ }\href {https://doi.org/10.1002/andp.201700043}
  {\bibfield  {journal} {\bibinfo  {journal} {Ann. Phys.-Berlin}\ }\textbf
  {\bibinfo {volume} {529}},\ \bibinfo {pages} {1700043} (\bibinfo {year}
  {2017})}\BibitemShut {NoStop}%
\bibitem [{\citenamefont {Narozhny}(2019)}]{Narozhny-2019}%
  \BibitemOpen
  \bibfield  {author} {\bibinfo {author} {\bibfnamefont {B.~N.}\ \bibnamefont
  {Narozhny}},\ }\bibfield  {title} {\bibinfo {title} {Electronic hydrodynamics
  in graphene},\ }\href
  {http://www.sciencedirect.com/science/article/pii/S0003491619302349}
  {\bibfield  {journal} {\bibinfo  {journal} {Ann. Phys.-New York}\ }\textbf
  {\bibinfo {volume} {411}},\ \bibinfo {pages} {167979} (\bibinfo {year}
  {2019})}\BibitemShut {NoStop}%
\bibitem [{\citenamefont {Moll}\ \emph {et~al.}(2016)\citenamefont {Moll},
  \citenamefont {Kushwaha}, \citenamefont {Nandi}, \citenamefont {Schmidt},\
  and\ \citenamefont {Mackenzie}}]{Moll-2016}%
  \BibitemOpen
  \bibfield  {author} {\bibinfo {author} {\bibfnamefont {P.~J.}\ \bibnamefont
  {Moll}}, \bibinfo {author} {\bibfnamefont {P.}~\bibnamefont {Kushwaha}},
  \bibinfo {author} {\bibfnamefont {N.}~\bibnamefont {Nandi}}, \bibinfo
  {author} {\bibfnamefont {B.}~\bibnamefont {Schmidt}},\ and\ \bibinfo {author}
  {\bibfnamefont {A.~P.}\ \bibnamefont {Mackenzie}},\ }\bibfield  {title}
  {\bibinfo {title} {Evidence for hydrodynamic electron flow in {PdCoO$_2$}},\
  }\href {https://science.sciencemag.org/content/351/6277/1061} {\bibfield
  {journal} {\bibinfo  {journal} {Science}\ }\textbf {\bibinfo {volume}
  {351}},\ \bibinfo {pages} {1061} (\bibinfo {year} {2016})}\BibitemShut
  {NoStop}%
\bibitem [{\citenamefont {Cook}\ and\ \citenamefont {Lucas}(2019)}]{Cook-2019}%
  \BibitemOpen
  \bibfield  {author} {\bibinfo {author} {\bibfnamefont {C.~Q.}\ \bibnamefont
  {Cook}}\ and\ \bibinfo {author} {\bibfnamefont {A.}~\bibnamefont {Lucas}},\
  }\bibfield  {title} {\bibinfo {title} {Electron hydrodynamics with a
  polygonal {Fermi} surface},\ }\href
  {https://doi.org/10.1103/PhysRevB.99.235148} {\bibfield  {journal} {\bibinfo
  {journal} {Phys. Rev. B}\ }\textbf {\bibinfo {volume} {99}},\ \bibinfo
  {pages} {235148} (\bibinfo {year} {2019})}\BibitemShut {NoStop}%
\bibitem [{\citenamefont {Gorbar}\ \emph {et~al.}(2018)\citenamefont {Gorbar},
  \citenamefont {Miransky}, \citenamefont {Shovkovy},\ and\ \citenamefont
  {Sukhachov}}]{Gorbar-2018}%
  \BibitemOpen
  \bibfield  {author} {\bibinfo {author} {\bibfnamefont {E.~V.}\ \bibnamefont
  {Gorbar}}, \bibinfo {author} {\bibfnamefont {V.~A.}\ \bibnamefont
  {Miransky}}, \bibinfo {author} {\bibfnamefont {I.~A.}\ \bibnamefont
  {Shovkovy}},\ and\ \bibinfo {author} {\bibfnamefont {P.~O.}\ \bibnamefont
  {Sukhachov}},\ }\bibfield  {title} {\bibinfo {title} {Hydrodynamic electron
  flow in a {Weyl} semimetal slab: Role of {Chern}-{Simons} terms},\ }\href
  {https://doi.org/10.1103/PhysRevB.97.205119} {\bibfield  {journal} {\bibinfo
  {journal} {Phys. Rev. B}\ }\textbf {\bibinfo {volume} {97}},\ \bibinfo
  {pages} {205119} (\bibinfo {year} {2018})}\BibitemShut {NoStop}%
\bibitem [{\citenamefont {Gorbar}\ \emph {et~al.}(2019)\citenamefont {Gorbar},
  \citenamefont {Miransky}, \citenamefont {Shovkovy},\ and\ \citenamefont
  {Sukhachov}}]{Gorbar-2019}%
  \BibitemOpen
  \bibfield  {author} {\bibinfo {author} {\bibfnamefont {E.~V.}\ \bibnamefont
  {Gorbar}}, \bibinfo {author} {\bibfnamefont {V.~A.}\ \bibnamefont
  {Miransky}}, \bibinfo {author} {\bibfnamefont {I.~A.}\ \bibnamefont
  {Shovkovy}},\ and\ \bibinfo {author} {\bibfnamefont {P.~O.}\ \bibnamefont
  {Sukhachov}},\ }\bibfield  {title} {\bibinfo {title} {Hydrodynamics of
  {Fermi} arcs: Bulk flow and surface collective modes},\ }\href
  {https://doi.org/10.1103/PhysRevB.99.155120} {\bibfield  {journal} {\bibinfo
  {journal} {Phys. Rev. B}\ }\textbf {\bibinfo {volume} {99}},\ \bibinfo
  {pages} {155120} (\bibinfo {year} {2019})}\BibitemShut {NoStop}%
\bibitem [{\citenamefont {Molenkamp}\ and\ \citenamefont
  {de~Jong}(1994)}]{Molenkamp-1994}%
  \BibitemOpen
  \bibfield  {author} {\bibinfo {author} {\bibfnamefont {L.~W.}\ \bibnamefont
  {Molenkamp}}\ and\ \bibinfo {author} {\bibfnamefont {M.~J.~M.}\ \bibnamefont
  {de~Jong}},\ }\bibfield  {title} {\bibinfo {title}
  {Electron-electron-scattering-induced size effects in a two-dimensional
  wire},\ }\href {https://doi.org/10.1103/PhysRevB.49.5038} {\bibfield
  {journal} {\bibinfo  {journal} {Phys. Rev. B}\ }\textbf {\bibinfo {volume}
  {49}},\ \bibinfo {pages} {5038} (\bibinfo {year} {1994})}\BibitemShut
  {NoStop}%
\bibitem [{\citenamefont {de~Jong}\ and\ \citenamefont
  {Molenkamp}(1995)}]{deJong-1995}%
  \BibitemOpen
  \bibfield  {author} {\bibinfo {author} {\bibfnamefont {M.~J.~M.}\
  \bibnamefont {de~Jong}}\ and\ \bibinfo {author} {\bibfnamefont {L.~W.}\
  \bibnamefont {Molenkamp}},\ }\bibfield  {title} {\bibinfo {title}
  {Hydrodynamic electron flow in high-mobility wires},\ }\href
  {https://doi.org/10.1103/PhysRevB.51.13389} {\bibfield  {journal} {\bibinfo
  {journal} {Phys. Rev. B}\ }\textbf {\bibinfo {volume} {51}},\ \bibinfo
  {pages} {13389} (\bibinfo {year} {1995})}\BibitemShut {NoStop}%
\bibitem [{\citenamefont {Lucas}\ and\ \citenamefont
  {Hartnoll}(2017)}]{Lucas-2017b}%
  \BibitemOpen
  \bibfield  {author} {\bibinfo {author} {\bibfnamefont {A.}~\bibnamefont
  {Lucas}}\ and\ \bibinfo {author} {\bibfnamefont {S.~A.}\ \bibnamefont
  {Hartnoll}},\ }\bibfield  {title} {\bibinfo {title} {Resistivity bound for
  hydrodynamic bad metals},\ }\href {https://doi.org/10.1073/pnas.1711414114}
  {\bibfield  {journal} {\bibinfo  {journal} {Proc. Natl. Acad. Sci. USA}\
  }\textbf {\bibinfo {volume} {114}},\ \bibinfo {pages} {11344} (\bibinfo
  {year} {2017})}\BibitemShut {NoStop}%
\bibitem [{\citenamefont {Schofield}(1999)}]{Schofield-1999}%
  \BibitemOpen
  \bibfield  {author} {\bibinfo {author} {\bibfnamefont {A.~J.}\ \bibnamefont
  {Schofield}},\ }\bibfield  {title} {\bibinfo {title} {Non-{Fermi} liquids},\
  }\href
  {https://www.tandfonline.com/doi/abs/10.1080/001075199181602?casa_token=PhAGlopVFpEAAAAA%3A-HRfUphzv6Ix_roE81znOjPoKGC3Wvk3LSSPmWnmubAxX3aF9IrowwEGFVx6DRbigezPAbCRqwvN&}
  {\bibfield  {journal} {\bibinfo  {journal} {Contemp. Phys.}\ }\textbf
  {\bibinfo {volume} {40}},\ \bibinfo {pages} {95} (\bibinfo {year}
  {1999})}\BibitemShut {NoStop}%
\bibitem [{\citenamefont {Zaanen}\ \emph {et~al.}(2015)\citenamefont {Zaanen},
  \citenamefont {Liu}, \citenamefont {Sun},\ and\ \citenamefont
  {Schalm}}]{Zaanen-2015holo}%
  \BibitemOpen
  \bibfield  {author} {\bibinfo {author} {\bibfnamefont {J.}~\bibnamefont
  {Zaanen}}, \bibinfo {author} {\bibfnamefont {Y.}~\bibnamefont {Liu}},
  \bibinfo {author} {\bibfnamefont {Y.-W.}\ \bibnamefont {Sun}},\ and\ \bibinfo
  {author} {\bibfnamefont {K.}~\bibnamefont {Schalm}},\ }\href@noop {} {\emph
  {\bibinfo {title} {Holographic duality in condensed matter physics}}}\
  (\bibinfo  {publisher} {Cambridge University Press},\ \bibinfo {address}
  {Cambridge},\ \bibinfo {year} {2015})\BibitemShut {NoStop}%
\bibitem [{\citenamefont {Crossno}\ \emph {et~al.}(2016)\citenamefont
  {Crossno}, \citenamefont {Shi}, \citenamefont {Wang}, \citenamefont {Liu},
  \citenamefont {Harzheim}, \citenamefont {Lucas}, \citenamefont {Sachdev},
  \citenamefont {Kim}, \citenamefont {Taniguchi}, \citenamefont {Watanabe},
  \citenamefont {Ohki},\ and\ \citenamefont {Fong}}]{Crossno-2016}%
  \BibitemOpen
  \bibfield  {author} {\bibinfo {author} {\bibfnamefont {J.}~\bibnamefont
  {Crossno}}, \bibinfo {author} {\bibfnamefont {J.~K.}\ \bibnamefont {Shi}},
  \bibinfo {author} {\bibfnamefont {K.}~\bibnamefont {Wang}}, \bibinfo {author}
  {\bibfnamefont {X.}~\bibnamefont {Liu}}, \bibinfo {author} {\bibfnamefont
  {A.}~\bibnamefont {Harzheim}}, \bibinfo {author} {\bibfnamefont
  {A.}~\bibnamefont {Lucas}}, \bibinfo {author} {\bibfnamefont
  {S.}~\bibnamefont {Sachdev}}, \bibinfo {author} {\bibfnamefont
  {P.}~\bibnamefont {Kim}}, \bibinfo {author} {\bibfnamefont {T.}~\bibnamefont
  {Taniguchi}}, \bibinfo {author} {\bibfnamefont {K.}~\bibnamefont {Watanabe}},
  \bibinfo {author} {\bibfnamefont {T.~A.}\ \bibnamefont {Ohki}},\ and\
  \bibinfo {author} {\bibfnamefont {K.~C.}\ \bibnamefont {Fong}},\ }\bibfield
  {title} {\bibinfo {title} {Observation of the {Dirac} fluid and the breakdown
  of the {Wiedemann}-{Franz} law in graphene},\ }\href
  {http://science.sciencemag.org/content/351/6277/1058/tab-pdf} {\bibfield
  {journal} {\bibinfo  {journal} {Science}\ }\textbf {\bibinfo {volume}
  {351}},\ \bibinfo {pages} {1058} (\bibinfo {year} {2016})}\BibitemShut
  {NoStop}%
\bibitem [{\citenamefont {Krishna~Kumar}\ \emph {et~al.}(2017)\citenamefont
  {Krishna~Kumar}, \citenamefont {Bandurin}, \citenamefont {Pellegrino},
  \citenamefont {Cao}, \citenamefont {Principi}, \citenamefont {Guo},
  \citenamefont {Auton}, \citenamefont {Shalom}, \citenamefont {Ponomarenko},
  \citenamefont {Falkovich}, \citenamefont {Watanabe}, \citenamefont
  {Taniguchi}, \citenamefont {Grigorieva}, \citenamefont {Levitov},
  \citenamefont {Polini},\ and\ \citenamefont {Geim}}]{Krishna-Kumar-2017}%
  \BibitemOpen
  \bibfield  {author} {\bibinfo {author} {\bibfnamefont {R.}~\bibnamefont
  {Krishna~Kumar}}, \bibinfo {author} {\bibfnamefont {D.}~\bibnamefont
  {Bandurin}}, \bibinfo {author} {\bibfnamefont {F.}~\bibnamefont
  {Pellegrino}}, \bibinfo {author} {\bibfnamefont {Y.}~\bibnamefont {Cao}},
  \bibinfo {author} {\bibfnamefont {A.}~\bibnamefont {Principi}}, \bibinfo
  {author} {\bibfnamefont {H.}~\bibnamefont {Guo}}, \bibinfo {author}
  {\bibfnamefont {G.}~\bibnamefont {Auton}}, \bibinfo {author} {\bibfnamefont
  {M.~B.}\ \bibnamefont {Shalom}}, \bibinfo {author} {\bibfnamefont {L.~A.}\
  \bibnamefont {Ponomarenko}}, \bibinfo {author} {\bibfnamefont
  {G.}~\bibnamefont {Falkovich}}, \bibinfo {author} {\bibfnamefont
  {K.}~\bibnamefont {Watanabe}}, \bibinfo {author} {\bibfnamefont
  {T.}~\bibnamefont {Taniguchi}}, \bibinfo {author} {\bibfnamefont {I.~V.}\
  \bibnamefont {Grigorieva}}, \bibinfo {author} {\bibfnamefont
  {L.}~\bibnamefont {Levitov}}, \bibinfo {author} {\bibfnamefont
  {M.}~\bibnamefont {Polini}},\ and\ \bibinfo {author} {\bibfnamefont
  {A.}~\bibnamefont {Geim}},\ }\bibfield  {title} {\bibinfo {title}
  {Superballistic flow of viscous electron fluid through graphene
  constrictions},\ }\href {https://www.nature.com/articles/nphys4240}
  {\bibfield  {journal} {\bibinfo  {journal} {Nat. Phys.}\ }\textbf {\bibinfo
  {volume} {13}},\ \bibinfo {pages} {1182} (\bibinfo {year}
  {2017})}\BibitemShut {NoStop}%
\bibitem [{\citenamefont {Ella}\ \emph {et~al.}(2019)\citenamefont {Ella},
  \citenamefont {Rozen}, \citenamefont {Birkbeck}, \citenamefont {Ben-Shalom},
  \citenamefont {Perello}, \citenamefont {Zultak}, \citenamefont {Taniguchi},
  \citenamefont {Watanabe}, \citenamefont {Geim}, \citenamefont {Ilani},\ and\
  \citenamefont {Sulpizio}}]{Ella-2019}%
  \BibitemOpen
  \bibfield  {author} {\bibinfo {author} {\bibfnamefont {L.}~\bibnamefont
  {Ella}}, \bibinfo {author} {\bibfnamefont {A.}~\bibnamefont {Rozen}},
  \bibinfo {author} {\bibfnamefont {J.}~\bibnamefont {Birkbeck}}, \bibinfo
  {author} {\bibfnamefont {M.}~\bibnamefont {Ben-Shalom}}, \bibinfo {author}
  {\bibfnamefont {D.}~\bibnamefont {Perello}}, \bibinfo {author} {\bibfnamefont
  {J.}~\bibnamefont {Zultak}}, \bibinfo {author} {\bibfnamefont
  {T.}~\bibnamefont {Taniguchi}}, \bibinfo {author} {\bibfnamefont
  {K.}~\bibnamefont {Watanabe}}, \bibinfo {author} {\bibfnamefont {A.~K.}\
  \bibnamefont {Geim}}, \bibinfo {author} {\bibfnamefont {S.}~\bibnamefont
  {Ilani}},\ and\ \bibinfo {author} {\bibfnamefont {J.~A.}\ \bibnamefont
  {Sulpizio}},\ }\bibfield  {title} {\bibinfo {title} {Simultaneous voltage and
  current density imaging of flowing electrons in two dimensions},\ }\href
  {https://www.nature.com/articles/s41565-019-0398-x} {\bibfield  {journal}
  {\bibinfo  {journal} {Nat. Nanotechnol.}\ }\textbf {\bibinfo {volume} {14}},\
  \bibinfo {pages} {480} (\bibinfo {year} {2019})}\BibitemShut {NoStop}%
\bibitem [{\citenamefont {Mackenzie}(2017)}]{Mackenzie-2017}%
  \BibitemOpen
  \bibfield  {author} {\bibinfo {author} {\bibfnamefont {A.~P.}\ \bibnamefont
  {Mackenzie}},\ }\bibfield  {title} {\bibinfo {title} {The properties of
  ultrapure delafossite metals},\ }\href
  {https://doi.org/10.1088/1361-6633/aa50e5} {\bibfield  {journal} {\bibinfo
  {journal} {Rep. Prog. Phys.}\ }\textbf {\bibinfo {volume} {80}},\ \bibinfo
  {pages} {032501} (\bibinfo {year} {2017})}\BibitemShut {NoStop}%
\bibitem [{\citenamefont {Nandi}\ \emph {et~al.}(2018)\citenamefont {Nandi},
  \citenamefont {Scaffidi}, \citenamefont {Kushwaha}, \citenamefont {Khim},
  \citenamefont {Barber}, \citenamefont {Sunko}, \citenamefont {Mazzola},
  \citenamefont {King}, \citenamefont {Rosner}, \citenamefont {Moll},
  \citenamefont {K\"{o}nig}, \citenamefont {Moore}, \citenamefont {Hartnoll},\
  and\ \citenamefont {Mackenzie}}]{Nandi-2018}%
  \BibitemOpen
  \bibfield  {author} {\bibinfo {author} {\bibfnamefont {N.}~\bibnamefont
  {Nandi}}, \bibinfo {author} {\bibfnamefont {T.}~\bibnamefont {Scaffidi}},
  \bibinfo {author} {\bibfnamefont {P.}~\bibnamefont {Kushwaha}}, \bibinfo
  {author} {\bibfnamefont {S.}~\bibnamefont {Khim}}, \bibinfo {author}
  {\bibfnamefont {M.~E.}\ \bibnamefont {Barber}}, \bibinfo {author}
  {\bibfnamefont {V.}~\bibnamefont {Sunko}}, \bibinfo {author} {\bibfnamefont
  {F.}~\bibnamefont {Mazzola}}, \bibinfo {author} {\bibfnamefont {P.~D.}\
  \bibnamefont {King}}, \bibinfo {author} {\bibfnamefont {H.}~\bibnamefont
  {Rosner}}, \bibinfo {author} {\bibfnamefont {P.~J.}\ \bibnamefont {Moll}},
  \bibinfo {author} {\bibfnamefont {M.}~\bibnamefont {K\"{o}nig}}, \bibinfo
  {author} {\bibfnamefont {J.}~\bibnamefont {Moore}}, \bibinfo {author}
  {\bibfnamefont {S.}~\bibnamefont {Hartnoll}},\ and\ \bibinfo {author}
  {\bibfnamefont {A.}~\bibnamefont {Mackenzie}},\ }\bibfield  {title} {\bibinfo
  {title} {Unconventional magneto-transport in ultrapure {PdCoO$_2$} and
  {PtCoO$_2$}},\ }\href {https://www.nature.com/articles/s41535-018-0138-8}
  {\bibfield  {journal} {\bibinfo  {journal} {npj Quantum Mater.}\ }\textbf
  {\bibinfo {volume} {3}},\ \bibinfo {pages} {66} (\bibinfo {year}
  {2018})}\BibitemShut {NoStop}%
\bibitem [{\citenamefont {Gooth}\ \emph {et~al.}(2018)\citenamefont {Gooth},
  \citenamefont {Menges}, \citenamefont {Kumar}, \citenamefont {S{\"u}$\beta$},
  \citenamefont {Shekhar}, \citenamefont {Sun}, \citenamefont {Drechsler},
  \citenamefont {Zierold}, \citenamefont {Felser},\ and\ \citenamefont
  {Gotsmann}}]{Gooth-2018}%
  \BibitemOpen
  \bibfield  {author} {\bibinfo {author} {\bibfnamefont {J.}~\bibnamefont
  {Gooth}}, \bibinfo {author} {\bibfnamefont {F.}~\bibnamefont {Menges}},
  \bibinfo {author} {\bibfnamefont {N.}~\bibnamefont {Kumar}}, \bibinfo
  {author} {\bibfnamefont {V.}~\bibnamefont {S{\"u}$\beta$}}, \bibinfo {author}
  {\bibfnamefont {C.}~\bibnamefont {Shekhar}}, \bibinfo {author} {\bibfnamefont
  {Y.}~\bibnamefont {Sun}}, \bibinfo {author} {\bibfnamefont {U.}~\bibnamefont
  {Drechsler}}, \bibinfo {author} {\bibfnamefont {R.}~\bibnamefont {Zierold}},
  \bibinfo {author} {\bibfnamefont {C.}~\bibnamefont {Felser}},\ and\ \bibinfo
  {author} {\bibfnamefont {B.}~\bibnamefont {Gotsmann}},\ }\bibfield  {title}
  {\bibinfo {title} {Thermal and electrical signatures of a hydrodynamic
  electron fluid in tungsten diphosphide},\ }\href
  {https://www.nature.com/articles/s41467-018-06688-y.pdf?origin=ppub}
  {\bibfield  {journal} {\bibinfo  {journal} {Nat. Comm.}\ }\textbf {\bibinfo
  {volume} {9}},\ \bibinfo {pages} {1} (\bibinfo {year} {2018})}\BibitemShut
  {NoStop}%
\bibitem [{\citenamefont {Jaoui}\ \emph {et~al.}(2018)\citenamefont {Jaoui},
  \citenamefont {Fauqu{\'e}}, \citenamefont {Rischau}, \citenamefont {Subedi},
  \citenamefont {Fu}, \citenamefont {Gooth}, \citenamefont {Kumar},
  \citenamefont {S{\"u}{\ss}}, \citenamefont {Maslov}, \citenamefont {Felser},\
  and\ \citenamefont {Behnia}}]{Jaoui-2018}%
  \BibitemOpen
  \bibfield  {author} {\bibinfo {author} {\bibfnamefont {A.}~\bibnamefont
  {Jaoui}}, \bibinfo {author} {\bibfnamefont {B.}~\bibnamefont {Fauqu{\'e}}},
  \bibinfo {author} {\bibfnamefont {C.~W.}\ \bibnamefont {Rischau}}, \bibinfo
  {author} {\bibfnamefont {A.}~\bibnamefont {Subedi}}, \bibinfo {author}
  {\bibfnamefont {C.}~\bibnamefont {Fu}}, \bibinfo {author} {\bibfnamefont
  {J.}~\bibnamefont {Gooth}}, \bibinfo {author} {\bibfnamefont
  {N.}~\bibnamefont {Kumar}}, \bibinfo {author} {\bibfnamefont
  {V.}~\bibnamefont {S{\"u}{\ss}}}, \bibinfo {author} {\bibfnamefont {D.~L.}\
  \bibnamefont {Maslov}}, \bibinfo {author} {\bibfnamefont {C.}~\bibnamefont
  {Felser}},\ and\ \bibinfo {author} {\bibfnamefont {K.}~\bibnamefont
  {Behnia}},\ }\bibfield  {title} {\bibinfo {title} {Departure from the
  {Wiedemann}--{Franz} law in {WP$_2$} driven by mismatch in {T}-square
  resistivity prefactors},\ }\href
  {https://www.nature.com/articles/s41535-018-0136-x} {\bibfield  {journal}
  {\bibinfo  {journal} {npj Quantum Mater.}\ }\textbf {\bibinfo {volume} {3}},\
  \bibinfo {pages} {64} (\bibinfo {year} {2018})}\BibitemShut {NoStop}%
\bibitem [{\citenamefont {Fu}\ \emph {et~al.}(2018)\citenamefont {Fu},
  \citenamefont {Scaffidi}, \citenamefont {Waissman}, \citenamefont {Sun},
  \citenamefont {Saha}, \citenamefont {Watzman}, \citenamefont {Srivastava},
  \citenamefont {Li}, \citenamefont {Schnelle}, \citenamefont {Werner},
  \citenamefont {Kamminga}, \citenamefont {Sachdev}, \citenamefont {Parkin},
  \citenamefont {Hartnoll}, \citenamefont {Felser},\ and\ \citenamefont
  {Gooth}}]{Fu-2018}%
  \BibitemOpen
  \bibfield  {author} {\bibinfo {author} {\bibfnamefont {C.}~\bibnamefont
  {Fu}}, \bibinfo {author} {\bibfnamefont {T.}~\bibnamefont {Scaffidi}},
  \bibinfo {author} {\bibfnamefont {J.}~\bibnamefont {Waissman}}, \bibinfo
  {author} {\bibfnamefont {Y.}~\bibnamefont {Sun}}, \bibinfo {author}
  {\bibfnamefont {R.}~\bibnamefont {Saha}}, \bibinfo {author} {\bibfnamefont
  {S.~J.}\ \bibnamefont {Watzman}}, \bibinfo {author} {\bibfnamefont {A.~K.}\
  \bibnamefont {Srivastava}}, \bibinfo {author} {\bibfnamefont
  {G.}~\bibnamefont {Li}}, \bibinfo {author} {\bibfnamefont {W.}~\bibnamefont
  {Schnelle}}, \bibinfo {author} {\bibfnamefont {P.}~\bibnamefont {Werner}},
  \bibinfo {author} {\bibfnamefont {M.~E.}\ \bibnamefont {Kamminga}}, \bibinfo
  {author} {\bibfnamefont {S.}~\bibnamefont {Sachdev}}, \bibinfo {author}
  {\bibfnamefont {S.~S.~P.}\ \bibnamefont {Parkin}}, \bibinfo {author}
  {\bibfnamefont {S.~A.}\ \bibnamefont {Hartnoll}}, \bibinfo {author}
  {\bibfnamefont {C.}~\bibnamefont {Felser}},\ and\ \bibinfo {author}
  {\bibfnamefont {J.}~\bibnamefont {Gooth}},\ }\bibfield  {title} {\bibinfo
  {title} {Thermoelectric signatures of the electron-phonon fluid in
  {PtSn$_4$}},\ }\href {https://arxiv.org/abs/1802.09468} {\bibfield  {journal}
  {\bibinfo  {journal} {arXiv:1802.09468}\ } (\bibinfo {year}
  {2018})}\BibitemShut {NoStop}%
\bibitem [{\citenamefont {Lea}\ \emph {et~al.}(1973)\citenamefont {Lea},
  \citenamefont {Birks}, \citenamefont {Lee},\ and\ \citenamefont
  {Dobbs}}]{Lea-1973}%
  \BibitemOpen
  \bibfield  {author} {\bibinfo {author} {\bibfnamefont {M.~J.}\ \bibnamefont
  {Lea}}, \bibinfo {author} {\bibfnamefont {A.~R.}\ \bibnamefont {Birks}},
  \bibinfo {author} {\bibfnamefont {P.~M.}\ \bibnamefont {Lee}},\ and\ \bibinfo
  {author} {\bibfnamefont {E.~R.}\ \bibnamefont {Dobbs}},\ }\bibfield  {title}
  {\bibinfo {title} {The dispersion of transverse zero sound in liquid helium
  3},\ }\href {http://stacks.iop.org/0022-3719/6/i=11/a=004} {\bibfield
  {journal} {\bibinfo  {journal} {J. Phys. C: Solid State}\ }\textbf {\bibinfo
  {volume} {6}},\ \bibinfo {pages} {L226} (\bibinfo {year} {1973})}\BibitemShut
  {NoStop}%
\bibitem [{\citenamefont {Roach}\ and\ \citenamefont
  {Ketterson}(1976)}]{Roach-1976}%
  \BibitemOpen
  \bibfield  {author} {\bibinfo {author} {\bibfnamefont {P.~R.}\ \bibnamefont
  {Roach}}\ and\ \bibinfo {author} {\bibfnamefont {J.~B.}\ \bibnamefont
  {Ketterson}},\ }\bibfield  {title} {\bibinfo {title} {Observation of
  transverse zero sound in normal $^{3}\mathrm{He}$},\ }\href
  {https://doi.org/10.1103/PhysRevLett.36.736} {\bibfield  {journal} {\bibinfo
  {journal} {Phys. Rev. Lett.}\ }\textbf {\bibinfo {volume} {36}},\ \bibinfo
  {pages} {736} (\bibinfo {year} {1976})}\BibitemShut {NoStop}%
\bibitem [{\citenamefont {Silin}(1957{\natexlab{a}})}]{Silin-1958a}%
  \BibitemOpen
  \bibfield  {author} {\bibinfo {author} {\bibfnamefont {V.~P.}\ \bibnamefont
  {Silin}},\ }\bibfield  {title} {\bibinfo {title} {Theory of a degenerate
  electron liquid},\ }\href@noop {} {\bibfield  {journal} {\bibinfo  {journal}
  {J. Exptl. Theoret. Phys. (U.S.S.R.)}\ }\textbf {\bibinfo {volume} {33}},\
  \bibinfo {pages} {495} (\bibinfo {year} {1957}{\natexlab{a}})},\ \bibinfo
  {note} {[\href{{http://www.jetp.ac.ru/cgi-bin/dn/e_006_02_0387.pdf}}{{Sov.
  Phys. JETP}} \textbf{6}, 387 (1958)]}\BibitemShut {NoStop}%
\bibitem [{\citenamefont {Silin}(1957{\natexlab{b}})}]{Silin-1958b}%
  \BibitemOpen
  \bibfield  {author} {\bibinfo {author} {\bibfnamefont {V.}~\bibnamefont
  {Silin}},\ }\bibfield  {title} {\bibinfo {title} {On the theory of the
  anomalous skin effect in metals},\ }\href@noop {} {\bibfield  {journal}
  {\bibinfo  {journal} {J. Exptl. Theoret. Phys. (U.S.S.R.)}\ }\textbf
  {\bibinfo {volume} {33}},\ \bibinfo {pages} {1282} (\bibinfo {year}
  {1957}{\natexlab{b}})},\ \bibinfo {note}
  {[\href{http://www.jetp.ac.ru/cgi-bin/dn/e_006_05_0985.pdf}{{Sov. Phys.
  JETP}} \textbf{6}, 985 (1958)]}\BibitemShut {NoStop}%
\bibitem [{\citenamefont {Beekman}\ \emph {et~al.}(2017)\citenamefont
  {Beekman}, \citenamefont {Nissinen}, \citenamefont {Wu},\ and\ \citenamefont
  {Zaanen}}]{Beekman-2017}%
  \BibitemOpen
  \bibfield  {author} {\bibinfo {author} {\bibfnamefont {A.~J.}\ \bibnamefont
  {Beekman}}, \bibinfo {author} {\bibfnamefont {J.}~\bibnamefont {Nissinen}},
  \bibinfo {author} {\bibfnamefont {K.}~\bibnamefont {Wu}},\ and\ \bibinfo
  {author} {\bibfnamefont {J.}~\bibnamefont {Zaanen}},\ }\bibfield  {title}
  {\bibinfo {title} {Dual gauge field theory of quantum liquid crystals in
  three dimensions},\ }\href {https://doi.org/10.1103/PhysRevB.96.165115}
  {\bibfield  {journal} {\bibinfo  {journal} {Phys. Rev. B}\ }\textbf {\bibinfo
  {volume} {96}},\ \bibinfo {pages} {165115} (\bibinfo {year}
  {2017})}\BibitemShut {NoStop}%
\bibitem [{\citenamefont {Jackson}(1962)}]{Jackson-1962}%
  \BibitemOpen
  \bibfield  {author} {\bibinfo {author} {\bibfnamefont {J.~D.}\ \bibnamefont
  {Jackson}},\ }\href@noop {} {\emph {\bibinfo {title} {Classical
  Electrodynamics}}}\ (\bibinfo  {publisher} {John Wiley and Sons},\ \bibinfo
  {address} {Hoboken},\ \bibinfo {year} {1962})\BibitemShut {NoStop}%
\bibitem [{\citenamefont {Dressel}\ and\ \citenamefont
  {Gr\"{u}ner}(2002)}]{Dressel-2001}%
  \BibitemOpen
  \bibfield  {author} {\bibinfo {author} {\bibfnamefont {M.}~\bibnamefont
  {Dressel}}\ and\ \bibinfo {author} {\bibfnamefont {G.}~\bibnamefont
  {Gr\"{u}ner}},\ }\href@noop {} {\emph {\bibinfo {title} {Electrodynamics of
  Solids}}}\ (\bibinfo  {publisher} {Cambridge University Press},\ \bibinfo
  {address} {Cambridge},\ \bibinfo {year} {2002})\BibitemShut {NoStop}%
\bibitem [{\citenamefont {Sondheimer}(2001)}]{Sondheimer-2001}%
  \BibitemOpen
  \bibfield  {author} {\bibinfo {author} {\bibfnamefont {E.~H.}\ \bibnamefont
  {Sondheimer}},\ }\bibfield  {title} {\bibinfo {title} {The mean-free path of
  electrons in metals},\ }\href {https://doi.org/10.1080/00018730152707225}
  {\bibfield  {journal} {\bibinfo  {journal} {Adv. Phys.}\ }\textbf {\bibinfo
  {volume} {50}},\ \bibinfo {pages} {499} (\bibinfo {year} {2001})}\BibitemShut
  {NoStop}%
\bibitem [{\citenamefont {Levitov}\ \emph {et~al.}(2013)\citenamefont
  {Levitov}, \citenamefont {Shtyk},\ and\ \citenamefont
  {Feigelman}}]{Levitov-2013}%
  \BibitemOpen
  \bibfield  {author} {\bibinfo {author} {\bibfnamefont {L.~S.}\ \bibnamefont
  {Levitov}}, \bibinfo {author} {\bibfnamefont {A.~V.}\ \bibnamefont {Shtyk}},\
  and\ \bibinfo {author} {\bibfnamefont {M.~V.}\ \bibnamefont {Feigelman}},\
  }\bibfield  {title} {\bibinfo {title} {Electron-electron interactions and
  plasmon dispersion in graphene},\ }\href
  {https://doi.org/10.1103/PhysRevB.88.235403} {\bibfield  {journal} {\bibinfo
  {journal} {Phys. Rev. B}\ }\textbf {\bibinfo {volume} {88}},\ \bibinfo
  {pages} {235403} (\bibinfo {year} {2013})}\BibitemShut {NoStop}%
\bibitem [{\citenamefont {Sun}\ \emph {et~al.}(2018)\citenamefont {Sun},
  \citenamefont {Basov},\ and\ \citenamefont {Fogler}}]{Sun-2018}%
  \BibitemOpen
  \bibfield  {author} {\bibinfo {author} {\bibfnamefont {Z.}~\bibnamefont
  {Sun}}, \bibinfo {author} {\bibfnamefont {D.~N.}\ \bibnamefont {Basov}},\
  and\ \bibinfo {author} {\bibfnamefont {M.~M.}\ \bibnamefont {Fogler}},\
  }\bibfield  {title} {\bibinfo {title} {Universal linear and nonlinear
  electrodynamics of a {Dirac} fluid},\ }\href
  {https://doi.org/10.1073/pnas.1717010115} {\bibfield  {journal} {\bibinfo
  {journal} {Proc. Natl. Acad. Sci. U. S. A.}\ }\textbf {\bibinfo {volume}
  {115}},\ \bibinfo {pages} {3285} (\bibinfo {year} {2018})}\BibitemShut
  {NoStop}%
\bibitem [{\citenamefont {Levchenko}\ and\ \citenamefont
  {Schmalian}(2020)}]{Levchenko-2020}%
  \BibitemOpen
  \bibfield  {author} {\bibinfo {author} {\bibfnamefont {A.}~\bibnamefont
  {Levchenko}}\ and\ \bibinfo {author} {\bibfnamefont {J.}~\bibnamefont
  {Schmalian}},\ }\bibfield  {title} {\bibinfo {title} {Transport properties of
  strongly coupled electron–phonon liquids},\ }\href
  {https://doi.org/https://doi.org/10.1016/j.aop.2020.168218} {\bibfield
  {journal} {\bibinfo  {journal} {Ann. Phys.}\ }\textbf {\bibinfo {volume}
  {419}},\ \bibinfo {pages} {168218} (\bibinfo {year} {2020})}\BibitemShut
  {NoStop}%
\bibitem [{\citenamefont {Khoo}\ \emph {et~al.}(2021)\citenamefont {Khoo},
  \citenamefont {Pientka},\ and\ \citenamefont
  {Sodemann}}]{Khoo-2021_preprint}%
  \BibitemOpen
  \bibfield  {author} {\bibinfo {author} {\bibfnamefont {J.~Y.}\ \bibnamefont
  {Khoo}}, \bibinfo {author} {\bibfnamefont {F.}~\bibnamefont {Pientka}},\ and\
  \bibinfo {author} {\bibfnamefont {I.}~\bibnamefont {Sodemann}},\ }\bibfield
  {title} {\bibinfo {title} {The universal shear conductivity of {Fermi}
  liquids and spinon {Fermi} surface states and its detection via spin qubit
  noise magnetometry},\ }\href {https://arxiv.org/abs/2103.05095} {\bibfield
  {journal} {\bibinfo  {journal} {arXiv:2103.05095}\ } (\bibinfo {year}
  {2021})}\BibitemShut {NoStop}%
\bibitem [{\citenamefont {Link}\ \emph {et~al.}(2016)\citenamefont {Link},
  \citenamefont {Orth}, \citenamefont {Sheehy},\ and\ \citenamefont
  {Schmalian}}]{Link-2015}%
  \BibitemOpen
  \bibfield  {author} {\bibinfo {author} {\bibfnamefont {J.~M.}\ \bibnamefont
  {Link}}, \bibinfo {author} {\bibfnamefont {P.~P.}\ \bibnamefont {Orth}},
  \bibinfo {author} {\bibfnamefont {D.~E.}\ \bibnamefont {Sheehy}},\ and\
  \bibinfo {author} {\bibfnamefont {J.}~\bibnamefont {Schmalian}},\ }\bibfield
  {title} {\bibinfo {title} {Universal collisionless transport of graphene},\
  }\href {https://doi.org/10.1103/PhysRevB.93.235447} {\bibfield  {journal}
  {\bibinfo  {journal} {Phys. Rev. B}\ }\textbf {\bibinfo {volume} {93}},\
  \bibinfo {pages} {235447} (\bibinfo {year} {2016})}\BibitemShut {NoStop}%
\bibitem [{\citenamefont {Delacr\'etaz}\ \emph {et~al.}(2017)\citenamefont
  {Delacr\'etaz}, \citenamefont {Gout\'eraux}, \citenamefont {Hartnoll},\ and\
  \citenamefont {Karlsson}}]{Delacretaz-2017}%
  \BibitemOpen
  \bibfield  {author} {\bibinfo {author} {\bibfnamefont {L.~V.}\ \bibnamefont
  {Delacr\'etaz}}, \bibinfo {author} {\bibfnamefont {B.}~\bibnamefont
  {Gout\'eraux}}, \bibinfo {author} {\bibfnamefont {S.~A.}\ \bibnamefont
  {Hartnoll}},\ and\ \bibinfo {author} {\bibfnamefont {A.}~\bibnamefont
  {Karlsson}},\ }\bibfield  {title} {\bibinfo {title} {Theory of hydrodynamic
  transport in fluctuating electronic charge density wave states},\ }\href
  {https://doi.org/10.1103/PhysRevB.96.195128} {\bibfield  {journal} {\bibinfo
  {journal} {Phys. Rev. B}\ }\textbf {\bibinfo {volume} {96}},\ \bibinfo
  {pages} {195128} (\bibinfo {year} {2017})}\BibitemShut {NoStop}%
\bibitem [{\citenamefont {Inkof}\ \emph {et~al.}(2020)\citenamefont {Inkof},
  \citenamefont {K{\"u}ppers}, \citenamefont {Link}, \citenamefont
  {Gout{\'e}raux},\ and\ \citenamefont {Schmalian}}]{Inkof-2019}%
  \BibitemOpen
  \bibfield  {author} {\bibinfo {author} {\bibfnamefont {G.~A.}\ \bibnamefont
  {Inkof}}, \bibinfo {author} {\bibfnamefont {J.~M.}\ \bibnamefont
  {K{\"u}ppers}}, \bibinfo {author} {\bibfnamefont {J.~M.}\ \bibnamefont
  {Link}}, \bibinfo {author} {\bibfnamefont {B.}~\bibnamefont
  {Gout{\'e}raux}},\ and\ \bibinfo {author} {\bibfnamefont {J.}~\bibnamefont
  {Schmalian}},\ }\bibfield  {title} {\bibinfo {title} {Quantum critical
  scaling and holographic bound for transport coefficients near {Lifshitz}
  points},\ }\href {https://link.springer.com/article/10.1007/JHEP11(2020)088}
  {\bibfield  {journal} {\bibinfo  {journal} {J. High Energy Phys.}\ }\textbf
  {\bibinfo {volume} {2020}}\bibinfo  {number} { (11)},\ \bibinfo {pages}
  {1}}\BibitemShut {NoStop}%
\bibitem [{\citenamefont {Forcella}\ \emph {et~al.}(2014)\citenamefont
  {Forcella}, \citenamefont {Zaanen}, \citenamefont {Valentinis},\ and\
  \citenamefont {van~der Marel}}]{FZVM-2014}%
  \BibitemOpen
\bibfield  {number} {  }\bibfield  {author} {\bibinfo {author} {\bibfnamefont
  {D.}~\bibnamefont {Forcella}}, \bibinfo {author} {\bibfnamefont
  {J.}~\bibnamefont {Zaanen}}, \bibinfo {author} {\bibfnamefont
  {D.}~\bibnamefont {Valentinis}},\ and\ \bibinfo {author} {\bibfnamefont
  {D.}~\bibnamefont {van~der Marel}},\ }\bibfield  {title} {\bibinfo {title}
  {Electromagnetic properties of viscous charged fluids},\ }\href
  {https://doi.org/10.1103/PhysRevB.90.035143} {\bibfield  {journal} {\bibinfo
  {journal} {Phys. Rev. B}\ }\textbf {\bibinfo {volume} {90}},\ \bibinfo
  {pages} {035143} (\bibinfo {year} {2014})}\BibitemShut {NoStop}%
\bibitem [{\citenamefont {Veselago}(1968)}]{Veselago-1968}%
  \BibitemOpen
  \bibfield  {author} {\bibinfo {author} {\bibfnamefont {V.~G.}\ \bibnamefont
  {Veselago}},\ }\bibfield  {title} {\bibinfo {title} {The electrodynamics of
  substances with simultaneously negative values of $\epsilon$ and $\mu$},\
  }\href {http://ufn.ru/en/articles/1968/4/d/} {\bibfield  {journal} {\bibinfo
  {journal} {Sov. Phys. Uspekhi}\ }\textbf {\bibinfo {volume} {10}},\ \bibinfo
  {pages} {509} (\bibinfo {year} {1968})}\BibitemShut {NoStop}%
\bibitem [{\citenamefont {Forcella}\ \emph {et~al.}(2017)\citenamefont
  {Forcella}, \citenamefont {Prada},\ and\ \citenamefont
  {Carminati}}]{Forcella-2016}%
  \BibitemOpen
  \bibfield  {author} {\bibinfo {author} {\bibfnamefont {D.}~\bibnamefont
  {Forcella}}, \bibinfo {author} {\bibfnamefont {C.}~\bibnamefont {Prada}},\
  and\ \bibinfo {author} {\bibfnamefont {R.}~\bibnamefont {Carminati}},\
  }\bibfield  {title} {\bibinfo {title} {Causality, nonlocality, and negative
  refraction},\ }\href {https://doi.org/10.1103/PhysRevLett.118.134301}
  {\bibfield  {journal} {\bibinfo  {journal} {Phys. Rev. Lett.}\ }\textbf
  {\bibinfo {volume} {118}},\ \bibinfo {pages} {134301} (\bibinfo {year}
  {2017})}\BibitemShut {NoStop}%
\bibitem [{\citenamefont {Veselago}\ \emph {et~al.}(2006)\citenamefont
  {Veselago}, \citenamefont {Braginsky}, \citenamefont {Shklover},\ and\
  \citenamefont {Hafner}}]{Veselago-2006}%
  \BibitemOpen
  \bibfield  {author} {\bibinfo {author} {\bibfnamefont {V.}~\bibnamefont
  {Veselago}}, \bibinfo {author} {\bibfnamefont {L.}~\bibnamefont {Braginsky}},
  \bibinfo {author} {\bibfnamefont {V.}~\bibnamefont {Shklover}},\ and\
  \bibinfo {author} {\bibfnamefont {C.}~\bibnamefont {Hafner}},\ }\bibfield
  {title} {\bibinfo {title} {Negative refractive index materials},\ }\href
  {https://doi.org/10.1166/jctn.2006.002} {\bibfield  {journal} {\bibinfo
  {journal} {J. Comput. Theor. Nanos.}\ }\textbf {\bibinfo {volume} {3}},\
  \bibinfo {pages} {189} (\bibinfo {year} {2006})}\BibitemShut {NoStop}%
\bibitem [{\citenamefont {Khoo}\ and\ \citenamefont
  {Sodemann~Villadiego}(2019)}]{Khoo-2019}%
  \BibitemOpen
  \bibfield  {author} {\bibinfo {author} {\bibfnamefont {J.~Y.}\ \bibnamefont
  {Khoo}}\ and\ \bibinfo {author} {\bibfnamefont {I.}~\bibnamefont
  {Sodemann~Villadiego}},\ }\bibfield  {title} {\bibinfo {title} {Shear sound
  of two-dimensional {Fermi} liquids},\ }\href
  {https://doi.org/10.1103/PhysRevB.99.075434} {\bibfield  {journal} {\bibinfo
  {journal} {Phys. Rev. B}\ }\textbf {\bibinfo {volume} {99}},\ \bibinfo
  {pages} {075434} (\bibinfo {year} {2019})}\BibitemShut {NoStop}%
\bibitem [{\citenamefont {Khoo}\ \emph {et~al.}(2020)\citenamefont {Khoo},
  \citenamefont {Chang}, \citenamefont {Pientka},\ and\ \citenamefont
  {Sodemann}}]{Khoo-2020}%
  \BibitemOpen
  \bibfield  {author} {\bibinfo {author} {\bibfnamefont {J.~Y.}\ \bibnamefont
  {Khoo}}, \bibinfo {author} {\bibfnamefont {P.-Y.}\ \bibnamefont {Chang}},
  \bibinfo {author} {\bibfnamefont {F.}~\bibnamefont {Pientka}},\ and\ \bibinfo
  {author} {\bibfnamefont {I.}~\bibnamefont {Sodemann}},\ }\bibfield  {title}
  {\bibinfo {title} {Quantum paracrystalline shear modes of the electron
  liquid},\ }\href {https://doi.org/10.1103/PhysRevB.102.085437} {\bibfield
  {journal} {\bibinfo  {journal} {Phys. Rev. B}\ }\textbf {\bibinfo {volume}
  {102}},\ \bibinfo {pages} {085437} (\bibinfo {year} {2020})}\BibitemShut
  {NoStop}%
\bibitem [{Note1()}]{Note1}%
  \BibitemOpen
  \bibinfo {note} {The dispersion relation can equivalently be written in the
  form $\omega (\protect \bm {q})$, as commonly done in solid-state textbooks.
  In this paper, we adopt the viewpoint of $\protect \bm {q}(\omega )$,
  consistently with the picture of an electromagnetic wave of real-valued
  frequency $\omega \in \protect \mdmathbb {R}$ which excites a Fermi-liquid
  shear mode with complex-valued momentum $\protect \bm {q} \in \protect
  \mdmathbb {C}$.}\BibitemShut {Stop}%
\bibitem [{\citenamefont {Bedell}\ and\ \citenamefont
  {Pethick}(1982)}]{Bedell-1982}%
  \BibitemOpen
  \bibfield  {author} {\bibinfo {author} {\bibfnamefont {K.}~\bibnamefont
  {Bedell}}\ and\ \bibinfo {author} {\bibfnamefont {C.~J.}\ \bibnamefont
  {Pethick}},\ }\bibfield  {title} {\bibinfo {title} {Viscoelastic behavior in
  a normal {Fermi} liquid},\ }\href {https://doi.org/10.1007/BF00681588}
  {\bibfield  {journal} {\bibinfo  {journal} {J. Low Temp. Phys.}\ }\textbf
  {\bibinfo {volume} {49}},\ \bibinfo {pages} {213} (\bibinfo {year}
  {1982})}\BibitemShut {NoStop}%
\bibitem [{\citenamefont {Gurzhi}(1963)}]{Gurzhi-1963}%
  \BibitemOpen
  \bibfield  {author} {\bibinfo {author} {\bibfnamefont {R.}~\bibnamefont
  {Gurzhi}},\ }\bibfield  {title} {\bibinfo {title} {Minimum of resistance in
  impurity-free conductors},\ }\href@noop {} {\bibfield  {journal} {\bibinfo
  {journal} {J .Exptl. Theoret. Phys. (U.S.S.R.)}\ }\textbf {\bibinfo {volume}
  {44}},\ \bibinfo {pages} {771} (\bibinfo {year} {1963})},\ \bibinfo {note}
  {[\href{http://jetp.ac.ru/cgi-bin/dn/e_017_02_0521.pdf}{{Sov. Phys. JETP}}
  \textbf{17}, 521 (1963)]}\BibitemShut {NoStop}%
\bibitem [{\citenamefont {Jaggi}(1991)}]{Jaggi-1990}%
  \BibitemOpen
  \bibfield  {author} {\bibinfo {author} {\bibfnamefont {R.}~\bibnamefont
  {Jaggi}},\ }\bibfield  {title} {\bibinfo {title} {Electron-fluid model for dc
  size effect},\ }\href {http://dx.doi.org/10.1063/1.347315} {\bibfield
  {journal} {\bibinfo  {journal} {J. Appl. Phys.}\ }\textbf {\bibinfo {volume}
  {69}},\ \bibinfo {pages} {816} (\bibinfo {year} {1991})}\BibitemShut
  {NoStop}%
\bibitem [{\citenamefont {Alekseev}(2016)}]{Alekseev-2016}%
  \BibitemOpen
  \bibfield  {author} {\bibinfo {author} {\bibfnamefont {P.~S.}\ \bibnamefont
  {Alekseev}},\ }\bibfield  {title} {\bibinfo {title} {Negative
  magnetoresistance in viscous flow of two-dimensional electrons},\ }\href
  {https://doi.org/10.1103/PhysRevLett.117.166601} {\bibfield  {journal}
  {\bibinfo  {journal} {Phys. Rev. Lett.}\ }\textbf {\bibinfo {volume} {117}},\
  \bibinfo {pages} {166601} (\bibinfo {year} {2016})}\BibitemShut {NoStop}%
\bibitem [{\citenamefont {Scaffidi}\ \emph {et~al.}(2017)\citenamefont
  {Scaffidi}, \citenamefont {Nandi}, \citenamefont {Schmidt}, \citenamefont
  {Mackenzie},\ and\ \citenamefont {Moore}}]{Scaffidi-2017}%
  \BibitemOpen
  \bibfield  {author} {\bibinfo {author} {\bibfnamefont {T.}~\bibnamefont
  {Scaffidi}}, \bibinfo {author} {\bibfnamefont {N.}~\bibnamefont {Nandi}},
  \bibinfo {author} {\bibfnamefont {B.}~\bibnamefont {Schmidt}}, \bibinfo
  {author} {\bibfnamefont {A.~P.}\ \bibnamefont {Mackenzie}},\ and\ \bibinfo
  {author} {\bibfnamefont {J.~E.}\ \bibnamefont {Moore}},\ }\bibfield  {title}
  {\bibinfo {title} {Hydrodynamic electron flow and {Hall} viscosity},\ }\href
  {https://doi.org/10.1103/PhysRevLett.118.226601} {\bibfield  {journal}
  {\bibinfo  {journal} {Phys. Rev. Lett.}\ }\textbf {\bibinfo {volume} {118}},\
  \bibinfo {pages} {226601} (\bibinfo {year} {2017})}\BibitemShut {NoStop}%
\bibitem [{\citenamefont {Reuter}\ and\ \citenamefont
  {Sondheimer}(1948)}]{Reuter-1948}%
  \BibitemOpen
  \bibfield  {author} {\bibinfo {author} {\bibfnamefont {G.~E.~H.}\
  \bibnamefont {Reuter}}\ and\ \bibinfo {author} {\bibfnamefont {E.~H.}\
  \bibnamefont {Sondheimer}},\ }\bibfield  {title} {\bibinfo {title} {The
  theory of the anomalous skin effect in metals},\ }\href
  {https://doi.org/10.1098/rspa.1948.0123} {\bibfield  {journal} {\bibinfo
  {journal} {Proc. R. Soc. A}\ }\textbf {\bibinfo {volume} {195}},\ \bibinfo
  {pages} {336} (\bibinfo {year} {1948})}\BibitemShut {NoStop}%
\bibitem [{\citenamefont {Khalatnikov}\ and\ \citenamefont
  {Abrikosov}(1957)}]{Khalatnikov-1958}%
  \BibitemOpen
  \bibfield  {author} {\bibinfo {author} {\bibfnamefont {I.~M.}\ \bibnamefont
  {Khalatnikov}}\ and\ \bibinfo {author} {\bibfnamefont {A.~A.}\ \bibnamefont
  {Abrikosov}},\ }\bibfield  {title} {\bibinfo {title} {Dispersion of sound in
  a {Fermi} liquid},\ }\href@noop {} {\bibfield  {journal} {\bibinfo  {journal}
  {J. Exptl. Theoret. Phys. (U.S.S.R.)}\ }\textbf {\bibinfo {volume} {33}},\
  \bibinfo {pages} {110} (\bibinfo {year} {1957})},\ \bibinfo {note}
  {[\href{http://www.jetp.ac.ru/cgi-bin/dn/e_006_01_0084.pdf}{{Sov. Phys.
  JETP}} \textbf{6}, 84 (1958)]}\BibitemShut {NoStop}%
\bibitem [{\citenamefont {Dupuis}(2019)}]{Dupuis-lect-2011}%
  \BibitemOpen
  \bibfield  {author} {\bibinfo {author} {\bibfnamefont {N.}~\bibnamefont
  {Dupuis}},\ }\href {https://www.lptmc.jussieu.fr/files/chap_fl.pdf} {\bibinfo
  {title} {Notes on the many-body problem - {Fermi}-liquid theory}} (\bibinfo
  {year} {2019})\BibitemShut {NoStop}%
\bibitem [{\citenamefont {Berthod}(2018)}]{Berthod-2018}%
  \BibitemOpen
  \bibfield  {author} {\bibinfo {author} {\bibfnamefont {C.}~\bibnamefont
  {Berthod}},\ }\href {https://doi.org/10.1088/978-0-7503-1741-2} {\emph
  {\bibinfo {title} {{Spectroscopic Probes of Quantum Matter}}}},\ 2053-2563\
  (\bibinfo  {publisher} {IOP Publishing},\ \bibinfo {address} {Bristol},\
  \bibinfo {year} {2018})\BibitemShut {NoStop}%
\bibitem [{Note2()}]{Note2}%
  \BibitemOpen
  \bibinfo {note} {In the hydrodynamic/collisional regime $\omega \tau _c \ll
  1$, the transverse collective mode is damped, i.e.\spacefactor \@m {}, it has
  a complex sound velocity $s=\omega /(q v_F^{*}) \in \protect \mdmathbb {C}$
  with equal real and imaginary parts \cite {Lea-1973}. In such conditions, the
  transverse mode is called \protect \emph {relaxational mode} in part of the
  literature, distinguishing this regime from the one of propagating transverse
  sound. Other references employ the label \protect \emph {damped transverse
  sound} in referring to damped transverse waves in the collisional regime. In
  this paper, we will employ the nomenclature \protect \emph {relaxational
  mode} and \protect \emph {damped transverse sound} as synonyms, meaning a
  transverse collective mode satisfying Eq.\spacefactor \@m {} (\ref {eq:Lea})
  in the regime $\omega \tau _c \ll 1$. Generic solutions of Eq.\spacefactor
  \@m {} (\ref {eq:Lea}), without specifying whether we are in
  collisional/collisionless regime, will be named \protect \emph {shear mode}
  or \protect \emph {transverse sound} in this paper.}\BibitemShut {Stop}%
\bibitem [{Note3()}]{Note3}%
  \BibitemOpen
  \bibinfo {note} {In this paper we will refer to real solutions of
  Eq.\spacefactor \@m {} (\ref {eq:Lea}) in the collisionless limit $\omega
  \tau _c \rightarrow +\infty $ as \protect \emph {transverse zero sound} or
  \protect \emph {propagating shear}, consistently with the literature \cite
  {Abrikosov-1959,Lea-1973,Roach-1976}}\BibitemShut {NoStop}%
\bibitem [{\citenamefont {Dobbs}(2000)}]{Dobbs-2000}%
  \BibitemOpen
  \bibfield  {author} {\bibinfo {author} {\bibfnamefont {R.}~\bibnamefont
  {Dobbs}},\ }\href@noop {} {\emph {\bibinfo {title} {Helium {Three}}}}\
  (\bibinfo  {publisher} {Oxford University Press},\ \bibinfo {address}
  {Oxford},\ \bibinfo {year} {2000})\BibitemShut {NoStop}%
\bibitem [{\citenamefont {Brooker}(1967)}]{Brooker-1967}%
  \BibitemOpen
  \bibfield  {author} {\bibinfo {author} {\bibfnamefont {G.~A.}\ \bibnamefont
  {Brooker}},\ }\bibfield  {title} {\bibinfo {title} {The acoustic impedance of
  liquid helium 3: theory},\ }\href
  {http://stacks.iop.org/0370-1328/90/i=2/a=310} {\bibfield  {journal}
  {\bibinfo  {journal} {P. Phys. Soc.}\ }\textbf {\bibinfo {volume} {90}},\
  \bibinfo {pages} {397} (\bibinfo {year} {1967})}\BibitemShut {NoStop}%
\bibitem [{\citenamefont {van~der Marel}\ \emph {et~al.}(2011)\citenamefont
  {van~der Marel}, \citenamefont {van Mechelen},\ and\ \citenamefont
  {Mazin}}]{vanderMarel-2011}%
  \BibitemOpen
  \bibfield  {author} {\bibinfo {author} {\bibfnamefont {D.}~\bibnamefont
  {van~der Marel}}, \bibinfo {author} {\bibfnamefont {J.~L.~M.}\ \bibnamefont
  {van Mechelen}},\ and\ \bibinfo {author} {\bibfnamefont {I.~I.}\ \bibnamefont
  {Mazin}},\ }\bibfield  {title} {\bibinfo {title} {Common {Fermi}-liquid
  origin of {T$^{2}$} resistivity and superconductivity in $n$-type
  {SrTiO}${}_{3}$},\ }\href {https://doi.org/10.1103/PhysRevB.84.205111}
  {\bibfield  {journal} {\bibinfo  {journal} {Phys. Rev. B}\ }\textbf {\bibinfo
  {volume} {84}},\ \bibinfo {pages} {205111} (\bibinfo {year}
  {2011})}\BibitemShut {NoStop}%
\bibitem [{\citenamefont {Coleman}(2015)}]{Coleman-2015mb}%
  \BibitemOpen
  \bibfield  {author} {\bibinfo {author} {\bibfnamefont {P.}~\bibnamefont
  {Coleman}},\ }\href@noop {} {\emph {\bibinfo {title} {{Introduction to
  Many-Body Physics}}}}\ (\bibinfo  {publisher} {Cambridge University Press},\
  \bibinfo {address} {Cambridge},\ \bibinfo {year} {2015})\BibitemShut
  {NoStop}%
\bibitem [{\citenamefont {Mahan}(2000)}]{Mahan-2000}%
  \BibitemOpen
  \bibfield  {author} {\bibinfo {author} {\bibfnamefont {G.~D.}\ \bibnamefont
  {Mahan}},\ }\href@noop {} {\emph {\bibinfo {title} {{Many-Particle
  Physics}}}}\ (\bibinfo  {publisher} {Springer US},\ \bibinfo {address} {New
  York},\ \bibinfo {year} {2000})\BibitemShut {NoStop}%
\bibitem [{\citenamefont {Bruus}\ and\ \citenamefont
  {Flensberg}(2004)}]{Bruus-2004mb}%
  \BibitemOpen
  \bibfield  {author} {\bibinfo {author} {\bibfnamefont {H.}~\bibnamefont
  {Bruus}}\ and\ \bibinfo {author} {\bibfnamefont {K.}~\bibnamefont
  {Flensberg}},\ }\href@noop {} {\emph {\bibinfo {title} {{Many-body Quantum
  Theory in Condensed Matter Physics: An Introduction}}}}\ (\bibinfo
  {publisher} {Oxford University Press},\ \bibinfo {address} {Oxford},\
  \bibinfo {year} {2004})\BibitemShut {NoStop}%
\bibitem [{\citenamefont {Baggioli}\ \emph {et~al.}(2020)\citenamefont
  {Baggioli}, \citenamefont {Gran},\ and\ \citenamefont
  {Torns{\"o}}}]{Baggioli-2020}%
  \BibitemOpen
  \bibfield  {author} {\bibinfo {author} {\bibfnamefont {M.}~\bibnamefont
  {Baggioli}}, \bibinfo {author} {\bibfnamefont {U.}~\bibnamefont {Gran}},\
  and\ \bibinfo {author} {\bibfnamefont {M.}~\bibnamefont {Torns{\"o}}},\
  }\bibfield  {title} {\bibinfo {title} {Transverse collective modes in
  interacting holographic plasmas},\ }\href
  {https://doi.org/10.1007/JHEP04(2020)106} {\bibfield  {journal} {\bibinfo
  {journal} {J. High Energy Phys.}\ }\textbf {\bibinfo {volume} {2020}}\bibinfo
   {number} { (4)},\ \bibinfo {pages} {1}}\BibitemShut {NoStop}%
\bibitem [{\citenamefont {Ledwith}\ \emph {et~al.}(2019)\citenamefont
  {Ledwith}, \citenamefont {Guo}, \citenamefont {Shytov},\ and\ \citenamefont
  {Levitov}}]{Ledwith-2019}%
  \BibitemOpen
\bibfield  {number} {  }\bibfield  {author} {\bibinfo {author} {\bibfnamefont
  {P.}~\bibnamefont {Ledwith}}, \bibinfo {author} {\bibfnamefont
  {H.}~\bibnamefont {Guo}}, \bibinfo {author} {\bibfnamefont {A.}~\bibnamefont
  {Shytov}},\ and\ \bibinfo {author} {\bibfnamefont {L.}~\bibnamefont
  {Levitov}},\ }\bibfield  {title} {\bibinfo {title} {Tomographic dynamics and
  scale-dependent viscosity in {2D} electron systems},\ }\href
  {https://doi.org/10.1103/PhysRevLett.123.116601} {\bibfield  {journal}
  {\bibinfo  {journal} {Phys. Rev. Lett.}\ }\textbf {\bibinfo {volume} {123}},\
  \bibinfo {pages} {116601} (\bibinfo {year} {2019})}\BibitemShut {NoStop}%
\bibitem [{\citenamefont {Berthod}\ \emph {et~al.}(2013)\citenamefont
  {Berthod}, \citenamefont {Mravlje}, \citenamefont {Deng}, \citenamefont
  {\ifmmode~\check{Z}\else \v{Z}\fi{}itko}, \citenamefont {van~der Marel},\
  and\ \citenamefont {Georges}}]{Berthod-2013}%
  \BibitemOpen
  \bibfield  {author} {\bibinfo {author} {\bibfnamefont {C.}~\bibnamefont
  {Berthod}}, \bibinfo {author} {\bibfnamefont {J.}~\bibnamefont {Mravlje}},
  \bibinfo {author} {\bibfnamefont {X.}~\bibnamefont {Deng}}, \bibinfo {author}
  {\bibfnamefont {R.}~\bibnamefont {\ifmmode~\check{Z}\else \v{Z}\fi{}itko}},
  \bibinfo {author} {\bibfnamefont {D.}~\bibnamefont {van~der Marel}},\ and\
  \bibinfo {author} {\bibfnamefont {A.}~\bibnamefont {Georges}},\ }\bibfield
  {title} {\bibinfo {title} {Non-{Drude} universal scaling laws for the optical
  response of local {Fermi} liquids},\ }\href
  {https://doi.org/10.1103/PhysRevB.87.115109} {\bibfield  {journal} {\bibinfo
  {journal} {Phys. Rev. B}\ }\textbf {\bibinfo {volume} {87}},\ \bibinfo
  {pages} {115109} (\bibinfo {year} {2013})}\BibitemShut {NoStop}%
\bibitem [{Note4()}]{Note4}%
  \BibitemOpen
  \bibinfo {note} {For electrons interacting with an ideal ionic Bravais
  lattice, momentum is conserved only up to a reciprocal lattice vector \cite
  {Ashcroft-1976}: although the \protect \emph {global} momentum of electrons
  and the lattice is conserved, the two individual momentum components for
  electrons and lattice vibrations are not \cite {Narozhny-2019}}\BibitemShut
  {NoStop}%
\bibitem [{\citenamefont {Pal}\ \emph {et~al.}(2012)\citenamefont {Pal},
  \citenamefont {Yudson},\ and\ \citenamefont {Maslov}}]{Pal-2012}%
  \BibitemOpen
  \bibfield  {author} {\bibinfo {author} {\bibfnamefont {H.}~\bibnamefont
  {Pal}}, \bibinfo {author} {\bibfnamefont {V.}~\bibnamefont {Yudson}},\ and\
  \bibinfo {author} {\bibfnamefont {D.}~\bibnamefont {Maslov}},\ }\bibfield
  {title} {\bibinfo {title} {Resistivity of non-{Galilean}-invariant
  {Fermi}-and non-{Fermi} liquids},\ }\href
  {http://www.lmaleidykla.lt/ojs/index.php/physics/article/view/2358}
  {\bibfield  {journal} {\bibinfo  {journal} {Lith. J. Phys.}\ }\textbf
  {\bibinfo {volume} {52}},\ \bibinfo {pages} {142} (\bibinfo {year}
  {2012})}\BibitemShut {NoStop}%
\bibitem [{\citenamefont {Maslov}\ and\ \citenamefont
  {Chubukov}(2017)}]{Maslov-2017}%
  \BibitemOpen
  \bibfield  {author} {\bibinfo {author} {\bibfnamefont {D.~L.}\ \bibnamefont
  {Maslov}}\ and\ \bibinfo {author} {\bibfnamefont {A.~V.}\ \bibnamefont
  {Chubukov}},\ }\bibfield  {title} {\bibinfo {title} {Optical response of
  correlated electron systems},\ }\href
  {http://stacks.iop.org/0034-4885/80/i=2/a=026503} {\bibfield  {journal}
  {\bibinfo  {journal} {Rep. Prog. Phys.}\ }\textbf {\bibinfo {volume} {80}},\
  \bibinfo {pages} {026503} (\bibinfo {year} {2017})}\BibitemShut {NoStop}%
\bibitem [{Note5()}]{Note5}%
  \BibitemOpen
  \bibinfo {note} {I am grateful to F.\spacefactor \@m {} Pientka for pointing
  out the difference between the factors $\protect \tilde {\beta }$ and $\beta
  $, with and without relaxation respectively.}\BibitemShut {Stop}%
\bibitem [{Note6()}]{Note6}%
  \BibitemOpen
  \bibinfo {note} {In Ref.\spacefactor \@m {} \protect \rev@citealp {FZVM-2014}
  an expression for $\nu (\omega )$ as a function of frequency $\omega $ and
  collision time $\tau _c$ was obtained through an analytical fit of the
  numerical dispersion for transverse sound in a neutral Fermi liquid,
  i.e.\spacefactor \@m {} Eq.\spacefactor \@m {} (\ref {eq:Lea}). The fitted
  functional behavior of $\nu (\omega )$ qualitatively agrees with the
  generalized shear modulus (\ref {eq:nu_q0_vel}), which is derived directly
  from the Fermi-liquid kinetic equation in this paper.}\BibitemShut {Stop}%
\bibitem [{\citenamefont {Gurzhi}(1968)}]{Gurzhi-1968}%
  \BibitemOpen
  \bibfield  {author} {\bibinfo {author} {\bibfnamefont {R.~N.}\ \bibnamefont
  {Gurzhi}},\ }\bibfield  {title} {\bibinfo {title} {Hydrodynamic effects in
  solids at low temperature},\ }\href
  {https://doi.org/10.1070/pu1968v011n02abeh003815} {\bibfield  {journal}
  {\bibinfo  {journal} {Sov. Phys. Uspekhi}\ }\textbf {\bibinfo {volume}
  {11}},\ \bibinfo {pages} {255} (\bibinfo {year} {1968})}\BibitemShut
  {NoStop}%
\bibitem [{\citenamefont {Lawrence}\ and\ \citenamefont
  {Wilkins}(1973)}]{Laurence-1973}%
  \BibitemOpen
  \bibfield  {author} {\bibinfo {author} {\bibfnamefont {W.~E.}\ \bibnamefont
  {Lawrence}}\ and\ \bibinfo {author} {\bibfnamefont {J.~W.}\ \bibnamefont
  {Wilkins}},\ }\bibfield  {title} {\bibinfo {title} {Electron-electron
  scattering in the transport coefficients of simple metals},\ }\href
  {https://doi.org/10.1103/PhysRevB.7.2317} {\bibfield  {journal} {\bibinfo
  {journal} {Phys. Rev. B}\ }\textbf {\bibinfo {volume} {7}},\ \bibinfo {pages}
  {2317} (\bibinfo {year} {1973})}\BibitemShut {NoStop}%
\bibitem [{\citenamefont {Venger}\ and\ \citenamefont
  {Piskovoi}(2004)}]{Venger-2004}%
  \BibitemOpen
  \bibfield  {author} {\bibinfo {author} {\bibfnamefont {E.~F.}\ \bibnamefont
  {Venger}}\ and\ \bibinfo {author} {\bibfnamefont {V.~N.}\ \bibnamefont
  {Piskovoi}},\ }\bibfield  {title} {\bibinfo {title} {Consistency of boundary
  conditions in crystal optics with spatial dispersion},\ }\href
  {https://doi.org/10.1103/PhysRevB.70.115107} {\bibfield  {journal} {\bibinfo
  {journal} {Phys. Rev. B}\ }\textbf {\bibinfo {volume} {70}},\ \bibinfo
  {pages} {115107} (\bibinfo {year} {2004})}\BibitemShut {NoStop}%
\bibitem [{\citenamefont {Halevi}\ and\ \citenamefont
  {Fuchs}(1984)}]{Halevi-1984}%
  \BibitemOpen
  \bibfield  {author} {\bibinfo {author} {\bibfnamefont {P.}~\bibnamefont
  {Halevi}}\ and\ \bibinfo {author} {\bibfnamefont {R.}~\bibnamefont {Fuchs}},\
  }\bibfield  {title} {\bibinfo {title} {Generalised additional boundary
  condition for non-local dielectrics. {I}. {Reflectivity}},\ }\href
  {https://doi.org/10.1088/0022-3719/17/21/017} {\bibfield  {journal} {\bibinfo
   {journal} {J. Phys. C: Solid State}\ }\textbf {\bibinfo {volume} {17}},\
  \bibinfo {pages} {3869} (\bibinfo {year} {1984})}\BibitemShut {NoStop}%
\bibitem [{\citenamefont {Henneberger}(1998)}]{Henneberger-1998}%
  \BibitemOpen
  \bibfield  {author} {\bibinfo {author} {\bibfnamefont {K.}~\bibnamefont
  {Henneberger}},\ }\bibfield  {title} {\bibinfo {title} {Additional boundary
  conditions: An historical mistake},\ }\href
  {https://doi.org/10.1103/PhysRevLett.80.2889} {\bibfield  {journal} {\bibinfo
   {journal} {Phys. Rev. Lett.}\ }\textbf {\bibinfo {volume} {80}},\ \bibinfo
  {pages} {2889} (\bibinfo {year} {1998})}\BibitemShut {NoStop}%
\bibitem [{\citenamefont {Chen}\ and\ \citenamefont
  {Nelson}(1993)}]{Chen-1993}%
  \BibitemOpen
  \bibfield  {author} {\bibinfo {author} {\bibfnamefont {B.}~\bibnamefont
  {Chen}}\ and\ \bibinfo {author} {\bibfnamefont {D.~F.}\ \bibnamefont
  {Nelson}},\ }\bibfield  {title} {\bibinfo {title} {Wave-vector-space method
  for wave propagation in bounded media},\ }\href
  {https://doi.org/10.1103/PhysRevB.48.15365} {\bibfield  {journal} {\bibinfo
  {journal} {Phys. Rev. B}\ }\textbf {\bibinfo {volume} {48}},\ \bibinfo
  {pages} {15365} (\bibinfo {year} {1993})}\BibitemShut {NoStop}%
\bibitem [{\citenamefont {Fuchs}\ and\ \citenamefont
  {Halevi}(1992)}]{Halevi-1992disp}%
  \BibitemOpen
  \bibfield  {author} {\bibinfo {author} {\bibfnamefont {R.}~\bibnamefont
  {Fuchs}}\ and\ \bibinfo {author} {\bibfnamefont {P.}~\bibnamefont {Halevi}},\
  }\href@noop {} {\emph {\bibinfo {title} {{Spatial Dispersion in Solids and
  Plasmas}}}}\ (\bibinfo  {publisher} {North-Holland},\ \bibinfo {address}
  {Amsterdam},\ \bibinfo {year} {1992})\BibitemShut {NoStop}%
\bibitem [{\citenamefont {Cocoletzi}\ and\ \citenamefont
  {Moch\'{a}n}(2005)}]{Cocoletzi-2005}%
  \BibitemOpen
  \bibfield  {author} {\bibinfo {author} {\bibfnamefont {G.~H.}\ \bibnamefont
  {Cocoletzi}}\ and\ \bibinfo {author} {\bibfnamefont {W.~L.}\ \bibnamefont
  {Moch\'{a}n}},\ }\bibfield  {title} {\bibinfo {title} {Excitons: from
  excitations at surfaces to confinement in nanostructures},\ }\href
  {https://doi.org/https://doi.org/10.1016/j.surfrep.2004.12.001} {\bibfield
  {journal} {\bibinfo  {journal} {Surf. Sci. Rep.}\ }\textbf {\bibinfo {volume}
  {57}},\ \bibinfo {pages} {1 } (\bibinfo {year} {2005})}\BibitemShut {NoStop}%
\bibitem [{\citenamefont {Melnyk}\ and\ \citenamefont
  {Harrison}(1968)}]{Melnyk-1968}%
  \BibitemOpen
  \bibfield  {author} {\bibinfo {author} {\bibfnamefont {A.~R.}\ \bibnamefont
  {Melnyk}}\ and\ \bibinfo {author} {\bibfnamefont {M.~J.}\ \bibnamefont
  {Harrison}},\ }\bibfield  {title} {\bibinfo {title} {Resonant excitation of
  plasmons in thin films by electromagnetic waves},\ }\href
  {https://doi.org/10.1103/PhysRevLett.21.85} {\bibfield  {journal} {\bibinfo
  {journal} {Phys. Rev. Lett.}\ }\textbf {\bibinfo {volume} {21}},\ \bibinfo
  {pages} {85} (\bibinfo {year} {1968})}\BibitemShut {NoStop}%
\bibitem [{\citenamefont {Melnyk}\ and\ \citenamefont
  {Harrison}(1970)}]{Melnyk-1970}%
  \BibitemOpen
  \bibfield  {author} {\bibinfo {author} {\bibfnamefont {A.~R.}\ \bibnamefont
  {Melnyk}}\ and\ \bibinfo {author} {\bibfnamefont {M.~J.}\ \bibnamefont
  {Harrison}},\ }\bibfield  {title} {\bibinfo {title} {Theory of optical
  excitation of plasmons in metals},\ }\href
  {https://doi.org/10.1103/PhysRevB.2.835} {\bibfield  {journal} {\bibinfo
  {journal} {Phys. Rev. B}\ }\textbf {\bibinfo {volume} {2}},\ \bibinfo {pages}
  {835} (\bibinfo {year} {1970})}\BibitemShut {NoStop}%
\bibitem [{\citenamefont {Fuchs}\ and\ \citenamefont
  {Kliewer}(1971)}]{Fuchs-1971}%
  \BibitemOpen
  \bibfield  {author} {\bibinfo {author} {\bibfnamefont {R.}~\bibnamefont
  {Fuchs}}\ and\ \bibinfo {author} {\bibfnamefont {K.~L.}\ \bibnamefont
  {Kliewer}},\ }\bibfield  {title} {\bibinfo {title} {Surface plasmon in a
  semi-infinite free-electron gas},\ }\href
  {https://doi.org/10.1103/PhysRevB.3.2270} {\bibfield  {journal} {\bibinfo
  {journal} {Phys. Rev. B}\ }\textbf {\bibinfo {volume} {3}},\ \bibinfo {pages}
  {2270} (\bibinfo {year} {1971})}\BibitemShut {NoStop}%
\bibitem [{\citenamefont {Bocquet}\ and\ \citenamefont
  {Barrat}(2007)}]{Bocquet-2007}%
  \BibitemOpen
  \bibfield  {author} {\bibinfo {author} {\bibfnamefont {L.}~\bibnamefont
  {Bocquet}}\ and\ \bibinfo {author} {\bibfnamefont {J.-L.}\ \bibnamefont
  {Barrat}},\ }\bibfield  {title} {\bibinfo {title} {Flow boundary conditions
  from nano- to micro-scales},\ }\href {https://doi.org/10.1039/B616490K}
  {\bibfield  {journal} {\bibinfo  {journal} {Soft Matter}\ }\textbf {\bibinfo
  {volume} {3}},\ \bibinfo {pages} {685} (\bibinfo {year} {2007})}\BibitemShut
  {NoStop}%
\bibitem [{\citenamefont {Pekar}(1957)}]{Pekar-1958}%
  \BibitemOpen
  \bibfield  {author} {\bibinfo {author} {\bibfnamefont {S.}~\bibnamefont
  {Pekar}},\ }\bibfield  {title} {\bibinfo {title} {The theory of
  electromagnetic waves in a crystal in which excitons are produced},\
  }\href@noop {} {\bibfield  {journal} {\bibinfo  {journal} {J. Exptl. Theoret.
  Phys. (U.S.S.R.)}\ }\textbf {\bibinfo {volume} {33}},\ \bibinfo {pages}
  {1022} (\bibinfo {year} {1957})},\ \bibinfo {note}
  {[\href{http://www.jetp.ac.ru/cgi-bin/dn/e_006_04_0785.pdf}{{Sov. Phys.
  JETP}} \textbf{6}, 785 (1958)]}\BibitemShut {NoStop}%
\bibitem [{\citenamefont {Ting}\ \emph {et~al.}(1975)\citenamefont {Ting},
  \citenamefont {Frankel},\ and\ \citenamefont {Birman}}]{Ting-1975}%
  \BibitemOpen
  \bibfield  {author} {\bibinfo {author} {\bibfnamefont {C.-S.}\ \bibnamefont
  {Ting}}, \bibinfo {author} {\bibfnamefont {M.}~\bibnamefont {Frankel}},\ and\
  \bibinfo {author} {\bibfnamefont {J.}~\bibnamefont {Birman}},\ }\bibfield
  {title} {\bibinfo {title} {Electrodynamics of bounded spatially dispersive
  media: The additional boundary conditions},\ }\href
  {https://doi.org/https://doi.org/10.1016/0038-1098(75)90688-2} {\bibfield
  {journal} {\bibinfo  {journal} {Solid State Commun.}\ }\textbf {\bibinfo
  {volume} {17}},\ \bibinfo {pages} {1285 } (\bibinfo {year}
  {1975})}\BibitemShut {NoStop}%
\bibitem [{\citenamefont {Kiselev}\ and\ \citenamefont
  {Schmalian}(2019)}]{Kiselev-2019}%
  \BibitemOpen
  \bibfield  {author} {\bibinfo {author} {\bibfnamefont {E.~I.}\ \bibnamefont
  {Kiselev}}\ and\ \bibinfo {author} {\bibfnamefont {J.}~\bibnamefont
  {Schmalian}},\ }\bibfield  {title} {\bibinfo {title} {Boundary conditions of
  viscous electron flow},\ }\href {https://doi.org/10.1103/PhysRevB.99.035430}
  {\bibfield  {journal} {\bibinfo  {journal} {Phys. Rev. B}\ }\textbf {\bibinfo
  {volume} {99}},\ \bibinfo {pages} {035430} (\bibinfo {year}
  {2019})}\BibitemShut {NoStop}%
\bibitem [{\citenamefont {Pippard}(1947{\natexlab{a}})}]{Pippard-1947a}%
  \BibitemOpen
  \bibfield  {author} {\bibinfo {author} {\bibfnamefont {A.~B.}\ \bibnamefont
  {Pippard}},\ }\bibfield  {title} {\bibinfo {title} {The surface impedance of
  superconductors and normal metals at high frequencies. {I}. {Resistance} of
  superconducting tin and mercury at 1200 {Mcyc}./sec.},\ }\href
  {https://royalsocietypublishing.org/doi/10.1098/rspa.1947.0121} {\bibfield
  {journal} {\bibinfo  {journal} {Proc. Roy. Soc. A}\ }\textbf {\bibinfo
  {volume} {191}} (\bibinfo {year} {1947}{\natexlab{a}})}\BibitemShut {NoStop}%
\bibitem [{\citenamefont {Pippard}(1947{\natexlab{b}})}]{Pippard-1947b}%
  \BibitemOpen
  \bibfield  {author} {\bibinfo {author} {\bibfnamefont {A.~B.}\ \bibnamefont
  {Pippard}},\ }\bibfield  {title} {\bibinfo {title} {The surface impedance of
  superconductors and normal metals at high frequencies. {II}. {The} anomalous
  skin effect in normal metals},\ }\href
  {http://doi.org/10.1098/rspa.1947.0122} {\bibfield  {journal} {\bibinfo
  {journal} {Proc. Roy. Soc. A}\ }\textbf {\bibinfo {volume} {191}},\ \bibinfo
  {pages} {385} (\bibinfo {year} {1947}{\natexlab{b}})}\BibitemShut {NoStop}%
\bibitem [{Note7()}]{Note7}%
  \BibitemOpen
  \bibinfo {note} {An exception to this is the transverse impedance in a
  neutral Fermi liquid, which was analyzed in Ref.\spacefactor \@m {} \protect
  \rev@citealp {Shahzamanian-2006}.}\BibitemShut {Stop}%
\bibitem [{\citenamefont {Chambers}(1952)}]{Chambers-1952}%
  \BibitemOpen
  \bibfield  {author} {\bibinfo {author} {\bibfnamefont {R.~G.}\ \bibnamefont
  {Chambers}},\ }\bibfield  {title} {\bibinfo {title} {The anomalous skin
  effect},\ }\href
  {https://royalsocietypublishing.org/doi/10.1098/rspa.1952.0226} {\bibfield
  {journal} {\bibinfo  {journal} {Proc. Roy. Soc. A}\ }\textbf {\bibinfo
  {volume} {215}},\ \bibinfo {pages} {481} (\bibinfo {year}
  {1952})}\BibitemShut {NoStop}%
\bibitem [{Note8()}]{Note8}%
  \BibitemOpen
  \bibinfo {note} {In Refs.\spacefactor \@m {} \protect \rev@citealp
  {Reuter-1948, Sondheimer-2001} a positive imaginary part of $Z(\omega )$ is
  obtained, since the convention $E(z,\omega ) \propto e^{i \omega t}$ is
  adopted.}\BibitemShut {Stop}%
\bibitem [{\citenamefont {Pippard}(1947{\natexlab{c}})}]{Pippard-1947c}%
  \BibitemOpen
  \bibfield  {author} {\bibinfo {author} {\bibfnamefont {A.~B.}\ \bibnamefont
  {Pippard}},\ }\bibfield  {title} {\bibinfo {title} {The surface impedance of
  superconductors and normal metals at high frequencies. {III}. the relation
  between impedance and superconducting penetration depth},\ }\href
  {http://doi.org/10.1098/rspa.1947.0123} {\bibfield  {journal} {\bibinfo
  {journal} {Proc. Roy. Soc. A}\ }\textbf {\bibinfo {volume} {191}},\ \bibinfo
  {pages} {399} (\bibinfo {year} {1947}{\natexlab{c}})}\BibitemShut {NoStop}%
\bibitem [{\citenamefont {Tinkham}(1996)}]{Tinkham-1996int}%
  \BibitemOpen
  \bibfield  {author} {\bibinfo {author} {\bibfnamefont {M.}~\bibnamefont
  {Tinkham}},\ }\href@noop {} {\emph {\bibinfo {title} {Introduction to
  {Superconductivity}}}}\ (\bibinfo  {publisher} {Courier Corporation},\
  \bibinfo {address} {North Chelmsford},\ \bibinfo {year} {1996})\BibitemShut
  {NoStop}%
\bibitem [{\citenamefont {Gurzhi}(1959)}]{Gurzhi-1959}%
  \BibitemOpen
  \bibfield  {author} {\bibinfo {author} {\bibfnamefont {R.}~\bibnamefont
  {Gurzhi}},\ }\bibfield  {title} {\bibinfo {title} {Mutual electron
  correlations in metal optics},\ }\href@noop {} {\bibfield  {journal}
  {\bibinfo  {journal} {Zh. Eksp. Teor. Fiz.}\ }\textbf {\bibinfo {volume}
  {35}},\ \bibinfo {pages} {965} (\bibinfo {year} {1959})},\ \bibinfo {note}
  {\href{http://www.jetp.ac.ru/cgi-bin/dn/e_008_04_0673.pdf}{{[Sov. Phys.
  JETP}} \textbf{8}, 673 (1959)]}\BibitemShut {NoStop}%
\bibitem [{\citenamefont {Valentinis}\ \emph {et~al.}(2021)\citenamefont
  {Valentinis}, \citenamefont {Zaanen},\ and\ \citenamefont {Van
  Der~Marel}}]{FZVM-new}%
  \BibitemOpen
  \bibfield  {author} {\bibinfo {author} {\bibfnamefont {D.}~\bibnamefont
  {Valentinis}}, \bibinfo {author} {\bibfnamefont {J.}~\bibnamefont {Zaanen}},\
  and\ \bibinfo {author} {\bibfnamefont {D.}~\bibnamefont {Van Der~Marel}},\
  }\bibfield  {title} {\bibinfo {title} {Propagation of shear stress in
  strongly interacting metallic {Fermi} liquids enhances transmission of
  terahertz radiation},\ }\href {https://doi.org/10.1038/s41598-021-86356-2}
  {\bibfield  {journal} {\bibinfo  {journal} {Sci. Rep.}\ }\textbf {\bibinfo
  {volume} {11}},\ \bibinfo {pages} {1} (\bibinfo {year} {2021})}\BibitemShut
  {NoStop}%
\bibitem [{\citenamefont {Dingle}(1953{\natexlab{a}})}]{Dingle-1953a}%
  \BibitemOpen
  \bibfield  {author} {\bibinfo {author} {\bibfnamefont {R.}~\bibnamefont
  {Dingle}},\ }\bibfield  {title} {\bibinfo {title} {The anomalous skin effect
  and the reflectivity of metals {I}},\ }\href
  {http://www.sciencedirect.com/science/article/pii/S0031891453800352}
  {\bibfield  {journal} {\bibinfo  {journal} {Physica}\ }\textbf {\bibinfo
  {volume} {19}},\ \bibinfo {pages} {311 } (\bibinfo {year}
  {1953}{\natexlab{a}})}\BibitemShut {NoStop}%
\bibitem [{\citenamefont {Dingle}(1953{\natexlab{b}})}]{Dingle-1953d}%
  \BibitemOpen
  \bibfield  {author} {\bibinfo {author} {\bibfnamefont {R.~B.}\ \bibnamefont
  {Dingle}},\ }\bibfield  {title} {\bibinfo {title} {The anomalous skin effect
  and the reflectivity of metals: {IV}. {Theoretical} optical properties of
  thin metallic films},\ }\href
  {http://www.sciencedirect.com/science/article/pii/S0031891453801369}
  {\bibfield  {journal} {\bibinfo  {journal} {Physica}\ }\textbf {\bibinfo
  {volume} {19}},\ \bibinfo {pages} {1187 } (\bibinfo {year}
  {1953}{\natexlab{b}})}\BibitemShut {NoStop}%
\bibitem [{\citenamefont {Ashcroft}\ and\ \citenamefont
  {Mermin}(1976)}]{Ashcroft-1976}%
  \BibitemOpen
  \bibfield  {author} {\bibinfo {author} {\bibfnamefont {N.~W.}\ \bibnamefont
  {Ashcroft}}\ and\ \bibinfo {author} {\bibfnamefont {N.~D.}\ \bibnamefont
  {Mermin}},\ }\href@noop {} {\emph {\bibinfo {title} {Solid State Physics}}}\
  (\bibinfo  {publisher} {Harcourt},\ \bibinfo {address} {Fort Worth},\
  \bibinfo {year} {1976})\BibitemShut {NoStop}%
\bibitem [{\citenamefont {Valentinis}(2017)}]{Valentinis-thesis}%
  \BibitemOpen
  \bibfield  {author} {\bibinfo {author} {\bibfnamefont {D.}~\bibnamefont
  {Valentinis}},\ }\emph {\bibinfo {title} {Electronic correlations in
  interacting quantum matter}},\ \href@noop {} {Ph.D. thesis},\ \bibinfo
  {school} {University of Geneva} (\bibinfo {year} {2017})\BibitemShut
  {NoStop}%
\bibitem [{\citenamefont {Gochan}\ \emph {et~al.}(2019)\citenamefont {Gochan},
  \citenamefont {Li},\ and\ \citenamefont {Bedell}}]{Gochan-2019}%
  \BibitemOpen
  \bibfield  {author} {\bibinfo {author} {\bibfnamefont {M.~P.}\ \bibnamefont
  {Gochan}}, \bibinfo {author} {\bibfnamefont {H.}~\bibnamefont {Li}},\ and\
  \bibinfo {author} {\bibfnamefont {K.~S.}\ \bibnamefont {Bedell}},\ }\bibfield
   {title} {\bibinfo {title} {Viscosity bound violation in viscoelastic {Fermi}
  liquids},\ }\href
  {https://iopscience.iop.org/article/10.1088/2399-6528/ab292b} {\bibfield
  {journal} {\bibinfo  {journal} {J. Phys. Commun.}\ }\textbf {\bibinfo
  {volume} {3}},\ \bibinfo {pages} {065008} (\bibinfo {year}
  {2019})}\BibitemShut {NoStop}%
\bibitem [{\citenamefont {Liao}\ and\ \citenamefont
  {Galitski}(2019)}]{Liao-2019}%
  \BibitemOpen
  \bibfield  {author} {\bibinfo {author} {\bibfnamefont {Y.}~\bibnamefont
  {Liao}}\ and\ \bibinfo {author} {\bibfnamefont {V.}~\bibnamefont
  {Galitski}},\ }\bibfield  {title} {\bibinfo {title} {Critical viscosity of a
  fluctuating superconductor},\ }\href
  {https://doi.org/10.1103/PhysRevB.100.060501} {\bibfield  {journal} {\bibinfo
   {journal} {Phys. Rev. B}\ }\textbf {\bibinfo {volume} {100}},\ \bibinfo
  {pages} {060501} (\bibinfo {year} {2019})}\BibitemShut {NoStop}%
\bibitem [{\citenamefont {Lavasani}\ \emph {et~al.}(2019)\citenamefont
  {Lavasani}, \citenamefont {Bulmash},\ and\ \citenamefont
  {Das~Sarma}}]{Lavasani-2019}%
  \BibitemOpen
  \bibfield  {author} {\bibinfo {author} {\bibfnamefont {A.}~\bibnamefont
  {Lavasani}}, \bibinfo {author} {\bibfnamefont {D.}~\bibnamefont {Bulmash}},\
  and\ \bibinfo {author} {\bibfnamefont {S.}~\bibnamefont {Das~Sarma}},\
  }\bibfield  {title} {\bibinfo {title} {{Wiedemann}-{Franz} law and {Fermi}
  liquids},\ }\href {https://doi.org/10.1103/PhysRevB.99.085104} {\bibfield
  {journal} {\bibinfo  {journal} {Phys. Rev. B}\ }\textbf {\bibinfo {volume}
  {99}},\ \bibinfo {pages} {085104} (\bibinfo {year} {2019})}\BibitemShut
  {NoStop}%
\bibitem [{\citenamefont {Shahzamanian}\ and\ \citenamefont
  {Yavary}(2007)}]{Shahzamanian-2006}%
  \BibitemOpen
  \bibfield  {author} {\bibinfo {author} {\bibfnamefont {M.~A.}\ \bibnamefont
  {Shahzamanian}}\ and\ \bibinfo {author} {\bibfnamefont {H.}~\bibnamefont
  {Yavary}},\ }\bibfield  {title} {\bibinfo {title} {The transverse impedance
  of normal liquid $^3${He}},\ }\href
  {https://doi.org/10.1142/S0217979207037429} {\bibfield  {journal} {\bibinfo
  {journal} {Int. J. Mod. Phys. B}\ }\textbf {\bibinfo {volume} {21}},\
  \bibinfo {pages} {2979} (\bibinfo {year} {2007})}\BibitemShut {NoStop}%
\end{thebibliography}
\end{document}